\documentclass[a4paper,10pt]{book}

\usepackage[T1]{fontenc}
\usepackage[utf8]{inputenc}
\usepackage[french,german,english]{babel}

\usepackage{lmodern}

\usepackage{setspace} 

\usepackage{graphicx,xcolor,wrapfig}
\graphicspath{{images/}}
\usepackage{subcaption}

\usepackage{booktabs}
\usepackage{lipsum}
\usepackage{microtype}
\usepackage{url}
\usepackage[final]{pdfpages}
\usepackage{multirow}
\usepackage{xfrac}
\usepackage{bbm}
\usepackage{braket}
\usepackage{enumitem}

\usepackage{fancyhdr}

\fancypagestyle{plain}{
	\fancyhf{}

	\fancyfoot[R]{\thepage}
	}
\fancypagestyle{addpagenumbersforpdfimports}{
	\fancyhead{}
	
	\fancyfoot{}
    \fancyfoot[R]{\thepage}
}

\usepackage{listings}
\usepackage{hyperref}
\hypersetup{pdfborder={0 0 0},
	colorlinks=true,
	linkcolor=black,
	citecolor=black,
	urlcolor=black}
\makeatletter
\def\cleardoublepage{\clearpage\if@twoside \ifodd\c@page\else
    \hbox{}
    \thispagestyle{empty}
    \newpage
    \if@twocolumn\hbox{}\newpage\fi\fi\fi}
\makeatother \clearpage{\pagestyle{plain}\cleardoublepage}

\usepackage{color}
\usepackage{tikz}
\usepackage[explicit]{titlesec}
\newcommand*\chapterlabel{}
\titleformat{\chapter}[display]  
	{\normalfont\bfseries\Huge} 
	{\gdef\chapterlabel{\thechapter\ }}     
 	{0pt} 
 	  {\begin{tikzpicture}[remember picture,overlay]
    \node[yshift=-8cm] at (current page.north west)
      {\begin{tikzpicture}[remember picture, overlay]
        \draw[fill=black] (0,0) rectangle(35.5mm,15mm);
        \node[anchor=north east,yshift=-7.2cm,xshift=34mm,minimum height=30mm,inner sep=0mm] at (current page.north west)
        {\parbox[top][30mm][t]{15mm}{\raggedleft $\phantom{\textrm{l}}$\color{white}\chapterlabel}};  
        \node[anchor=north west,yshift=-7.2cm,xshift=37mm,text width=\textwidth,minimum height=30mm,inner sep=0mm] at (current page.north west)
              {\parbox[top][30mm][t]{\textwidth}{\color{black}#1}};
       \end{tikzpicture}
      };
   \end{tikzpicture}
   \gdef\chapterlabel{}
  } 

\titlespacing*{\chapter}{0pt}{50pt}{30pt}
\titlespacing*{\section}{0pt}{13.2pt}{*0}  
\titlespacing*{\subsection}{0pt}{13.2pt}{*0}
\titlespacing*{\subsubsection}{0pt}{13.2pt}{*0}

\newcounter{myparts}
\newcommand*\partlabel{}
\titleformat{\part}[display]  
	{\normalfont\bfseries\Huge} 
	{\gdef\partlabel{\thepart\ }}     
 	{0pt} 
 	  {\setlength{\unitlength}{20mm}
	  \addtocounter{myparts}{1}
	  \begin{tikzpicture}[remember picture,overlay]
    \node[anchor=north west,xshift=-65mm,yshift=-6.9cm-\value{myparts}*20mm] at (current page.north east) 
      {\begin{tikzpicture}[remember picture, overlay]
        \draw[fill=black] (0,0) rectangle(62mm,20mm);   
        \node[anchor=north west,yshift=-6.1cm-\value{myparts}*20mm,xshift=-60.5mm,minimum height=30mm,inner sep=0mm] at (current page.north east)
        {\parbox[top][30mm][t]{55mm}{\raggedright \color{white}Part \partlabel $\phantom{\textrm{l}}$}};  
        \node[anchor=north east,yshift=-6.1cm-\value{myparts}*20mm,xshift=-63.5mm,text width=\textwidth,minimum height=30mm,inner sep=0mm] at (current page.north east)
              {\parbox[top][30mm][t]{\textwidth}{\raggedleft \color{black}#1}};
       \end{tikzpicture}
      };
   \end{tikzpicture}
   \gdef\partlabel{}
  } 

\usepackage{amsmath}
\makeatletter
\def\resetMathstrut@{%
  \setbox\z@\hbox{%
    \mathchardef\@tempa\mathcode`\(\relax
      \def\@tempb##1"##2##3{\the\textfont"##3\char"}%
      \expandafter\@tempb\meaning\@tempa \relax
  }%
  \ht\Mathstrutbox@1.2\ht\z@ \dp\Mathstrutbox@1.2\dp\z@
}
\makeatother

\usepackage{subfiles}

\renewcommand{\thechapter}{\Roman{chapter}}

\newcommand*{\del}{\mathop{\mathrm{{}\partial}}\mathopen{}}
\newcommand*{\setC}{\ensuremath{\mathbb{C}}}
\newcommand*{\setN}{\ensuremath{\mathbb{N}}}

\newcommand*{\setR}{\ensuremath{\mathbb{R}}}

\usepackage{acronym}
\acrodef{nlo}[\textsc{nlo}]{next-to-leading-order}
\acrodef{lg}[\textsc{lg}]{Landau--Ginzburg}
\acrodef{ccwz}[\textsc{ccwz}]{Callan--Coleman--Wess--Zumino~\cite{Coleman:1969sm,Callan:1969sn}}
\acrodef{hw}[\textsc{hw}]{Hanany--Witten~\cite{Hanany:1996ie}}
\acrodef{ahw}[\textsc{ahw}]{Affleck--Harvey--Witten~\cite{Affleck:1982as}}
\acrodef{kss}[\textsc{kss}]{Kutasov--Schwimmer--Seiberg~\cite{Kutasov:1995ss}}
\acrodef{nsvz}[\textsc{nsvz}]{Novikov--Shifman--Vainshtein--Zakharov~\cite{Novikov:1983uc}}
\acrodef{kk}[\textsc{kk}]{Kaluza--Klein}
\acrodef{sym}[\textsc{sym}]{super Yang--Mills}
\acrodef{vev}[\textsc{vev}]{vacuum expectation value}
\acrodef{gk}[\textsc{gk}]{Giveon--Kutasov~\cite{Giveon:2008zn}}
\acrodef{cs}[\textsc{cs}]{Chern--Simons}
\acrodef{fi}[\textsc{fi}]{Fayet--Iliopoulos}
\acrodef{kp}[\textsc{kp}]{Kim and Park in~\cite{Kim:2013cma}}
\acrodef{scft}[\textsc{scft}]{super-conformal field theory}
\acrodefplural{scft}[\textsc{scft}s]{super-conformal field theories}
\acrodef{qft}[\textsc{qft}]{quantum field theory}
\acrodefplural{qft}[\textsc{qft}s]{quantum field theories}
\acrodef{cft}[\textsc{cft}]{conformal field theory}
\acrodefplural{cft}[\textsc{cft}s]{conformal field theories}
\acrodef{eft}[\textsc{eft}]{effective field theory}
\acrodef{ope}[\textsc{ope}]{operator product expansion}
\acrodef{dof}[\textsc{dof}]{degree of freedom}
\acrodef{qm}[\textsc{qm}]{quantum mechanics}
\acrodefplural{dof}[\textsc{dof}]{degrees of freedom}
\acrodef{uv}[\textsc{uv}]{ultraviolet}
\acrodef{lsm}[\textsc{lsm}]{linear sigma model}
\acrodef{nlsm}[\textsc{nlsm}]{non-linear sigma model}
\acrodef{qcd}[\textsc{qcd}]{quantum chromodynamics}
\acrodef{aha}[\textsc{aha}]{Aharony~\cite{Aharony:1997gp}}
\acrodef{dsz}[\textsc{dsz}]{Dirac--Schwinger--Zwanziger}
\acrodef{bps}[\textsc{bps}]{Bogomol'nyi--Prasad--Sommerfield}
\acrodef{Wline}[\textsc{w} line]{Wilson line}
\acrodef{Hline}[\textsc{h} line]{'t~Hooft line}
\acrodef{WH}[\textsc{wh}]{Wilson--'t~Hooft}
\acrodef{ir}[\textsc{ir}]{infrared}
\acrodef{uv}[\textsc{uv}]{ultraviolet}
\acrodef{eom}[\textsc{eom}]{equations of motion}
\acrodef{MC}[\textsc{mc}]{Monte Carlo}
\acrodef{lhs}[\textsc{lhs}]{left-hand side}
\acrodef{rn}[\textsc{rn}]{Reissner--Nordström}
\acrodef{adsrn}[\textsc{ads-rn}]{anti-de Sitter--Reissner--Nordström}
\acrodef{adscft}[\textsc{ads-cft}]{\textsc{ads-cft}}
\acrodef{adseft}[\textsc{ads-eft}]{\textsc{ads-eft}}
\acrodef{gb}[\textsc{gb}]{Gauss--Bonnet}
\acrodef{adm}[\textsc{adm}]{Arnowitt--Deser--Misner~\cite{Arnowitt:1959ah}}
\acrodef{wgc}[\textsc{wgc}]{weak gravity conjecture}
\acrodef{ehm}[\textsc{ehm}]{Einstein--Hilbert--Maxwell}
\acrodef{sct}[\textsc{sct}]{special conformal transformation}
\acrodef{rg}[\textsc{rg}]{renormalization group}
\acrodef{bdg}[\textsc{bdg}]{Bogoliubov--de Gennes}
\acrodef{dimreg}[\textsc{dimreg}]{dimensional regularization}
\acrodef{wkb}[\textsc{wkb}]{Wentzel--Kramers--Brillouin}

\acrodef{nrcft}[\textsc{nrcft}]{nonrelativistic conformal field theory}
\acrodefplural{nrcft}[\textsc{nrcft}s]{nonrelativistic conformal field theories}
\acrodef{bcs}[\textsc{bcs}]{Bardeen-Cooper-Schrieffer}
\acrodef{bec}[\textsc{bec}]{Bose-Einstein condensate}

\acrodef{ssb}[\textsc{ssb}]{spontaneous symmetry breaking}
\acrodef{ic}[\textsc{ic}]{initial conditions}
\acrodef{lo}[\textsc{lo}]{leading order}
\acrodef{nlo}[\textsc{nlo}]{next-to-leading order}
\acrodef{nnlo}[\textsc{nnlo}]{next-to-next-to-leading order}

\usepackage{standalone}
\usepackage{stmaryrd}
\newcommand*{\Op}[2]{\mathcal{O}^{(#1,#2)}_{\text{bulk}}}
\newcommand*{\Z}[1]{Z^{#1}_{\text{edge}}}
\newcommand*{\Ecasimir}[1]{E_{\text{Casimir}}^{#1}}
\newcommand*{\mudim}[1]{\left[#1\right]_\mu}

\newcommand{\vect}[1]{\mathbf{#1}}
\newcommand{\proptime}{\tau}
\newcommand{\Kdiag}{K}
\newcommand{\Koff}{\tilde{K}}
\newcommand{\bbdiag}{\mathbb{B}}
\newcommand{\bboff}{\tilde{\mathbb{B}}}

\newcommand*{\deltaop}{\mathop{\hat \delta}\nolimits}

\newcommand*\Laplacian{\mathop{}\!\mathbin\Delta}
\newcommand*{\Lag}{\mathcal{L}}
\DeclareMathOperator{\Id}{\mathbbm{1}}
\DeclareDocumentCommand\Gammaop{}{\do@opbraces{\operatorname{\Gamma}}}
\newcommand{\cev}[1]{\reflectbox{\ensuremath{\vec{\reflectbox{\ensuremath{#1}}}}}}

\RequirePackage{xparse}
\newcommand{\poisson}[1]{\left\{ #1 \right\}}
\newcommand{\ppoisson}[1]{\left\{\left\{ #1 \right\}\right\}}
\newcommand{\GammaF}[1]{\Gamma\left(#1\right)}
\newcommand{\pqtybis}[1]{\left( #1 \right)}

\usepackage{cancel}

\newcommand{\pFq}[5]{\prescript{}{#1}{F}_{#2}\left[ \begin{matrix} #3 \\ #4 \end{matrix}; #5 \right]}

\DeclareUnicodeCharacter{2212}{\ensuremath{-}}

\usepackage{amsmath,amssymb,mathrsfs,mathtools,physics,bbold}
\usepackage[color=lightgray]{todonotes}

\newcommand*\expo [1] {\mathop{}\!\mathrm{e}^{#1}}

\usepackage{csquotes} 
\usepackage[backend=biber,style=numeric-comp,sorting=none]{biblatex}
\begin{document}
\frontmatter
\begin{titlepage}
\includegraphics[width=4cm]{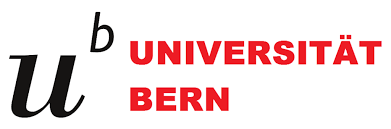}

\begin{center}
\large
\sffamily
\vfill

{\huge\textsc{The Large-Charge Expansion in}} \\ \vspace{3mm}
{\huge\textsc{Nonrelativistic Conformal}} \\ \vspace{3mm}
{\huge\textsc{Field Theories}}
    
\vfill\vfill

Inaugural dissertation \\
of the Faculty of Science, \\
University of Bern \\
~\\
presented by \\
{\Large \textbf{Vito Pellizzani}} \\
from Neuchâtel

\vfill\vfill

Supervisor of the doctoral thesis: \\
\textbf{Prof. Dr. Susanne Reffert} \\
Institute for Theoretical Physics \\
Albert Einstein Center for Fundamental Physics \\
University of Bern

\vfill
\includegraphics[width=0.10\textwidth]{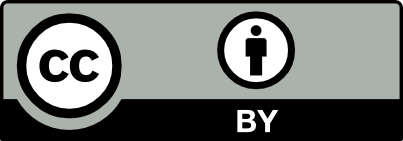} \\
\footnotesize{This work is licensed under a \\ Creative Commons Attribution 4.0 International License \\ \url{https://creativecommons.org/licenses/by/4.0/}}
\end{center}
\end{titlepage}

\begin{titlepage}
\includegraphics[width=4cm]{images/logo}

\begin{center}
\large
\sffamily
\vfill

{\huge\textsc{The Large-Charge Expansion in}} \\ \vspace{3mm}
{\huge\textsc{Nonrelativistic Conformal}} \\ \vspace{3mm}
{\huge\textsc{Field Theories}}
    
\vfill\vfill

Inaugural dissertation \\
of the Faculty of Science, \\
University of Bern \\
~\\
presented by \\
{\Large \textbf{Vito Pellizzani}} \\
from Neuchâtel

\vfill\vfill

Supervisor of the doctoral thesis: \\
\textbf{Prof. Dr. Susanne Reffert} \\
Institute for Theoretical Physics \\
Albert Einstein Center for Fundamental Physics \\
University of Bern

\vfill

Accepted by the Faculty of Science.
\end{center}

\begingroup
\large
\sffamily
\begin{minipage}[c]{5cm}
    Bern, September 19, 2024 \\
    ~
\end{minipage}
\hfill
\begin{minipage}[c]{6cm}
    The Dean, \\
    Prof. Dr. Jean-Louis Reymond
\end{minipage}
\endgroup
\end{titlepage}
\setcounter{page}{0}
\chapter*{Acknowledgements}
~\newline~\newline~
\markboth{Acknowledgements}{Acknowledgements}
\addcontentsline{toc}{chapter}{Acknowledgements}

The year 2020 coincided with three major events that impacted my life: the beginning of my PhD, the covid-19 pandemic, and the birth of my first daughter. The latter two shaped the way I experienced the former. It was overall a fantastic adventure.\vspace{2mm}

I therefore want to express my gratitude to my advisor, Susanne Reffert, for offering me the chance to join her group. Early on, she encouraged me to publish results, even on my own, and involved me into an international collaboration which, by now, can be regarded as successful and very fruitful for my career. It also lowered the impact of the pandemic on my work as I got used to collaborating remotely, but also to thinking about my own independent research. Her support beyond research was also invaluable, as she has always been enthusiastic and patient regarding my new life as a father. Naturally, my gratitude also goes to Domenico Orlando, not just for the many stories he likes to tell about the small---and sometimes weird---community we belong to, but also and foremost for his scientific enthusiasm when collaborating. Moreover, his incredible ability to communicate with Simeon Hellerman is still second to none. Joke aside, this is a good place to also thank Simeon and Ian Swanson for accepting to work with me, despite the very uncomfortable time differences that we had to deal with. I learned a lot from the many online meetings that we had all together, during which the laggy connection was of great help for me to think twice about the deep insights that everyone else had.\vspace{2mm}

I also want to thank the friends that I made within the group, namely, Rafael Moser and Ioannis Kalogerakis---my fellow PhD students---as well as Nicola Dondi, Donald Huber-Youmans, Fabio Apruzzi and Giacomo Sberveglieri---the tireless postdocs who patiently answered all of my ignorant questions about physics. A special thank goes to Nicola for his feedback on parts of this thesis. The prospect of collaborating in the future with some of you fills me with excitement. It was also a pleasure to interact with the bright Master students who temporarily joined the group: Daniil, Jan, Rebecca, Lara, Tim, Omar and Tabea. My gratitude also goes, beyond Bern, to the PhD students and postdocs who made my journey more enjoyable, in particular, Adrian, Pietro, Rahel, Eren and Francesco. I am also grateful to Adrien, Rahel, Donald and Sara for co-organizing the first and second editions of the \emph{Advanced Lectures in Physics in Switzerland}, which promise to become a long and successful series of events. As mentioned, part of my time as a PhD student in Bern overlapped with the pandemic, and I perhaps never tried hard enough to engage with the people outside of my research group once the confinement was over. My gratitude goes also to them, in particular for the many lively coffee breaks.\vspace{2mm}

I did not repeat this mistake when I arrived in Oxford for a visit that lasted almost a year. There, I received a very warm welcome, and I want to thank the Mathematical Physics group for this: Andrea, Enrico, Beppe, Palash, Lea, Marta, Clément, Tabea, Alice, Alison, Maria, Adam, Romain, Akshay, Matteo, Pieter, Chris, Christian, Julius, Gabriel and all the others. I am also deeply grateful to Mark Mezei for his guidance, past and future. I am very excited to the prospect of interacting with all of you two additional years.\vspace{2mm}

Finally, I want to acknowledge the tremendous and unconditional support of my wife. Thank you for your patience and resilience. For looking in the same direction as me, although I had no idea what I was looking at and you knew it. I also thank my parents for their continuous support. The last words here go to my daughters: thank you for reminding that what really matters is the ability to marvel at the smallest things.

\bigskip

\noindent\textit{Neuchâtel, \today}
\hfill V.~P.


\cleardoublepage
\chapter*{Abstract}
~\newline~\newline~
\addcontentsline{toc}{chapter}{Abstract} 


Conformal field theories are associated with critical phenomena and phase transitions and also play an essential role in string theory. Solving a \ac{cft} is an extremely constrained problem due to conformal invariance---the task essentially reduces to the computation of a set of numbers called the \ac{cft} data---yet it remains highly nontrivial. In fact, \acp{cft} usually are strongly coupled and thus require new tools that do not rely on perturbation theory. For these reasons, in the past few decades, they became one of the most active fields of research.\vspace{1mm}

A natural extension of these ideas with far-reaching implications for condensed matter systems---in which relativistic effects are not manifest---is to replace the Poincaré group by the Galilean group, thereby opening the way to a precise formulation of a whole new class of critical phenomena in terms of \acp{nrcft}. Many of the aforementioned tools developed to gain a better formal understanding of \acp{cft} can then be adapted to gain a strong predictive power in systems with direct experimental relevance.~\vspace{1mm}

In recent years, a new powerful tool has emerged: the large-charge expansion. It allows to systemically uncover part of the \ac{cft} data of theories with global symmetries, thereby revealing profound and universal features of these systems. It complements the set of analytical tools designed for the study of \acp{cft}, such as the various incarnations of the conformal bootstrap program. Using an exact, non-perturbative property known as the state-operator correspondence, most of the computations are further reduced to the evaluation of energy levels in a finite-density system described by a large-charge effective field theory, usually associated with a conformal superfluid phase. The large-charge dynamics in the nonrelativistic case is however richer---and therefore, more challenging---than its relativistic counterpart. In this thesis, we discuss recent progress which opened the door to many new exciting large-charge applications.



\chapter*{Foreword}
~\newline~\newline~
\markboth{Foreword}{Foreword}
\addcontentsline{toc}{chapter}{Foreword}

This thesis is based on the publications~\cite{Orlando:2020idm,Pellizzani:2021hzx,Hellerman:2021qzz,Hellerman:2023myh} by the author and his collaborators. The overarching topic is the large-charge expansion in \acp{nrcft}. The structure of the thesis is the following.
\begin{enumerate}[itemsep=5pt,partopsep=0pt,topsep=0pt,parsep=0pt]
    \item Chapter~\ref{chap:intro} gives an overview of the topic, starting from its motivation and a discussion of the context in which it fits. We propose a condensed matter invitation to understand the relevance of certain real-life critical phenomena to the formal study of \acp{cft}. We then discuss the main ingredients underlying the large-charge program in relativistic \acp{cft}, and present a comprehensive overview of the current status of research in this field. This is used as a reference point to present the state of the art in the nonrelativistic case, which allows us to clearly formulate how the content of this thesis fits into this landscape. In passing, we discuss the quantization of \acp{nrcft} in a way that sheds light on some of its surprising properties.
    \item Chapter~\ref{chap:NR_EFT} gives a coherent presentation of the results obtained in~\cite{Orlando:2020idm,Pellizzani:2021hzx,Hellerman:2021qzz}. These contributions marked a significant development in the nonrelativistic large-charge program. Indeed, large-charge techniques are based on the \ac{eft} framework, expressing in this context the emergence of simple collective behaviours---such as superfluidity---when the number of particles in the system, that is, the charge, becomes large. The \ac{eft} used previously was a correct, albeit incomplete description of the large-charge dynamics. The prescription to obtain the complete large-charge \ac{eft} and the corresponding renormalization procedure is the central object of the second chapter. In turn, this allows to prove the universality of the leading quantum correction, which we compute in $(2+1)$ and $(3+1)$ dimensions. Along the way, we discuss various equivalent prescriptions and schemes, account for a soft breaking of conformal invariance and propose a clear physical picture underlying the general large-charge expansion that we obtain. 
   \item Chapter~\ref{chap:ExplicitReal} is based on~\cite{Hellerman:2023myh}. It provides the first explicit verification of the nonrelativistic large-charge \ac{eft} predictions starting from a controlled microscopic description. The model that we study is relevant to the description of certain real-life critical phenomena, thereby echoing the discussion of the first chapter. We also include a small list of further microscopic models which are expect to be captured by the large-charge \ac{eft}.
   \item Finally, in the outlook, we critically discuss our results and summarize the various suggestions for future research that we make along the way.
\end{enumerate}
In the large-charge context, we hope to convey the idea that the content of this thesis lays the foundation for many important research projects in the future, as certain aspects of \acp{nrcft} have been underexplored so far, partly due to the technical challenges that we have overcome.

\tableofcontents

\setlength{\parskip}{1em}

\mainmatter
\chapter{Introduction} \label{chap:intro}
\setcounter{section}{0}

Quantum Field Theory (\acs{qft}) was originally developed as the theoretical framework for the description of interactions between fundamental particles and other problems in high-energy physics. It has now evolved into a powerful way to phrase problems across most areas of modern physics, from condensed matter and statistical physics to cosmology.

The main computational tool of \acs{qft} is the path integral, which expresses correlation functions as an average over all allowed field configurations, weighted by a phase factor given by the action:
\begin{equation}
    \ev{\mathcal{O}_n(x_n) \ldots \mathcal{O}_1(x_1)} = \frac{\int D\Phi \, \mathcal{O}_n(x_n) \ldots \mathcal{O}_1(x_1) \expo{iS[\Phi]}}{\int D\Phi \, \expo{iS[\Phi]}}.
\end{equation}
In the absence of additional assumptions, analytic access to this quantity is too much to ask for in general. However, from a formal perspective, this expression---or rather, its Euclidean version where the phase factor becomes $\expo{-S_E[\Phi]}$---makes it clear that when there exists a leading trajectory around which fluctuations are suppressed by some large factor, then one can consistently perform a saddle-point approximation. Most useful applications of \acs{qft} realize this scenario\footnote{Under additional constraints, some powerful tools can go much beyond the saddle-point approximations. Some typically require supersymmetry. Examples are localization \cite{Pestun:2016zxk}, integrability \cite{Beisert:2010jr} and dualities broadly defined (including holography \cite{Aharony:1999ti}), see for instance \cite{Seiberg:2016gmd}.}, which roughly fall in one of four categories.
\begin{itemize} [itemsep=5pt,partopsep=0pt,topsep=0pt,parsep=0pt]
    \item The action is that of a known theory deformed by interaction terms whose coupling constants are small. The saddle-point approximation is then controlled by $\hbar^{-1}$, thereby giving rise to \emph{perturbation theory}.  The most common example is when the leading trajectory is Gaussian---in which case the theory is said to be weakly coupled---as best illustrated by the successful match between the theory of quantum electrodynamics and the corresponding experimental data. But this also includes conformal perturbation theory as a way to explore the dynamics of a system near criticality~\cite{zamolodchikov1987renormalization}.
    \item The theory is strongly coupled, \emph{i.e.} the coupling constants are not small, but there exists an energy range in which observables are essentially not sensitive to field configurations with large quantum fluctuations. The latter can therefore be integrated out, at least formally. In practice, this is achieved by building an effective theory in terms of the reduced number of effective degrees of freedom in this energy range. The corresponding path integral is dominated by a leading trajectory with fluctuations suppressed by some large energy ratio controlling the saddle-point approximation. This is typically a low-energy phenomenon, \emph{e.g.} the pion Lagrangian, where the pions are the effective degrees of freedom resulting from the strong quark interactions below a certain energy scale ($\sim$ 1 GeV).
    \item A physical parameter is given---somewhat artificially---a large value that can be factored out of the action, thereby controlling the corresponding saddle-point approximation. This is for instance the famous 't~Hooft limit~\cite{tHooft:1973alw} or, more generally, the large-$N$ limit~\cite{Moshe:2003xn}, where $N$ is often the rank of a given global or gauge symmetry. Unlike the other scenarios, this is a modification of theory itself, not just a statement about a subset of observables.
    \item The path integral is given some boundary conditions that correspond to preparing a specific state carrying a large number of quanta, that is, a state that belongs to a large irreducible representation. The corresponding large quantum number can for instance be the spin \cite{Alday:2007mf,Fitzpatrick:2012yx,Komargodski:2012ek} or a global charge \cite{Hellerman:2015nra,Monin:2016jmo,Gaume:2020bmp}, and this framework is now referred to as the \emph{large quantum number expansion}. A toy model for this is the hydrogen atom, whose Hamiltonian $H=\frac{\vec{p}^2}{2m_e} - \frac{\hbar c \alpha}{r}$ has energy levels given by $E_n = \frac{E_0}{n^2}$. Here, $m_e$ is the mass of the electron, $\alpha$ is the fine-structure constant and $E_0=-\frac{\alpha^2 m_e c^2}{2}$ is the ground-state energy. The corresponding wave-functions carry three quantum numbers: $n\in\mathbb{N}^*$, $l\in\{0,\ldots,n-1\}$ and $m\in\{-l,\ldots,l\}$. The latter labels the projection of the angular momentum on the third axis. Of all the states with a fixed value of $m$, the one with the lowest energy has $m=n-1$, which yields $E_{m+1}=\frac{E_0}{(m+1)^2}$. On the other hand, the classical minimization of the energy $E = \frac{1}{2} m_e v^2 - \frac{\hbar c \alpha}{r}$ under the constraint of fixed angular momentum $\vec{r} \times \vec{p} = \hbar m$ gives $E_{classic}(m) = \frac{E_0}{m^2}$. One might argue that there is no reason to trust such a result, but it turns out to coincide with the quantum result in the large-$m$ limit. This is no coincidence. It is due to the fact that, in this regime, the de Broglie wavelength $\lambda \sim \frac{r}{m}$ varies slowly in the radial direction, indicating that the system 'classicalizes' and a treatment based on the WKB approximation is justified. Equivalently, this can be phrased in the path-integral language, where the constrained path integral is dominated by the classical trajectory at large $m$, see also \cite{Monin:2016jmo} for an expanded discussion of the so-called rigid rotor. While this is a simple quantum mechanical example, there exists a quantum field theory avatar of it which is a cornerstone of the modern large quantum number expansion program: the effective string theory description of the meson at large angular momentum $J$, where the meson is modelled by a long quickly rotating string. This is to be thought of as the flux tube connecting a quark/antiquark pair. At large $J$, the mass of the meson is found to be of the form \cite{Hellerman:2013kba,Hellerman:2016hnf}:
    \begin{equation} \label{eq:EST}
        M^2 = \# J + \# J^\frac{1}{4} - \frac{1}{12} + \order{J^{-1}}.
    \end{equation}
    Not only does it reproduce the famous Regge trajectory relation $M^2 \sim J$ at leading-order, but it also predicts a \emph{universal} Regge intercept, namely, the $J^0$ term.
\end{itemize}
Note that the second category, the one describing strongly-coupled theories admitting a particular coarse-grained description, typically refers to processes taking place in the vacuum. This is different from the last category, in which operators carrying large quantum numbers shift the saddle away from the vacuum, and the theory is therefore semiclassically described in terms of a certain \emph{phase of matter}. The latter features some emergent collective behaviour characterized by a reduced number of degrees of freedom. So the two situations are in fact very similar, and are both captured by the \emph{effective field theory} framework\footnote{The formal tools needed for \ac{eft} constructions were first developed in the context of particle physics and we shall therefore borrow some of the jargon associated with this field although, in the situations we will face, there will be no 'low-energy' regime in the RG sense, but merely a coarse-grained description of a system developing some collective behaviour. We hope this will be clear from the context.}. In the large quantum number expansion, this description is made in terms of the hydrodynamic Goldstone bosons associated with the breaking of some spacetime and sometimes also internal symmetries, caused by the finite density state. For instance, in the effective string theory example mentioned above, the relation Eq.~\eqref{eq:EST} is indeed obtained by carefully constructing the \ac{eft} for the Goldstone bosons associated with the breaking of translations transverse to the string.

\subsubsection{Outline of the chapter}

This thesis concerns general features of critical systems that are described in terms of a simple phases of matter when a global internal charge---to be thought of as the number of particles in the system (more on this later)---becomes large. This is the \emph{large-charge expansion} \cite{Hellerman:2015nra,Monin:2016jmo,Gaume:2020bmp} mentioned above. In particular, we will focus on conformal superfluid states in \acp{nrcft}.

A good place to start the investigation of the possible emergence of collective behaviours describing a large-charge phase is condensed matter. Indeed, unlike standard high-energy computations done in the vacuum, it describes finite density phases of matter which emerge as effective descriptions of the (often complicated) underlying quantum many-body system. This is what we propose to do in Section~\ref{sec:Intro_Invit}, where we give a standard---though slightly biased---description of the phenomenology and the phase diagram of the cold Fermi gas, making no reference to the large-charge expansion. Instead, we want to use these results to ask '\emph{what does a high-energy theoretical physicist learn from this condensed matter example?}'.

As we advocate later in Section~\ref{sec:Intro_LargeQ_rel}, the distinction between vacuum state and finite density state disappears in critical systems thanks to a powerful exact relation known as the state-operator correspondence. The phenomenology of the cold Fermi gas thus serves as a precious guide towards the understanding of certain properties of vacuum correlators in critical systems when some operator insertions carry a large global charge. Although cold Fermi gases are intrinsically nonrelativistic, this approach triggered an important series of formal developments in the context of \emph{relativistic} conformal field theories, which we review.

In Section~\ref{sec:Intro_LargeQ_nonrel}, we finally discuss the fate of the large-charge expansion in \acp{nrcft}. The purpose of this thesis is to convey the idea that \acp{nrcft} are in fact very well-suited for this program, for at least three reasons:
\begin{itemize} [itemsep=5pt,partopsep=0pt,topsep=0pt,parsep=0pt]
    \item \emph{every} \ac{nrcft} possesses a global $U(1)$ symmetry, associated with the number of particles in the system;
    \item the state-operator correspondence establishes an astonishing connection with ultracold atom experiments \cite{Nishida:2007pj,Nishida:2010tm,Goldberger:2014hca}, in which the number of particles is always large, meaning that theoretical predictions and experimental data can be compared directly;
    \item the kinematic constraints imposed by nonrelativistic conformal invariance \cite{Goldberger:2014hca} are somewhat weaker than their relativistic counterpart, but the large-charge expansion is equally powerful and even reveals a richer structure.
\end{itemize}
With this, we will be ready to enter the technical part of the thesis in Chapters~2 and 3.

\pagebreak
\section{A condensed matter invitation: cold atoms in a trap}
\label{sec:Intro_Invit}

We discuss some aspects of cold Fermi gases, see \emph{e.g.} \cite{altland_simons_2010,zwerger2011bcs,Schmitt:2014eka}.\footnote{\textbf{Note added:} the author was informed that the term \emph{condensed matter} may be misused in this context.}

One of the major theoretical successes in quantum many-body physics was the description of pairing in a cold Fermi gas by Bardeen, Cooper and Schrieffer almost 70 years ago~\cite{bardeen1957microscopic,bardeen1957theory}, which has been known since then as \acs{bcs} theory. The mechanism is simple and highly generic: at low temperatures and in the presence of an attractive potential between the fermions, even very weak, it is energetically favourable for the fermions close to the Fermi surface to pair up, thereby forming \emph{Cooper pairs}. The Fermi surface is therefore seen to be unstable at low temperatures against any form of attractive interaction due to Cooper pairing and the ground state of the system exhibits either superconductivity or superfluidity. Of course, Cooper pairing is not robust against large thermal fluctuations, so this is an emergent phenomenon below a critical temperature $T_c$. This said, at low enough temperatures, Cooper pairs can proliferate, causing a strong overlap between them. They form a stable collective condensate: breaking this phase requires to break all Cooper pairs, not just individual ones. This is the \acs{bcs} energy gap. Perhaps the most striking aspect of \acs{bcs} theory is the fact that it does not depend on the underlying mechanism causing the attractive potential between the fermions, therefore entailing a sense of universality.

In a metallic superconductor, the weakly attractive interaction can be seen as the consequence of the local distortion of the ionic crystal (the 'lattice of atoms') by a freely moving electron. This causes a local excess of positive charge which takes a characteristic time inversely proportional to the Debye frequency $\omega_D$ to relax back to equilibrium. During relaxation, another electron with opposite spin may get attracted, and the two of them pair up in momentum space. The characteristic time needed for the first electron to travel across a lattice spacing, on the other hand, is inversely proportional to the Fermi energy, typically much larger than $\omega_D$. Hence, the resulting pairing between the two electrons takes place over long distances with respect to the lattice spacing (typically, a factor of $10^3$). In quantum mechanical terms, the interaction is mediated by phonons, the quasi-particle accounting for the vibrations of the crystal. Once a large number of Cooper pairs have formed, they condense and the electrons can flow collectively without resistance, therefore achieving superconductivity.

Similarly, fermionic superfluidity is achieved through Bose-Einstein condensation (\acs{bec}) of the Cooper pairs in a cold gas of fermionic atoms, resulting in neutral frictionless motion. The first experimental realization of this phenomenon was with atoms of ${}^3$He \cite{osheroff1972evidence}. More recently (about 20 years ago), quantized vortices---a hallmark of superfluidity, see Figure~\ref{fig:vortices}---were observed in a gas of atoms of ${}^6$Li in an optical trap \cite{zwierlein2005vortices}, thereby confirming the underlying \acs{bcs} mechanism. The \acs{bcs} regime is however typically associated with a weakly-interacting phase, which raises the question of whether there exists a \emph{strong-coupling} (not necessarily attractive) mechanism allowing for the formation of fermionic superfluid states. In this case, one would imagine fermions with opposite spin to pair up into short-distance di-fermionic molecules, also called dimers, which could then condense. As we shall see, not only is the answer positive, but this can be achieved experimentally using \emph{e.g.} the lithium setup mentioned just above by departing from a \acs{bcs} state and increasing the interaction strength in a way that preserves superfluidity: this is the famous \acs{bcs}-\acs{bec} crossover that we discuss next. Of course, although the whole crossover is technically in a \acs{bec} phase, the '\acs{bec} side' of the crossover is really understood as the condensation of the \emph{tightly} bound pairs of fermions.

\subsection{BCS-BEC crossover and the unitary Fermi gas}

We refer to \cite{ketterle2008making} for a thorough review of the experimental preparation of ultracold Fermi gases, and to \cite{Strinati:2018wdg} for a general review and a report of the relevance of this setup to nuclear systems.\footnote{Inded, dilute nuclear matter features a \acs{bcs}--\acs{bec} crossover similar to that of cold Fermi gases, which can be described by the same microscopic models, though the interactions must have a finite range, and the crossover is realized by varying the density, as opposed to the scattering length. The experimental realization of the unitary Fermi gas in ultracold atomic gases raised the interest of the nuclear physics community as the neutron-neutron scattering length is known to be 'accidentally large' and should therefore approximately share the universal properties of the unitary Fermi gas. An exciting application of this is the possibility to develop a better understanding of neutron stars, whose inner crust is expected to host a dilute gas of neutrons in an almost critical superfluid state.}

\subsubsection{Experimental setup and phase diagram}

The different regimes of atomic systems are characterized by the inter-atomic interaction range $r_0$, the inter-atomic spacing $\rho^{-\frac{1}{3}}\sim k_F^{-1}$, where $\rho$ is the density at the center of the system and $k_F$ is the Fermi momentum\footnote{We always strip off a factor of $\hbar^{-1}$ to make momentum have the units of an inverse distance. So $k_F$ is in fact the Fermi wave-vector.}, the system size $V^\frac{1}{3}$ and the thermal wavelength $\lambda_T = \sqrt{\frac{2\pi\hbar^2}{mT}}$.\footnote{Throughout this thesis, we set the Boltzmann constant $k_B = 1$.} An atomic gas is characterized by the following hierarchy:
\begin{equation} \label{eq:QuantumGasHierarchy}
    r_0 \ll k_F^{-1} \ll V^\frac{1}{3}.
\end{equation}
The first inequality characterizes \emph{diluteness}, while the second requires the number of particles in the system to be large.
One further distinguishes
\begin{equation}
\begin{cases}
    r_0 \gg \lambda_T \quad & \text{quasi-classical collisions}, \\
    r_0 \ll \lambda_T \quad & \text{ultracold collisions}.
\end{cases}
\end{equation}
A \emph{quantum gas} satisfies the hierarchy \eqref{eq:QuantumGasHierarchy} and the second regime above (ultracold collisions).\footnote{Echoing the previous discussion, we note that we really are interested in the emergence of a semiclassical description of the \emph{quantum} regime. Of course, at high enough temperature, the system classicalizes in a somewhat trivial fashion which will not be relevant for us.} A \emph{degenerate} quantum gas satisfies instead \eqref{eq:QuantumGasHierarchy} and the stronger condition $k_F^{-1} \ll \lambda_T$, which implies ultracold collisions. 

The diluteness condition for a quantum gas has an important and almost generic consequence. Let us consider a generic spherically symmetric inter-atomic potential $U(r)$ with range $r_0$. A wave-function with quantum numbers $(n,l,m)$ is then subject to an effective potential of the form $U_{eff}(r) = U(r) + \frac{\hbar^2}{2m}\frac{l(l+1)}{r^2}$. Roughly speaking, a wave-packet with momentum $\sim k_F$ moving towards the origin of this potential will encounter a classical turning point when $r$ equals $r^* := \sqrt{l(l+1)} k_F^{-1}$ ($l\neq0$). By the diluteness condition, we have $r_0 \ll r^*$, meaning that the wave-packet bounces back even before feeling the atomic potential if $l\neq0$, and there are no collisions. Therefore, only $s$-wave scattering matters in a quantum gas.\footnote{A caveat is when there are so-called \emph{shape resonances} for $l>0$.}. In this case, the $S$-matrix is simply given by $S(k)=1+2ikf_s(k)$, where the $s$-wave scattering amplitude as function of the momentum transfer $k$ is given by
\begin{equation} \label{eq:Swave_Scatt}
    f_s(k) = -\frac{1}{a^{-1} + k[i + \order{r_0 k}]}.
\end{equation}
Since $k \leq k_F$, the $\order{r_0 k}$ correction is indeed small by the diluteness condition. The parameter $a=-f_s(0)$ is the $s$-wave scattering length. It can be thought of as the strength of an effective two-body contact interaction between the particles. However, if there is a regime where $a$ diverges with respect to $k$, then we simply get $f_s(k)=\frac{i}{k}$ and $S(k)=-1$, which saturates the unitarity bound. We therefore call this regime the \emph{unitary} limit, where the nature of the gas and the details of the inter-atomic potential no longer enter the description of the system: this is a nonrelativistic quantum critical phase. From a theoretical viewpoint, there is an emergent invariance under nonrelativistic conformal symmetries, known as the Schrödinger group.

\begin{figure}[!p]
	\centering
	\includegraphics[width=0.65\textwidth]{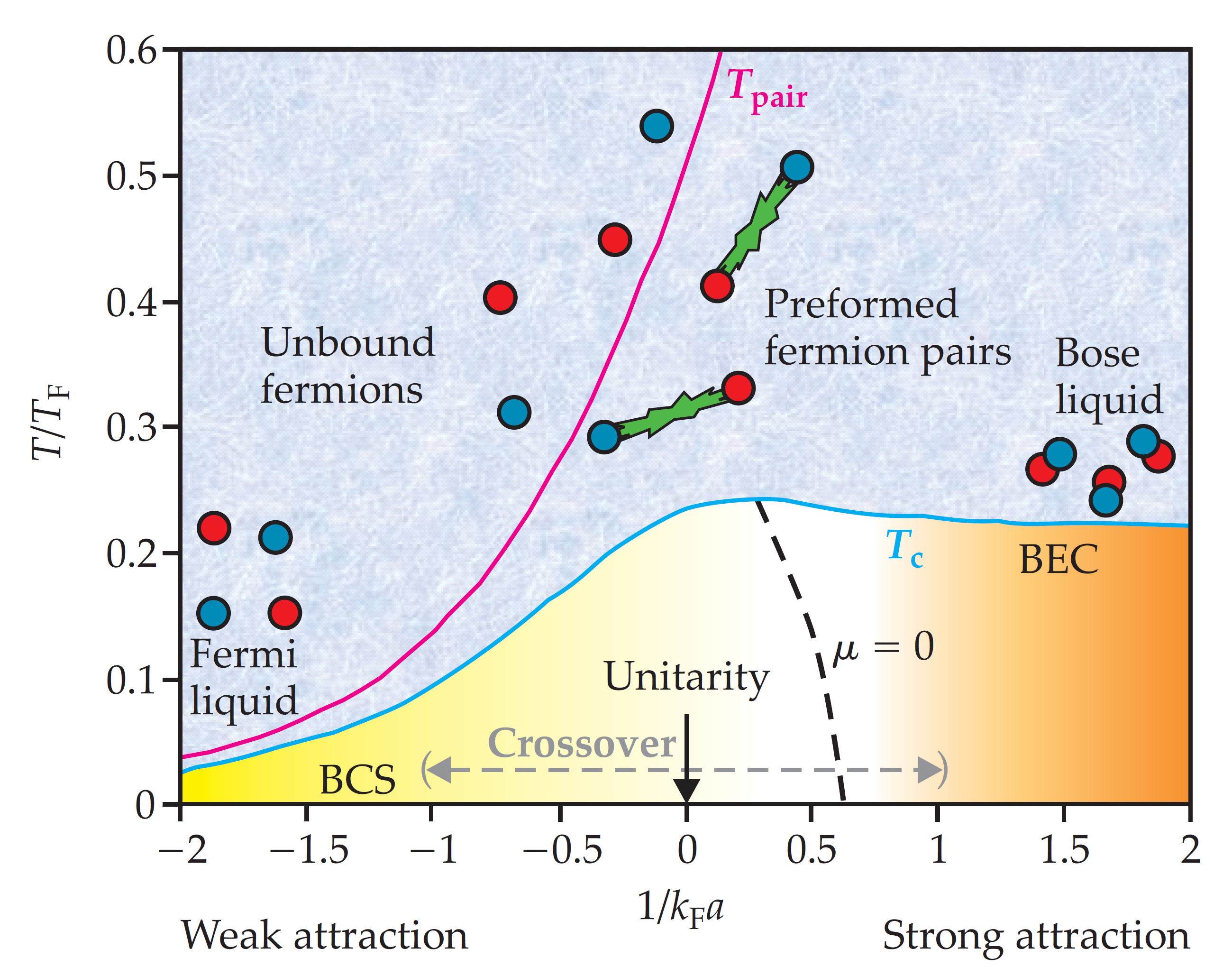}
	\caption{{Putative phase diagram of a cold Fermi gas (picture taken from \cite{sa2008fermions}). The value and even the sign of the dimensionless scattering parameter $\frac{1}{k_F a}$---where $k_F$ is the Fermi wave-vector and $a$ the $s$-wave scattering length---can be experimentally tuned. At very low-temperatures in the \acs{bcs} regime $\frac{1}{k_F a} \ll -1$, the interaction is weakly attractive and fermions form Cooper pairs, while for $\frac{1}{k_F a} \gg 1$, the interaction is strong and tightly binds pairs of fermions together. The latter system is effectively described by a weakly interacting bosonic gas of such molecules (also called dimers), i.e. a \acs{bec}. Both regimes are known to exhibit superfluidity and no phase transition occurs in between, indicating a smooth crossover that preserves superfluidity for all values of $k_F a$. This is particularly relevant for the strongly interacting crossover region $\frac{1}{k_F a} \in [-1, 1]$, centered around the resonant case $\frac{1}{k_F a} = 0$ known as the unitary limit. Above the critical temperature $T_c$ (the blue line), the gap closes and superfluidity is lost. At unitarity and in the \acs{bec} regime, this happens at an $\order{T_F}$ temperature, above which fermion pairing survives up to very high temperatures but is not able to condense, while $T_c$ is exponentially small in $1/k_F\abs{a}$ in the \acs{bcs} regime, where the loss of superfluidity essentially coincides with the breaking of fermion pairs.}}
	\label{fig:crossover}
\end{figure}

Fermionic quantum gases are typically prepared with atoms of ${}^{40}$K or ${}^6$Li confined by an optical (approximately harmonic) trap. The most remarkable experimental achievement in this setup is the ability to tune the strength of the interaction via a controllable magnetic field $B$ as
\begin{equation}
    a(B) = a_0 \left( 1 - \frac{\delta B}{B - B_c} \right).
\end{equation}
The unitary limit is attained at the so-called (Fano-)Feshbach resonance, occurring when $B$ approaches the critical value $B_c$ from above or from below, and $\delta B$ is called the width of the resonance.\footnote{There is a distinction between broad and narrow resonances, but the diluteness condition essentially corresponds to the resonance being broad~\cite{ketterle2008making}.} In ${}^6$Li, the resonance occurs around $B_c \approx 834$ G. The resulting nonrelativistic quantum critical phase is called the \emph{unitary Fermi gas}. This is a strongly interacting system, making it challenging to describe theoretically. This said, upon tuning the magnetic field $B$ one can reach either the \acs{bcs} or the \acs{bec} regime while preserving superfluidity. This is the famous \acs{bcs}-\acs{bec} crossover, see Figure~\ref{fig:crossover}. A complete theoretical description of the crossover is still lacking, although some qualitative features are appropriately captured by mean-field theory.

\subsubsection{Quantized vortices}

\begin{figure}[p]
    \centering
    \includegraphics[width=0.7\linewidth]{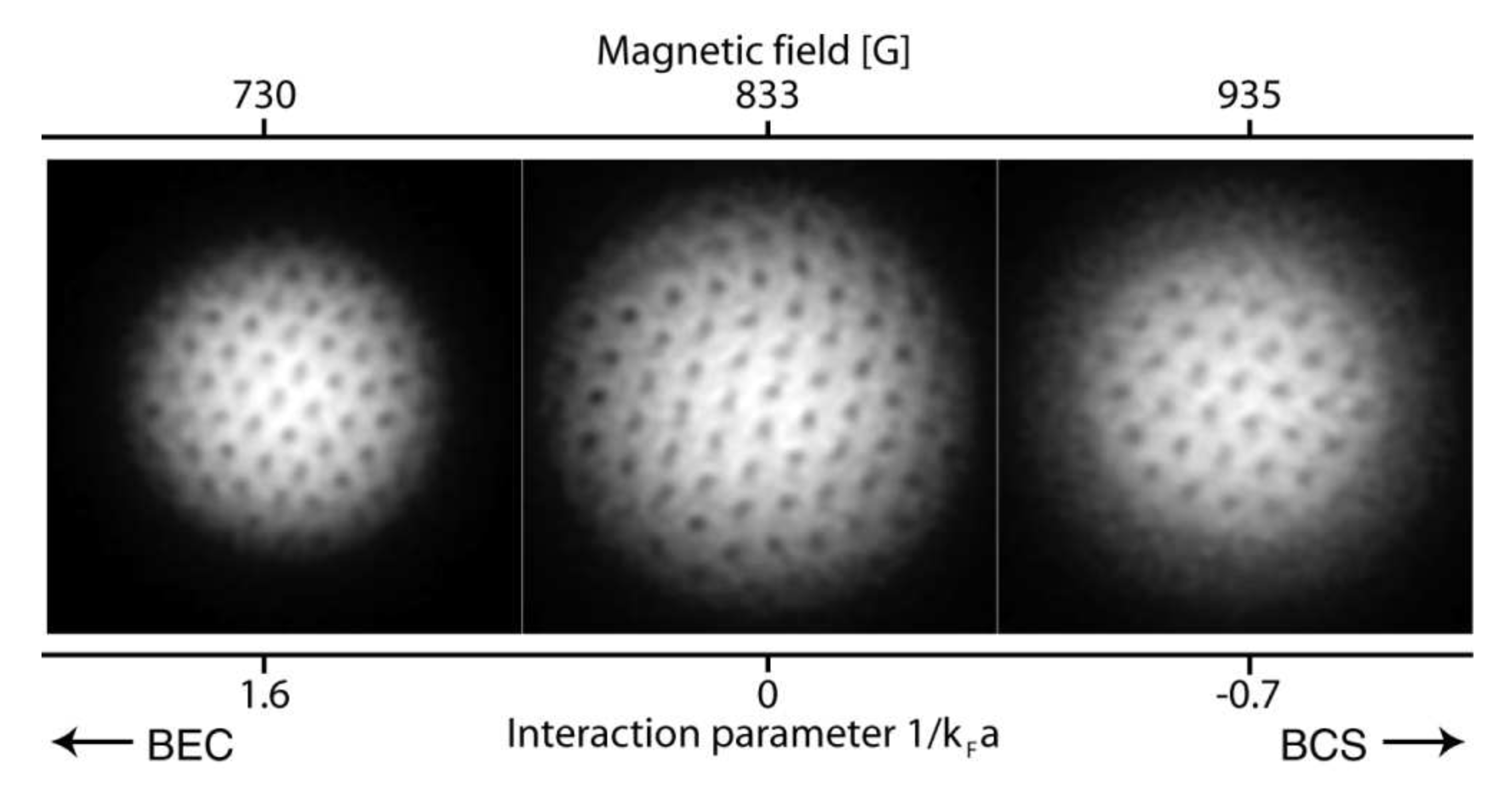}
    \caption{Picture taken from~\cite{ketterle2008making}, which is an updated realization of the original result presented in~\cite{zwierlein2005vortices}. Quantized vortices are clearly visible across the \acs{bec}-\acs{bcs} crossover, with the unitary regime shown in the middle, around 833~G. This is realized in a ultracold dilute gas of ${}^6$Li, and each picture is roughly 1mm wide.}
    \label{fig:vortices}
\end{figure}

An important feature mentioned earlier is the existence of a gap in the spectrum of single-particle fermionic excitations, thereby protecting the superfluid phase at low temperatures against thermal fluctuations. In the \acs{bcs} regime, mean-field theory predicts that the gap is of the form
\begin{equation}
    \Delta_{gap}(T) \sim T_F \expo{-\frac{\pi}{2k_F\abs{a}}} \sqrt{1-\frac{cst \cdot T}{T_F \expo{-\frac{\pi}{2k_F\abs{a}}}}},
\end{equation}
The gap is therefore seen to close as $T$ increases towards a critical value $T_c \sim T_F \expo{-\frac{\pi}{2k_F\abs{a}}} \sim \Delta_{gap}(0)$, which is exponentially small in the dimensionless parameter $\frac{1}{k_F\abs{a}}$. This is believed to be the correct qualitative picture. As the scattering length is tuned away from the \acs{bcs} regime, the critical temperature increases significantly to the order of the Fermi temperature $T_F$.

This means that there can be a variety of low-energy \emph{bosonic} excitations in the superfluid which do not spoil superfluidity. These are typically associated with the hydrodynamic Goldstone mode. Again, these vibration modes are associated with phonon excitations. We shall discuss the \ac{eft} for this mode in the critical superfluid state shortly. However, we want to point out that parametrically larger excitations can lead to the formation of quantized vortices and other superfluid phases.

Indeed, when a superfluid is given a large amount of angular momentum, the most energetically favourable way to accommodate this excitation is to develop one or many quantized vortices. Vortices are characterized by the fact that the order parameter drops to zero at their core, thereby locally breaking the superfluid state and restoring symmetry. In an optical trap, the vortices---whose size is of the order of the so-called healing length $\xi_h$ in the \acs{bec} regime, or $k_F^{-1}$ in the \acs{bcs} one---are too small to be observed directly. This can however be circumvented by relaxing the condensate, which thus undergoes a rapid expansion. The vortices survive this process by virtue of angular momentum conservation, which allows the observation of vortex structures as in Figure~\ref{fig:vortices}.

From a microscopic perspective, the phase $\chi(t,\vect{r})$ of the superfluid order parameter,
\begin{equation}
    \chi =
    \begin{cases}
        -\arg(\psi) & \text{(bosonic)} \\
        -\frac{1}{2} \arg\left(\ev{\psi_\downarrow\psi_\uparrow}\right) \qquad & \text{(fermionic)}
    \end{cases}
\end{equation}
(where $\ev{\psi_\downarrow\psi_\uparrow}$ is the Cooper pair) is related to the so-called superfluid velocity
\begin{equation}
    \vect{v} := \frac{\hbar}{m_b} \vect{\nabla} \chi,
\end{equation}
where $m_b$ is the effective mass of the bosonic degree of freedom forming the superfluid. The famous Onsager-Feynman quantization condition then reads
\begin{equation}
    \oint \dd{\vect{l}} \cdot \vect{v} = \frac{2\pi\hbar n}{m_b},
    \qquad (n\in\mathbb{Z}).
\end{equation}
The integral is a closed loop inside the superfluid. If it does not enclose a vortex, then $n=0$. A vortex carrying $\abs{n}>1$ tends to decay into multiple vortices with unit charge.

In viscous fluids, classical vortices may form momentarily due to some external stress but are generally metastable and eventually decay into rigid-body motion. However, they may sometimes be long lived. Instead, there are two properties that really distinguish them from superfluid vortices: the fact that the latter carry a \emph{quantized} charge, and the fact that, in a superfluid, the ground state with some large angular momentum is always a state with vortices. Interestingly, vortex nucleation in a trap or container necessarily arises from external \emph{edge} excitations, where the order parameter vanishes. If that was not the case, their spontaneous formation in the bulk of the droplet of particles would imply a discontinuity. This is indeed verified experimentally where, after inputting some large angular momentum into the superfluid, the edge of the droplet of particles undergoes some turbulence that eventually leads to the formation of vortices that quickly navigate towards the bulk of the droplet to adopt the configuration of minimal energy. It was predicted and then verified that they form the so-called Abrikosov hexagonal lattice~\cite{Abrikosov1956MagneticPO} to minimize the total energy. Further experimental~\cite{Coddington_2003} and theoretical~\cite{Watanabe_2008} works showed that the lattice planes of this hexagonal arrangement can undergo shear distortion due to the so-called Tkachenko modes~\cite{Tkachenko1966OnVL,1966JETP23.1049T,Tkachenko1969ElasticityOV,Moroz:2018noc,Moroz:2019qdw}, which were recently addressed in the context of fracton physics~\cite{Nguyen:2020yve}.

However, in the presence of a large amount of vortices, the vortex cores may start to overlap, thereby presumably undergoing a phase transition to a new state: a giant vortex~\cite{Kasamatsu_2002,Fischer_2003,Guo_2020}. This is a configuration where the order parameter vanishes everywhere except on an external annulus of quickly rotating superfluid.

\subsubsection{Effective field theory}

As mentioned, a quantitatively accurate microscopic description of the crossover is lacking. We shall however study one such qualitative model in Section~\ref{sec:UFG_largeN}. On the other hand, regardless of what the correct microscopic theory is, this system is subject to the emergence of a collective behaviour below the critical temperature, namely, superfluidity. There is a unique low-energy degree of freedom describing such a phase. This is the hydrodynamic Goldstone mode: the phonon $\pi$. It is the fluctuation around the superfluid ground-state profile for the phase of the condensate:
\begin{equation}
    \chi(t,\vect{r}) = \mu t + \pi(t, \vect{r}),
\end{equation}
where $\mu$ is the chemical potential associated with the $U(1)$ particle number symmetry.\footnote{See \cite{Watanabe:2013uya,Brauner:2020zov} for enlightening discussions about Goldstone theorems at finite chemical potential.} It is therefore natural to imagine a low-energy effective description in terms of $\chi$ only, constrained by the symmetries of the system. This approach is particularly interesting as the symmetries get enhanced at the unitarity point: the \ac{eft} is that of a conformal superfluid. The nonrelativistic conformal group---the Schrödinger group as already mentioned---is of the form
\begin{equation}
    U(1) \times \left[ (SL(2,\mathbb{R}) \times SO(d)) \ltimes (\mathbb{R}^d \times \mathbb{R}^d) \right],
\end{equation}
where the components, from left to right, correspond to the particle number symmetry; the generators $(H,D,C)$ associated with time translations, dilatations and special conformal transformations; spatial rotations; Galilean boosts; spatial translations. We will investigate the properties of \acp{nrcft} more in detail later, but at this stage, we want to point out that the spontaneous symmetry breaking pattern that a nonrelativistic conformal superfluid realizes is
\begin{equation}
    U(1) \times \left[ (SL(2,\mathbb{R}) \times SO(d)) \ltimes (\mathbb{R}^d \times \mathbb{R}^d) \right]
    \stackrel{SSB}{\longrightarrow}
    (\bar{H} \times SO(d)) \ltimes (\mathbb{R}^d \times \mathbb{R}^d),
\end{equation}
where $\bar{H}$ denotes time translations as generated by the operator $H-\hbar\mu \mathcal{Q}$, and $\mathcal{Q}$ is the particle number generator. Correctly implementing this symmetry breaking pattern in the presence of a background potential (representing the harmonic trapping potential applied in experiments) turns out to be a rather non-trivial task. This was done in \cite{Son:2005rv} by carefully analysing the constraints imposed by general coordinate invariance.

At this point, it is not clear what the scales controlling the derivative expansion in this \ac{eft} are. We postpone the answer to later. At leading order, however, the \ac{eft} should match the standard low-energy descriptions of the unitary Fermi gas, which are the Thomas-Fermi approximation and superfluid hydrodynamics. They are characterized by a single Wilsonian coefficient $\xi$---the so-called Bertsch parameter---whose approximate value is known from experiments~\cite{Ku_2012} and Monte-Carlo simulations~\cite{Carlson:2011kv,Endres:2012cw} to be around $\xi=0.37$. It is defined as the ratio between the ground-state energy of the Fermi gas at unitarity and that of the noninteracting Fermi gas.

The leading-order effective Lagrangian for a nonrelativistic superfluid, conformal or not, in the presence of an external potential $A_0(\vect{r})$ is a function
\begin{equation}
    \mathcal{L}_{LO}(\chi) = P(U)
\end{equation}
of the invariant building-block\footnote{Here, $m$ corresponds to the effective mass of the bosonic degree of freedom forming the superfluid, denoted $m_b$ earlier. We drop the subscript for convenience.}
\begin{equation}
    U(t,\vect{r}) = \dot\chi - A_0(\vect{r}) - \frac{\hbar}{2m} (\vec\nabla \chi)^2.
\end{equation}
Note that the superfluid density is given by $\rho=P'(U)$. Upon turning off the background potential $A_0(\vect{r})$ and evaluating this relation in the superfluid ground state, one finds $\rho_0=P'(\mu)$, which must reproduce the equation of state, so $P$ is interpreted at zero temperature as the pressure.
Finally, the equation of motion is the continuity equation:
\begin{equation}
    \dot{\rho} - \vect{\nabla} \cdot (\rho \vect{v}) = 0,
\end{equation}
where $\vect{v}$ is the superfluid density introduced earlier.

If we now specialize to the unitary regime, there is a unique choice for the function $P(\mu)$ as dictated by dimensional analysis:
\begin{equation}
    P(\mu) = c_0 \sqrt{\frac{m^3}{\hbar}} \mu^\frac{5}{2}.
\end{equation}
In turn, this fixes completely the leading-order Lagrangian, and it is easy enough to verify that the Bertsch parameter is related to the Wilsonian coefficient $c_0$ as
\begin{equation} \label{eq:C0_VS_Bertsch}
	c_0 = \frac{2^{5/2}}{15\pi^2\xi^{3/2}}.
\end{equation}
The phonon dispersion relation is fixed at leading order by solving the continuity equation at linear order in the fluctuations $\pi$, namely\footnote{For later convenience, we wrote the result in general spatial dimension $d$, in which case the leading-order Lagrangian at unitarity is proportional to $U^{\frac{d}{2}+1}$.}
\begin{equation}
    \ddot{\pi} - \frac{2\hbar}{md} \left(\mu - A_0(\vect{r}) \right) + \frac{\hbar}{m} \vec\nabla A_0 \cdot \vec\nabla\pi = 0.
\end{equation}
The case of interest for us is that of a harmonic potential $A_0(\vect{r}) = \frac{m\omega^2}{2\hbar} \vect{r}^2$, in which case this linearized equation of motion reduces to a hypergeomtric equation, which yields the following spectrum of phonon excitations~\cite{Kravec:2018qnu}:\footnote{These excitations are actually associated with either internal motion ('breathing mode') or center-of-mass excitations.}
\begin{equation} \label{eq:PhonoDispRel}
    E_{n,l}^d = \hbar\omega \sqrt{\frac{4}{d} n^2 + 4 n + \frac{4}{d} l n - \frac{4}{d} n + l}.
\end{equation}
We shall come back to this dispersion relation much later.

We are now reaching the end of this condensed matter detour. There is one final observation we would like to make which is of central importance for the philosophy underlying this thesis, namely, the fate of the controlling parameter in this \ac{eft} description. Going back to $d=3$, the next-to-leading order Lagrangian contains two additional Wilsonian coefficients~\cite{Son:2005rv}:
\begin{equation}\label{eq:eff-action-nlo}
	\Lag_{eff}(\chi) = c_0 \sqrt{\frac{m^3}{\hbar}} U^{5/2} + c_1 \sqrt{\hbar m} \frac{(\vec\nabla U)^2}{U^\frac{1}{2}} + c_2 \sqrt{\frac{\hbar^3}{m}} \left[\left(\vec\nabla^2 \chi\right)^2 - 9 \frac{m}{\hbar} \vec\nabla^2 A_0(\vect{r})\right] U^\frac{1}{2},
\end{equation}
Corrections to the phonon dispersion relation, among others, could be computed from there. This was done in~\cite{Son:2005rv} when the external potential is set to zero. Instead, we quote here the result for the ground-state energy of the system trapped in a harmonic potential as a function of the total number of particles $Q$:
\begin{equation} \label{eq:trap-energy-Wilson-parameters}
	\frac{E(Q)}{\hbar\omega} = \frac{\sqrt{\xi}}{4} (3Q)^\frac{4}{3} - \sqrt{2}\pi^2 \xi \left(c_1 - \frac{9}{2}c_2\right) (3Q)^\frac{2}{3} + \order{Q^\frac{5}{9}}.
\end{equation}
This is the central result. While this \ac{eft} for a nonrelativistic conformal superfluid was simply organized as a derivative expansion, it now appears clearly that the controlling parameter is the number of particles in the system, assumed to be large from the beginning. Of course, this might have been anticipated from the discussion around Eq.~\eqref{eq:QuantumGasHierarchy} defining quantum gases. It will soon become clear that this could be regarded as the first modern application of the large-charge program. However, there is still a missing ingredient: theories constrained by conformal invariance are usually expressed in terms of their so-called \acs{cft} data. The result established above may or may not be related in a simple manner to a piece of \ac{nrcft} data. The answer to this question is surprising and will be given later.

At this point, we simply want to advocate the following philosophy: on their quest to collect pieces of \acs{cft} data, high-energy physicists may reasonably expect a semiclassical picture such as the one we have just discussed to emerge in extreme regimes where some quantum numbers take very large values. In the case of global internal charges---akin to the particle number $Q$ above---this semiclassical description would correspond to a phase of matter, of which the conformal superfluid is the most natural example. In fact, large \emph{classes} of \acp{cft} may feature the same emergent collective behaviour, and the corresponding statements about their \acs{cft} data would be highly generic. What is more, the whole spectrum of excitations---which for a conformal superfluid corresponds to phonons and the various vortex phases described earlier---may be conjectured to describe operators carrying even larger quantum numbers. The phenomenology of the unitary Fermi gas and nonrelativistic condensed matter systems in general may be regarded as the main source of inspiration for such a program. This is what we discuss next.
\pagebreak
\section{High-energy perspective: large-charge CFT data}
\label{sec:Intro_LargeQ_rel}

Perhaps the most central paradigm in modern theoretical physics is the abstract formulation of quantum field theories as points in an infinite space, related to one another by the action of the Renormalization Group, thereby forming \emph{flows} of theories. Under reasonable assumptions, these trajectories start and end on fixed points where the spacetime symmetries---i.e. the Poincaré group---of the theory get enhanced to the conformal group, thus becoming (relativistic) \acp{cft}.
Conformal field theories are immensely constrained---they are fully characterized by their so-called \ac{cft} data---yet difficult to solve; they have direct experimental relevance, most notably in condensed matter; and they have ramifications in almost all areas of modern theoretical physics, from statistical physics to quantum gravity. The task of understanding this class of theories is of such importance that the development of new specific techniques has been one of the main endeavors of the community in the past few decades. Examples include:
\begin{itemize} [itemsep=5pt,partopsep=0pt,topsep=0pt,parsep=0pt]
    \item Monte-Carlo simulations: this is a set of non-perturbative numerical techniques that relies on repeated random sampling. In the context of \acp{cft}, some of the most celebrated results concern the 3D Ising model, see~\cite{gupta1996critical,Ferrenberg_2018}.
    \item Fuzzy-sphere regularization~\cite{Zhu:2022gjc,Hu:2023xak,Han:2023yyb,Zhou:2023qfi,Hu:2023ghk,Hofmann:2023llr,Han:2023lky,Zhou:2023fqu,Hu:2024pen,Zhou:2024dbt,Dedushenko:2024nwi,Cuomo:2024psk}\footnote{To date (September 2024), this list of references is essentially exhaustive! This is to say how recent this topic is (in the context of conformal theories at least).}: this non-perturbative method is a new exciting way to efficiently extract \ac{cft} data numerically. Interestingly, it is implemented on the cylinder frame, $\mathbb{R} \times S_R^{d-1}$, which is, as we discuss next, the natural starting point for the large-charge program too. In $d>2$, the compact component of the cylinder geometry is curved, and it can therefore only embed a finite number of regular lattices, containing very few lattice points, making lattice simulations on the cylinder geometry not very efficient. Fuzzy spheres regularization is an alternative way to 'discretize' the sphere, which has proven very powerful in recent years.
    \item Analytic techniques: for instance, the usual perturbative small-$\epsilon$ and large-$N$ expansions, or exact methods such as integrability or supersymmetric localization.
\end{itemize}
Most other modern approaches exploit \ac{cft} axioms to encode theory-independent constraints on the \ac{cft} data. They broadly fall under the name \emph{bootstrap}. The modern version of the conformal bootstrap program was initiated in~\cite{Rattazzi:2008pe}, which aims to exploit the power of crossing symmetry, unitary and the \ac{ope} to systematically carve out the space of allowed \ac{cft} data, both numerically~\cite{Poland:2022qrs} and analytically~\cite{Hartman:2022zik}. One of the most successful instances of the analytic conformal bootstrap is the lightcone bootstrap~\cite{Fitzpatrick:2012yx,Komargodski:2012ek,Simmons-Duffin:2016wlq}, which investigates the consequences of a specific kinematic limit of the crossing equation, namely, the lightcone limit. One such consequence, valid for \emph{any} \ac{cft} in $d>2$, is that, given two primaries with twists\footnote{The twist is simply the difference between the conformal dimension $\Delta$ and the spin $J$: $\tau=\Delta-J$.} $\tau_1$ and $\tau_2$ and any $n\in\mathbb{N}$, there exists an infinite family of operators, dubbed doubled-twist operators, whose twist behaves as a function of the spin $J$ as
\begin{equation} \label{eq:LargeSpin}
    \lim_{J\to\infty} \tau(J) = \tau_1 + \tau_2 + 2n.
\end{equation}
Consequently, for a fixed large value of $J$, the lightest operator with such spin always has a conformal dimension with a Regge-like behaviour:
\begin{equation} \label{eq:LargeSpinRegge}
    \Delta_J = J + \ldots
\end{equation}
In the holographic context, this is associated with a simple semiclassical effective description in the bulk of AdS in terms of weakly interacting probe particles rotating far away from each other. Unsurprisingly, other approaches based on effective field theory were subsequently developed to study emergent universal features of \ac{cft} data, such as the \emph{large-charge expansion}~\cite{Hellerman:2015nra,Monin:2016jmo,Gaume:2020bmp} or the thermal effective action \cite{Benjamin:2023qsc,Benjamin:2024kdg}. These methods differ from other bootstrap techniques as their primary goal is not to place sharp bounds on the \ac{cft} data, but rather identify common patterns across large classes of \acp{cft}.

Indeed, as an \ac{eft}-based approach, much of the excitement about the large-charge expansion stems from its ability to deduce structural properties of \acp{cft} with global symmetries without referring to a specific model, thereby classifying them into \emph{large-charge universality classes}, the most prominent one being associated with a conformal superfluid \ac{eft} in the cylinder frame, which realizes the simplest example of a \emph{macroscopic limit} \cite{Jafferis:2017zna}, as we shall review. Of course, this \ac{eft} closely resembles the one discussed in the case of the unitary Fermi gas, so the phenomenology of the latter is expected to be an invaluable guiding principle for the exploration of the various phases with both macroscopic charge and macroscopic spin.

This is what we want to establish in this section. We proceed pedantically by reviewing some basic but important notions of conformal field theory and then moving on to a discussion of (relativistic) large-charge universality classes. In the next section, we mimic this structure to present the state of the art in the case of \emph{nonrelativistic} \acp{nrcft}, which is the focus of this thesis.

\subsection{Harnessing CFT data from the cylinder frame}

We therefore start with a discussion of correlation functions in conformal field theories, assuming prior basic knowledge. For conventions, we refer to~\cite{Rychkov:2016iqz}. Any local unitary conformal field theory in flat space---by which we mean Euclidean spacetime $\mathbb{R}^d$---has a traceless stress tensor, $T_\mu^\mu=0$, which implies that it is invariant under Weyl rescaling of its metric: $\dd{s}^2 = \Omega(x)^2 \dd{s}^2_{\mathbb{R}^d}$. This maps the flat-space theory to an equivalent theory on a possibly curved manifold $\mathcal{M}_d$, which is said to be conformally flat. This is a powerful statement. Indeed, while \acp{cft} are commonly regarded as theories without any dimensionful parameter, $\mathcal{M}_d$ can actually have compact directions, so it is in fact more illuminating to state that \emph{conformal field theories can be expressed in terms of an arbitrary length scale $R$ in various ways (associated with various conformally flat manifolds), although any physical observable is, crucially, independent of $R$.} In situations where another dimensionful quantity comes into play, like a charge density, one can employ dimensional analysis to full power. We will establish this result more precisely in what follows.

But first, we recall that a general result valid for any conformally flat manifold $\mathcal{M}_d$, with metric $\dd{s}^2 = \Omega_{\mathcal{M}_d}(x)^2 \dd{s}_{\mathbb{R}^d}^2$, is the following. Insertion of a local scalar primary operator $\mathcal{O}$ with conformal dimension $\Delta$ in flat space gets mapped to an insertion of the operator
\begin{equation}
    \mathcal{O}_{\mathcal{M}_d}(\theta) := \Omega_{\mathcal{M}_d}(x(\theta))^{-\Delta} \mathcal{O}(x(\theta))
\end{equation}
on $\mathcal{M}_d$, where $\theta$ denotes the coordinate on $\mathcal{M}_d$ related to $x(\theta)$ by the Weyl mapping. Therefore, correlation functions of these new operators are related to those in flat space as
\begin{equation}
    \frac{\ev{\mathcal{O}_{\mathcal{M}_d}(\theta_n) \ldots \mathcal{O}_{\mathcal{M}_d}(\theta_1)}_{\mathcal{M}_d}}{\ev{1}_{\mathcal{M}_d}} = \left(\prod_{i=1}^n \Omega_{\mathcal{M}_d}(x(\theta_i))^{-\Delta_i}\right)
    \frac{\ev{\mathcal{O}(x(\theta_n)) \ldots \mathcal{O}(x(\theta_1))}_{\mathbb{R}^d}}{\ev{1}_{\mathbb{R}^d}}.
\end{equation}
The denominators account for a possible Weyl anomaly on $\mathcal{M}_d$, which is known to occur in even dimensions. We will however ignore this issue for now and drop these terms.

As alluded to earlier, there is a specific choice of conformally flat manifold $\mathcal{M}_d$ that will be of central interest for us: the cylinder $\mathbb{R} \times S_R^{d-1}$. But let us first start in flat space and derive later the consequences on that particular manifold. In particular, the two-point function reads
\begin{equation} \label{eq:CFT_2pt}
\begin{aligned}
    \ev{\mathcal{O}(x_2) \mathcal{O}(x_1)}_{\mathbb{R}^d}
    & = \frac{1}{\abs{x_2-x_1}^{2\Delta}} \\
    & = \frac{1}{\abs{x_2}^{2\Delta}} \left[ 1 - 2\frac{x_1\cdot x_2}{x_2^2} + \abs{\frac{x_1}{x_2}}^2 \right]^{-\Delta} \\
    & = \frac{1}{\abs{x_2}^{2\Delta}} \sum_{n=0}^\infty C_n^{(\Delta)}\left(\frac{x_1\cdot x_2}{\abs{x_1}\abs{x_2}}\right) \abs{\frac{x_1}{x_2}}^n,
\end{aligned}
\end{equation}
where the Gegenbauer polynomial $C_n^{(\Delta)}(z)$ is a specific hypergeometric function:
\begin{equation}
\begin{aligned}
    C_n^{(\Delta)}(z) = \pFq{2}{1}{-n,2\Delta+n}{\Delta+\frac{1}{2}}{\frac{1-z}{2}}.
\end{aligned}
\end{equation}
This motivates the usual prescription for sending an insertion to infinity, namely,
\begin{equation}
    \mathcal{O}(\infty) := \lim_{x_2\to\infty} \abs{x_2}^{2\Delta} \mathcal{O}(x_2),
\end{equation}
such that $\ev{\mathcal{O}(\infty) \mathcal{O}(0)}_{\mathbb{R}^d} = 1$. This is however a bit abrupt, as this specific configuration does not seem to capture any piece of \ac{cft} data. We are going to show that, on the cylinder, the corresponding configuration with the limit taken appropriately actually computes the conformal dimension $\Delta$. Moreover, the series in Eq.~\eqref{eq:CFT_2pt} can be seen to resum the contributions of the descendants, in a sense that we shall make precise.

In order to do so, we need to pick a specific choice of quantization. This is a partly arbitrary prescription which amounts to foliating spacetime in a way compatible with the spacetime symmetries of the system. The canonical choice is \emph{equal-time} quantization, where parallel codimension one surfaces foliate spacetime, and the Hamiltonian implements evolution from one such slice to another. \acp{cft} are instead most conveniently quantized with respect to the dilatation operator or any other operator in its conjugacy class. Once a given quantization prescription has been chosen, it becomes straightforward to establish a fundamental property of \acp{cft} that we shall invoke extensively, namely, the state-operator correspondence. In what follows, we shall discuss two such prescriptions.

\subsubsection{Radial quantization and the state-operator correspondence}

We choose to quantize the flat-space theory with respect to the dilatation operator $D$ and choose an arbitrary length scale $R$, usually set to one. This is the so-called radial quantization, in which an insertion of a local primary operator at the origin prepares a state living on the radial slice with zero radius, \emph{cf.} Figure~\ref{fig:RadQuant}. The state is thus defined as
\begin{equation}
    \ket{\mathcal{O}} := R^\Delta\, \mathcal{O}(0) \ket{0} = R^\Delta \expo{-iP\cdot x} \mathcal{O}(x) \ket{0} .
\end{equation}
We used\footnote{We always think in terms of Euclidean time $\tau:=ix^0$, and we work with the mostly minus signature.} $\mathcal{O}(x) = \expo{iP\cdot x} \mathcal{O}(0) \expo{-iP\cdot x}$ and the fact that the vacuum $\ket{0}$ is invariant under any conformal transformation. Moreover, every choice of quantization comes with a corresponding rule for hermitian conjugation. In radial quantization, we have\footnote{We define the reflection (or inversion) $\mathcal{R}x := R^2\frac{x^\mu}{x^2}$, and we use $r=\abs{x}$.} $\mathcal{O}(x)^\dagger = \left(\frac{R}{r}\right)^{2\Delta} \mathcal{O}\left(\mathcal{R}x\right)$ and $P^\dagger_\mu = K_\mu$, in particular. Therefore,
\begin{equation}
    \bra{\mathcal{O}} = \bra{0} \mathcal{O}(\infty) R^{-\Delta} = \bra{0} \mathcal{O}(x)^\dagger \expo{iK\cdot x} R^\Delta .
\end{equation}
This construction also implies that the insertion of an operator (primary or not) anywhere in $\mathbb{R}^d$ at a distance $r$ from the origin prepares a state on the radial slice of radius $r$. This state, which is generically a linear combination of primary states and their descendants, evolves 'backward' or 'forward' with respect to $D$. Conversely, a state prepared on an arbitrary slice may be evolved back to the slice of zero radius, which is equivalent to a (generically non-primary) operator insertion at the origin. Of course, this reasoning can also be used to insert an operator at infinity. This establishes the state-operator correspondence in radial quantization.

Using spherical coordinates $x=\abs{x}\hat{n}$ and parametrizing $\abs{x}=R\expo{\frac{\tau}{R}}$, the insertion of $\mathcal{O}\left(R\expo{\frac{\tau_1}{R}}\hat{n}_1\right)$ on the 'far right' of a (radially ordered) correlation function acts as (we multiply by $R^\Delta$ for convenience)
\begin{equation} \label{eq:OflatKet}
\begin{aligned}
    R^\Delta \mathcal{O}\left(R\expo{\frac{\tau_1}{R}}\hat{n}_1\right)\ket{0}
    & = \expo{i(P\cdot\hat{n}_1)R\expo{\frac{\tau_1}{R}}} \ket{\mathcal{O}} \\
    & = \expo{-\frac{\tau_1}{R}\Delta} \sum_{n=0}^\infty \frac{i^n}{n!} \expo{\frac{\tau_1}{R}D} \ket{n,\hat{n}_1},
\end{aligned}
\end{equation}
where we introduced the linear combination of level-$n$ descendants $\ket{n,\hat{n}_1} := R^n (P\cdot\hat{n}_1)^n\ket{\mathcal{O}}$ with conformal dimension $\Delta+n$, and we purposely extracted an exponential term in front of the sum in the last line. Similarly, insertion of $\mathcal{O}$ on the far left acts as
\begin{equation}
\begin{aligned}
    \bra{0} \mathcal{O}\left(R\expo{\frac{-\tau_2}{R}}\hat{n}_2\right)^\dagger R^\Delta \expo{-\frac{\tau_2}{R}2\Delta}
    = \expo{-\frac{\tau_2}{R}\Delta} \sum_{n=0}^\infty \frac{(-i)^n}{n!} \bra{n,\hat{n}_2} \expo{-\frac{\tau_2}{R}D},
\end{aligned}
\end{equation}
where $\bra{n,\hat{n}}=\bra{\mathcal{O}}(K\cdot\hat{n}_2)^nR^n$. For instance, the flat-space two-point function becomes
\begin{equation} \label{eq:TwoPointFunction_expanded}
    \ev{\mathcal{O}\left(R\expo{\frac{\tau_2}{R}}\hat{n}_2\right) \mathcal{O}\left(R\expo{\frac{\tau_1}{R}}\hat{n}_1\right)}_{\mathbb{R}^d} = \left(R^2\expo{\frac{\tau_1+\tau_2}{R}}\right)^{-\Delta} \sum_{n=0}^\infty \frac{1}{(n!)^2} \mel{n,\hat{n}_2}{\expo{-\frac{\tau_2-\tau_1}{R}D}}{n,\hat{n}_1}.
\end{equation}
It is an interesting exercise---and a non-trivial check of this construction---to verify that
\begin{equation}
    \frac{1}{(n!)^2} \braket{n,\hat{n}_2}{n,\hat{n}_1} = C_n^{(\Delta)}(\hat{n}_1\cdot\hat{n}_2),
\end{equation}
so as to match the series expansion in Eq.~\eqref{eq:CFT_2pt}. We went through this simple exercise to illustrate the fact that the dilatation operator really is the evolution operator in radial quantization, and the insertions of $\mathcal{O}(x_2)$ and $\mathcal{O}(x_1)$ correspond to states $\bra{\mathcal{O},x_2}$ and $\ket{\mathcal{O},x_1}$ living on \emph{different} slices. They thus need to be evolved backward or forward to evaluate their overlap, which is what the operator $\expo{-\frac{\tau_2-\tau_1}{R}D}$ does.

\subsubsection{Conformal dimensions from the cylinder frame}

As mentioned, we are interested in mapping the theory on the cylinder $\mathbb{R} \times S_R^{d-1}$. The Weyl map that achieves this is a local rescaling of the radial slices to spheres of fixed radius $R$, which get stacked up on top of each other in the $\mathbb{R}$ direction, see Figure~\ref{fig:CylinderQuant}. The slices with zero and infinite radius get mapped to the 'extremities' of this cylinder. The evolution operator in radial quantization generates translations along the non-compact direction, which is therefore identified with the action of the Hamiltonian of the theory on the cylinder. Coordinates on the cylinder will be denoted by $(\tau,\hat{n})$, which corresponds to the point $R\expo{\frac{\tau}{R}} \hat{n} \in \mathbb{R}^d$. Conversely, $(\tau,\hat{n})=\left(R\log\frac{\abs{x}}{R},\frac{x^\mu}{\abs{x}}\right)$, and the Weyl factor is therefore
\begin{equation}
    \Omega_{cyl.}(x) = \frac{R}{\abs{x}} = \expo{-\frac{\tau}{R}}.
\end{equation}
Correspondingly, $\mathcal{O}_{cyl.}(\tau,\hat{n}) = \expo{\Delta\frac{\tau}{R}} \mathcal{O}\left(R\expo{\frac{\tau}{R}} \hat{n}\right)$.

We now look at the fate of Eq.~\eqref{eq:OflatKet} in terms of cylinder operators. The exponential that we purposely extracted out is simply the Weyl factor, such that
\begin{equation}
    R^\Delta \mathcal{O}_{cyl}(\tau_1, \hat{n}_1)\ket{0}
    = \sum_{n=0}^\infty \frac{i^n}{n!} \expo{\frac{\tau_1}{R}D} \ket{n,\hat{n}_1}
    \quad\stackrel{\tau_1\to-\infty}{\longrightarrow}\quad \expo{\frac{\tau_1}{R}\Delta} \ket{\mathcal{O}} =: \ket{\mathcal{O},in},
\end{equation}
and similarly for the hermitian conjugate, defining $\bra{\mathcal{O},out}$. We then have
\begin{equation}
\begin{aligned}
    \ev{\mathcal{O}_{cyl.}(\tau_2,\hat{n}_2) \mathcal{O}_{cyl.}(\tau_1,\hat{n}_1)}_{cyl.}
    & = \frac{1}{R^{2\Delta}} \sum_{n=0}^\infty \frac{1}{(n!)^2} \mel{n,\hat{n}_2}{\expo{-(\tau_2-\tau_1)H_{cyl.}}}{n,\hat{n}_1},
\end{aligned}
\end{equation}
where we identified the dilatation operator $\frac{D}{R}$ acting in flat space with the cylinder Hamiltonian. In particular, the spectrum of $D$ is seen to be completely equivalent to the energy spectrum of the cylinder theory via the relation $\Delta = R\,E$. Upon sending the insertions to $\tau_1\to-\infty$ and $\tau_2\to\infty$, we obtain
\begin{equation} \label{eq:DeltaFromCylinder}
    \braket{\mathcal{O},out}{\mathcal{O},in} = \lim_{T\to\infty}
    \ev{\expo{-T H_{cyl.}}}{\mathcal{O}} = \expo{-\frac{T}{R}\Delta},
\end{equation}
Of course, strictly speaking, this gives 0 which, up to rescaling, is equivalent to the flat-space two-point function with an insertion at the origin and another one at infinity. The more precise way to express the result is instead
\begin{equation} \label{eq:DeltaFromCylinderBis}
\boxed{
    \Delta = -\lim_{T\to\infty} \frac{R}{T} \log \ev{\expo{-T H_{cyl.}}}{\mathcal{O}}
}
\end{equation}
This amounts to computing the partition function---or rather, the free energy---of the system on the cylinder prepared in a given state $\mathcal{O}$. In fact, in practice, it is often convenient to consider the thermal cylinder theory, that is, the the theory on $S_\beta^1 \times S_R^{d-1}$ with inverse temperature $\beta$, in which case the zero-temperature limit, $\beta\to\infty$, projects the constrained partition function onto the ground state, just as the large-separation limit did:
\begin{equation}
    \Delta = -\lim_{\beta\to\infty} \frac{R}{\beta} \log \ev{\expo{-\beta H_{cyl.}}}{\mathcal{O}}.
\end{equation}
In situations where the explicit computation of $\Delta$ is hard, this suggests an alternative avenue where one needs to analytically harness the corresponding path integral. Echoing the discussion at the beginning of the chapter, one may hope to find configurations where this can be done \emph{semiclassically}. When this can indeed be achieved, the computation of correlators of the form
\begin{equation}
    \mel{\mathcal{O}',out}{\mathcal{O}_{\delta_n}(\tau_n,\hat{n}_n) \ldots \mathcal{O}_{\delta_1}(\tau_1,\hat{n}_1)}{\mathcal{O},in}
\end{equation}
and their corresponding \ac{cft} data will be controlled by the \emph{same} semiclassical trajectory. Here, the cylinder operators\footnote{We omit the \emph{cyl.} subscript for convenience.} $\mathcal{O}_{\delta_i}$ are light operators and $n$ is not too big. The \emph{in} and \emph{out} states may be different so as to make the quantum numbers match in the correlator, but both should give rise to the same semiclassical description. A natural scenario is when they carry some large quantum numbers. In this case, Eq.~\eqref{eq:DeltaFromCylinderBis} suggests that \emph{one does not even need to specify explicitly} what the underlying operator $\mathcal{O}$ is. Indeed, given a non-exhaustive set of quantum numbers, the conformal dimension that this formula computes will be the one associated with the operator carrying these quantum numbers that minimizes the energy on the cylinder. This far-reaching observation indicates the possibility to make \emph{theory-independent} statements, potentially valid across large classes of theories. Before discussing this in details, we shall quickly comment on a different, albeit equivalent, choice of quantization.

\subsubsection{NS-quantization and the state-operator correspondence}

It is illuminating to show the equivalence between radial quantization and another choice of quantization. In fact, there are good reasons to consider the one that we shall discuss here, commonly dubbed \emph{NS-quantization}~\cite{Rychkov:2016iqz}. Our motivation to introduce it here is that it is the natural starting point to quantize \acp{nrcft}, as we shall discuss in the next section.

Let $g$ be an element of the conformal group, and let us denote the generators of the conformal group (rotations, translations, dilatations and \acp{sct}) by $J_a$. Acting with $\text{Ad}_g$ on each of them, \emph{viz.}
\begin{equation}
    \bar{J}_a := g J_a g^{-1},
\end{equation}
preserves the commutation relations, that is, it defines an automorphism of the conformal algebra. Now, let us choose $g = \expo{-i\frac{\pi}{4}(RP_0+\frac{K_0}{R})}$, in which case
\begin{equation} \label{eq:LuscherMach_hamil}
    \bar{D} = \frac{1}{2} \left( R P_0 - \frac{K_0}{R} \right).
\end{equation}
This is commonly known as the Lüscher-Mack Hamiltonian~\cite{luscher1975global}. Quantization with respect to $\bar{D}$ is called NS-quantization.

\begin{figure} [!ht]
\centering
\begin{subfigure}{0.3\textwidth}
    \includegraphics[height=4cm]{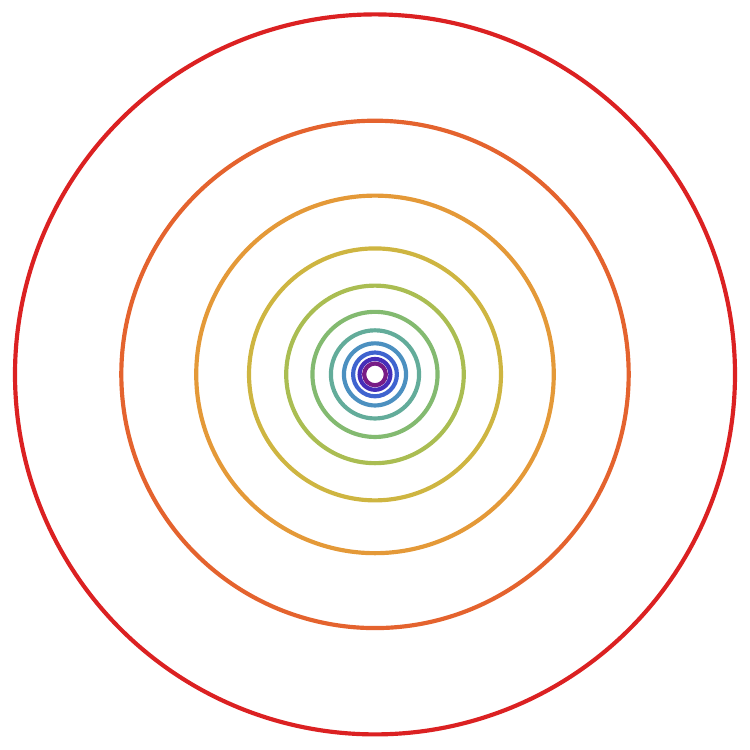}
    \caption{Radial quantization on $\mathbb{R}^d$.}
    \label{fig:RadQuant}
\end{subfigure}
\hfill
\begin{subfigure}{0.3\textwidth}
    \includegraphics[height=7cm]{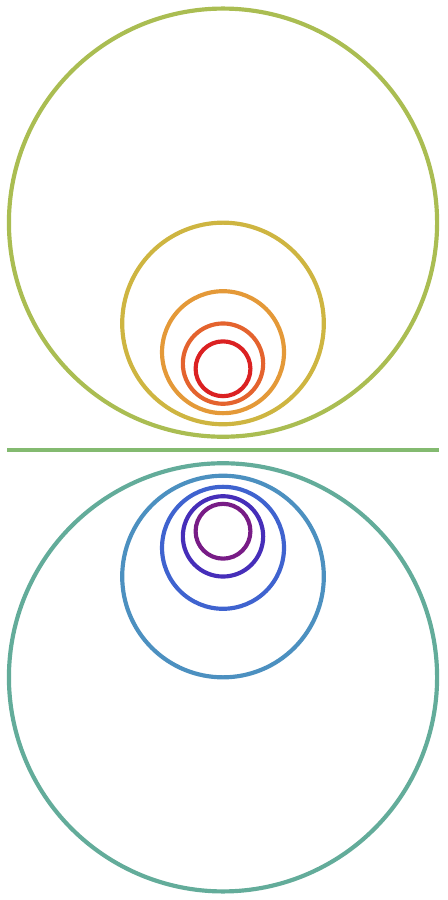}
    \caption{NS-quantization on $\mathbb{R}^d$.}
    \label{fig:NSQuant}
\end{subfigure}
\hfill
\begin{subfigure}{0.3\textwidth}
    \includegraphics[height=7cm]{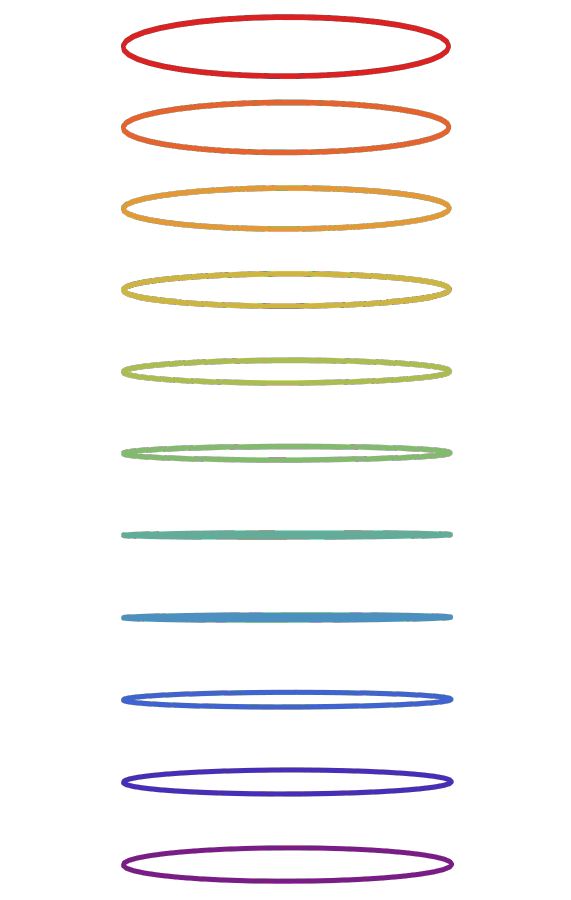}
    \caption{Quantization on $\mathbb{R}\times S_R^{d-1}$.}
    \label{fig:CylinderQuant}
\end{subfigure}

\caption{Quantization on $\mathbb{R}^d$ can be done with respect to any element of the conformal algebra, seen as a conformal Killing vector. The most natural choice is to take $D$, in which case $\mathbb{R}^d$ gets foliated into radial slices centered at the origin, see (a). States living on a slice of radius $R\expo{\frac{\tau_1}{R}}$ can be evolved into states living on a slice of radius $R\expo{\frac{\tau_2}{R}}$ by means of the evolution operator $\expo{-\frac{\tau_2-\tau_1}{R}D}$. However, any other operator in the conjugacy class of $D$ is an equally good choice, \emph{e.g.} the Lüscher-Mack Hamiltonian $\frac{1}{2}(RP_0-K_0/R)$, thereby slicing $\mathbb{R}^d$ as shown in (b). In this case, the slices contract around the points $(\pm R, \vec{0})$. Inserting a local primary operator $\bar{\mathcal{O}}$ at one of these points prepares the corresponding state $\ket{\mathcal{O}}$ (or $\bra{\mathcal{O}}$). Conversely, shrinking a state corresponds to a local operator insertion. This is the state-operator correspondence in NS-quantization. Any other choice of quantization would also have two contraction points (one being possibly at infinity), and insertions of operators at these points establish the state-operator correspondence. On the cylinder, see (c), there is a canonical choice of quantization where one chooses the generator of translations along the non-compact direction as the evolution operator, which naturally gets interpreted as the Hamiltonian on the cylinder. We insist on the fact that this foliation of the cylinder is equivalent to any conformal foliation of $\mathbb{R}^d$, not just the radial one. The colours of the slices indicate how they get mapped across these three frames.}
\label{fig:Quant}
\end{figure}

Going from radial quantization to the new one simply amounts to going from $x\in(\mathbb{R}^d)_{rad.}$ to $y\in(\mathbb{R}^d)_{NS}$ as dictated by the action of $g$:
\begin{equation}
   y^\mu(x) = \left(R\frac{\abs{x}^2-R^2}{R^2+2Rx^0+\abs{x}^2} \, , \, R\frac{2R x^i}{R^2+2Rx^0+\abs{x}^2}\right).
\end{equation}
Conversely,
\begin{equation}
   x^\mu(y) = \left(R\frac{R^2-\abs{y}^2}{R^2-2Ry^0+\abs{y}^2} \, , \, R\frac{2R y^i}{R^2-2Ry^0+\abs{y}^2}\right).
\end{equation}
The points $x=0$, respectively $x=\infty$, get mapped to $y^\mu=(-R,\vec{0})$, respectively $y^\mu=(R,\vec{0})$. Seen as a conformal Killing vector, $\bar{D}$ looks like a magnetic field joining the two poles of a magnet, hence the name (NS stands for North-South). Spherical slices perpendicular to this field correspond to the new foliation of spacetime, see Figure~\ref{fig:NSQuant}. This particular choice of quantization puts the time coordinate $y^0$ on a slightly different footing, which is appropriate \emph{e.g.} for analytic continuation to Lorentzian signature~\cite{Kravchuk:2018htv}. As mentioned, this will also be the ideal starting point for the quantization of \acp{nrcft}.

Primary operators are defined as $\bar{\mathcal{O}}(y)=g \mathcal{O}(x(y)) g^{-1}$ (with $g$ in field representation), such that they satisfy for instance $[\bar{D},\bar{\mathcal{O}}(-R,\vec{0})] = -i\Delta\bar{\mathcal{O}}(-R,\vec{0})$. The relationship between $\mathcal{O}$ and $\bar{\mathcal{O}}$ is therefore\footnote{In fact, we absorb a factor of $2^\Delta$ in the definition of $\bar{\mathcal{O}}$.}
\begin{equation}
    \mathcal{O}(x) = \left(\frac{R^2}{R^2+2Rx^0+\abs{x}^2}\right)^\Delta \bar{\mathcal{O}}(y(x)).
\end{equation}
In particular, $\mathcal{O}(0) = \bar{\mathcal{O}}(-R,\vec{0})$
, such that
\begin{equation} \label{eq:StateOpMap_NS}
    \ket{\mathcal{O}}
    = R^\Delta \bar{\mathcal{O}}(-R,\vec{0}) \ket{0}
    = R^\Delta \expo{-R P_0} \bar{\mathcal{O}}(0) \ket{0}.
\end{equation}
Hermitian conjugation in this quantization simply amounts to sending $y^0 \to -y^0$, such that
\begin{equation}
    \bra{\mathcal{O}}
    = \bra{0} \bar{\mathcal{O}}(R,\vec{0}) R^\Delta
    = \bra{0} \bar{\mathcal{O}}(0) \expo{-RP_0} R^\Delta.
\end{equation}
The state-operator correspondence in NS-quantization is therefore established by inserting local operators at the points $(\pm R, \vec{0})$ and evolving them using $\expo{-\frac{\tau}{R}\bar{D}}$. The fact that this is indeed the correct evolution operator could be seen by writing the equivalent of Eq.~\eqref{eq:TwoPointFunction_expanded} for NS-quantization. Conversely, states prepared on arbitrary slices may be evolved to either of the zero-radius slices, corresponding to local operator insertions at $(\pm R, \vec{0})$.

The foliation with respect to $\bar{D}$ can also be mapped to the cylinder just as was done in radial quantization. This is most conveniently expressed from the cylinder coordinate $(\tau,\hat{n})$ to the flat-space one $y^\mu \in (\mathbb{R}^d)_{NS}$, which gives
\begin{equation} \label{eq:CylinderToNS_coord}
    y^\mu(\tau,\hat{n}) = \left(R\frac{\sinh\frac{\tau}{R}}{\cosh\frac{\tau}{R}+n^0} \, , \, R\frac{n^i}{\cosh\frac{\tau}{R}+n^0} \right),
\end{equation}
so the Weyl factor is given by $\bar\Omega_{cyl.}(\tau,\hat{n}) = \cosh\frac{\tau}{R}+n^0$, and primary operators on the cylinder are related to those in flat-space as
\begin{equation}
    \mathcal{O}_{cyl.}(\tau,\hat{n}) = 2^\Delta \left(\cosh\frac{\tau}{R}+n^0\right)^{-\Delta} \bar{\mathcal{O}} \left(R\frac{\sinh\frac{\tau}{R}}{\cosh\frac{\tau}{R}+n^0} \, , \, R\frac{n^i}{\cosh\frac{\tau}{R}+n^0} \right).
\end{equation}

As we shall see later, quantization of \acp{nrcft} is similar to the relativistic case discussed here with $d=1$\footnote{See \emph{e.g.}~\cite{Sen:2011cn} for a more detailed discussion of the $d=1$ case.}, in which case the cylinder is just a line, with coordinate that we choose to denote by $\tilde\tau:=\frac{\tau}{2}$ instead of $\tau$. Moreover, let us denote the coordinate $y^0$ on $(\mathbb{R})_{NS}$ by $\bar\tau$ instead, and define $\omega:=\frac{1}{R}$. Then, Eq.~\eqref{eq:CylinderToNS_coord} becomes
\begin{equation} \label{eq:NStoCylinder1D}
    \omega\bar\tau = \tanh\omega\tilde\tau,
\end{equation}
and
\begin{equation} \label{eq:Operators_FromNStoCylinder1D}
\begin{cases}
    \bar{\mathcal{O}}(\bar\tau) & = (\cosh\omega\tilde\tau)^{2\Delta} \, \mathcal{O}_{cyl.}(\tilde\tau), \\
    \mathcal{O}_{cyl.}(\tilde\tau) & = \left(1-(\omega\bar\tau)^2\right)^\Delta \, \bar{\mathcal{O}}(\bar\tau).
\end{cases}
\end{equation}
Accordingly, the cylinder two-point function simplifies to
\begin{equation}
    \ev{\mathcal{O}_{cyl.}(\tilde\tau_2) \mathcal{O}_{cyl.}(\tilde\tau_1)}_{cyl.} = \frac{\omega^{2\Delta}}{\sinh(\omega(\tilde\tau_2-\tilde\tau_1))^{2\Delta}},
\end{equation}
which behaves like $\sim \expo{-\frac{\tau_2-\tau_1}{R}\Delta}$ when $(\tau_2-\tau_1)\to\infty$, to be compared to Eq.~\eqref{eq:DeltaFromCylinder}.

Although we restricted here to the $d=1$ case for convenience and for its relevance to the nonrelativistic discussion, this illustrates the complete equivalence between radial quantization, NS-quantization and quantization on the cylinder, which of course holds in general dimension.

\subsection{Towards large-charge universality classes: the macroscopic limit}

We mentioned earlier that the limit of large separation of external insertions in the cylinder frame is suggestive of a simple computational algorithm to extract \ac{cft} data. Indeed, the computation of $\Delta$ and other pieces of \ac{cft} data boils down to a path-integral computation on the cylinder which, supposedly, acquires a semiclassical description in some 'extreme' regime. Let us be more precise.

Consider the state $\ket{\mathcal{O}} =: \ket{\vec{Q},a}$ carrying some quantum numbers $\vec{Q}$ (the Cartan charges) with respect to a global symmetry $G$, as well as other quantum numbers (like conformal dimension and spin) denoted by $a$. It is generically possible to have several primary states carrying the same Cartan charges $\vec{Q}$ but different $a$. Nonetheless, for large values of the Cartan charges, the corresponding partition function is dominated by semiclassical configurations associated with some specific spontaneous symmetry breaking pattern~\cite{Monin:2016jmo}\footnote{Bear in mind that no spontaneous symmetry breaking can occur on the cylinder because the spatial slice has finite volume. This is therefore a statement about the phase that the system would enter into upon sending the radius of the cylinder to infinity, thereby recovering flat space. We shall discuss this in detail shortly.}. In the large-separation limit on the cylinder, of all operators having the same $\vec{Q}$, only the one associated with the lowest-action trajectory survives.\footnote{In fact, if the operator is not a singlet under the global symmetry, these states may be degenerate.} For a given set of Cartan charges $\vec{Q}$, such an operator has the lowest conformal dimension. We shall call this operator the \emph{large-charge operator} $\mathcal{O}_{\vec{Q}}$, with conformal dimension $\Delta_{\vec{Q}}$. Insertions of the large-charge operator at $0$ and $\infty$ in radial quantization respectively break translations $P_\mu$ and \acp{sct} $K_\mu$. Rotations may or may not be preserved, depending on whether $\mathcal{O}_{\vec{Q}}$ carries some spin and/or is not in a symmetric traceless representation of $G$. Similarly, dilatations and the global internal symmetry $G$ may or may not be spontaneously broken.

In this thesis, we shall be mostly concerned with the simple case of a global $U(1)$ symmetry, although part of the general discussion that we present next may easily be extended to non-abelian global symmetries.

\subsubsection{Macroscopic limit}

In essence, the large-charge \ac{cft} data, including $\Delta_Q$, can be probed \emph{at leading order} in $Q$ upon taking $Q\to\infty$ in an appropriate fashion on the cylinder, which is called the \emph{macroscopic limit}~\cite{Jafferis:2017zna,Komargodski:2021zzy,Cuomo:2024fuy}. The large-charge program is therefore concerned with the operators of a given \ac{cft} that yield a non-trivial macroscopic limit, as defined below. This can be phrased in a 'kinematic' way---similar in spirit to the lightcone bootstrap---with no reference to the underlying breaking pattern. It thus allows to identify generic features of all possible large-charge behaviours. Of course, various breaking patterns may give rise to the same macroscopic limit, so one then needs to refine this analysis by making explicit assumptions about the large-charge state.

For simplicity, consider a \ac{cft} (in $d>2$) with a global $U(1)$ symmetry, and let $\left\{\ket{Q}\right\}_Q$ denote the set of large-charge states---that is, the states with the lowest energy on the cylinder for a given $Q\gg1$---labelled by the allowed (quantized\footnote{In fact, the argument requires to treat $Q$ as a continuous parameter, thereby assuming some notion of analyticity of the dependence of these large-charge families on the charge $Q$. The equivalent statement for the spin is a well-established fact in the \ac{cft} literature \cite{Caron-Huot:2017vep}. Here, we have to make the assumption that analyticity is valid at least at leading order in $Q$.}) values of $Q$. We then choose a fixed set of light operators $\mathcal{O}_{\delta_i,q_i}$ (their conformal dimensions $\delta_i$ and charges $q_i$ are of order one), and define a \emph{large-charge family} of correlators as the set of correlators
\begin{equation} \label{eq:MacroLimit_correl}
    \mel{Q+\sum_i \delta_i,out}{\mathcal{O}_{\delta_n,q_n}(\tau_n,\hat{n}_n) \ldots \mathcal{O}_{\delta_1,q_1}(\tau_1,\hat{n}_1)}{Q,in}
\end{equation}
for all allowed values of $Q$. The central assumption is that there must exist a set of light operators and a constant $\gamma\in\mathbb{R}_+$ such that, as we take the limit
\begin{itemize} [itemsep=5pt,partopsep=0pt,topsep=0pt,parsep=0pt]
    \item $Q\to\infty$, $R\to\infty$ with $\nu:=\frac{Q^\frac{1}{\gamma}}{R}$ kept fixed, as well as
    \item the relative distances in flat-space $\abs{x(\tau_i,\hat{n}_i)-x(\tau_j,\hat{n}_j)}$ kept fixed,
\end{itemize}
the resulting normalized flat-space correlator
\begin{equation}
    \nu^{-(\delta_1+\ldots+\delta_n)} \times \mel{flat}{\mathcal{O}_{\delta_n,q_n}(x(\tau_n,\hat{n}_n)) \ldots \mathcal{O}_{\delta_1,q_1}(x(\tau_1,\hat{n}_1))}{flat}
\end{equation}
evaluated in the resulting non-trivial flat-space state $\ket{flat}$ associated with the scale $\nu$, is finite and non zero. This is what we call a non-trivial macroscopic limit. Taking the reverse logic, namely, assuming the existence of a non-trivial macroscopic limit without making assumptions about the nature of $\ket{Q}$ (or $\ket{flat}$) allows to extract the behaviour of large-charge data at leading order in $Q$ as a function of $\gamma$. For instance, large-charge \ac{ope} coefficients must scale like
\begin{equation}
    \lambda_{Q,\mathcal{O}_{\delta,q},Q+q} \sim Q^\frac{\delta}{\gamma}
\end{equation}
at leading order in $Q$. The fate of higher-point functions may be analysed in a similar fashion, where the corresponding cross-ratios become functions of $Q$ via the second condition above (fixed relative distances). The existence of a macroscopic limit corresponds to the requirement that on a patch of the cylinder whose typical size is much smaller than $R$, the dynamics is well approximated by the flat-space state. This can also be phrased in terms of the original flat-space theory---with no reference to the cylinder frame---by saying that, in the presence of large-charge insertions at the origin and infinity, the correlators of light operator insertions far away from the large-charge insertions may be effectively computed as if they were in the state $\ket{flat}$.

\subsubsection{A simple case}

However, determining the value of $\gamma$ is generically very hard. A particularly natural assumption that \emph{implies} the existence of a macroscopic limit for a precise value of $\gamma$ is the following. Suppose that the large-charge operator $\mathcal{O}_Q$ does not carry a spin that grows with $Q$, implying that the corresponding state on the cylinder is approximately homogeneous, and suppose that $\Delta_Q \sim Q^\alpha$ for large enough $Q$. We further assume that the energy density and the charge density on the cylinder, which are of the form
\begin{equation} \label{eq:MacroLim_Simple}
\begin{aligned}
    \text{energy density } & \sim \frac{Q^\alpha}{R^d}, \\
    \text{charge density } & \sim \frac{Q}{R^{d-1}},
\end{aligned}
\end{equation}
remain finite in the limit $Q,R\to\infty$. This corresponds to the relation
\begin{equation}
    \text{energy density} \sim Q^{\alpha-\frac{d}{d-1}} \times \text{(charge density)}^\frac{d}{d-1}.
\end{equation}

For $\alpha>\frac{d}{d-1}$, the charge density vanishes, suggesting that the corresponding flat-space state is a generic heavy state, not a large-charge one. The limit where the energy density is kept fixed while $R\to\infty$ is known as the \emph{thermodynamic limit} \cite{Lashkari:2016vgj}. A general expectation, known as the eigenstate thermalization hypothesis (ETH)~\cite{Srednicki:1994mfb,DAlessio:2015qtq}, is that correlators with heavy external insertions thermalize at late times. It relies upon some 'genericity' and 'typicality' assumptions, valid for most heavy states. However, large-charge states are very atypical in this sense, so these really are different limits, and the case $\alpha > \frac{d}{d-1}$ does not yield a macroscopic limit. 

For $\alpha<\frac{d}{d-1}$, either the charge density is infinite, which is incompatible with the existence of a macroscopic limit, or the energy density must be zero, indicating the presence of a flat direction. We therefore reach the following powerful general result \cite{Cuomo:2024fuy}: a theory that has both $\lim_{Q\to\infty} \Delta_Q Q^{-\frac{d}{d-1}} = 0$ and a non-trivial a macroscopic limit necessarily admits a moduli space of vacua.\footnote{That is, conformal symmetry is spontaneously broken.} The mechanism causing the emergence of a moduli space is the absence of a potential for the dilaton. This requires a high level of fine tuning or symmetry protection. In fact, all known local interacting \acp{cft} with this property are supersymmetric and the global symmetry is the $R$-symmetry. In this case, zero-energy states in flat space cause some scalars charged under the $R$-symmetry to acquire nontrivial vacuum expectation values which parametrize the moduli space. The operators $\mathcal{O}_Q$ are BPS operators, whose scaling dimensions are enforced by supersymmetry to be $\Delta_Q \sim Q$ without corrections (so $\alpha=1$). The sufficient condition established above for the existence of a moduli space is part of a bigger program to understand under which circumstances a \ac{cft} can undergo spontaneous symmetry breaking of conformal symmetry~\cite{Caetano:2023zwe,Cuomo:2024fuy,Ivanovskiy:2024vel}.

The case $\alpha=\frac{d}{d-1}$ is therefore the most natural behaviour in the absence of supersymmetry or other strong constraints. The corresponding non-trivial macroscopic limit has
\begin{equation}
    \gamma = d-1.
\end{equation}
This behaviour has been established in the large-charge sector of a variety of systems, corresponding to conformal superfluid states, Fermi spheres, extremal Reissner-Nordström black-holes, etc. Below, we review the state of the art of research in this field.

Importantly, the macroscopic limits described here do not capture subleading corrections in $Q$ in any trivial way. For instance, a more detailed analysis based on the conformal superfluid \ac{eft} on the cylinder shows that a characteristic feature of the conformal superfluid state is the presence of a universal $Q^0$ or $Q^0 \log Q$ term in the large-charge expansion of $\Delta_Q$. Whether such a term is also present in other phases with $\alpha=\frac{d}{d-1}$ remains to be shown.

\pagebreak
\subsection{Large-charge universality classes: state of the art}

We have identified various, albeit not many, admissible large-charge behaviours. This is a powerful statement: while there is no truly universal large-charge behaviour in generic \acp{cft} with global symmetries, their large-charge \ac{cft} data can exhibit only few characteristic features. This provides a \emph{classification tool} for \acp{cft}, where each theory belongs to a given large-charge universality class.

\subsubsection{Monopole operators}

The large-charge program in its most modern version was initiated in~\cite{Hellerman:2015nra}, see also~\cite{Gaume:2020bmp} for a review. Historically, some of the ingredients of the large-charge expansion---most notably the state-operator correspondence and the cylinder frame---had already been employed as computational tools in the study of monopole operators in \acp{cft}~\cite{Borokhov:2002ib,Metlitski:2008dw,Dyer:2013fja,Karthik:2018rcg}. The large-charge machinery was subsequently applied explicitly to this class of operators in~\cite{Dyer:2015zha,Chester:2017vdh,delaFuente:2018qwv,Dupuis:2019uhs,Dupuis:2021yej,Dupuis:2019xdo,Dupuis:2021flq,Chester:2022wur,Boyack:2023uml}, where evidence was found that they belong to the superfluid universality class, which we describe in details below.

\subsubsection{Moduli universality class}

Another important early line of development that we alluded to before is the universality class of theories with a moduli space, mostly (or perhaps exclusively) populated by supersymmetric theories for which the large-charge operators are \acs{bps} operators. The seminal work of Berenstein, Maldacena and Nastase~\cite{Berenstein:2002jq}, quickly followed by Gross, Mikhailov and Roiban~\cite{Gross:2002su}, among others, used a mix of analytic techniques including the large $R$-charge limit, the 't Hooft limit and the constraints of $\mathcal{N}=4$ supersymmetry. Subsequently, many supersymmetric models were extensively studied in light of the newly developed large-charge methods~\cite{Hellerman:2017veg,Hellerman:2017sur,Hellerman:2018xpi,Hellerman:2020sqj,Baume:2020ure,Hellerman:2021yqz,Hellerman:2021duh,Bourget:2018obm,Beccaria:2018xxl,Cremonesi:2022fvg,Paul:2023rka,Giombi:2021zfb,Giombi:2022anm,Baume:2022cot,Brown:2023why,Brown:2024yvt,Caetano:2023zwe,Ivanovskiy:2024vel,Grassi:2019txd,Grassi:2024bwl,Heckman:2024erd}. Of course, the object of interest is usually not the conformal dimension of the lightest operator charged under a fixed $R$-charge $R$, as the latter is protected by supersymmetry to be exactly linear in $R$. However, by virtue of supersymmetry, other important pieces of \ac{cft} data can be extracted and the large-charge expansion may even sometimes be resummed. Supersymmetric models therefore provide a wealth of examples were analytic results can be obtained and even compared to those obtained using different techniques.

For instance, consider an $\mathcal{N}=2$ \acs{scft} in four dimensions. It has an $SU(2) \times U(1)$ $R$-symmetry and a moduli space of degenerate vacua. The subspace of the moduli space which is described at low energies by a free photon and its superpartners is called the Coulomb branch. This low-energy description is the result of the spontaneous breaking of conformal symmetries and the $U(1)$ $R$-symmetry. The order parameters for this breaking are the superconformal primaries annihilated by all $\bar{Q}$ supercharges and their vacuum expectation values parametrize the Coulomb branch vacua. They carry a $U(1)$ $R$-charge and their conformal dimensions are given by $\frac{R}{2}$. They are called chiral (or Coulomb branch) primaries. A central feature of these operators is that their \ac{ope} is regular, leading to a ring structure. In particular, in rank-1 theories, the Coulomb branch is generated by a single operator with dimension $\Delta_\mathcal{O}$, so the rest of the operators are simply labelled by an integer $n\in\mathbb{N}$ and their ring structure reads
\begin{equation}
    \lim_{y\to x} \mathcal{O}^n(x) \mathcal{O}^m(y) = \mathcal{O}^{n+m}(x)
\end{equation}
as an operator identity. A particularly simple class of correlators are the so-called extremal correlators, which have a single anti-chiral operator and are fixed to be of the form
\begin{equation}
    \ev{\mathcal{O}^{n_1}(x_1) \ldots \mathcal{O}^{n_k}(x_k)\bar{\mathcal{O}}^m(y)} = G_{n_1,\ldots,n_k;m}(\tau,\bar\tau) \prod_{i=1}^k \frac{1}{\abs{x_i-y}^{2n_i\Delta_\mathcal{O}}},
\end{equation}
with $m = \sum_{i=1}^k n_i$ and $\tau$ is the exactly marginal coupling. The function $G_{n_1,\ldots,n_k;m}(\tau,\bar\tau)$ is not fixed by symmetries and it therefore encodes the dynamics of extremal correlators at a given point of the moduli space. Upon fusing the first $k$ insertions, \emph{i.e.} $x_1,\ldots,x_k\to x$, the computation of this function boils down to extracting the normalization of the resulting two-point function, and it therefore only depends on the total charge $m$, not the single $n_i$'s. We shall therefore denote it $G_{2m}(\tau,\bar\tau)$ instead. When $m$ is large, a large-charge \ac{eft} can be implemented to compute this function. In fact, its large-charge expansion can be completely resummed, up to exponential corrections~\cite{Hellerman:2017sur,Bourget:2018obm,Hellerman:2018xpi,Grassi:2019txd}, resulting in
\begin{equation}
    G_{2n}(\tau,\bar\tau) = \left[ \expo{nA(\tau,\bar\tau)+B(\tau,\bar\tau)} + \order{\expo{-\sqrt{n}}} \right] \Gamma(n\Delta_\mathcal{O} + \alpha + 1).
\end{equation}
Here, $A(\tau,\bar\tau)$ and $B(\tau,\bar\tau)$ are not determined within the large-charge \ac{eft}, and $\alpha=2(a_{CFT}-a_{free})$ is the variation of the $a$-anomaly between the free and the interacting theories. This result is in very good agreement with bootstrap results~\cite{Gimenez-Grau:2020jrx} for $n=1$ and, in $\mathcal{N}=2$ SQCD, it matches the numerical estimates from localization~\cite{Gerchkovitz:2016gxx}. Very recently, this result has been generalized to higher-rank theories~\cite{Grassi:2024bwl}. The above result could have been derived qualitatively by observing that extremal correlators can be regarded as large-charge families admitting a non-trivial macroscopic limit~\cite{Cuomo:2024fuy}.

\subsubsection{Fermi-sphere universality class}

We also mentioned another striking example of a large-charge phase of matter sharing the same type of macroscopic limit as the superfluid universality class (namely, $\gamma=d-1$), although their semiclassical description is fundamentally different: the Fermi sphere state~\cite{Dondi:2022zna}. Interacting systems with a Fermi surface are expected to be described at low energies by a Fermi liquid \ac{eft}\footnote{There is also the possibility to have a non-Fermi liquid, that is, a gapless phase with no fermionic quasiparticles around the Fermi surface. However, in the context of large charge, no such example is known to date.} akin to the one described in~\cite{Polchinski:1992ed}. This effective description is notoriously harder to use compared to the superfluid one. Moreover, in the nonrelativistic case, there is a symmetry obstruction to improving this \ac{eft} to the conformal case~\cite{Rothstein:2017niq}. In~\cite{Dondi:2022zna}, the Fermi sphere was obtained in the Gross-Neveu model at large-$N$. Whether this configuration is a truly stable large-charge state beyond the strict $N\to\infty$ limit remains to be shown. In an effort to address the problem of the emergence of a Fermi sphere state at large charge from \ac{eft} considerations, it would most likely be more appropriate to start from the newly proposed approach to the Fermi surface \ac{eft} of~\cite{Delacretaz:2022ocm}.

\subsubsection{Swampland and holography}

Recently, another research direction has been developed in connection with the swampland program~\cite{Vafa:2005ui,Palti:2019pca} in holographic theories (see also~\cite{Harlow:2018tng} for an extensive discussion of the fate of global symmetries in quantum gravity). The so-called (abelian) charge-convexity conjecture states that, in any \ac{cft} in $d>2$ with a $U(1)$ symmetry, the following convexity condition is satisfied~\cite{Aharony:2021mpc,Antipin:2021rsh,Sharon:2023drx,Moitra:2023yyc,Aharony:2023ike}:
\begin{equation}
    \Delta_{(n+m)Q_0} \geq \Delta_{n Q_0} + \Delta_{m Q_0}, \qquad \forall n,m \in \mathbb{N}^*.
\end{equation}
Here, $Q_0$ is the charge associated with the onset of convexity in $\Delta_Q$, which denotes as before the conformal dimension of the lightest operator of fixed charge $Q$. Originally, $Q_0$ was proposed to always be of order one, although it was quickly shown to be wrong~\cite{Sharon:2023drx}. However, relaxing this condition and allowing $Q_0$ to be arbitrarily large is compatible with all known results in the literature so far. For \acp{cft} with a holographic dual description, this conjecture is roughly equivalent to the famous weak-gravity conjecture~\cite{Arkani-Hamed:2006emk}. Whether generic field theoretic arguments such as the existence of a macroscopic limit can partly prove such conjectures is a very exciting open question, \emph{cf.}~\cite{Nakayama:2015hga,Cuomo:2022kio,Cuomo:2024fuy}.

Still in the context of holography, the so-called holographic superconductors~\cite{Hartnoll:2008vx,Hartnoll:2008kx,Hartnoll:2009sz,Esposito:2016ria,delaFuente:2020yua} may be regarded as a large-charge universality class on their own. Similarly, cases where the large-charge state would be an extremal Reissner-Nordström black holes, which have a parametrically large degeneracy of operators close to the lightest state, could be regarded as a new universality class. However, these states are believed to be unstable. If this is true, there probably is a large-charge argument on the \ac{cft} side---a generalization of~\cite{Jafferis:2017zna}---that would rule them out. See also~\cite{Ishii:2024pnh}, and~\cite{Brown:2024ajk} for a broader discussion of the fate of charged black holes.

\subsubsection{Superfluid universality class}

And finally, there is the superfluid universality class, which is the main focus of this thesis (albeit in the nonrelativistic case, see next section). We mentioned that large-charge families of correlators are described semiclassically by a path-integral trajectory associated with a given symmetry-breaking pattern on the cylinder, although, strictly speaking, there is no spontaneous symmetry breaking in finite volume. The fate of the dilation operator and the global internal symmetry group $G$ is theory-dependent. However, a breaking pattern commonly realized in nature is the one that breaks both of them while preserving a linear combination. We have already discussed this in the description of the nonrelativistic conformal superfluid. To be more precise, the relativistic breaking pattern describing a conformal superfluid state is given by
\begin{equation}
    SO(d+1,1) \times G \stackrel{SSB}{\longrightarrow} SO(d) \times \bar{D} \times G',
\end{equation}
where $\bar{D} = D + \vec{\mu} \cdot \vec{Q}$ is the unbroken combination of the dilation operator and the Cartan charges $Q_I$. The coefficients $\mu_I$ are interpreted as chemical potentials, and $G'$ is the unbroken subgroup of $G$. As far as microscopic realizations of this breaking pattern are concerned, it seems that the large-charge dynamics of \acp{cft} is vastly dominated by this configuration in the absence of extra constraints like supersymmetry or higher-spin symmetry. The corresponding effective action $S_{eff}[\chi_I,\pi_A]$ is the most general action realizing the above symmetry breaking pattern in terms of the Goldstone fields $\chi_I$ and possible extra degrees of freedom $\pi_A$. It gets complemented by a term imposing boundary conditions so as to reproduce the correct charges of the \emph{in} and \emph{out} states. For instance, in the absence of extra light operator insertions, one can compute Eq.~\eqref{eq:DeltaFromCylinder} semiclassically from
\begin{equation}
    \mel{\vec{Q},out}{\expo{-H_{cyl.}T}}{\vec{Q},in} = \int\prod D\chi_I \prod D\pi_A \exp\left[ -S_{eff}[\chi_I,\pi_A] - \frac{iQ_I}{\Omega_d R^{d-1}} \int_{-T}^T \dd{\tau} \int \dd{S} \dot\chi_I \right]
\end{equation}
with $T\to\infty$ and $\Omega_d = \frac{2\pi^\frac{d}{2}}{\Gamma\left(\frac{d}{2}\right)}$. The action contains higher and higher derivative terms (accompanied by new Wilsonian coefficients) which are more and more suppressed by the large charges. The saddle-point approximation is performed around the lowest energy configuration (the superfluid ground state) and quantum corrections, \emph{i.e.} loop effects, typically kick in at order $Q^0$ in the expansion. In the case $G=U(1)$\footnote{In the non-Abelian case such as the $O(N)$ model, only large-charge operators in the completely symmetric traceless representation are associated with a homogeneous superfluid ground state. For a discussion of inhomogeneous ground states, \emph{cf.}~\cite{Alvarez-Gaume:2016vff,Hellerman:2017efx,Hellerman:2018sjf,Cuomo:2020gyl} and the numerical results~\cite{Banerjee:2021bbw,Rong:2023owx}. Even in the simple homogeneous case, the fluctuations above the ground state contain type-II (or type-B) Goldstone modes and their gapped partners~\cite{Gaume:2020bmp}, which have nonrelativistic dispersion relations. They describe operators in slightly asymmetric representations.}, the superfluid \ac{eft} takes the following simple form~\cite{Monin:2016jmo}:
\begin{equation}
    S_{eff}[\chi] = \int\limits_{\mathbb{R}\times S_R^{d-1}}\dd[d]{x} \sqrt{g} \abs{\partial\chi}^d \left[ c_0 + c_1 \left( \frac{\mathcal{R}}{(\partial\chi)^2} + (d-1)(d-2) \frac{(\nabla_\mu\abs{\partial\chi})^2}{(\partial\chi)^4} \right) + \ldots \right]
\end{equation}
to be thought of as the relativistic version of Eq.~\eqref{eq:eff-action-nlo} in general \emph{spacetime} dimension $d$ ($\mathcal{R}$ is the Ricci curvature),
and it is now well established that this results in $\Delta_Q$ having the following structure (\emph{cf.} for instance~\cite{Dondi:2022wli})
\begin{equation}
\begin{aligned}
    \Delta_Q & = Q^\frac{d}{d-1} \left[ a_1 + \frac{a_2}{Q^\frac{2}{d-1}} + \frac{a_3}{Q^\frac{4}{d-1}} + \ldots  \right] + \sum_{l=1}^\infty Q^{\frac{d}{d-1}(1-l)} \left[ \beta_{l,1} + \frac{\beta_{l,2}}{Q^\frac{2}{d-1}} + \frac{\beta_{l,3}}{Q^\frac{4}{d-1}} + \ldots \right]    
\end{aligned}
\end{equation}
The first series is the ground state (or 'classical') contribution, in which the coefficients $a_i$ are associated with new Wilsonian coefficients in the effective action, \emph{e.g.}~\cite{Cuomo:2020rgt}
\begin{equation}
    a_1 = \frac{c_0(d-1)\Omega_{d-1}}{(c_0d\Omega_{d-1})^\frac{d}{d-1}},
    \qquad
    a_2 = \frac{c_1(d-1)(d-2)\Omega_{d-1}}{(c_0d\Omega_{d-1})^\frac{d-2}{d-1}},
    \qquad
    \text{etc.}
\end{equation}
The second set of contributions, on the other hand, denotes the $l$-loop effects. The coefficients $\beta_{l,i}$ generically depend on previous Wilsonian coefficients, and the contributions may get logarithmic enhancements in even dimensions. However, for a given dimension, the $\beta_{1,1}$ term is universal across the superfluid universality class. The canonical examples are
\begin{equation} \label{eq:DeltaQ_rel}
    \Delta_Q =
    \begin{cases}
        a_1 Q^\frac{3}{2} + a_2 Q^\frac{1}{2} - 0.0937\ldots + \order{Q^{-\frac{1}{2}}}, \qquad & (d=3), \\
        a_1 Q^\frac{4}{3} + a_2 Q^\frac{2}{3} - \frac{1}{48\sqrt{3}} \log Q + \order{Q^0}, \qquad & (d=4).
    \end{cases}
\end{equation}
The phonon excitations above the superfluid ground state have a dispersion relation of the form
\begin{equation} \label{eq:Phonon_rel}
    \omega_J = \frac{1}{R} \sqrt{\frac{J(J+d-2)}{d-1}} + \order{Q^{-\frac{2}{d-1}}},
\end{equation}
which gives rise to the $Q^0$ terms above. These fluctuations describe large-charge states with small spin. For large values of the spin, the simple conformal superfluid \ac{eft} description breaks down as the ground-state configuration transitions to a superfluid state with vortices, which requires a new effective description. The description of various phases that the system enters into as we vary the charge and the spin is entirely guided by our phenomenological discussion of the last section and is somewhat conjectural.

\pagebreak
Here is a brief summary for $d=3$, based on~\cite{Cuomo:2017vzg,Cuomo:2019ejv,Cuomo:2022kio}.\footnote{\textbf{Note added:} in the window $Q^\frac{3}{2} \ll J \ll Q^2$, a state of lower energy that the one described here was proposed in~\cite{Choi:2025tql}. It effectively corresponds to a charged rotating normal fluid, which allows to establish some qualitative connection to black hole physics via the fluid-gravity correspondence.}
\begin{itemize} [itemsep=5pt,partopsep=0pt,topsep=0pt,parsep=0pt]
    \item $0 \leq J \ll Q^\frac{1}{2}$: the spectrum of large-charge spinning operators with small spin $J$ is simply given by Eq.~\eqref{eq:DeltaQ_rel} above, with the addition of a phonon excitation of spin $J$, that is,
    \begin{equation}
        \Delta_{Q,J} = a_1 Q^\frac{3}{2} + a_2 Q^\frac{1}{2} + \frac{1}{\sqrt{2}} \sqrt{J(J+1)} - 0.09372 + \ldots
    \end{equation}
    \item $Q^\frac{1}{2} \ll J \ll Q$: in this range, the superfluid develops a single vortex/anti-vortex pair rotating around the $S_R^2$, and we find
    \begin{equation}
        \Delta_{Q,J} = a_1 Q^\frac{3}{2} + \frac{1}{6a_1} Q^\frac{1}{2} \log \frac{J(J+1)}{Q} + \ldots
    \end{equation}
    As the spin increases, the vortices move towards the poles.
    \item $Q \ll J \ll Q^\frac{3}{2}$: after reaching the poles, new vortices need to form in order to accommodate more angular momentum. The resulting lattice of vortices rotating around the sphere exhibits an effective rigid body motion, with
    \begin{equation}
        \Delta_{Q,J} = a_1 Q^\frac{3}{2} + \frac{1}{2a_1} \frac{J^2}{Q^\frac{3}{2}} + \ldots
    \end{equation}
    As $J\sim Q^\frac{3}{2}$, the rotation speed exceeds the speed of sound and the system transitions to a new phase.
    \item $Q^\frac{3}{2} \ll J \ll Q^2$: in this regime, the vortex cores overlap and collapse into the giant vortex state mentioned earlier. The order parameter is non-zero only in some small region around the equator, and we have
    \begin{equation}
        \Delta_{Q,J} = J + \frac{9a_1^2}{4\pi} \frac{Q^3}{J} + \ldots
    \end{equation}
    The fact that the leading term is linear in $J$ is indicative of the fact that the system is reaching the universal Regge regime mentioned earlier.
    \item $Q^2 \ll J$: when the spin is the dominating quantum number, the leading-order dynamics is truly universal (not just within the superfluid universality class), and must match the large-spin predictions. The picture is that of $Q$ weakly interacting 'partons'---that is, unit charge spinless quanta---spinning around the equator, whose conformal dimensions we denote by $\Delta_\Phi$. They carry unit global charge, such that (compare to Eq.~\eqref{eq:LargeSpinRegge})
    \begin{equation}
        \Delta_{Q,J} = J + \Delta_\Phi Q + \ldots.
    \end{equation}
\end{itemize}

\begin{figure}[h]
	\centering
	\includegraphics[width=0.75\textwidth]{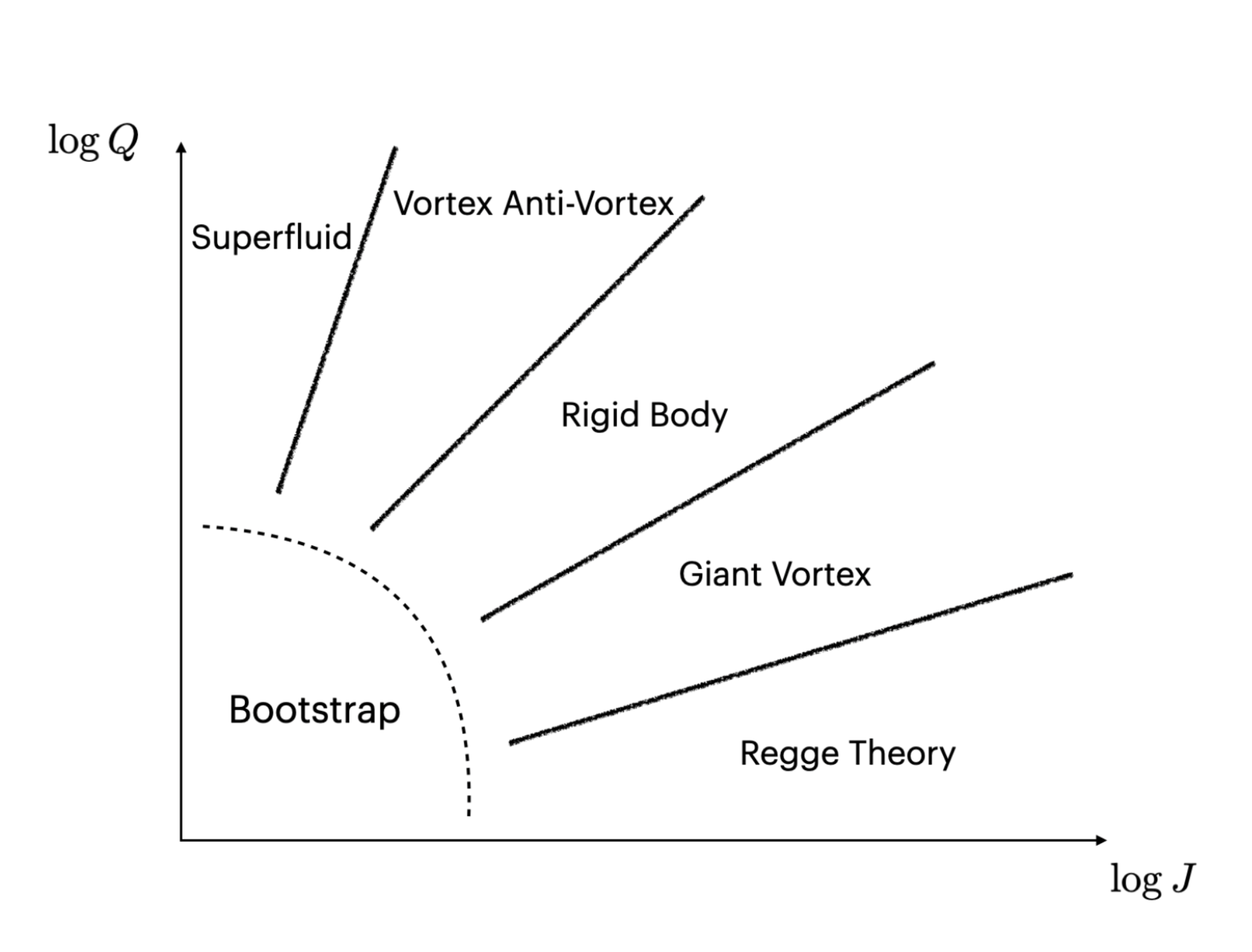}
	\caption{\emph{Picture taken from \cite{Cuomo:2022kio}. Proposed phase diagram at large charge $Q$ versus large spin $J$ within the superfluid universality class.}}
	\label{fig:largeQJ}
\end{figure}

See Figure~\ref{fig:largeQJ} for a schematic representation of the large-charge versus large-spin phase diagram, and~\cite{Fardelli:2024heb} for a recent discussion of some of these phases from a holographic perspective. It remains to be understood how, starting from the Regge regime, one would fall into this specific superfluid phase diagram or the one associated with another large-charge universality class. It is also not clear what the nature of the phase transitions from one regime to the next is. It would be interesting to start from a microscopic (\emph{e.g.} large-$N$) description and explicitly identify the leading path-integral trajectory in one of the intermediate phases.

This is the state of the art for the computation of conformal dimensions in the simple conformal superfluid case. What concerns the computation of other pieces of \ac{cft} data, there is by now a very good understanding of the fate of correlators with one or two light operator insertions~\cite{Hellerman:2015nra,Monin:2016jmo,Jafferis:2017zna,Cuomo:2020rgt,Cuomo:2021ygt,Dondi:2022wli}. They match the generic predictions of the macroscopic limit analysis in the 'simple case' analysed previously, with $\alpha=\frac{d}{d-1}$. Correlators involving three large-charge operators were investigated in~\cite{Cuomo:2021ygt}, which is presumably the first example of a universal heavy-heavy-heavy correlator without supersymmetry in $d>2$.

Variations and extensions of the simple conformal superfluid setup have been investigated too, \emph{e.g.} in parity-violating theories~\cite{Cuomo:2021qws} and theories with boundaries/defects~\cite{Cuomo:2021cnb}.\footnote{It is worth mentioning here that semiclassical techniques inspired by large-charge methods have recently been developed in the context of conformal defects carrying large quantum numbers (large spin in particular)~\cite{Cuomo:2022xgw,Krishnan:2024wrc}.} We also mention that some 'bootstrap' techniques can be applied on top of the large-charge construction. Indeed, it is for instance possible to place bounds on the Wilsonian coefficients appearing in the superfluid \acp{eft} by invoking positivity arguments~\cite{Creminelli:2022onn}.

When it comes to microscopic checks of the \ac{eft} predictions, one usually has to employ double-scaling limits, be it in the $\epsilon$ or large-$N$ expansions~\cite{Alvarez-Gaume:2019biu,Watanabe:2019pdh,Badel:2019oxl,Badel:2019khk,Arias-Tamargo:2019xld,Arias-Tamargo:2019kfr,Arias-Tamargo:2020fow,Antipin:2020abu,Antipin:2020rdw,Antipin:2021akb,Antipin:2021jiw,Giombi:2020enj,Giombi:2022gjj,Orlando:2021usz,Moser:2021bes,Dondi:2024vua}. In these controlled setups, it becomes possible to analyse the analytic structure of the asymptotic large-charge expansion and apply the tools of resurgence theory, as was done in~\cite{Dondi:2021buw,Antipin:2022dsm}. Finally, one can also non-perturbatively test the \ac{eft} predictions using numerics, \emph{cf.} in particular~\cite{Banerjee:2017fcx,Banerjee:2019jpw,Banerjee:2021bbw,Cuomo:2023mxg,Rong:2023owx}, where excellent agreement is found.

This concludes this overview of the \emph{relativistic} large-charge landscape. We shall now move on to the description of the nonrelativistic large-charge program, thereby facing the striking fact that a lot remains to be done.
\vfill


\pagebreak
\section{Large-charge sector of nonrelativistic CFTs: an overview}
\label{sec:Intro_LargeQ_nonrel}


To the high-energy physicists developing the large-charge methods, it came as a surprise that \acp{nrcft} were the perfect ground on which to run the large-charge program. As mentioned, this is due to the fact that every \ac{nrcft} has a global $U(1)$ symmetry counting the number $Q$ of particles in the system, and the state-operator correspondence, which we are going to discuss shortly establishes an astonishing connection with the ultracold atom experiments mentioned earlier, in which the charge $Q$ is always large. Much of the large-charge methodology can therefore be adapted to this class of theories, and the corresponding superfluid universality class is expected, on physical grounds, to be the most relevant large-charge behaviour. Although the condensed matter community does necessarily focus on the computation of large-charge \ac{cft} data, it is only fair to acknowledge the fact that many results, especially when it comes to \ac{eft}-building for various phases of matter, predate the modern large-charge program. One such example is the \ac{eft} for a trapped nonrelativistic conformal superfluid introduced earlier, and we shall close the loop at the end of this section by coming back to it and discuss how it provides a direct way to extract large-charge \ac{cft} data. From now on, $d$ denotes the number of \emph{spatial} dimensions, and we set $\hbar=m=1$ in this section.

\subsection{Nonrelativistic conformal field theories in a nutshell}

Very much like the conformal group is the maximal kinematical group that leaves the free Maxwell or free Klein-Gordon equations invariant, the Schrödinger group is the maximal kinematical group that leaves the free Schrödinger equation invariant~\cite{niederer1972maximal,Hagen:1972pd}\footnote{However, the maximal symmetry \emph{algebra} is in fact enhanced to the infinite \emph{Weyl algebra}~\cite{Valenzuela:2009gu,Bekaert_2012} in a way similar to the higher-spin symmetry algebra of the free Klein-Gordon equation~\cite{Giombi:2016ejx}.}. And very much like the former is an extension of the Poincaré group, the latter is an extension of the (centrally extended) Galilean group. In fact, as already noted in~\cite{niederer1972maximal} more than 50 years ago, the main structural difference between the Galilean and Schrödinger groups is that time translations are enhanced to a $SL(2, \mathbb{R})$ group generated by time translations, dilatations and special conformal transformations. This has profound consequences which we shall discuss.

\subsubsection{Schrödinger group and algebra}

The Schrödinger algebra has been studied~\emph{e.g.} in \cite{niederer1972maximal,Hagen:1972pd,Henkel:1993sg,Mehen:1999nd,Nishida:2007pj,Fuertes:2009ex,Volovich:2009yh,Golkar:2014mwa,Goldberger:2014hca,Pal:2018idc,Karananas:2021bqw}.\footnote{We follow the conventions of~\cite{Goldberger:2014hca} in particular.} It consists of the Galilean algebra, that is, time translations $H : (t, \vect{r}) \to (t + a, \vect{r})$, space translations $P_i : (t, \vect{r}) \to (t, \vect{r} + \vect{a})$, Galilean boosts $K_i : (t, \vect{r}) \to (t, \vect{r} + \vect{v} t)$ and spatial rotations $M_{ij} = -M_{ji} : (t, \vect{r}) \to (t, M \vect{r})$ ($i,j = 1, \ldots, d$), together with its central extension $\mathcal{Q}$ associated with a $U(1)$ particle number symmetry, as well as dilatations $D : (t, \vect{r}) \to (\lambda^2 t, \lambda \vect{r})$ and a single special conformal transformation (sometimes called expansion) $C = I H I : (t, \vect{r}) \to \left(\frac{t}{1 + a t}, \frac{\vect{r}}{1 + a t}\right)$, where $I : (t, \vect{r}) \to \left(\frac{1}{t}, \frac{\vect{r}}{t}\right)$ is an inversion.

\begin{table}[t]
\begin{tabular}{|c|c|c|c|c|c|c|c|} \hline
    $X \backslash Y$ & $M_{kl}$ & $P_k$ & $K_k$ & $H$ & $C$ & $D$ & $\mathcal{Q}$ \\ \hline
    $M_{ij}$ & \begin{tabular}{@{}c@{}} $i(\delta_{ik} M_{jl} + \delta_{jl} M_{ik}$ \\ $- \delta_{jk} M_{il} - \delta_{il} M_{jk})$\end{tabular} & $i(\delta_{ik} P_j - \delta_{jk} P_i)$ & $i(\delta_{ik} K_j - \delta_{jk} K_i)$ & 0 & 0 & 0 & 0 \\ \hline
    $P_i$ & $i(\delta_{il} P_k - \delta_{ik} P_l)$ & 0 & $-i\delta_{ik} \mathcal{Q}$ & 0 & $-iK_i$ & $-iP_i$ & 0 \\ \hline
    $K_i$ & $i(\delta_{il} K_k - \delta_{ik} K_l)$ & $i\delta_{ik} \mathcal{Q}$ & 0 & $iP_i$ & 0 & $iK_i$ & 0 \\ \hline
    $H$ & 0 & 0 & $-iP_k$ & 0 & $-iD$ & $-2iH$ & 0 \\ \hline
    $C$ & 0 & $iK_k$ & 0 & $iD$ & 0 & $2iC$ & 0 \\ \hline
    $D$ & 0 & $iP_k$ & $-iK_k$ & $2iH$ & $-2iC$ & 0 & 0 \\ \hline
    $\mathcal{Q}$ & 0 & 0 & 0 & 0 & 0 & 0 & 0 \\ \hline
\end{tabular}
\caption{The full Schrödinger algebra $[X, Y]$. Note in particular the subalgebra $[D, H] = 2 i H$, $[H, C] = -i D$ and $[C, D] = 2 i C$ generating an $SL(2, \mathbb{R})$ subgroup.}
\end{table}

Its action on time and $d$-dimensional space can be compactly written as
\begin{equation}
    t \longrightarrow \frac{a t + b}{c t + d},
    \hspace{10mm} \text{and} \hspace{10mm}
    \vect{r} \longrightarrow \frac{M \vect{r} + \vect{v} t + \vect{a}}{c t + d},
\end{equation}
where $a d - b c = 1$. The subgroup generated upon setting $M = \mathbb{1}$, $\vect{v} = 0$ and $\vect{a} = 0$ is isomorphic to $SL(2, \mathbb{R})$. The group can therefore be written as
\begin{equation}
    U(1) \times \left[ (SL(2,\mathbb{R}) \times SO(d)) \ltimes (\mathbb{R}^d \times \mathbb{R}^d) \right].
\end{equation}
The easiest way to establish the group law is to use the following faithful matrix representation:
\begin{equation}
    \mathcal{L}_g =
    \begin{pmatrix}
        M & \vect{v} & \vect{a} \\
        0 & a & b \\
        0 & c & d
    \end{pmatrix},
\end{equation}
which satisfies $\mathcal{L}_{g} \mathcal{L}_{g'} = \mathcal{L}_{gg'}$, from which $gg'$ can easily be read off.

In a nonrelativistic quantum field theory, that is, a theory invariant under the Galilean group, if the 'tracelessness' condition
\begin{equation}
    2 H = T_{ii}
\end{equation}
is satisfied, the spacetime symmetry group gets enhanced to the Schrödinger group. Here, $T_{ij}$ is the stress tensor. When the theory is local, the currents associated with each of the generators can then be built out of the charge density $\rho(\vect{r})$ and the charge current $\vect{j}(\vect{r})$ as follows:\footnote{Charge and momentum conservation yield, respectively, $\partial_t \rho + \partial_i j_i = 0$ and $\partial_t j_i + \partial_j \Pi_{ij}=0$, where $\Pi_{ij}$ is the stress tensor density.}
\begin{equation} \label{eq:GeneratorCurrents}
\begin{cases}
    M_{ij} & = \int\dd{\vect{r}} (r_i j_j(\vect{r}) - r_j j_i(\vect{r})), \\
    P_i & = \int\dd{\vect{r}} j_i(\vect{r}), \\
    K_i & = \int\dd{\vect{r}} r_i \rho(\vect{r}), \\
\end{cases}
\quad
\begin{cases}
    C & = \frac{1}{2} \int\dd{\vect{r}} \vect{r}^2 \rho(\vect{r}), \\
    D & = \int\dd{\vect{r}} \vect{r} \cdot \vect{j}(\vect{r}), \\
    \mathcal{Q} & = \int\dd{\vect{r}} \rho(\vect{r}).
\end{cases}
\end{equation}

\subsubsection{Unitary irreducible representations}

We quickly summarize the notion of primary operators and the corresponding irreducible representations of the Schrödinger group because there is an important difference with respect to the relativistic case that we want to point out. As usual, we define a local operator as an operator that depends on a position in time and space $(t, \vect{r})$ such that
\begin{equation}
    \mathcal{O}(t, \vect{r}) = \expo{i(t H - \vect{r} \cdot \vect{P})} \mathcal{O}(0, \vect{0}) \expo{-i(t H - \vect{r} \cdot \vect{P})},
\end{equation}
which is equivalent to the statement that time and space translations act on it as
\begin{equation} \label{eq:NRCFT_localOp}
\begin{aligned}
    [H, \mathcal{O}(t, \vect{r})] & = -i \partial_t \mathcal{O}(t, \vect{r}), \\
    [P_i, \mathcal{O}(t, \vect{r})] & = i \partial_i \mathcal{O}(t, \vect{r}).
\end{aligned}
\end{equation}
A local operator is said to have a well-defined scaling dimension $\Delta$, spin\footnote{Saying that $\mathcal{O}$ has spin $J$ means it belongs to a finite-dimensional irreducible representation of the rotation algebra $o(d)$, $\Sigma_{ij}$, in which case $J$ refers to a collection of Young tableaux labels characterizing uniquely the representation, as these are not just labeled by half-integers when $d > 3$~\cite{Bekaert_2012}.} $J$ and particle number $Q$ if it is an eigenvector of $D$, $M_{ij}$ and $\mathcal{Q}$ at the origin, respectively, in the sense that
\begin{equation} \label{eq:NRCFT_localOpQuantumNumbers}
\begin{aligned}
    [D, \mathcal{O}] & = i \Delta \mathcal{O}, \\
    [M_{ij}, \mathcal{O}^a] & = i {(\Sigma_{ij})^a}_b \mathcal{O}^b, \\
    [\mathcal{Q}, \mathcal{O}] & = Q \mathcal{O},
\end{aligned}
\end{equation}
where here and in what follows, $\mathcal{O} = \mathcal{O}(0, \vect{0})$ refers to an operator at the origin. The set of all local operators $\mathcal{O}_\alpha(t, \vect{r})$ form a representation of the Schrödinger algebra at the origin, \emph{i.e.} for any element $A$ of the algebra,
\begin{equation}
    [A, \mathcal{O}_\alpha] = A_{\alpha\beta} \mathcal{O}_\beta.
\end{equation}
We are interested in constructing the unitary irreducible representations of the Schrödinger algebra. We note that, given Eq.~\eqref{eq:NRCFT_localOpQuantumNumbers}, we have
\begin{equation} \label{eq:NRCFT_ScalingLadder}
\begin{aligned}
    [D,[K_i, \mathcal{O}]] & = i (\Delta - 1) [K_i, \mathcal{O}], \\
    [D,[C, \mathcal{O}]] & = i (\Delta - 2) [C, \mathcal{O}], \\
    [D,[P_i, \mathcal{O}]] & = i (\Delta + 1) [P_i, \mathcal{O}], \\
    [D,[H, \mathcal{O}]] & = i (\Delta + 2) [H, \mathcal{O}]. \\
\end{aligned}
\end{equation}
In a unitary theory, the spectrum of scaling dimensions is positive and bounded from below. Therefore, one can build lowest-weight representations starting from local \emph{primary operators}, that is, operators satisfying Eqs~\eqref{eq:NRCFT_localOp}, \eqref{eq:NRCFT_localOpQuantumNumbers} and
\begin{equation} \label{eq:NRCFT:PrimaryDef}
    [K_i, \mathcal{O}] = [C, \mathcal{O}] = 0
	\rlap{\hspace{5mm} \text{(primary operator)}}.
\end{equation}
For such an operator, one then has
\begin{equation} \label{eq:KC_on_Primary}
\begin{aligned}
    [K_i,[P_j, \mathcal{O}]] & = i \delta_{ij} Q \mathcal{O}, \\
    [C, [P_j, \mathcal{O}]] & = 0.
\end{aligned}
\end{equation}
In the so-called massive sector $Q \neq 0$, the operator $[P_j, \mathcal{O}]$ is therefore not a primary operator, and we call it a \emph{descendant operator} of the primary operator $\mathcal{O}$. It has scaling dimension $\Delta + 1$ as can be seen from Eq.~\eqref{eq:NRCFT_ScalingLadder}. Similarly, since
\begin{equation} 
\begin{aligned}
    [K_i,[H, \mathcal{O}]] & = i [P_i, \mathcal{O}], \\
    [C, [H, \mathcal{O}]] & = -\Delta \mathcal{O},
\end{aligned}
\end{equation}
the operator $[H, \mathcal{O}]$ is not a primary operator, and we also call it a descendant of $\mathcal{O}$ with scaling dimension $\Delta + 2$. The tower of operators obtained by acting repeatedly with $P_i$ and $H$ on a given primary operator with $Q \neq 0$ forms a (massive) unitary irreducible representation of the Schrödinger group.

The situation is however more subtle in the 'massless' sector. Indeed, if a primary operator $\mathcal{O}$ has $Q = 0$, then Eq.~\eqref{eq:KC_on_Primary} tells us that $[P_j, \mathcal{O}]$ is also a primary operator. Similarly, if an operator $\mathcal{O}$ is such that $[K_j, \mathcal{O}]$ is a primary, then $\mathcal{O}$ cannot be a descendant of $[K_j, \mathcal{O}]$ in the sense that it cannot be obtained upon acting with either $P_i$ or $H$ on it. Dealing with the $Q=0$ sector of theory requires some care~\cite{Golkar:2014mwa,Bekaert_2012,Pal:2018idc}, and we shall assume $Q>0$ from now on. The situation is therefore qualitatively different from relativistic \acp{cft}, where there is no such distinction.


This said, we can now give the action of the Schrödinger algebra away from the origin:
\begin{equation} \label{eq:NRCFT_WardId}
\begin{aligned}
    [\mathcal{Q}, \mathcal{O}(t, \vect{r})] & = Q \mathcal{O}(t, \vect{r}), \\
    [D, \mathcal{O}(t, \vect{r})] & = i (2 t \partial_t + r_i \partial_i + \Delta) \mathcal{O}(t, \vect{r}), \\
    [C, \mathcal{O}(t, \vect{r})] & = \left(-i t^2 \partial_t - i t r_i \partial_i - i t \Delta + \frac{\vect{r}^2}{2} Q \right) \mathcal{O}(t, \vect{r}), \\
    [K_i, \mathcal{O}(t, \vect{r})] & = (-i t \partial_i + r_i Q)\mathcal{O}(t, \vect{r}), \\
\end{aligned}
\end{equation}
and rotations act as usual. This also sets the canonical units of the generators.

\subsubsection{Correlation functions}

Equipped with the Ward identities of Eq.~\eqref{eq:NRCFT_WardId}, kinematic constraints can be imposed to fix the form of correlators, which we denote by
\begin{equation}
    G_n(\tau_1,\vect{r}_1;\ldots;\tau_n,\vect{r}_n) := \ev{\mathcal{O}_1(\tau_1,\vect{r}_1) \ldots \mathcal{O}_n(\tau_n,\vect{r}_n)}.
\end{equation}
Note that we only discuss scalar operators for simplicity. The simplest Ward identity is the one imposed by $\mathcal{Q}$, namely, charge conservation $\sum_{i=1}^n Q_i = 0$. If this is not satisfied, the correlator must vanish. In what follows, we assume charge conservation. Two-point functions are then completely fixed, up to normalization, to be
\begin{equation}
    G_2(t_1, \vect{r}_1; t_2, \vect{r}_2) = \expo{-i \frac{Q_1}{2} \frac{\vect{r}_{12}^2}{t_{12}}} \times \frac{c \, \delta_{\Delta_1,\Delta_2}}{t_{12}^{\Delta_1}}.
\end{equation}
On the other hand, $n$-point functions with $n>2$ depend on $n^2-3n+1$ invariant cross-ratios through theory-dependent functions~\cite{Volovich:2009yh}. For instance, three-point functions are of the form
\begin{equation}
    G_3(t_1, \vect{r}_1; t_2, \vect{r}_2; t_3, \vect{r}_3) = \expo{-i \frac{Q_1}{2} \frac{\vect{r}_{13}^2}{t_{13}} - i \frac{Q_2}{2} \frac{\vect{r}_{23}^2}{t_{23}}} \prod_{i < j} t_{ij}^{\Delta/2 - \Delta_i - \Delta_j} \times F(v_{123}),
\end{equation}
where the cross-ratio is given by
\begin{equation}
    v_{ijn} = \frac{(\vect{r}_{in} t_{jn} - \vect{r}_{jn} t_{in})^2}{2 t_{ij} t_{in} t_{jn}}
\end{equation}
and $\Delta:=\sum_{i=1}^3 \Delta_i$. The unfixed function satisfies $F(0)=\text{cst}$. Observe that upon placing the insertions along the time axis (\emph{i.e.} $\vect{r}_i = \vect{0}$), these correlators take the form dictated by $SL(2,\mathbb{R})$ invariance of \acp{cft} in one spacetime dimension, with the only modification that time scales twice as much as 'standard' coordinates under dilatations.

\subsubsection{Quantization and the state-operator correspondence}

When one encounters the nonrelativistic state-operator correspondence~\cite{Nishida:2007pj,Goldberger:2014hca} for the first time, it might seem a bit mysterious or surprising.\footnote{To the best of our knowledge, the first person to fully appreciate the connection between the quantization of \acp{nrcft} and NS-quantization of \acp{cft} in one dimension is Sridip Pal~\cite{sridipDiablerets} (see also some of his preliminary results in~\cite{Pal:2018idc}). This connection is also established in~\cite{Karananas:2021bqw} in a slightly less transparent way. See~\cite{Baiguera:2024vlj} for a recent discussion in a related context.} In light of our previous discussion of the quantization of relativistic \acp{cft}, we present here a natural explanation. Because of the $SL(2,\mathbb{R})$ subgroup, \acp{nrcft} are naturally quantized with respect to any element in the conjugacy class of $D$ within this subgroup. Since the conformal group in one dimension is precisely $SL(2,\mathbb{R})$, quantization of \acp{nrcft} closely resembles the relativistic one in one dimension. The main difficulty is to keep track of sign conventions and factors of two coming from the fact that the time coordinate scales twice as much under dilatations as it would in the relativistic case.

In particular, we choose to quantize the theory with respect to
\begin{equation} \label{eq:HarmHamil}
    \expo{-i\frac{\pi}{4}\left(\frac{H}{\omega}-\omega C\right)} \,(\omega D)\, \expo{i\frac{\pi}{4}\left(\frac{H}{\omega}-\omega C\right)} = H + \omega^2 C,
\end{equation}
to be compared with Eq.~\eqref{eq:LuscherMach_hamil}. Here, we include an overall factor of $\omega$ for convenience, and we call this generator the harmonic Hamiltonian $H_{harm.} := H + \omega^2 C$. We shall come back to it shortly, but let us first discuss the state-operator correspondence in this quantization.

Seen as a conformal Killing vector field, $H_{harm.}$ gives rise to a foliation similar to that of NS-quantization, albeit slightly deformed due to the unusual scaling property of the time component. The slices contract around the points $\left(\pm\frac{1}{\omega},\vect{r}\right)$. The state-operator correspondence is thus established upon inserting operators at one of these points and, conversely, evolving a state prepared on an arbitrary slice until it reaches the slice of vanishing size. To be concrete, consider a scalar primary operator\footnote{Note that in nonrelativistic theories, there are no antiparticles and operators thus either create or annihilate excitations. Operators with a dagger denote the former. Note also that we refrain from adopting the barred notation used in NS-quantization of relativistic \acp{cft} for simplicity.} $\mathcal{O}^\dagger(\tau, \vect{r})$ in Euclidean time, with charge $Q > 0$ and conformal dimension $\Delta$. Then, similarly to Eq.~\eqref{eq:StateOpMap_NS}, we define the state
\begin{equation}
    \ket{\mathcal{O}} := \omega^\frac{\Delta}{2} \expo{-\frac{H}{\omega}} \mathcal{O}^\dagger(0) \ket{0},
\end{equation}
where $\ket{0}$ is the Schrödinger invariant vacuum state, and
\begin{equation}
    \bra{\mathcal{O}} = \bra{0} \mathcal{O}(0) \expo{-\frac{H}{\omega}} \omega^\frac{\Delta}{2}.
\end{equation}
Of course, it satisfies
\begin{equation}
\begin{aligned}
    \mathcal{Q} \ket{\mathcal{O}} & = Q \ket{\mathcal{O}} \\
    H_{harm.} \ket{\mathcal{O}} & = \omega \Delta \ket{\mathcal{O}},
\end{aligned}
\end{equation}
showing explicitly that the spectrum of $H_{harm.}$ is the same as that of $D$, as befits two operators in the same conjugacy class.

\pagebreak
The rest of the generators are obtained upon applying the same rotation as in Eq.~\eqref{eq:HarmHamil} to the original generators. For instance, $P_i,K_i,H,C$ become, respectively,
\begin{equation}
\begin{aligned}
    P^\pm_i & := \frac{1}{\sqrt{2 \omega}} P_i \pm i \sqrt{\frac{\omega}{2}} K_i \\
    L^\pm & := \frac{1}{2} \left( \frac{1}{\omega} H - \omega C \pm i D \right).
\end{aligned}
\end{equation}
They act as raising/lowering operators for the eigenvalues of $H_{harm.}$:
\begin{equation}
\begin{aligned}
    [H_{harm.}, P^\pm_i] & = \pm \omega P^\pm_i, \\
    [H_{harm.}, L^\pm] & = \pm 2 \omega L^\pm,
\end{aligned}
\end{equation}
thereby reproducing Eq.~\eqref{eq:NRCFT_ScalingLadder} in terms of the new generators. Of course, since we choose $\mathcal{O}$ to be a primary operator, $\ket{\mathcal{O}}$ is a lowest-weight state:
\begin{equation}
    P^-_i \ket{\mathcal{O}} = L^- \ket{\mathcal{O}} = 0.
\end{equation}
The corresponding representation is generated upon acting with $P_i^+$ and/or $L^+$ one or many times on $\ket{\mathcal{O}}$, which gives a tower of descendant states to be identified with the descendant operators previously introduced. From these considerations, one can also establish the \ac{ope} for \acp{nrcft}, although we refer to~\cite{Golkar:2014mwa,Goldberger:2014hca} for further details.





\subsubsection{Conformal dimensions from the harmonic frame}

From the viewpoint of the $SL(2,\mathbb{R})$ subgroup, we can then map the theory on the 'cylinder'. In fact, this just maps the time axis to itself. However, it also rearranges the slices (which extend through time and space) into codimension one planes perpendicular to the time axis. In Euclidean signature, this covers all of spacetime and $H_{harm.}$ is thus seen to generate translation along the time axis: this is the 'cylinder' Hamiltonian. The term 'cylinder frame' is however not very enlightening here, and we shall call it 'harmonic frame' instead for reasons that will become clear shortly. The map from $(\tau,\vect{r})\in (\mathbb{R}_\tau \times \mathbb{R}^d)_{NS}$ to $(\tilde\tau,\vect{\tilde{r}})\in (\mathbb{R}_\tau \times \mathbb{R}^d)_{harm.}$ is given by
\begin{equation}
\begin{cases}
    \omega\tau = \tanh(\omega\tilde\tau), \\
    \vect{r} = \frac{1}{\cosh{\omega\tilde\tau}} \vect{\tilde{r}},
\end{cases}
\qquad
\begin{cases}
    \omega\tilde\tau = \text{arctanh}(\omega\tau), \\
    \vect{\tilde{r}} = \frac{1}{\sqrt{1-(\omega\tau)^2}} \vect{r},
\end{cases}
\end{equation}
to be compared with Eq.~\eqref{eq:NStoCylinder1D}, which we obtained upon introducing by hand some factors of two in anticipation of this result.

Correlation functions can be obtained in the harmonic frame from the original ones using the following relations:
\begin{equation}
\begin{cases}
    \mathcal{O}(\tau,\vect{r}) & = (\cosh{\omega\tilde\tau})^\frac{\Delta}{2} \expo{-\frac{Q\omega}{2} \tanh(\omega\tilde\tau) \vect{\tilde{r}}^2} \, \tilde{\mathcal{O}}(\tilde\tau,\vect{\tilde{r}}), \\
    \tilde{\mathcal{O}}(\tilde\tau,\vect{\tilde{r}}) & = \left(1-(\omega\tau)^2\right)^\frac{\Delta}{2} \expo{\frac{Q\omega^2}{2} \frac{\tau\vect{r}^2}{1-(\omega\tau)^2}} \, \mathcal{O}(\tau,\vect{r}),
\end{cases}
\end{equation}
which again resembles the one-dimensional relativistic case, see Eq.~\eqref{eq:Operators_FromNStoCylinder1D}. The exponentials here and in the correlation functions before stem from the fact that the global $U(1)$ charge enters the Schrödinger algebra as a central extension, and the operators therefore belong to projective representations.

The two-point function in the harmonic frame with insertions along the time axis takes the following simple form:
\begin{equation}
    \ev{\mathcal{O}_{harm.}(\tilde\tau_1,\vect{0}) \mathcal{O}_{harm.}^\dagger(\tilde\tau_2,\vect{0})} = \frac{\omega^{\Delta}}{\sinh(\omega(\tilde\tau_2-\tilde\tau_1))^{\Delta}},
\end{equation}
which behaves like $\sim \expo{-\omega(\tilde\tau_2-\tilde\tau_1)\Delta}$ when $(\tau_2-\tau_1)\to\infty$. The fact that we set $\vect{\tilde{r}}_1=\vect{\tilde{r}}_2=\vect{0}$ does not matter in this limit, and one can compute $\Delta$ just as in the relativistic case:
\begin{equation}\boxed{
    \Delta = -\lim_{T\to\infty} \frac{1}{\omega T} \log \ev{\expo{-T H_{harm.}}}{\mathcal{O}}
}.\end{equation}
The rest of the discussion therefore applies verbatim, and one expects a semiclassical treatment to be equally valid when the operator $\mathcal{O}$ carries large quantum numbers.

The last crucial observation concerns the term 'harmonic' which we have not yet commented on. Looking back at Eq.~\eqref{eq:GeneratorCurrents}, we see that
\begin{equation}
    H_{harm.} = H + \int\dd{\vect{r}} \, \frac{\omega^2}{2} \vect{r}^2 \, \rho(\vect{r}).
\end{equation}
This is the Hamiltonian of the original theory \emph{coupled to an external harmonic potential}, akin to those used in ultracold atom experiments. The experimentally measured ground-state energy at criticality is therefore in exact correspondence with the large-charge conformal dimension $\Delta_Q$, which can be directly computed semiclassically from the \ac{eft} of the trapped nonrelativistic conformal superfluid! We quoted the result in $(3+1)$ dimensions in Eq.~\eqref{eq:trap-energy-Wilson-parameters}, which is now seen to be equivalent to
\begin{equation} \label{eq:DeltaQ_NR_prelim}
	\Delta_Q = \frac{\sqrt{\xi}}{4} (3Q)^\frac{4}{3} - \sqrt{2}\pi^2 \xi \left(c_1 - \frac{9}{2}c_2\right) (3Q)^\frac{2}{3} + \order{Q^\frac{5}{9}}.
\end{equation}

\subsection{Nonrelativistic large-charge landscape} \label{sec:NR_largeQ_landscape}

The fate of the macroscopic limit in \acp{nrcft} has not yet been investigated systematically. There is however a natural scenario akin to the 'simple case' \eqref{eq:MacroLim_Simple} discussed earlier\footnote{When comparing the two situations, keep in mind the fact that $d$ now denotes the number of spatial dimensions.}, namely, when the
\begin{equation}
\begin{aligned}
    \text{energy density } & \sim \omega^{\frac{d}{2}+1} Q^{\alpha-\frac{1}{2}} \\
    \text{charge density } & \sim \omega^\frac{d}{2} \, Q^\frac{1}{2}
\end{aligned}
\end{equation}
are fixed as we take the limit $\omega\to0$, $Q\to\infty$, for some $\alpha\in\mathbb{R}_+$. Here, the densities are taken at the center of the trapping potential.\footnote{We shall discuss this technicality at length in due time.} The second relation tells us to think of the volume of the system as $\sim\omega^{-\frac{d}{2}} Q^\frac{1}{2}$, and the first relation then implies $\Delta_Q \sim Q^\alpha$ at leading order. The two relations combined give
\begin{equation}
    \text{energy density} \sim Q^{\alpha-\frac{d+1}{d}} \times \text{(charge density)}^\frac{d+2}{d}.
\end{equation}
This suggests that $\alpha=\frac{d+1}{d}$ is again the most natural scaling behaviour in the absence of strong additional symmetry constraints. Note that it indeed matches Eq.~\eqref{eq:DeltaQ_NR_prelim}. The above relation however also suggests that a sufficient condition for the spontaneous breaking of conformal symmetry is the same as the one established in the relativistic case. This is expected to be relevant in the context of nonrelativistic supersymmetric theories, see~\cite{Meyer:2017zfg} for a related discussion.

This said, to the best of our knowledge, there has not been any work at large charge in supersymmetric \acp{nrcft}.\footnote{There is an important and fascinating exception to this statement, which we briefly describe in Subsection~\ref{sec:anyons}.} It would be interesting to investigate the relevance of the large-charge program in this class of theories both from a condensed matter perspective and from a 'high-energy' one. Similarly, \ac{nrcft} in the context of holography is not as well understood as its relativistic counterpart, see however~\cite{Balasubramanian:2008dm,Son:2008ye,Taylor:2008tg,Bekaert_2012,Jensen:2014aia,Taylor:2015glc,Kolekar:2018sba}. Additionally, there is a symmetry obstruction to enhancing the nonrelativistic Fermi liquid \ac{eft} to the conformal case~\cite{Rothstein:2017niq}.\footnote{\textbf{Note added:} this statement was recently invalidated in~\cite{Delacretaz:2025ifh}.}

Altogether, this puts the conformal superfluid state at the center of the stage as the \emph{only} nonrelativistic large-charge universality class explored to date. A correct understanding of this class is however more challenging than its relativistic counterpart, which is the subject of this thesis. The state-of-the-art understanding of \acp{nrcft} in sectors of large charge is based on the work of the author and his collaborators~\cite{Hellerman:2020eff,Orlando:2020idm,Pellizzani:2021hzx,Hellerman:2021qzz,Hellerman:2023myh}, building upon important earlier results \cite{Son:2005rv,schroedinger,Kravec:2018qnu,Kravec:2019djc}. Let us review previous achievements first.

\subsubsection{Previous achievements}

Since the nonrelativistic large-charge program is deeply rooted in the study of the unitary Fermi gas, it is fair to say that several adjacent results exist in the condensed literature. Perhaps the two most relevant works are the Monte-Carlo simulations performed in~\cite{Chang_2007,Blume_2007,Endres:2012cw,Yin_2015} to evaluate $\Delta_Q$ non-perturbatively. They predate the modern large-charge program, so it would be interesting to refine these simulations and make contact with those established in the relativistic case. This is also a good place to mention the seminal works of Shina Tan~\cite{Tan_2008a,Tan_2008b,Tan_2008c}, which marked a major development in the description of many-body systems with short-range interaction, including cold Fermi gases. In these works, a set of universal relations connecting short-range two-body correlations to thermodynamics were established in terms of the so-called \emph{contact} $C$. These results were phrased in the language of many-body physics, but subsequent works used \acs{qft} methods such as the \acs{ope} to reproduce and sometimes generalize them~\cite{Braaten:2008uh,Braaten:2008bi} (see also~\cite{Werner_2006,Mehen:2007dn} for important earlier developments). We shall not discuss these results here, and refer the reader to this beautiful series of papers.

When it comes to the study of the large-charge \ac{eft} \emph{per se}, a major obstacle that one faces stems from the coupling to an external harmonic potential required by the nonrelativistic state-operator correspondence. While this is what establishes a direct relation with the experimental setup used in ultracold atoms experiments, it also implies that one has to study an inhomogeneous superfluid configuration supported on a ball at the edge of which the particle density vanishes due to the energy barrier. The fundamental problem that this causes is the breakdown of the naive EFT, which is intrinsically controlled by the charge density. The effective description of such systems was therefore cursed by edge divergences---as first observed in the seminal work~\cite{Son:2005rv}---and further progress was obstructed.

\pagebreak
This said, the first formal applications of large-charge ideas appeared in \cite{schroedinger,Kravec:2018qnu,Kravec:2019djc}, where some progress was made to uncover the structural properties of the conformal dimension of the lightest primary operator at large charge, $\Delta_Q$. The $d=3$ case reproduced the results of~\cite{Son:2005rv}, while new results were found in general $d$ and $d=2$. Subleading computations were however still out of reach due to edge divergences.

Instead, other \emph{leading-order} results were established. The authors of \cite{Kravec:2018qnu,Kravec:2019djc} derived the spectrum of phonon excitations, Eq.~\eqref{eq:PhonoDispRel}, and identified the same first two regimes featuring vortices describing operators with large spin. Explicitly, they found in $(2+1)$ dimensions the following hierarchy
\begin{equation}
    \Delta_{Q,J} =
    \begin{cases}
        J^\frac{1}{2} + \Delta_Q + \ldots & 0 \leq J \ll Q^\frac{1}{3} \\
        \sqrt{\frac{\pi c_0}{2}} J^\frac{1}{2} \log J + \Delta_Q + \ldots \qquad & Q^\frac{1}{3} \ll J \ll Q \\
        \sqrt{\frac{9\pi c_0}{2}} \frac{J^2}{Q^\frac{2}{3}} + \Delta_Q + \ldots & Q \ll J \ll Q^\frac{3}{2},
    \end{cases}
\end{equation}
where $\Delta_Q = \frac{2}{3\sqrt{2\pi c_0}} Q^\frac{3}{2}$. As in the relativistic case, the first case corresponds to a single phonon excitation carrying an angular momentum $J$, then a single vortex is created, followed by the formation of many vortices exhibiting rigid body motion. We refer to the original publication for the $(3+1)$ case.\footnote{In $d=3$, an intermediate regime---between the first and the second---is present and associated with the formation of a vortex string at the edge of the cloud of particles. The authors of~\cite{Kravec:2019djc} could not reliably apply \ac{eft} techniques in this regime, but merely make estimates. It would be interesting to revisit this problem in light of the results of this thesis.} Regimes with even larger spin have not been discussed so far. It is reasonable to expect a giant vortex to emerge here as well. However, the Regge regime was established from a bootstrap argument, namely, the lightcone limit of the crossing equation, which seems to be relativistic in nature. Whether a simple kinematic argument can be found to establish the parametric dependence of the conformal dimension on the spin when the latter is the dominant quantum number is an open question.\footnote{As general fact, the study of formal properties of \acp{nrcft} is critically underdeveloped.} They also computed the universal behaviour of the simplest large-charge three-point function with a single light primary operator $\mathcal{O}_{\delta,q}$ at position $y$ in the harmonic frame, which goes at leading order like\footnote{The constant $\zeta$ is related to the leading Wilsonian coefficient $c_0$. The explicit relation will be given in the next chapter.}
\begin{equation}
    F(v=i\omega y^2) \propto Q^\frac{\delta}{2d} \left(1 - \frac{\omega y^2}{2 \zeta Q^\frac{1}{d}} \right)^\frac{\delta}{2} \expo{-\frac{1}{2} q\omega y^2}.
\end{equation}
It would be interesting to obtain this structure from a macroscopic-limit argument. Recently, this result was vastly generalized to many light operator insertions in~\cite{Beane:2024kld}, where the computation is performed in the theory without harmonic trap. This is consistent by virtue of the macroscopic limit, as long as one is merely interested in leading-order results, which is common practice in the large-charge literature when it comes to computing pieces of \ac{cft} data other than the large-charge conformal dimension. 

\pagebreak
\subsubsection{State of the art}

At this point, the main challenge was therefore to overcome the difficulty caused by edge divergences to access subleading corrections, which possibly contained universal quantum corrections, just as in the relativistic case. As mentioned, their evaluation was technically and conceptually obstructed by the breakdown of the large-charge \ac{eft} close to the edge of the cloud of particles. It was nevertheless possible to compute the leading quantum correction to the large-charge expansion in 2+1 dimensions based on the known spectrum of phonon excitations. It was reasonable to assume that it was a universal term, although it was yet to be proven. This contribution was computed in~\cite{Orlando:2020idm}, where a soft breaking of scale invariance caused by a slightly massive dilaton was accounted for, thereby shedding light on the near-conformal phase of the theory along the lines of~\cite{Orlando:2019skh,Orlando:2020yii}.\footnote{In \cite{Beane:2024kld}, a microscopic description of the unitary Fermi gas without trap is considered, on top of which a relevant deformation related to the presence of a finite $s$-wave scattering length is added (together with an irrelevant deformation associated with effective range corrections). It would be interesting to contrast these two options to explore the near-critical regime. See also~\cite{Escobedo:2009bh,Pavaskar:2024pqf}.} Simultaneously, it was realized that the problem of edge divergences is in fact cured by the presence of edge operators complementing the original, naive \ac{eft}~\cite{Hellerman:2020eff}. This is similar to theories with sharp boundaries, although the situation here is more subtle to implement. This improved recipe helped uncovering the complete structure of the lowest conformal dimension at large charge in~\cite{Pellizzani:2021hzx}, thereby proving the universality of the quantum correction in 2+1 dimensions and opening the avenue for many more investigations. These results were later reformulated and refined in~\cite{Hellerman:2021qzz}, with the important added value of the computation of the universal quantum correction in 3+1 dimensions using a nontrivial $\zeta$-function regularization. All of these results are presented in Chapter~\ref{chap:NR_EFT}. Subsequently, the first-ever microscopic verification of the nonrelativistic large-charge predictions, including edge effects, was performed in~\cite{Hellerman:2023myh} using a large-$N$ description of the unitary Fermi gas---akin to mean-field theory---and perfect agreement was found. We present these computations in Chapter~\ref{chap:ExplicitReal}.

This concludes the overview of the nonrelativistic large-charge landscape. We mentioned several open questions, and we shall mention several more at the end of the thesis, in the hope to trigger ideas for future research.

\chapter{Nonrelativistic large-charge effective field theory} \label{chap:NR_EFT}
\setcounter{section}{0}
This chapter is based on references~\cite{Orlando:2020idm,Pellizzani:2021hzx,Hellerman:2021qzz} by the author and collaborators. The introduction sets the stage and presents in a coherent way the interplay between these works. Section~\ref{sec:NR_EFT_simple} is based on~\cite{Orlando:2020idm}, Section~\ref{sec:NR_EFT_systematicsI} on~\cite{Pellizzani:2021hzx}, and Section~\ref{sec:NR_EFT_systematicsII} on~\cite{Hellerman:2021qzz}. Appears in Section~\ref{sec:NR_EFT_quantum} the computation of quantum corrections done in~\cite{Orlando:2020idm} and \cite{Hellerman:2021qzz}.

\vspace{-5mm}
\section{Introduction} \label{sec:NR_EFT_intro}

As argued in the previous chapter, quantum critical points in nonrelativistic systems are phenomenologically rich and theoretically challenging. Like relativistic critical points, they are generically strongly coupled and therefore do not fit in the framework of standard perturbation theory. One
of the few analytical tools available to study them is to analyse the collective behaviour of a large number of particles in the system, treating the inverse particle number as a small
parameter on which to build a suitable and systematic perturbation theory.

To correctly understand the context, we recall that the phenomenological observation of the emergence of collective behaviours---namely, superfluidity, phonon excitations and vortex formation---in trapped cold atom systems was used as a guide to explore the various phases of the large-charge versus large-spin phase diagram of the superfluid universality class of relativistic \acp{cft}, culminating in the work \cite{Cuomo:2022kio}. However, these experimental setups are intrinsically nonrelativistic, and one would therefore expect that the latter observations may be even more relevant to the abstract study of \ac{nrcft} data. The surprise here comes from the fact that, more than a mere guiding principle, the connection is \emph{exact}, as established by the nonrelativistic correspondence: experimental data such as energy levels or density profiles of the trapped system \emph{must be} reproduced by the nonrelativistic large-charge \ac{eft}. Therefore, better experimental data---\emph{e.g.} close but not exactly at the critical point---may help building a sharper theoretical description, and vice versa, where \ac{eft} predictions can help improving the numerical fits and extract universal coefficients beyond the Bertsch parameter.

Our theoretical journey in this field of research therefore revolves around \acp{nrcft}, for which the state-operator correspondence establishes a bijective mapping between the spectrum of conformal dimensions of (positively charged) local operators and the energy spectrum of states in an external spherical harmonic trap~\cite{Nishida:2007pj,Goldberger:2014hca},
\begin{equation}
    A_0(\vect{r}) = \frac{m \omega^2}{2 \hbar} \vect{r}^2,
\end{equation}
as already explained. In particular, we study the structure of the conformal dimension $\Delta_Q$ of the lowest operator of fixed charge $Q \gg 1$ in an expansion in $1/Q$, which can be accessed via the ground-state energy of the trapped system with $Q$ particles confined to a spherical cloud. We work in general spatial dimension $d$.

\subsubsection{Outline of the chapter}

We study the structural dependence on the charge via \ac{eft} techniques, assuming that the corresponding state in the harmonic trap realizes superfluid spontaneous symmetry breaking.\footnote{Recall that spontaneous symmetry breaking only occurs in infinite volume, so symmetry breaking here has to be understood in terms of the microscopic limit: in the harmonic frame, the large-charge state is well-approximated by a state $\ket{flat}$ which would cause spontaneous breaking in the infinite-volume limit. This justifies using a semiclassical description, which we sometimes refer to as a 'low-energy' description by abuse of language.} We also assume that there are no additional light degrees of freedom beside the hydrodynamic Goldstone boson $\chi$. Under this assumptions, we use two different \ac{eft} descriptions.

The first employs a dilaton-dressing prescription \emph{à la} Coleman \cite{Coleman:1988aos} that implements conformal invariance. We shall refer to this prescription as \emph{dilaton \acp{eft}} and illustrate it by including just a few terms in Section~\ref{sec:NR_EFT_simple}. This construction contains all the necessary ingredients to compute the leading-order large-charge contributions to $\Delta_Q$ and to discuss the intrinsic breaking of the \ac{eft} due to the vanishing of the particle density close to the edge of the cloud of particles.

In Section~\ref{sec:NR_EFT_systematicsI}, we perform a much more systematic analysis of this \ac{eft}-building prescription and classify all types of 'classical' large-charge contributions to the conformal dimension $\Delta_Q$, building upon \cite{Hellerman:2020eff}. This is achieved thanks to a careful discussion of the renormalization of divergences associated with the aforementioned breakdown of the \ac{eft}, in a physical-cutoff procedure that will be referred to as $\delta$-layer regularization. Large-charge dilaton \acp{eft}, however, introduce a gapped radial mode which is not strictly necessary from a 'low-energy' perspective (see e.g. \cite{Brauner:2014aha} for a discussion on gapped Goldstone bosons). We do so because the description of the system in terms of the dilaton mode allows us to explore the near-conformal regime of the theory by introducing a small dilaton mass, still following Coleman \cite{Coleman:1988aos}, based on the general expectation detailed in \cite{PhysRevD.103.105026,PhysRevD.101.065018} that a dilaton-like mode appears near a smooth quantum phase transition. Note that, while the concept of conformal dimension becomes ill-defined---or rather, scheme dependent---away from criticality, the corrections induced by the dilaton mass to the ground-state energy of the trapped system are interesting \emph{per se}. In this work, one can therefore think of $\Delta_Q$ as the ground-state energy, which exactly corresponds to the conformal dimension of the lowest operator of charge $Q$ only when the dilaton mass vanishes.

After identifying the signature of this soft breaking of conformal invariance, we switch it off in Section~\ref{sec:NR_EFT_systematicsII} and move on to the second \ac{eft} description, which is an action for the Goldstone boson only. We show that the two descriptions are equivalent. 
More importantly, we reevaluate the problem of the renormalization of edge divergences using a simpler, though less physical, scheme: dimensional regularization. This is of course in perfect agreement with the previous results, but it also clarifies and makes more precise the classification achieved using $\delta$-layer regularization.

Finally, in Section~\ref{sec:NR_EFT_quantum}, we discuss the theory-independent one-loop correction to the conformal dimension $\Delta_Q$ and compute it explicitly in $d=2$ and $d=3$.
\pagebreak
\section{Warm-up: a simple class of large-charge dilaton EFTs} 

In this section, we consider a simple class of $(d+1)$-dimensional effective theories invariant under the Schrödinger group. A more general construction will be discussed in the next section, but the purpose here is to illustrate certain important properties and highlight some issues related to edge divergences that we will have to deal with subsequently. Moreover, we purposely choose to keep track of the constants $\hbar$, $m$ and the strength of the potential $\omega$. In the next two sections, we shall set them to one, so this serves as a reference point for restoring these constants, \emph{e.g.} if one wants to compare with some results in the condensed matter literature.


As mentioned, in order to build this class of theories, we apply a simple procedure pointed out by Coleman in~\cite{Coleman:1988aos}, which promotes a Lorentz-invariant theory to a conformally invariant one using a dilaton-dressing prescription. This can be straightforwardly adapted to nonrelativistic theories, where one starts with a theory invariant under the Galilean group and promotes it to the Schrödinger group\footnote{\emph{Cf.} \cite{Arav:2017plg,Argurio:2020jcq} for a similar discussion.}. For a second, let us discuss both of these cases at the same time by introducing the so-called dynamical exponent $z$, in terms of which a scale transformation reads ($\alpha\in\mathbb{R}$)
\begin{equation}
    (t, \vect{r}) \to (e^{z\alpha} t, e^\alpha \vect{r}).
\end{equation}
Obviously, $z=1$ (resp. $z=2$) corresponds to relativistic (resp. nonrelativistic) scale transformations. The dilaton is then a real scalar field transforming non-linearly under the latter, namely,
\begin{equation}
	\sigma(t, \vect{r}) \longrightarrow \sigma(e^{z\alpha} t, e^\alpha \vect{r}) + \frac{d + z - 2}{2 f} \alpha,
\end{equation}
where the dimensionful parameter $f$ may be regarded as the dilaton (inverse) decay constant, see \emph{e.g.} \cite{Komargodski:2011vj}. Any operator $\mathcal{O}_\Delta$ of dimension $\Delta$ can then be dressed with an appropriate power of $\sigma$ to become marginal:
\begin{equation} \label{eq:simple_DressingRule}
	\mathcal{O}_\Delta(t, \vect{r}) \stackrel{\text{dressing}}{\longrightarrow} e^{\frac{2 (\Delta - d - z)}{d+z-2} f \sigma(t, \vect{r})} \mathcal{O}_\Delta(t, \vect{r}).
\end{equation}
In particular, the identity operator should be dressed as $\mathbb{1} \to e^{-\frac{2(d+z)}{d+z-2} f \sigma}$.

We are interested in the unified description of the class of critical theories featuring a global $U(1)$ symmetry, as is the case by default in the Schrödinger group\footnote{As a reminder, the generators of the Galilean algebra are given by: time and spatial translations, rotations and Galilean boosts, that is, uniform motion with velocity $\vect{v}$. It admits a $U(1)$ central extension associated with the particle number generator. Together with the nonrelativistic scale transformation,
\begin{equation}\label{eq:nonrela_dila_transfo}
    (t,\vect{r})  \rightarrow (t',\vect{r}') = (e^{2\alpha}t,e^{\alpha} \vect{r}),
\end{equation}
where $\alpha$ is a real parameter, and nonrelativistic \ac{sct}, 
\begin{equation}\label{eq:nonrela_SCT} 
    (t,\vect{r})  \rightarrow (t', \vect{r}') = \left(\frac{t}{1 + \lambda t}, \frac{\vect{r}}{1 + \lambda t}\right),
\end{equation}
with $\lambda$ a real parameter, they form the Schrödinger algebra.}, under the assumption that they realize the superfluid spontaneous breaking pattern at large charge. The task is simple enough: first construct the most general \ac{eft} for the Goldstone mode $\chi$ invariant under Galilean or Lorentz symmetry that nonlinearly realizes the $U(1)$ symmetry, and then appropriately dress the operators with the dilaton $\sigma$ so as to make them marginal. Note that in general, these two fields can then be conveniently recast as\footnote{We use the letter $\psi$ in analogy with the usual literature on Schrödinger-invariant theories, although it is a bosonic field here. In Chapter~\ref{chap:ExplicitReal}, however, we will study a model of interacting fermions that belong to this superfluid large-charge universality class.}\textsuperscript{,}\footnote{Note that this definition excludes $\psi = 0$, but keep in mind that this field is meant to be expanded around a non-zero expectation value in the charged ground state.}
\begin{equation} \label{eq:simple_repackage}
	\psi = \frac{1}{f} e^{-f \sigma - i \chi},
\end{equation}
and we shall therefore refer to $a \equiv |\psi| = \frac{1}{f} e^{-f \sigma}$, with dimension $[a]=\frac{d+z-2}{2}$, as the \emph{radial mode}. The most general large-charge dilaton \ac{eft} thus contains every possible operator obtained from dilaton-dressed Lorentz/Galilean-invariant expressions---infinitely many of which exist---as well as derivative terms for the dilaton itself. In the \ac{eft} jargon, every such term comes with a  model-dependent \emph{Wilsonian coefficient}. \acp{eft} are naturally equipped with \ac{uv} and \ac{ir} cutoffs, $\Lambda_{UV}$ and $\Lambda_{IR}$, beyond which the effective description breaks down. The utility of \acp{eft} therefore relies on the notion of large separation of scales $\Lambda_{UV} \gg \Lambda_{IR}$ which, as we shall soon see, is implemented in our case by the large-charge limit.

At large charge, the previously introduced radial mode gets gapped with mass of order $\Lambda_{UV}$ under spontaneous breaking of the $U(1)$ symmetry and thus decouples. Upon integrating it out, one would recover an equivalent large-charge effective action for the Goldstone $\chi$ alone in the form of what is sometimes referred to as a \ac{nlsm}, as opposed to dilaton \acp{eft} which may be referred to as \acp{lsm}. Various methods, e.g. the coset construction \cite{Monin:2016jmo, Kravec:2018qnu}, allow for a systematic construction of this kind of \acp{eft}. Section~\ref{sec:NR_EFT_systematicsII} discusses this description further. Here, we stick to the dilaton \ac{eft} in view of including a small dilaton mass deformation in Section~\ref{sec:NR_EFT_systematicsI}.

\vspace{-3mm}
\label{sec:NR_EFT_simple}
\subsection{The action}

We now go back specifically to nonrelativistic theories, hence $z=2$. The main building block for the construction of a Galilean-invariant action for the Goldstone mode $\chi$ in the presence of an external harmonic trap is
\begin{equation}
	U = \del_t \chi - A_0(\vect{r}) - \frac{\hbar}{2m} \del_i \chi \del_i \chi,
\end{equation}
where $m$ is the mass of the particles, and the presence of $A_0(\vect{r})$ requires some notion of general coordinate invariance, as discussed in \cite{Son:2005rv}. As we shall see in Section~\ref{sec:NR_EFT_systematicsI}, derivatives of this operator, as well as other operators featuring more derivatives of the Goldstone mode $\chi$ contribute to the effective action but, for now, let us ignore them and focus on the power series in $U$,
\begin{equation}
		\mathcal{L}(\chi) = -k_0 + \sum_{n=1}^\infty k_n U^n + \ldots,
\end{equation}
where the $k_i$'s are dimensionful Wilsonian coefficients in units of $M L^{2-d} T^{i-2}$. We now apply the dressing rule, Eq.~\eqref{eq:simple_DressingRule}, to each of these terms, which we conveniently write in terms of the radial mode. Since\footnote{The units of $f^{-1}$ and $\sigma$ are $M^\frac{1}{2} L^\frac{2-d}{2} T^{-\frac{1}{2}}$.} $[a] = \frac{d}{2}$ and $[U] = 2$, we simply obtain the following new Lagrangian:
\begin{equation} \label{eq:lagr_series}
		\mathcal{L}(\chi, a) = -\frac{\hbar \kappa}{2 m} e^{-2 f \sigma} (\del_i \sigma)^2 - k_0 a^{2+\frac{4}{d}} + a^{2+\frac{4}{d}} \sum_{n=1}^\infty k_n \cdot \left( \frac{U}{a^\frac{4}{d}} \right)^n + \ldots,
\end{equation}
in which we included a kinetic term for the dilaton, together with a new coefficient $\kappa$ with $[\kappa] = [k_1]$. Note that time derivatives for $\sigma$ would break boost invariance~\cite{Arav:2017plg}, so the first term is really the canonical nonrelativistic kinetic term for the dilaton.

From here on, we shall somewhat arbitrarily truncate the tower of operators to $n=1$, although we shall discuss in more details how to justify this in Section~\ref{sec:NR_EFT_systematicsI}. The simple class of dilaton \acp{eft} alluded to earlier corresponds to Eq.~\eqref{eq:lagr_series} with three Wilsonian coefficients: $\kappa$, $k_0$ and $k_1$. In fact, for reasons that will become clear shortly, we expect the dilaton kinetic term to contribute to \ac{nlo} in the large-charge expansion of any observable, so there are enough ingredients here to illustrate the \ac{lo} computations and discuss the breakdown of the naive \ac{eft}. Repackaging the theory in terms the scalar complex field of Eq.~\eqref{eq:simple_repackage}, we obtain a (partly) familiar model for the Schrödinger particle:
\begin{equation} \label{eq:Schroedinger_lag}
	\mathcal{L}(\psi) = c_0 \left[ \frac{i}{2} ( \psi^*D_t \psi - \psi D_t \psi^*) - \frac{\hbar}{2m} \del_i \psi^* \del_i \psi \right] - \frac{\hbar c_1}{2 m} e^{-2 f \sigma} (\del_i \sigma)^2 - \frac{\hbar d}{2(d+2) m \hbar^\frac{2}{d}} g (\psi^*\psi)^{\frac{2}{d}+1},
\end{equation}
where $D_t:=\partial_t-iA_0(\vect{r})$ and the dimensionless Wilsonian coefficients $c_i$ depend on the previous coefficients through
\begin{align}
	c_0 &= f^2 k_1, & c_1 &= \kappa - f^2 k_1, & g &= \frac{2(d+2)m}{d\hbar^2} (\hbar f^2)^{\frac{2}{d}+1} k_0.
\end{align}

Note that we have not discussed whether the dilaton-dressing prescription implements invariance not only under scale transformations, but also \ac{sct}. It is however straightforward to verify that this is indeed the case under the transformation laws~\cite{Arav:2017plg, schroedinger}
\begin{equation}
\begin{cases}
	\sigma(t, \vect{r}) & \stackrel{SCT}{\longrightarrow}~ \sigma(t, \vect{r}) - \frac{d}{2 f} \ln(1 + \lambda t) \\
	\chi(t, \vect{r}) & \stackrel{SCT}{\longrightarrow}~ \chi(t, \vect{r}) + \frac{m}{2 \hbar} \frac{\lambda \vect{r}^2}{1 + \lambda t}.
\end{cases}	
\end{equation}
The second line corresponds to the usual transformation law for $\psi$ in the Schrödinger model \cite{Hagen:1972pd, schroedinger}. Note that invariance of the $c_0$-term is ensured by the fact that the time derivative transforms under Eq.~\eqref{eq:nonrela_SCT} as: $\del_t \to (1 + \lambda t)^2 \del_t + \lambda (1 + \lambda t) r_i \partial_i$. Then,
\begin{equation}
\begin{aligned}
		\frac{i}{2} (\psi^* \dot\psi - \psi \dot\psi^*) \to (1 + \lambda t)^{d + 2} & \left[ \frac{i}{2} (\psi^* \dot\psi - \psi \dot\psi^*) + \frac{m}{2 \hbar} \frac{\lambda^2 \vect{r}^2}{(1 + \lambda t)^2} \psi^* \psi \right. \\
		& \quad \left. + \frac{i \lambda r_i}{2 (1 + \lambda t)} (\psi^* \del_i \psi - \psi \del_i \psi^*) \right]
\end{aligned}
\end{equation}
and
\begin{equation}
\begin{aligned}
		\frac{\hbar}{2m} |\del_i \psi|^2 \to (1 + \lambda t)^{d+2} & \left[ \frac{\hbar}{2m} |\del_i \psi|^2 + \frac{m}{2 \hbar} \frac{\lambda^2 \vect{r}^2}{(1 + \lambda t)^2} \psi^* \psi \right. \\
		& \quad \left. + \frac{i \lambda r_i}{2 (1 + \lambda t)} (\psi^* \del_i \psi - \psi \del_i \psi^*) \right].
\end{aligned}
\end{equation}
Subtracting the second expression from the first shows invariance under \ac{sct}. The last two terms of Eq.~\eqref{eq:Schroedinger_lag} are readily seen to be invariant as they do not contain time derivatives. For instance, invariance of the $c_1$-term is shown as follows:
\begin{equation}
	e^{-2 f \sigma} (\del_i \sigma)^2 \to e^{-2 f \sigma + d \ln(1 + \lambda t)} [(1 + \lambda t) \del_i \sigma]^2 = (1 + \lambda t)^{d+2}  e^{-2 f \sigma} (\del_i \sigma)^2.
\end{equation}
This being said, we are ready to move on to the computation of the scaling dimension $\Delta_Q$.

\subsection{Large-charge conformal dimension}%
\label{sec:semiclass}

We start from the Lagrangian constructed above, expressed in terms of $\chi$, the radial mode $a$ and the dimensionless Wilsonian coefficients $c_i$:
\begin{equation}\label{eq:LSMLagU}
	\mathcal{L}(\chi, a) = c_0 a^2 U - \frac{\hbar d}{2(d+2) m \hbar^\frac{2}{d}} g a^{2+\frac{4}{d}} - \frac{c_1 \hbar}{2 m} (\partial_i a)^2,
\end{equation}
so that the \ac{eom} read
\begin{equation}\label{eq:eom}
\begin{cases}
	a(t,\vect{r})^{\frac{4}{d}+1} & = \frac{c_0 \hbar^\frac{2}{d}}{g} \left[ \frac{2 m}{\hbar} a(t,\vect{r}) U(t,\vect{r}) + \frac{c_1}{c_0} \nabla^2 a(t,\vect{r}) \right], \\
	0 & = \del_t \rho(t,\vect{r}) + \del_i j_i(t,\vect{r}),
\end{cases}
\end{equation}
where the charge density and the current are given by
\begin{align}
	\rho(t,\vect{r}) &= \frac{c_0}{\hbar} \psi^* \psi = \frac{c_0}{\hbar} a^2, & j_i(t,\vect{r}) &= -\frac{\hbar}{m} \rho  \del_i \chi.
\end{align}
The second \ac{eom} is usually referred to as the continuity equation. Note also that the charge density $\rho$ has units of $L^{-d}$. This makes the total charge dimensionless as it should be. Finally, the Hamiltonian density is given by
\begin{equation} \label{energy_density}
	\mathcal{E} = \frac{\hbar}{2 m} \left[ c_0 a^2 (\partial_i \chi)^2 + \frac{2 m c_0}{\hbar} A_0(\vect{r}) a^2 + c_1 (\partial_i a)^2 + \frac{d g}{(d + 2) \hbar^\frac{2}{d}} a^{\frac{4}{d}+2} \right].
\end{equation}


\subsubsection{Ground-state solution and scales}

As mentioned, we work with the underlying assumption that the ground state of the trapped system at fixed large charge realizes spontaneous breaking of the $U(1)$ symmetry. This requires us to look for a time-dependent solution to the \acp{eom} for $\chi$ and a spherically symmetric one for the radial mode:
\begin{equation} \label{eq:GS_profile}
  \begin{cases}
    \ev{a} = v(r), \\
    \ev{\chi} = \mu \cdot t.
  \end{cases}
\end{equation}
In this sense, $\mu$ is interpreted as a chemical potential sourcing the fixed $U(1)$ global charge $Q$, and it is fixed, together with $v(r)$, by solving
\begin{equation} \label{eq:GS_EoM}
  \begin{cases}
    v(r)^{\frac{4}{d}+1} = \frac{c_0 \hbar^\frac{2}{d}}{g} \left[ \frac{2 m}{\hbar} v(r)  U_0(r) + \frac{c_1}{c_0}  \left( \del_r^2 v(r) + \frac{d - 1}{r} \del_r v(r) \right) \right], \\
    Q = \int \dd{\vect{r}} \rho_0(r) = \frac{c_0}{\hbar} \int \dd{\vect{r}} v(r)^2.
  \end{cases}
\end{equation}
Here, \(U_0(r)\) is the expectation value of $U$ in this background, namely,
\begin{equation}
  U_0(r) := \ev{U} = \mu - \frac{m \omega^2}{2 \hbar} r^2, 
\end{equation}
and the charge density of the ground state is denoted by $\rho_0(r) := \frac{c_0}{\hbar} v(r)^2$. Importantly, the latter is not a constant in space due to the presence of the harmonic potential. Although we are going to be more explicit shortly, in the regime where large separation of scales is realized in the \ac{eft}, the derivative terms in the first equation of \eqref{eq:GS_EoM} contribute at a parametrically lower order compared to the other terms. This is guaranteed by construction of the \ac{eft}, where derivatives necessarily introduce factors of $\Lambda_{IR}$ and, in turn, their contributions to any observables is suppressed by powers of $\epsilon := \frac{\Lambda_{IR}}{\Lambda_{UV}}$ with respect to terms with less derivatives. Therefore, to leading order, $\rho_0(r)$ has the same profile as $U_0(r)^\frac{d}{2}$, and they both vanish at a distance that will be referred to as the {\em cloud radius} \(R_{\text{cl}}\), given by
\begin{equation}
    R_{\text{cl}} := \sqrt{\frac{2\hbar\mu}{m\omega^2}}.
\end{equation}
This is the \ac{ir} cutoff of our effective description, that is, the longest fluctuation that can be captured by the \ac{eft}. Note that this is equivalent to a turning point analysis, and the cloud of particles has radius $R_{cl}$ only in a classical sense: the boundary undergoes small fluctuations that do not affect our effective description, see \cite{Hellerman:2020eff} for more details. In order to simplify the computations, we introduce the dimensionless coordinate
\begin{equation}
    z := 1 - \frac{r^2}{R_{cl}^2},
\end{equation}
which measures the distance from the classical boundary of the cloud. Since spherical symmetry is preserved by the superfluid ground state, it will be convenient to express every expectation value as a function of $z$.

Additionally, the chemical potential itself sets a length scale, which we normalize as
\begin{equation}
    R_\mu := \sqrt{\frac{2\hbar}{m\mu}}.
\end{equation}
It may be regarded as the \ac{uv} cutoff. This is also clear from the fact that the radial mode fluctuations around the superfluid background \eqref{eq:GS_profile} acquire a quadratic term upon expanding $a^{2+\frac{4}{d}}$, corresponding to a 'mass' inversely proportional to $R_\mu$. Our expansion parameter is therefore given by
\begin{equation}
    \epsilon = \frac{R_\mu}{R_{cl}} = \frac{\omega}{\mu}.
\end{equation}
By consequence, large separation of scales is achieved by taking $\mu$ large in units of $\omega$. Expectedly, this will in turn correspond to the large-charge limit, $Q\gg1$.

It is now straightforward to solve the \ac{eom} for $v$ (the first equation in \eqref{eq:GS_EoM}) order by order in $\epsilon$, which takes the suggestive form
\begin{equation}
	\left( \frac{v(z)}{v_{hom}} \right)^\frac{4}{d} = z + \epsilon^2 \frac{c_1}{c_0} \frac{(1-z) v'' - \frac{d}{2} v'}{v}.
\end{equation}
Primes correspond to derivatives with respect to $z$, and
\begin{equation}
    v_{hom} = \sqrt{\hbar} \left[ \frac{4 c_0}{g R_\mu^2} \right]^\frac{d}{4}  
\end{equation}
is the expectation value of the radial mode in the homogeneous system, that is, without the harmonic trap (\emph{i.e.} $\omega \to 0$, which implies $R_{cl} \to \infty$ and $z \to 1$).
From this, it is fairly easy to conclude that the ground-state configuration that solves this equation to \ac{nlo} is given by
\begin{equation} \label{eq:radial_vev}
	v(z) = v_{hom} z^\frac{d}{4} \left[ 1 - \epsilon^2 \frac{c_1}{c_0} \frac{d^2}{64} \left\{ \frac{4 - d}{z^3} + \frac{3 d - 4}{z^2} \right\} + \mathcal{O}(\epsilon^4) \right].
\end{equation}
As anticipated, the term proportional to $c_1$ gives a subleading correction. However, this only true away from the edge of the cloud. As we approach it, $z^3$ in the denominator (the most divergent term) starts competing with $\epsilon^2$ and triggers the breakdown of the \ac{eft} when
\begin{equation}
  \frac{\epsilon^2}{z^3} \sim \order{1}.
\end{equation}
The renormalization of these edge divergences requires a careful treatment and is the subject of Sections~\ref{sec:NR_EFT_systematicsI} and \ref{sec:NR_EFT_systematicsII}. For now, we shall content ourselves with an estimate of the subleading contributions one would obtain after a proper renormalization. This is achieved by simply cutting off any divergent integral over $z$ by restricting them to the interval $[\delta, 1]$, with
\begin{equation}
    \delta = \epsilon^\frac{2}{3}.
\end{equation}
This---or rather, the more precise version thereof presented in Section~\ref{sec:NR_EFT_systematicsI}---is referred to as $\delta$-layer regularization.

\subsubsection{The charge and the ground-state energy density}

With this somewhat incomplete prescription in mind, we move on to relating the charge and the chemical potential, thereby showing that separation of scales is indeed achieved by the large-charge condition.

The ground-state charge density is related to the chemical potential $\mu$ through $\rho_0 = \frac{c_0}{\hbar} v^2$, \emph{i.e.}
\begin{equation}
	\rho_0(z) = \rho_{hom}  z^\frac{d}{2} \left[ 1 - \epsilon^2 \frac{c_1}{c_0} \frac{d^2}{32} \left\{ \frac{4 - d}{z^3} + \frac{3 d - 4}{z^2} \right\} + \mathcal{O}(\epsilon^4) \right],
\end{equation}
where the homogeneous profile
\begin{equation}
  \rho_{hom} = c_0 \left[ \frac{2 c_0 m}{\hbar g} \mu \right]^{\frac{d}{2}}
\end{equation}
is the ground-state charge density configuration in the absence of the harmonic trap. Conceptually, it merely remains to integrate $\rho_0(z)$ over the volume of the cloud, that is, over $z\in[0,1]$. But it should be clear by now that this introduces divergences, which we regulate as
\begin{equation} \label{eq:charge_div}
	Q = \frac{c_0 \Omega_d}{\hbar} \int_0^{R_{cl}} \dd{r} r^{d-1}  v(r)^2 = Q_{LO} \left[ 1 - I_{\text{div}} + \mathcal{O}(\epsilon^4) \right],
\end{equation}
where the leading term $Q_{LO}$ is defined through
\begin{equation}
	Q_{LO} = \frac{1}{(\xi  \epsilon)^d} = \frac{1}{\xi^d}  \left( \frac{\mu}{\omega} \right)^d,
\end{equation}
and $\xi = \sqrt{\frac{g}{4 \pi c_0}} \left[ \frac{2 \Gamma(d)}{c_0 \Gamma\left( \frac{d}{2} \right)} \right]^\frac{1}{d}$ is a dimensionless constant, generically of order one.
The divergent part $I_{\text{div}}$ that needs to be regularized is given by
\begin{equation} \label{eq:I_div}
	I_{\text{div}} = \epsilon^2 \frac{c_1}{c_0} \frac{d^2 \Gamma(d)}{16 \Gamma^2 \left( \frac{d}{2} \right)} \int_\delta^1 \dd{z}~(1-z)^{\frac{d}{2}-1} \left\{ \frac{4 - d}{z^{3-\frac{d}{2}}} + \frac{3 d - 4}{z^{2-\frac{d}{2}}} \right\}
    \sim
    \order{\epsilon^\frac{d+2}{3}},
\end{equation}
where we indicated its leading contribution after removing the $\delta$-layer. This is of course much smaller than one and corresponds to a subleading correction in Eq.~\eqref{eq:charge_div}, as it should. One can therefore invert the relationship between $Q$ and $\mu$ to obtain 
\begin{equation}
  \mu = (\omega \xi) Q_{LO}^{\frac{1}{d}} = (\omega \xi) Q^{\frac{1}{d}} \left[ 1 + \order{Q^{-\frac{d+2}{3d}}} \right].
\end{equation}
It is evident that large separation of scales is indeed achieved in the large-charge regime, $Q\gg1$.

Similarly, in the superfluid ground-state given in Eq.~\eqref{eq:GS_profile}, the energy density (Eq.~\eqref{energy_density}) becomes
\begin{equation}
	\mathcal{E}_0(z) = \frac{d}{d + 2} \frac{2 c_0 \hbar}{m R_\mu^2} v^2 \left[ \left( 1 + \epsilon^2 \frac{c_1}{c_0} \left( \frac{v'}{v} \right)^2 \right) \left( 1 + \frac{2}{d} \right)  (1-z) + \left( \frac{v}{v_{hom}} \right)^\frac{4}{d} \right],
\end{equation}
into which we can now plug Eq.~(\ref{eq:radial_vev}) for $v(z)$ to get
\begin{equation} \label{eq:E0_final}
	\mathcal{E}_0(z) = \mathcal{E}_{hom} z^\frac{d}{2} \left[ 1 + \frac{2 (1-z)}{d} - \epsilon^2 \frac{c_1}{c_0} \frac{d (d + 2)}{32} \left\{ \frac{4 - d}{z^3} + \frac{3 d - 6}{z^2} + \frac{2}{z} \right\} + \mathcal{O}(\epsilon^4) \right],
\end{equation}
where the leading-order term is given by
\begin{equation}
  \mathcal{E}_{hom} = \frac{d}{d + 2} \frac{g \hbar^2}{2 m} \left[ \frac{4 c_0}{g R_\mu^2} \right]^{\frac{d}{2}+1}.
\end{equation}
The term proportional to $c_1$ again signals the breakdown of the perturbative expansion close to the edge of the cloud, and $\delta$-layer regularization is required to perform the integration over the volume of the cloud.

\subsubsection{Conformal dimension}

We are finally ready for the last computation of this section: the conformal dimension of the lowest operator of charge $Q\gg1$, corresponding to the superfluid ground-state energy of the trapped system via the relation
\begin{equation}
	\Delta_Q = \frac{E_0}{\hbar \omega},
\end{equation}
where $E_0$ is the integral over the cloud of the ground-state energy density Eq.~(\ref{eq:E0_final}). Expressing the result in terms of $Q$, we find
\begin{equation}
	\Delta_Q = \frac{d}{d + 1} \xi Q^\frac{d+1}{d} + Q^\frac{d-1}{d}  \frac{c_1}{\xi c_0}  \frac{d^2 \Gamma(d)}{8 \Gamma^2\left( \frac{d}{2} \right)} \int_0^1 dz~\frac{(1-z)^\frac{d}{2}}{z^{2-\frac{d}{2}}} + \ldots.
\end{equation}
The leading-order dependence of the conformal dimension on $Q$ is thus exactly as in the relativistic case. The second term is  a priori suppressed by a factor of $Q^{-\frac{2}{d}}$ with respect to the first one, but gives in fact
\begin{equation}
    Q^\frac{d-1}{d}  \frac{c_1}{\xi c_0}  \frac{d^2 \Gamma(d)}{8 \Gamma^2\left( \frac{d}{2} \right)} \int_0^1 dz~\frac{(1-z)^\frac{d}{2}}{z^{2-\frac{d}{2}}} =
    \begin{cases}
        \frac{c_1}{6 \xi c_0} \sqrt{Q} \log Q + \order{\sqrt{Q}} \qquad & (d=2), \\
        \frac{d^3}{8(d-2)} \frac{c_1}{\xi c_0} Q^\frac{d-1}{d} & (d > 2).
    \end{cases}
\end{equation}
To obtain the above result, we had to remove the $\delta$-layer from the integral in $d=2$, unlike in $d=3$ where it was convergent. It is important to note at this stage that multiplying $\delta$ by an arbitrary $\order{1}$ number and implementing a proper renormalization of the divergences cannot affect the coefficient of the logarithm\footnote{This was first argued in this context in \cite{Kravec:2018qnu}.} in $d=2$. Moreover, we mentioned in the first chapter that it was established in the seminal work~\cite{Son:2005rv} that in $d=3$, the \acs{nnlo} corrections are of order $Q^\frac{5}{9}$, \emph{cf.} Eq.~\eqref{eq:trap-energy-Wilson-parameters}. This will be re-derived and generalized in Section~\ref{sec:NR_EFT_systematicsI}, but for now, we have reached the following conclusion:
\begin{equation}
  \label{eq:conformal-dimensions}
\Delta_Q =
\begin{cases}
	\frac{2}{3} \xi Q^\frac{3}{2} + \frac{c_1}{6 \xi c_0} \sqrt{Q} \log Q + \order{\sqrt{Q}} \qquad & (d=2), \\
	\frac{3}{4} \xi Q^\frac{4}{3} + \frac{27}{8} \frac{c_1}{\xi c_0} Q^\frac{2}{3} + \mathcal{O}(Q^\frac{5}{9}) & (d = 3).
\end{cases}
\end{equation}
These results were discussed in~\cite{Kravec:2018qnu} using the \ac{nlsm} obtained via the coset construction.

\subsection{Lessons from this simple class of theories}

This section was devoted to the study of a simple, though non-trivial, class of large-charge dilaton \acp{eft}. The goal was to illustrate how to exploit the nonrelativistic state-operator correspondence to compute the smallest conformal dimension $\Delta_Q$ in a fixed large-charge sector by computing the corresponding superfluid ground-state energy in the trap. Somewhat surprisingly, the leading dependence of $\Delta_Q$ on the charge $Q\gg1$ is exactly the same as in the relativistic case, namely,
\begin{equation}
    \Delta_Q = a_1 Q^\frac{d+1}{d} + \ldots.
\end{equation}
We also included just enough ingredients in this construction to explore the first correction to this expression, and it hints at the fact that there might be a tower of contributions suppressed by integer powers of $Q^\frac{2}{d}$ with respect to the leading term, as in the relativistic case again. However, the presence of the $\sqrt{Q} \log Q$ term in $d=2$ and the further $\order{Q^{-\frac{5}{9}}}$ correction in $d=3$ (that we were not able to see in this simple example) have no counterparts in the relativistic setup. The fact that the state-operator correspondence requires to couple the system to an \emph{inhomogeneous} external potential introduces these extra structures in the large-charge expansion. Note also that, while kinematical constraints are somewhat weaker in \acp{nrcft}, the large-charge expansion is equally powerful. For these reasons, the nonrelativistic large-charge expansion may be regarded as richer than its relativistic counterpart.

It however remains to promote the analysis performed in this section to a fully systematic and sharp construction. The breakdown of an \ac{eft} in the presence of an 'extended object' like the edge of the cloud is not exotic, though: in the presence of a boundary or a defect, an effective theory is doomed to break down in the region close to it even without the inhomogeneity caused by the external potential, and an effective theory of operators living on this extend object must complement the bulk one in order for the whole description to be consistent. Implementing this in the context of the large-charge dilaton \ac{eft} is the subject of the next section.
\pagebreak
\section{Systematic analysis I: large-charge dilaton EFT}
\label{sec:NR_EFT_systematicsI}

Let us set the stage again by going through the same steps as before and filling the gaps. In this section, however, we set $\hbar=m=\omega=1$. The external harmonic potential now simply reads
\begin{equation}
		A_0(\vect{r}) = \frac{1}{2} \vect{r}^2,
\end{equation}
and restricts the support of the (classical) theory to a ball of finite radius, i.e., a cloud---or droplet---of particles at the edge of which the particle density drops. Unlike in the relativistic case where the state-operator correspondence maps the theory to a fixed background, the cloud is a dynamical object whose boundary undergoes quantum fluctuations. Even without doing the computations of the last section explicitly, it is very intuitive to understand that the \ac{eft} is not designed to describe the short-distance physics close to the edge. This was addressed already in \cite{Son:2005rv}, and discussed for the first time in the large-charge literature in \cite{Kravec:2018qnu}. The consequence is the breakdown of the effective description in this region, as manifested by edge divergences occurring at tree-level, that is, even before doing any sort of loop calculation. A physical sharp cutoff procedure was discussed in these two references, where the so-called $\delta$-layer is removed at the edge in order to regularize the theory, as illustrated in the previous section. However, a complete \emph{renormalization} prescription in terms of counterterms that yield scheme-independent results was still lacking. Recently, the authors of \cite{Hellerman:2020eff} initiated a thorough discussion on how to cure this situation within the effective theory framework: an additional edge \ac{eft} complements the naive 'bulk' \ac{eft} discussed so far. This section, based on \cite{Pellizzani:2021hzx} by the author, reviews this prescription in the language of the large-charge dilaton \ac{eft} and broadly extends the results of \cite{Hellerman:2020eff}. It also includes a discussion of the near-conformal window via the introduction of a small dilaton mass.

\subsection{Leading-order Lagrangian} \label{sec:LO_lagr}

Recall that we started from the observation that the leading-order term in the nonrelativistic large-charge dilaton \ac{eft} can be obtained by dressing powers of $U \equiv \dot\chi - A_0(r) - \frac{1}{2} (\del_i \chi)^2$ with the dilaton or, equivalently, the radial mode:
\begin{equation} \label{eq:lagr_series_bis}
		\mathcal{L}(\chi, a) = -k_0 a^{2+\frac{4}{d}} + a^{2+\frac{4}{d}} \sum_{n=1}^\infty k_n \cdot \left( \frac{U}{a^\frac{4}{d}} \right)^n.
\end{equation}
In the superfluid ground state, the $U(1)$ and conformal symmetries are spontaneously broken, where the expectation value of the Goldstone mode is $\langle \chi \rangle = \mu \cdot t$, with $\mu$ the chemical potential. The equation of motion for $a$ imposes that the ratio $\frac{U}{a^\frac{4}{d}}$ is necessarily a constant at leading order. Correspondingly, their expectation value are of the form
\begin{equation} \label{eq:a_and_U}
	\langle a \rangle^\frac{4}{d} \sim \langle U \rangle = \mu \left( 1 - \frac{r^2}{2 \mu} \right).
\end{equation}
From the form of the action, Eq.~(\ref{eq:lagr_series_bis}), one readily sees that fluctuations around $\ev{a}$ acquire a 'mass' $m_a^2 \sim \langle a \rangle^\frac{4}{d} \sim \mu$, which sets a \ac{uv} scale $R_\mu \equiv \sqrt{\frac{2}{\mu}}$. Moreover, the ground-state charge density $\rho_0 \propto \langle a \rangle^2$ is supported on the interval $r \in [0, R_{cl}]$ where $R_{cl} \equiv \sqrt{2 \mu}$ defines the classical radius of the cloud, and sets an \ac{ir} length-scale. Upon integrating the charge density over this region, one thus finds that the total charge scales, at leading-order, like
\begin{equation}
	Q \sim \mu^d.
\end{equation}
The effective theory description is under perturbative control when there is a large separation of scales,
\begin{equation}
		R_{cl} \gg R_{\mu},
\end{equation}
which amounts to requiring that the controlling parameter $\frac{R_\mu}{R_{cl}} = \frac{1}{\mu} \sim Q^{-\frac{1}{d}}$ be small. This, in turn, is achieved in the large-charge regime $Q \gg 1$.

It should be clear by now that keeping track of the massive mode $a$ in the low-energy description is redundant in that we have to account for towers of operators, as in Eq.~(\ref{eq:lagr_series_bis}), that give the same leading-order contribution to observables. Roughly speaking, integrating the radial mode out in Eq.~(\ref{eq:lagr_series_bis}) gives a single leading-order term $U^{1+\frac{d}{2}}$ in the \ac{nlsm}, and trading $a^\frac{4}{d}$ for $U$ is therefore equivalent at the level of the \ac{nlsm}. The working prescription is thus to minimally include the dilaton by selecting the simplest term in each of these towers, such that Wilsonian coefficients can be matched across the two different types of \acp{eft}. Typically,  the leading-order Lagrangian becomes
\begin{equation} \label{eq:lagr_LO}
	\mathcal{L}_{LO}(\chi, a) = c_0 a^2 U - \frac{d}{2 (d + 2)} g a^{2+\frac{4}{d}},
\end{equation}
where we renamed and rescaled the Wilsonian coefficients for convenience. Correspondingly, the ground-state energy---and therefore, the conformal dimension of the lowest operator of charge $Q\gg1$ in the system without trap---is given by (\emph{cf.} Section~\ref{sec:NR_EFT_simple})
\begin{equation} \label{eq:Delta_LO}
	\Delta_Q = \frac{d}{d + 1} \zeta Q^\frac{d+1}{d},
\end{equation}
where $\zeta = \sqrt{\frac{g}{4 \pi c_0}} \left[ \frac{2 \Gamma(d)}{c_0 \Gamma\left(\frac{d}{2}\right)} \right]^\frac{1}{d}$ is a dimensionless constant, in accordance with the \ac{nlsm} result \cite{Kravec:2018qnu}.

Recall also that we introduced the dimensionless coordinate $z \equiv 1 - \frac{r^2}{R_{cl}^2} = 1 - \frac{r^2}{2 \mu}$, which measures the distance from the classical boundary of the cloud. Since spherical symmetry is preserved by the superfluid ground state, it will be convenient to express every expectation value as a function of $z$. Useful properties are
\begin{equation} \label{eq:z_stuff}
\begin{aligned}
	& (\del_i f(\vect{r})) (\del_i g(\vect{r})) = \frac{2(1-z)}{\mu} f'(z) g'(z), \\
	& \nabla^2 f(\vect{r}) = \frac{2}{\mu} \left[ (1-z) f''(z) - \frac{d}{2} f'(z) \right], \\
	& \int_{cloud} \dd{\vect{r}} \, f(\vect{r}) = \frac{(2 \pi \mu)^\frac{d}{2}}{\Gamma\left(\frac{d}{2}\right)} \int_0^1 \dd{z} \, (1 - z)^\frac{d-2}{2} f(z),
\end{aligned}
\end{equation}
where primes refer to derivatives with respect to $z$ and $f, g$ are spherically invariant functions. Note that spatial derivatives of operators a priori make their contributions to the conformal dimension $\Delta_Q$ parametrically smaller due to the division by $\mu \sim Q^\frac{1}{d}$.

At this stage, let us point out that Eq.~(\ref{eq:lagr_LO}) in $d = 3$ corresponds to the Thomas-Fermi approximation of the unitary Fermi gas, and yields, among other things, the known expression for the doubly-integrated density experimentally measured in \cite{PhysRevLett.92.120401}, namely
\begin{equation} \label{eq:profile_LO}
\begin{aligned}
	n(r_3) & = \iint \dd{r_1} \dd{r_2} \, \rho(\vect{r}) \\
	& = \frac{2 \pi g}{5} \left[ \frac{2 c_0 \mu}{g} \left( 1 - \frac{r_3^2}{2 \mu} \right) \right]^\frac{5}{2},
\end{aligned}
\end{equation}
where the $r_i$ take values in the cloud. We shall discuss corrections to this expression soon.

\subsection{Dressing operators} \label{sec:dressing_rules}

The presence of the dilaton field, via the radial mode $a(t, \vect{r}) = \frac{1}{f} e^{-f \sigma(t, \vect{r})}$, allows for the dressing of operators to marginality, as discussed previously. At the same time, the breakdown of the effective theory near the edge of the cloud is associated with the fact that the particle density $\rho$---and hence $a$ too---vanishes \cite{Son:2005rv, Kravec:2018qnu, Orlando:2020idm}. From a large-charge perspective, this means that the dressing rule based on $a$ is only appropriate when edge effects are negligible, i.e. in the \emph{bulk} of the cloud (to be defined more precisely later). Close to the boundary, the effective theory stemming from this 'bulk dressing rule' alone fails to appropriately describe the system, and another dressing rule based on a nonvanishing, nonsingular operator needs to take over \cite{Hellerman:2020eff}. In this way, the naive bulk \ac{eft} comes together with a complementary edge \ac{eft} such that edge divergences get renormalized away.

In order to identify the correct dressing operators, we must think in terms of \ac{eft} building for systems that non-linearly realize the $U(1)$ and conformal symmetries, \emph{cf.} \cite{Hellerman:2020eff}. Consider the set of all scalar primaries $\{\mathcal{O}_i\}_i$ with nonzero finite expectation value in the charged ground state, and let ($\Delta_i$, $\mu_i$) denote their conformal dimension and their leading power-law dependence on $\mu$ when evaluated in the superfluid background. Then, the dressing operators (usually, just one) are the ones that maximize the ratio $\frac{\mu_i}{\Delta_i}$. From a microscopic point of view, this comes from integrating out the heavy modes, whose masses, in turn, coincide with the scale set by the chemical potential after spontaneous breaking of the $U(1)$ and conformal symmetries. In our dilaton-\ac{eft} approach, the heavy mode is introduced by construction as the radial mode, and the dressing operators must be built out of it. A crucial aspect to consider when applying this argument is the fact that expectation values may vanish or diverge depending on the location within the cloud of particles.

Therefore, let us consider an operator involving powers of $a$ and its derivatives $(\del_i a)^2$, \emph{i.e.}
\begin{equation}
	\mathcal{D}_{b,c} \equiv \left[ a^{2b} (\del_i a)^{2c} \right]^\frac{2}{d \cdot (b + c) + 2 c},
\end{equation}
where $b, c$ can be any positive numbers for now, and the overall power is chosen such that its dimension is fixed:
\begin{equation}
	[\mathcal{D}_{b,c}] = 2.
\end{equation}
Indeed, note that $[a^2] = d$ and $[(\del_i a)^2] = d+2$. If $\mathcal{D}_{b,c}$ was to be regarded as a dressing operator, it is straightforward to see that, for a scalar primary $\mathcal{O}_\Delta$ with dimension $\Delta$ to be dressed to marginality, the rule would be
\begin{equation}
	\mathcal{O}_{dressed} := \mathcal{O}_\Delta \cdot \mathcal{D}_{b,c}^{\frac{d + 2 - \Delta}{2}}.
\end{equation}
However, $b$ and $c$ can only take specific values as dictated by our previous discussion. For this, we need to assess the power-law dependence on $\mu$ of the expectation value of $\mathcal{D}_{b,c}$ in the superfluid ground state. We find, to leading-order,
\begin{equation} \label{eq:mu_Dab}
	\langle \mathcal{D}_{b,c} \rangle \sim \mu^{1 - \frac{4 c}{d \cdot (b + c) + 2 c}} \cdot z^{1 - \frac{6 c}{d \cdot (b + c) + 2 c}},
\end{equation}
since $\langle a \rangle \sim (\mu \cdot z)^\frac{d}{4}$ and $(\del_i \langle a \rangle)^2 \sim \mu^\frac{d-2}{2} \cdot z^\frac{d-4}{2}$.

In the bulk of the cloud, we can consider $z$ to be of order 1, and the bulk dressing operator---which we recall must maximize its $\mu$-scaling---corresponds to $c=0$. This yields the natural dressing rule used in the previous section, namely,
\begin{equation} \label{eq:dress_bulk}
	\mathcal{D}_{bulk} \equiv a^\frac{4}{d},
\end{equation}
consistent with the fact that the radial mode is gapped at the cutoff scale set by $\mu$.

At the edge, however, the main constraint comes from the fact that the appropriate dressing operator can neither vanish nor diverge. In short, its leading-order dependence on $\mu$ in the ground state should feature neither positive nor negative powers of $z$: it is a constant. We must therefore set $d \cdot (b + c) = 4 c$ and conclude that
\begin{equation} \label{eq:dress_edge}
	\mathcal{D}_{edge} \equiv \left[ a^{\frac{8}{d}-2} (\del_i a)^2 \right]^\frac{1}{3}.
\end{equation}
This operator is proportional to $\left| \del_i \left(a^\frac{4}{d}\right) \right|^\frac{2}{3}$ which is equivalent to the edge dressing rule originally discussed in \cite{Hellerman:2020eff} upon trading $a^\frac{4}{d}$ for $U$ at the level of the \ac{nlsm}.


\subsection{The complete large-charge dilaton EFT}

Now that we are equipped with these dressing operators, let us discuss what the corresponding dressing rules are. The main requirement is that, given any bulk scalar primary operator $\mathcal{O}_\Delta$ allowed by symmetries and with a non-vanishing expectation value in the charged ground state, the dressed operator must be marginal. Indeed, although conformal symmetry is non-linearly realized in the superfluid \ac{eft}, no dimensionful parameter can enter its construction, so only marginal operators are allowed.

The bulk dressing rule is simple enough (and already known): the dressed operator entering the bulk \ac{eft} is given by
\begin{equation} \label{eq:BulkDressingRule}
    \mathcal{O}_{bulk} := \mathcal{O}_\Delta \cdot a^\frac{2(d+2-\Delta)}{d}.
\end{equation}
The edge dressing rule, on the other hand, comes with an extra ingredient: every edge \ac{eft} operator is the projection of a bulk operator $\mathcal{O}_\Delta$ on the codimension-one (fluctuating, as opposed to sharp) boundary of the cloud and is thus incorporated into the action by including an operator-valued Dirac $\delta$-function $\delta(\mathcal{D}_{bulk})$ of dimension $-2$, which is then dressed to marginality using Eq.~(\ref{eq:dress_edge}). The dressed operators entering the edge \ac{eft} are therefore of the form
\begin{equation} \label{eq:edge_dressing}
\begin{aligned}
	\mathcal{O}_{edge} & := \mathcal{O}_\Delta \cdot \delta(\mathcal{D}_{bulk}) \cdot \mathcal{D}_{edge}^\frac{d+4-\Delta}{2} \\
	&~ = \mathcal{O}_\Delta \cdot \delta(a^\frac{4}{d}) \cdot \left[ a^{\frac{8}{d}-2} (\del_i a)^2 \right]^\frac{d+4-\Delta}{6}.
\end{aligned}
\end{equation}
Given that the expectation value of the radial mode is of the form (cf. Eq.~(\ref{eq:a_and_U}))
\begin{equation}
	v(z) = v_{hom} \cdot z^\frac{d}{4} \cdot [1 + \text{subleading}],
\end{equation}
where we recall tha $v_{hom}$ is the superfluid ground-state solution of Eq.~(\ref{eq:lagr_LO}) in the homogeneous case (i.e. without trap), the Dirac $\delta$-functional becomes
\begin{equation}
	\delta(v^\frac{4}{d}) = \delta\left( \frac{v^\frac{4}{d}}{v_{hom}^\frac{4}{d}} \right) \frac{1}{v_{hom}^\frac{4}{d}} = \frac{\delta(z)}{v_{hom}^\frac{4}{d}}.
\end{equation}
This illustrates how to make sense of the Dirac $\delta$-functional within the large-charge \ac{eft}, showing that edge \ac{eft} operators are correctly located at $z = 0$.

With these two dressing rules in mind, let us discuss the candidate operators $\mathcal{O}_\Delta$ to be dressed. The general construction of operators as constrained by general coordinate invariance was investigated in \cite{Son:2005rv} where it was shown that, besides $U$, $(\del_i U)^2$, $a$ and $(\del_i a)^2$, another allowed building block is
\begin{equation}
	Z \equiv \nabla^2 A_0 - \frac{1}{d^2} \left( \nabla^2 \chi \right)^2,
\end{equation}
with dimension $[Z] = 4$ and expectation value in the superfluid ground state given by\footnote{The fact that its expectation value does not scale with $\mu$ allows us to conclude a posteriori that $Z$ could not enter the dressing rules that we have just established.} $\langle Z \rangle = d$. From there, it is easy to convince oneself that any other operator with more derivatives of $\chi$ would have a vanishing contribution in the superfluid ground state. Hence, any operator $\mathcal{O}_\Delta$ used to build \ac{eft} operators is necessarily a composite operator made out of integer powers of $U$, $(\del_i U)^2$, $a$, $(\del_i a)^2$ and $Z$. For the same reason as explained in the construction of the leading-order Lagrangian, however, it is enough for the purpose of dilaton-\ac{eft} building to discard operators made out of $U$ and $\partial_i U$ as long as the corresponding operators involving the radial mode are minimally included in the action. The \ac{nlsm} is blind to this, and the Wilsonian coefficients can be matched across the two descriptions\footnote{In turn, discarding $U$ is very convenient because it then only appears in the Lagrangian density as given in Eq.~(\ref{eq:lagr_LO}), and the Legendre transform required to pass to the Hamiltonian $\mathcal{H} = \frac{\del \mathcal{L}}{\del U} \dot\chi - \mathcal{L}$ is trivial: an operator made out solely of $a$ and $\partial_i a$ simply gets its sign flipped.}. Finally, we can strip off powers of $a$ as they will be appropriately restored by the dressing. This dramatically reduces the possible building blocks for the \ac{eft} as we merely need to dress operators of the form
\begin{equation} \label{eq:Omn}
		\mathcal{O}^{(m,n)} \equiv (\del_i a)^{2m} Z^n,
\end{equation}
with $m, n$ two positive integers. Its dimension is $[\mathcal{O}^{(m,n)}] = (d+2)m + 4n$, and the bulk dressing rule, Eq.~\eqref{eq:BulkDressingRule}, instructs us how to turn it into a marginal operator entering the bulk \ac{eft}, namely,
\begin{equation} \label{eq:Omn_bulk}
\begin{aligned}
	\mathcal{O}^{(m,n)}_{bulk} & \equiv (\del_i a)^{2m} Z^n \cdot a^{\frac{2}{d}((d+2)(1-m)-4n)} \\
	& = \left( \frac{(\del_i a)^2}{a^{\frac{4}{d}+2}} \right)^m \left( \frac{Z}{a^\frac{8}{d}} \right)^n a^{\frac{4}{d}+2}.
\end{aligned}
\end{equation}
The construction of edge \ac{eft} operators is even simpler as we can strip off powers of $(\del_i a)^2$ as well, because they then get restored by the edge dressing rule. Hence, we only have to care about dressing the operators $Z^p$ ($p\in\mathbb{N}$), which results in
\begin{equation} \label{eq:Zn_edge}
	Z^p_{edge} \equiv Z^p \cdot \delta(a^\frac{4}{d}) \cdot \left[ a^{\frac{8}{d}-2} (\del_i a)^2 \right]^\frac{d+4(1-p)}{6}.
\end{equation}
The complete superfluid dilaton \ac{eft} therefore includes the leading-order piece, Eq.~\eqref{eq:lagr_LO}, and every operator given in Eqs~\eqref{eq:Omn_bulk} and \eqref{eq:Zn_edge}:
\begin{equation} \label{eq:FullDilatonEFT}
    \mathcal{L}(\chi,a) = c_0 a^2 U
    + a^{\frac{4}{d}+2} \sum_{m,n} c_{m,n} \, \left( \frac{(\del_i a)^2}{a^{\frac{4}{d}+2}} \right)^m \left( \frac{Z}{a^\frac{8}{d}} \right)^n
    + \delta(a^\frac{4}{d}) \sum_p \kappa_p \, Z^p \left[ a^{\frac{8}{d}-2} (\del_i a)^2 \right]^\frac{d+4(1-p)}{6},
\end{equation}
where the $c_{m,n}$'s and the $\kappa_p$'s are real dimensionless coefficients (Wilsonian coefficients). We also set $c_{0,0} = -\frac{d}{2 (d + 2)} g$ to match the previous notation.\footnote{Starting from a standard microscopic model, say, weakly-coupled, one would expect derivatives of the dilaton to enter the bulk action with $m=1$ only, stemming from the presence of a canonical kinetic term in the \ac{uv}. As we shall soon see, the contributions of the operators $\mathcal{O}^{(m,n)}_{bulk}$ to the large-charge expansion do not depend on $m$, hence, we shall discuss examples featuring $m=1$ only, as was already the case in Section~\ref{sec:NR_EFT_simple}.}

We can now start to understand how the presence of the edge \ac{eft} cures the problem of edge divergences faced in the previous section: operators at the edge act as counterterms by absorbing divergences in their Wilsonian coefficients. It simply remains to choose a regularization scheme.

\subsection{\texorpdfstring{$\delta$}{}-layer and renormalization of edge divergences} \label{sec:counterterms}

A physically motivated scheme is to effectively cut off a small spatial layer close the classical boundary of the cloud, as discussed in \cite{Son:2005rv, Kravec:2018qnu, Orlando:2020idm} and in the previous section. Following \cite{Hellerman:2020eff}, this regularization procedure can be made slightly more precise in terms of the dressing operators we have just constructed. Indeed, this layer is located where the bulk dressing rule starts to break down and requires the edge one to take over. This is characterized by the region where $\mathcal{D}_{bulk} \sim \mathcal{D}_{edge}$, \emph{i.e.}
\begin{equation} \label{eq:regul_condition}
	(\del_i a)^2 \sim a^{\frac{4}{d}+2},
\end{equation}
which, in the charged ground state, is satisfied when $z \sim \mu^{-\frac{2}{3}}$, and we thus define the $\delta$-layer\footnote{To contrast with the discussion in Section~\ref{sec:NR_EFT_simple}, we put a subscript $\alpha$ to indicate that $\alpha$ encodes the scheme dependence.}
\begin{equation}
		\delta_\alpha \equiv \frac{\alpha}{\mu^\frac{2}{3}} \sim \mathcal{O}\left( Q^{-\frac{2}{3d}} \right),
\end{equation}
for an arbitrary constant $\alpha \sim \mathcal{O}(1)$. We thereby define the bulk of the cloud as the region covered by the interval $\delta_\alpha \lesssim z \leq 1$.

The renormalization procedure\footnote{At this stage, it is worth mentioning another situation where an adapted version of this construction appears: effective string theory~\cite{Hellerman:2016hnf}. We mentioned it in the first chapter as one of the early applications of modern large-charge ideas. Effective string theory uses the framework of open relativistic strings with freely moving endpoints to characterize the dynamics of a confined quark/anti-quark pair. While the "bulk" of the string is an effective description of the flux tube connecting the two probes, the endpoints---effectively characterizing the probes---must be equipped with an edge \ac{eft} which corrects and enriches the effective predictions, \emph{cf.} Eq.~\eqref{eq:EST}.} thus consists in
\begin{enumerate} [itemsep=5pt,partopsep=0pt,topsep=0pt,parsep=0pt]
    \item complementing the bulk \ac{eft} with the edge \ac{eft} as discussed above,
    \item regularizing the divergent integrals of operators dressed in the bulk by removing the $\delta_\alpha$-layer from the domain of integration, wherein the regulator $\alpha$ then appears in logarithms and denominators, and thus serves to diagnose divergences, and
    \item using the edge \ac{eft} operators as counterterms whose coefficients get renormalized so as to absorb the resulting $\alpha$-dependence.
\end{enumerate}

\subsection{Large-charge conformal dimension} \label{sec:subleading}

We now put all of these ingredients together to push the large-charge expansion of $\Delta_Q$ beyond all previously known results. By construction, an \ac{eft} generically contains infinitely many terms, see in particular Eq.~\eqref{eq:FullDilatonEFT}. However, in the regime where large separation of scales is achieved---by taking $Q\gg1$ in our case---any observable computed from the \ac{eft} takes the form of an expansion in the ratio between the \ac{ir} and the \ac{uv} scales. It is therefore self-consistent to truncate the infinite series of operators entering the \ac{eft} to any desired order. In our case, observables take the form of an expansion in inverse power of the charge which receives contributions from both the bulk and the edge \acp{eft}, so the truncation must be performed in both of these sectors in a consistent way. For this, we need to evaluate to which order in the large-charge expansion every \ac{eft} operator contributes to observables. For this purpose, it is enough to perform this analysis on the conformal dimension $\Delta_Q$, which is what we turn to now.

Let us therefore consider an operator $\mathcal{O}_\Delta$ as before. We already know that it is enough to consider operators of the form $\mathcal{O}^{(m,n)}$ as defined in Eq.~\eqref{eq:Omn}, but let us make some general remarks first. The contribution of $\mathcal{O}_\Delta$ to the ground-state energy $\Delta_Q$ is obtained by integrating its expectation value, conveniently expressed as function of $z$, over the volume of the cloud using Eq.~(\ref{eq:z_stuff}). Depending on the nature of the resulting integrand, the integral may need to be regularized upon removing the $\delta_\alpha$-layer. More specifically, let $\mu[\mathcal{O}_\Delta]$ and $z[\mathcal{O}_\Delta]$ be such that the expectation value of the operator $\mathcal{O}_\Delta$ to leading-order in $\mu$ takes the form
\begin{equation}
	\langle\mathcal{O}_\Delta\rangle \sim \mu^{\mu[\mathcal{O}_\Delta]} \cdot z^{z[\mathcal{O}_\Delta]} + \mathrm{(subleading)}.
\end{equation}
The dressed operator in the bulk, Eq.~(\ref{eq:BulkDressingRule}), then has an expectation value that scales to leading order in $\mu$ as
\begin{equation}
	\langle \mathcal{O}_{bulk} \rangle \sim \mu^{\mu[\mathcal{O}_\Delta]+\frac{d+2-\Delta}{2}} \cdot z^{z[\mathcal{O}_\Delta] + \frac{d+2-\Delta}{2}} + \text{subleading}.
\end{equation}
It is now straightforward to analyze the leading contribution of this operator to $\Delta_Q$. Indeed, if
\begin{equation}
	z[\mathcal{O}_\Delta] + \frac{d+2-\Delta}{2} \leq -1,
\end{equation}
a divergence occurs when integrating over $z \in [0, 1]$. In particular, a logarithmic divergence appears when the above inequality is saturated. We thus regularize these divergences by removing the $\delta_\alpha$-layer and, accounting for the factor of $\mu^\frac{d}{2} \sim \sqrt{Q}$ from the measure (\emph{cf.} Eq.~(\ref{eq:z_stuff})), we find that $\Delta_Q$ receives contributions of the form
\begin{equation} \label{eq:DeltaQ_bulk_gen}
	\Delta_Q \ni
	\begin{cases}
		Q^{\frac{d+1}{d}-\frac{\Delta-2\mu[\mathcal{O}_\Delta]}{2d}} & \text{if} \quad \Delta < d +4 + 2 z[\mathcal{O}_\Delta] \\
		Q^{\frac{d+1}{d}-\frac{\Delta-2\mu[\mathcal{O}_\Delta]}{2d}} \cdot \log \frac{Q}{\alpha^{3d/2}} & \text{if} \quad \Delta = d +4 + 2 z[\mathcal{O}] \\
		\frac{Q^{\frac{2}{3}-(\Delta+4z[\mathcal{O}_\Delta]-6\mu[\mathcal{O}_\Delta]-2)/(6d)}}{\alpha^{(\Delta-2z[\mathcal{O}_\Delta]-d-4)/2}} & \text{if} \quad \Delta > d +4 + 2 z[\mathcal{O}_\Delta].
	\end{cases}
\end{equation}
The last two cases---the divergent ones---are $\alpha$-dependent and thus need to be renormalized using edge operators as counterterms.

Let us now be more explicit and analyze the fate of the operators $\mathcal{O}^{(m,n)}$, whose conformal dimension, leading power-law dependence on $\mu$ and corresponding $z$-scaling are given, respectively, by
\begin{equation}
\begin{aligned}
	& [\mathcal{O}^{(m,n)}] = (d+2)m + 4n, \\
	& \mu[\mathcal{O}^{(m,n)}] = \frac{d-2}{2} m, \\
	& z[\mathcal{O}^{(m,n)}] = \frac{d-4}{2} m.
\end{aligned}
\end{equation}
This immediately leads to the following contributions to $\Delta_Q$:
\begin{equation} \label{eq:DeltaQ_bulk}
	\Delta_Q \ni
	\begin{cases}
		Q^\frac{d+1-2(m+n)}{d} & \text{if} \quad 6m + 4n < d +4 \\
		Q^{\frac{2 d - 1}{3 d}-\frac{2n}{3d}} \cdot \log \frac{Q}{\alpha^{3d/2}} & \text{if} \quad 6m + 4n = d +4 \\
		\frac{Q^{\frac{2 d - 1}{3 d}-\frac{2n}{3d}}}{\alpha^{\frac{1}{2} (6m+4n-d-4)}} & \text{if} \quad 6m + 4n > d +4.
	\end{cases}
\end{equation}
Note that the divergent cases (the last two) give a contribution whose $Q$-scaling is independent of $m$. For instance, the set of operators $(\del_i a)^{2 m}$ with $m \geq 2$ in $d = 2$ all give $Q^\frac{1}{2}$-contributions after $\delta_\alpha$-regularization, but they have not yet appeared in the literature so far. The same holds for $(\del_i a)^{2 m} Z$ with $m \geq 1$ in $d = 2$, yielding $Q^\frac{1}{6}$-contributions, and similarly in $d = 3$ where sets of equally contributing operators give terms of order $Q^\frac{5}{9}$, $Q^\frac{1}{3}$, $Q^\frac{1}{9}$, etc. Again, in the context of dilaton-\ac{eft} building, minimal inclusion of such tower of operators is always sufficient, as we shall illustrate in Sec. \ref{sec:examples} by including one operator of each set.

It is fairly straightforward to perform the same analysis for the edge \ac{eft} operators of Eq.~\eqref{eq:Zn_edge} to find that their contribution to $\Delta_Q$ is of the form:
\begin{equation} \label{eq:DeltaQ_edge}
		\Delta_Q \ni Q^{\frac{2 d-1}{3d}-\frac{2p}{3d}}.
\end{equation}
This is thus seen to match the scheme-dependent contributions identified in Eq.~(\ref{eq:DeltaQ_bulk}), and one can thus conclude---somewhat expectedly---that the renormalization procedure is self-consistent. This analysis of divergences and counterterms provides us with a simple prescription to consistently truncate the effective action given in Eq.~\eqref{eq:FullDilatonEFT}. In Sec.~\ref{sec:examples}, we carry out the computation of $\Delta_Q$ in $d=2$ and $d=3$ up to corrections that scale with negative powers of the charge, but we first make some general observations.

\subsection{Properties} \label{sec:properties}
\subsubsection{Equation of motion}

Consider the leading-order Lagrangian Eq.~(\ref{eq:lagr_LO}). The \ac{eom} with respect to the radial mode $a$ then fixes $(v(z)/v_{hom})^\frac{4}{d} = z$ in the superfluid ground state, where $v_{hom} \equiv (2 \mu c_0/g)^\frac{d}{4}$ is the ground-state solution in the homogeneous case, i.e. when the trap is turned off. As we include subleading bulk and edge operators in the action, given in Eq.~(\ref{eq:Omn_bulk}) and Eq.~(\ref{eq:Zn_edge}), the \ac{eom} gets more and more complicated, although it can always be put in the form
\begin{equation} \label{eq:eom_d}
	\left( \frac{v(z)}{v_{hom}} \right)^\frac{4}{d} = z [1 + B(z, v, v', v'')]
\end{equation}
in the ground state. In the bulk, $B(z, v, v', v'') \ll 1$ and one can solve this equation order by order in an expansion in $\frac{1}{\mu}$. For example, adding the first subleading operator $-\frac{c_{1,0}}{2} \mathcal{O}^{(1,0)}_{bulk}$ yields
\begin{equation} \label{eq:v_NLO}
	\left( \frac{v(z)}{v_{hom}} \right)^\frac{4}{d} = z \left[ 1 - \frac{d}{16} \frac{c_{1,0}}{c_0} \frac{(4 - d) + (3 d - 4) z}{\mu^2 z^3} + \mathcal{O}\left( \frac{1}{\mu^4} \right) \right],
\end{equation}
which reproduces the simple result of Section~\ref{sec:NR_EFT_simple}.

\subsubsection{Chemical potential}

For a given perturbative solution $v(z)$ to the equation of motion, the chemical potential can then be expressed as a function of the charge $Q$ by inverting
\begin{equation}
	Q = \int_{cloud} \dd{\vect{r}} \, \rho(\vect{r}),
\end{equation}
with $\rho = c_0 v(z)^2$. Using Eq.~(\ref{eq:v_NLO}) and removing the $\delta_\alpha$-layer to regularize the divergent part, we find
\begin{equation}
	Q = \left( \frac{\mu}{\zeta} \right)^d \left[ 1 + \mathcal{O}\left( \mu^{-\frac{d+2}{3}} \right) \right],
\end{equation}
where $\zeta$ has already been found to be $\zeta = \sqrt{\frac{g}{4 \pi c_0}} \left[ \frac{2 \Gamma(d)}{c_0 \Gamma\left( \frac{d}{2} \right)} \right]^\frac{1}{d}$. Therefore,
\begin{equation} \label{eq:mu_generic}
	\mu = \zeta Q^\frac{1}{d} \left[ 1 + \mathcal{O}\left( Q^{-\frac{d+2}{3d}} \right) \right].
\end{equation}
It is tempting to give an explicit expression for the correction in the square bracket based on the solution found above for $v(z)$, but some remarks are in order. Pushing the expansion further in Eq.~(\ref{eq:v_NLO}), one would actually face terms of the form $\frac{c_{1,0}^k}{(\mu^2 z^3)^k}$ ($k \in \mathbb{N}$), which all become of order one close to the edge (i.e. where $z \approx \delta_\alpha \sim \mu^{-\frac{2}{3}}$), and contribute to the $Q^{-\frac{d+2}{3d}}$-correction in the chemical potential, which makes it hard to express it in closed form. A reasonable choice, though, is to focus on first-order perturbations in the Wilsonian coefficients, as was originally done in \cite{Hellerman:2020eff}. Similarly, the operator $-\frac{c_{2,0}}{4} \mathcal{O}^{(2,0)}_{bulk}$ also ends up contributing to the next-to-leading order in the chemical potential for exactly the same reason, and so does any operator $\mathcal{O}^{(m,0)}_{bulk}$, although we will not need to consider $m > 2$. So, to first-order in the perturbations, we get
\begin{equation}
    \mu = \zeta Q^\frac{1}{d} \left[ 1 + \frac{d^2 \Gamma(d)}{8 c_0 \Gamma\left(\frac{d}{2}\right)^2} \frac{1}{Q^\frac{d+2}{3d}} \left\{ \frac{c_{1,0}}{\alpha^\frac{4-d}{2}} + \frac{3 d^2}{32} \frac{c_{2,0} g}{c_0} \frac{1}{\alpha^\frac{10-d}{2}} \right\} + \mathcal{O}\left( Q^{-\frac{d+4}{3d}} \right) \right].
\end{equation}
Counterterm contributions would allow us to renormalize this expression and get rid of the $\alpha$-dependence, but for practical purposes, we shall do this and fix the renormalized coefficients such that they cancel all divergences only at the very end of the computation of $\Delta_Q$.

\subsubsection{Structure of the expansion}

We now elaborate on the structure of the expansion of $\Delta_Q$ by first noting that the contribution of $\mathcal{O}^{(m,n)}_{bulk}$ is itself an expansion. Indeed, using the leading-order solution $v(z) \sim (\mu \cdot z)^\frac{d}{4}$ (corrections do not change the argument) and Eq.~(\ref{eq:z_stuff}), we have
\begin{equation}
\begin{aligned}
	\int_{cloud} \dd{\vect{r}} \, \langle \mathcal{O}^{(m,n)}_{bulk} \rangle & \sim \mu^{d+1-2(m+n)} \int_0^1 \dd{z} \, \frac{(1 - z)^{\frac{d}{2}-1+m}}{z^{3m+2n-1-\frac{d}{2}}}.
\end{aligned}
\end{equation}
The integral on the right-hand side either converges and corresponds to the first case of Eq.~(\ref{eq:DeltaQ_bulk}), or needs to be regularized by setting the lower bound to $\delta_\alpha$, yielding the last two cases of this classification. In the latter situation, however, the upper bound of the integral always gives a finite result, thus continuing the expansion in $Q^{-\frac{2}{d}}$ starting at $Q^\frac{d+1}{d}$. For concreteness, consider
\begin{equation}
	\mathcal{O}^{(1,1)}_{bulk} = \frac{(\del_i a)^2 Z}{a^\frac{8}{3}}
\end{equation}
in $d = 3$. We obtain
\begin{equation}
\begin{aligned}
	\int_{cloud} \dd{\vect{r}} x \, \langle \mathcal{O}^{(1,1)}_{bulk} \rangle & \sim \int_{\delta_\alpha}^1 \dd{z} \, \frac{(1 - z)^\frac{3}{2}}{z^\frac{5}{2}} \\
	& = \pi + \frac{2}{3 \delta_\alpha^\frac{3}{2}} - \frac{3}{\sqrt{\delta_\alpha}} + \mathcal{O}\left( \sqrt{\delta_\alpha} \right) \\
	& = \frac{2 \zeta}{3 \alpha^\frac{3}{2}} Q^\frac{1}{3} - \frac{3 \zeta^\frac{1}{3}}{\sqrt{\alpha}} Q^\frac{1}{9} + \pi + \mathcal{O}\left( Q^{-\frac{1}{9}} \right),
\end{aligned}
\end{equation}
where $\pi$ comes from the upper bound and is suppressed by a factor of $(Q^{-\frac{2}{3}})^2$ with respect to the leading-order term. The rest is an expansion in $\delta_\alpha$ whose dependence on $\alpha$ should subsequently be absorbed by counterterms. All in all, the expansion of $\Delta_Q$ reads
\begin{equation} \label{eq:DeltaQ_result}
\begin{aligned}
	\Delta_Q & = Q^\frac{d+1}{d} \left[ a_1 + a_ 2 Q^{-\frac{2}{d}} + a_3 Q^{-\frac{4}{d}} + \ldots \right] \\
	& + Q^\frac{2d-1}{3d} \left[ b_1 + b_ 2 Q^{-\frac{2}{3d}} + b_3 Q^{-\frac{4}{3d}} + \ldots \right] \\
	& + Q^\frac{d-5}{3d} \left[ c_{1,0} + c_ 2 Q^{-\frac{2}{3d}} + c_3 Q^{-\frac{4}{3d}} + \ldots \right] + \ldots
\end{aligned}
\end{equation}
The first line is completely analogous to the relativistic case, while the rest is specific to \ac{nrcft}. The last line arises when $\mu$ is replaced by $Q$, according to Eq.~(\ref{eq:mu_generic}). When $d$ is even, it can be absorbed in the second line, where some of the $b_i$'s contain $\log Q$-terms---cf. again Eq.~(\ref{eq:DeltaQ_bulk}) for the classification.

\subsection{Small dilaton mass}

We now implement a small explicit breaking of conformal symmetry, as originally proposed by Coleman~\cite{Coleman:1988aos}. It is relatively straightforward to adapt to capture both the relativistic ($z = 1$) and the nonrelativistic ($z = 2$) cases at once. Referring to the beginning of Section~\ref{sec:NR_EFT_simple}, we recall that the dilaton is a field transforming under scale transformations as
\begin{equation}
	\sigma(t, \vect{r}) \longrightarrow \sigma(e^{z\alpha} t, e^\alpha \vect{r}) + \frac{d + z - 2}{2 f} \alpha,
\end{equation}
where $f$ is a dimensionful constant, as required by the relation $a = \frac{1}{f} e^{-f \sigma}$. Dressing the identity operator in the bulk is an easy operation, which we express in terms of $\sigma$ rather than $a$:
\begin{equation}
    \mathbb{1}_{bulk} := \frac{1}{f^{\eta}} \expo{-\eta f \sigma(t, \vect{r})},
\end{equation}
with $\eta := \frac{2(d + z)}{d + z - 2}$. This corresponds to the marginal interaction term for the radial mode that we have used so far---the term whose Wilsonian coefficient was denoted by $g$. Coleman's potential is a small modification of this term which consists in adding
\begin{equation} \label{eq:Uc}
\begin{aligned}
	U_C & = \frac{m_\sigma^2}{\eta^2 f^2} \left[ e^{-\eta f \sigma} + \eta f \sigma - 1 \right] \\
	& = \frac{m_\sigma^2}{\eta^2 f^2} \left[ (f a)^{\eta} - \eta \log(f a) - 1 \right].
\end{aligned}
\end{equation}
to the action, where $m_\sigma$ is a small parameter that we will refer to as the \emph{dilaton mass}. Indeed, to quadratic order,
\begin{equation}
	U_C \approx \frac{1}{2} m_\sigma^2 \sigma^2.
\end{equation}
It must be "small" in the following sense:
\begin{equation}
    m_\sigma^2 f^{\eta-2} \ll g,
\end{equation}
where $g$ is generically of $\order{1}$. In Eq.~\eqref{eq:Uc}, the linear piece in $\sigma$ is needed to eliminate the tadpole hidden in the exponential, and will appear to be responsible for the signature of the soft breaking of conformal symmetry in the computation of $\Delta_Q$ (see \cite{PhysRevD.101.065018} for the relativistic case). The dilaton mass $m_\sigma$ is therefore a finely-tuned small parameter encoding deviations from the critical point.

Finally, let us comment on the trace of the stress tensor. Although the breaking of scale invariance is very explicit due to the last two terms in Coleman's potential, we can somewhat quantify {\em how much} we break it. Indeed, the trace of the stress tensor does no longer vanish but equals\footnote{For more details on the tracelessness of the stress tensor in nonrelativistic theories, we refer the reader to \cite{Hagen:1972pd}.}
\begin{equation}
	T = \frac{d + z - 2}{2} \frac{m^2_\sigma}{f} \sigma.
\end{equation}

\subsection{Examples} \label{sec:examples}

We are now going to see the whole machinery in action through two examples, so let us repeat the recipe. We are interested in the conformal dimension of the lowest operator of charge $Q \gg 1$ in the theory without trap, $\Delta_Q$, which is given by the ground-state energy of the trapped system. After choosing at which order in $Q$ we want to truncate the expansion of $\Delta_Q$ (in what follows, we go up to $Q^0$), we select the corresponding terms in the complete large-charge dilaton \ac{eft} given in Eq.~\eqref{eq:FullDilatonEFT} based on their contributions to $\Delta_Q$ given in Eqs~\eqref{eq:DeltaQ_bulk} and \eqref{eq:DeltaQ_edge}. We also account for the small dilaton mass deformation that we have just discussed.


\subsubsection{The \texorpdfstring{$d=2$}{} case} \label{sec:ex_2D}

Typically, \acp{nrcft} in two spatial dimensions are relevant for the description of anyons \cite{Wilczek:1982wy,Nishida:2007pj}, which themselves are at the origin of the Aharonov-Bohm effect \cite{Bergman:1993kq} and whose existence has recently been confirmed in the context of the fractional quantum Hall effect \cite{Bartolomei_2020,nakamura2020direct}. Based on the previous discussion, we include the subleading operators $\mathcal{O}^{(m,n)}_{bulk}$ of Eq.~(\ref{eq:Omn_bulk}) for $(m, n) \in \{ (1, 0), (0, 1), (2, 0), (1, 1) \}$, as well as the edge \ac{eft} operators $Z^0_{edge}$ and $Z^1_{edge}$ constructed in Eq.~(\ref{eq:Zn_edge}). The Lagrangian thus reads\footnote{We introduce for convenience some extra normalization for the Wilsonian coefficients compared to Eq.~\eqref{eq:FullDilatonEFT}.}
\begin{equation}
\begin{aligned}
	\mathcal{L} & = c_0 a^2 U - \frac{g}{4} a^4 - \frac{m^2_\sigma}{16 f^2} \left[ (f a)^4 - 4 \log (f a) - 1 \right] \\
	& - \frac{c_{1,0}}{2} (\del_i a)^2 - \frac{c_{0,1}}{2} Z - \frac{c_{2,0}}{4} \frac{(\del_i a)^4}{a^4} - \frac{c_{1,1}}{4} \frac{(\del_i a)^2}{a^4} Z \\
	& + \delta(a^2) \frac{1}{2} \left[ \kappa_0 \, a^2 (\del_i a)^2 + \kappa_1 \left( \frac{a^2 (\del_i a)^2}{2} \right)^\frac{1}{3} Z \right].
\end{aligned}
\end{equation}
It is convenient to introduce $\tilde{g} := g + \frac{m_\sigma^2 f^2}{4}$. Using $\langle \chi \rangle = \mu \cdot t$, $\langle a \rangle = v(z)$ and Eq.~(\ref{eq:z_stuff}), we get
\begin{equation} \label{eq:2D_H0}
\begin{aligned}
	\mathcal{H}_0 & = c_0 v^2 \mu (1 - z) + \frac{\tilde g}{4} v^4 - \frac{m^2_\sigma}{16 f^2} \left[ 4 \log (f v) + 1 \right] \\
	& + \frac{(1 - z) v'^2}{\mu} \left[ c_{1,0} + \frac{c_{1,1}}{v^4} \right] + c_{0,1} + \frac{c_{2,0} (1 - z)^2}{\mu^2} \left( \frac{v'}{v} \right)^4 \\
	& - \delta(v^2) \left[ \frac{\kappa_0 (1 - z)}{\mu} (v v')^2 + \frac{\kappa_1 (1 - z)^\frac{1}{3}}{\mu^\frac{1}{3}} (v v')^\frac{2}{3} \right].
\end{aligned}
\end{equation}
The expectation value $v(z)$ is the solution of the \ac{eom} (cf. Eq.~(\ref{eq:eom_d}))
\begin{equation} \label{eq:2D_EoM}
	\left( \frac{v(z)}{v_{hom}} \right)^2 = z \left[ 1 + B(z,v,v',v'') \right],
\end{equation}
with $v_{hom} \equiv \sqrt{2 \mu c_0 / \tilde g}$, and $B(z,v,v',v'')$ is given by
\begin{equation}
\begin{aligned}
	B(z,v,v',v'') & = \frac{m_\sigma^2}{8 c_0 f^2 \mu z} \frac{1}{v^2} + \frac{c_{1,0}}{c_0 \mu^2 z} \frac{(1 - z) v'' - v'}{v} \\
	& + \frac{6 c_{2,0} (1 - z)}{c_0 \mu^3 z} \frac{{v'}^2}{v^4} \frac{(1 - z) (v v'' - {v'}^2) - v v'}{v^2} \\
	& + \frac{c_{1,1}}{c_0 \mu^2 z} \frac{(1 - z) (v v'' - 2 {v'}^2) - v v'}{v^6}
\end{aligned}
\end{equation}
in the bulk. Referring to the previous section, we find that the chemical potential is related to the charge as
\begin{equation}
\begin{aligned}
	\mu & = \zeta \sqrt{Q} \left[ 1 + \frac{1}{2 c_0} \frac{1}{Q^\frac{2}{3}} \left\{ \frac{c_{1,0}}{\alpha} + \frac{3}{8} \frac{c_{2,0} \tilde g}{c_0} \frac{1}{\alpha^4} \right\} + \mathcal{O}\left( Q^{-1} \right) \right],
\end{aligned}
\end{equation}
where $\zeta = \sqrt{\tilde g/(2 \pi c_0^2)}$. Note that the dilaton mass $m_\sigma$ modifies the expression of the chemical potential only beyond next-to-leading order. We are now in position to integrate Eq.~(\ref{eq:2D_H0}), removing the $\delta_\alpha$-layer when necessary. To first-order perturbations in the Wilsonian coefficients, we find
\begin{equation}
\begin{aligned}
	\Delta_Q^{(d=2)} & = \frac{2}{3} \zeta Q^\frac{3}{2} + \left[ \frac{c_{1,0}}{6 \zeta c_0} - \frac{\pi \zeta m_\sigma^2}{8 f^2} \right] \sqrt{Q} \log Q \\
	& - \frac{\kappa_0^{ren.}}{2 \zeta c_0} \sqrt{Q} - \left( 2 \pi^4 c_0 \right)^\frac{1}{3} \zeta \kappa_1^{ren.} \cdot Q^\frac{1}{6} + \order{Q^0},
\end{aligned}
\end{equation}
The renormalized couplings are
\begin{equation}
\begin{aligned}
	\kappa_0^{ren.} & = \kappa_0 + c_{1,0} \log \alpha - \frac{\tilde{g}}{24 c_0} \frac{c_{2,0}}{\alpha^3} + \mathrm{(finite)}, \\
	\kappa_1^{ren.} & = \kappa_1 + \left[ \left(\frac{c_0}{\tilde{g}}\right)^\frac{1}{3} c_{2,0} - \left(\frac{\tilde{g}}{c_0}\right)^\frac{2}{3} c_{1,1} \right] \frac{1}{8 \alpha^2} + \mathrm{(finite)}.
\end{aligned}
\end{equation}
Note that we have absorbed the contribution of $c_{0,1}$, which is finite, as well as a finite correction due to $m_\sigma$ into the finite part of $\kappa_0^{ren.}$. We thus see that the effect of the small dilaton mass deformation is a mere shift in the coefficients of the $\sqrt{Q} \log Q$ term, so there is essentially no signature of the small breaking of conformal symmetry in this scenario.

Let us mention that anyons are not invariant under parity and it would thus be interesting to extend this study to the case of parity-violating theories (cf. \cite{Kravec:2018qnu} for suggestions), in the spirit of \cite{Cuomo:2021qws} in the relativistic case.

\subsubsection{The \texorpdfstring{$d=3$}{} case} \label{sec:er_3D}

As mentioned in the introduction, the case of \acp{nrcft} in three spatial dimensions is relevant for the description of the unitary Fermi gas. In order to build the large-charge \ac{eft}, we again use the leading-order Lagrangian Eq.~(\ref{eq:lagr_LO}), to which we add Coleman's potential Eq.~(\ref{eq:Uc}), the operators given in Eq.~(\ref{eq:Omn_bulk}) with $(m, n) \in \{ (1, 0), (0, 1), (2, 0), (1, 1), (0, 2) \}$ and the edge operators, Eq.~(\ref{eq:Zn_edge}), constructed from $Z^0$, $Z^1$, and $Z^2$. We thus consider the following Lagrangian density:
\begin{equation}
\begin{aligned}
	\mathcal{L} & = c_0 a^2 U - \frac{3 \tilde g}{10} a^\frac{10}{3} + \frac{9 m^2_\sigma}{100 f^2} \left[ \frac{10}{3} \log (f a) + 1 \right] \\
	& - \frac{c_{1,0}}{2} (\del_i a)^2 - \frac{c_{0,1}}{3} a^\frac{2}{3} Z - \frac{c_{2,0}}{4} \frac{(\del_i a)^4}{a^\frac{10}{3}} - \frac{c_{1,1}}{6} \frac{(\del_i a)^2}{a^\frac{8}{3}} Z - \frac{c_{0,2}}{9} \frac{Z^2}{a^2} \\ 
	& + \delta(a^\frac{4}{3}) \left[ \kappa_0 \left( \frac{a^\frac{2}{3} (\del_i a)^2}{2} \right)^\frac{7}{6} + \frac{\kappa_1}{3} \left( \frac{a^\frac{2}{3} (\del_i a)^2}{2} \right)^\frac{1}{2} Z + \frac{\kappa_2}{9} \left( \frac{2}{a^\frac{2}{3} (\del_i a)^2} \right)^\frac{1}{6} Z^2 \right]
\end{aligned}
\end{equation}
where $\tilde g \equiv g \left( 1 + 3 m_\sigma^2 f^\frac{4}{3} / (10 g) \right)$. The ground-state energy density then reads
\begin{equation} \label{eq:3D_H0}
\begin{aligned}
	\mathcal{H}_0 & = c_0 \mu v^2 (1 - z) + \frac{3 \tilde g}{10} v^\frac{10}{3} - \frac{9 m^2_\sigma}{100 f^2} \left[ \frac{10}{3} \log (f v) + 1 \right] \\
	& + \frac{(1 - z) {v'}^2}{\mu} \left[ c_{1,0} + \frac{c_{1,1}}{v^\frac{8}{3}} \right] + c_{0,1} v^\frac{2}{3} + \frac{c_{2,0} (1 - z)^2}{\mu^2} \frac{{v'}^4}{v^\frac{10}{3}} + \frac{c_{0,2}}{v^2} \\
	& - \delta(v^\frac{4}{3}) \left[ \kappa_0 \left( \frac{1-z}{\mu} v^\frac{2}{3} {v'}^2 \right)^\frac{7}{6} + \kappa_1 \left( \frac{1-z}{\mu} v^\frac{2}{3} {v'}^2 \right)^\frac{1}{2} + \kappa_2 \left( \frac{1-z}{\mu} v^\frac{2}{3} {v'}^2 \right)^{-\frac{1}{6}} \right].
\end{aligned}
\end{equation}
Moreover, the expectation value of the radial mode $v(z)$ satisfies the equation of motion Eq.~(\ref{eq:eom_d}) with $v_{hom} \equiv (2 \mu c_0 / \tilde g)^\frac{3}{4}$ and
\begin{equation}
\begin{aligned}
	B(z, v, v', v'') & = \frac{3 m_\sigma^2}{20 c_0 f^2 \mu z} \frac{1}{v^2} + \frac{c_{1,0}}{2 c_0 \mu^2 z} \frac{2 (1 - z) v'' - 3 v'}{v} - \frac{c_{0,1}}{3 c_0 \mu z} \frac{1}{v^\frac{4}{3}} + \frac{c_{0,2}}{c_0 \mu z} \frac{1}{v^4} \\
	& + \frac{c_{2,0} (1 - z)}{c_0 \mu^3 z} \frac{{v'}^2}{v^\frac{10}{3}} \frac{(1 - z) (6 v v'' - 5 {v'}^2) - 9 v v'}{v^2} \\
    & + \frac{c_{1,1}}{6 c_0 \mu^2 z} \frac{(1 - z) (6 v v'' - 8 {v'}^2) - 9 v v'}{v^\frac{14}{3}}
\end{aligned}
\end{equation}
in the bulk. Consequently, the chemical potential reads
\begin{equation}
\begin{aligned}
	\mu & = \zeta Q^\frac{1}{3} \left[ 1 + \frac{9}{\pi c_0} \frac{1}{Q^\frac{5}{9}} \left\{ \frac{c_{1,0}}{\sqrt{\alpha}} + \frac{27 c_{2,0} \tilde{g}}{32 c_0} \frac{1}{\alpha^\frac{7}{2}} \right\} + \mathcal{O}\left( Q^{-\frac{7}{9}} \right) \right].
\end{aligned}
\end{equation}
Proceeding as before, we can conveniently express the result in terms of $\gamma^2 := \frac{18 \tilde{g}}{c_0}$ as
\begin{equation} \label{eq:Delta_Q_3D}
\begin{aligned}
	\Delta_Q^{(d=3)} & = \frac{3}{4} \zeta Q^\frac{4}{3} + \left[ \frac{27}{8 \zeta c_0} c_{1,0} + \frac{\tilde{g}}{\zeta c_0^2} c_{0,1} \right]  \cdot Q^\frac{2}{3} - 3^5 \sqrt{2} \pi (\zeta^5 \gamma^{-8})^\frac{1}{3} \kappa_0^{ren.} \cdot Q^\frac{5}{9} \\
	& - \frac{\sqrt{2} \pi m_\sigma^2 \zeta^\frac{3}{2}}{5 f^2} \cdot \sqrt{Q} \left[ \log Q + \left( \log \frac{12^6 \zeta^3 f^4}{\gamma^6} - 8 \right) \right] \\
	& - 3 \sqrt{2} \pi \zeta \kappa_1^{ren.} \cdot Q^\frac{1}{3} - \frac{\sqrt{2} \pi}{27} (\zeta \gamma^8)^\frac{1}{3} \kappa_2^{ren.} \cdot Q^\frac{1}{9} + \order{Q^0} \\
\end{aligned}
\end{equation}
where the renormalized couplings are given by
\begin{equation}
\begin{aligned}
	\kappa_0^{ren.} & = \kappa_0 - \frac{\gamma^\frac{5}{3}}{80} \frac{c_{2,0}}{\alpha^\frac{5}{2}} + \mathrm{(finite)} \\ 
	\kappa_1^{ren.} & = \kappa_1 + \left[ \frac{135}{8 \gamma} c_{2,0} - \frac{\gamma}{3} c_{1,1} \right] \frac{1}{4 \alpha^\frac{3}{2}} + \mathrm{(finite)} \\
	\kappa_2^{ren.} & = \kappa_2 - \left[ \frac{3^5}{2^3} \frac{c_{1,1}}{\gamma^\frac{5}{3}} + \frac{3^8 \cdot 5}{2^7} \frac{c_{2,0}}{\gamma^\frac{11}{3}} + \gamma^\frac{1}{3} c_{0,2} \right] \frac{1}{\sqrt{\alpha}} + \mathrm{(finite)}.
\end{aligned}
\end{equation}
The dilaton mass deformation $m_\sigma$ is responsible for the second line of Eq.~\eqref{eq:Delta_Q_3D} above, that is, the presence of the $\sqrt{Q} \log Q$ and $\sqrt{Q}$-terms: this is a testable signature of the soft breaking of conformal invariance. Moreover, there is a mixed $Q^0$-contribution that we have not displayed as it merely serves as a counterterm for the divergent one-loop Casimir energy to be computed in Section~\ref{sec:NR_EFT_quantum}, where a universal $Q^0 \log Q$-term arises as a consequence of this.

Among the many quantities that can be given corrections based on this construction (cf. also \cite{Son:2005rv}), let us come back to the doubly integrated density mentioned earlier, whose leading-order expression is given in Eq.~(\ref{eq:profile_LO}). We solve the equation of motion as in Eq.~(\ref{eq:v_NLO}), but we only account for the corrections caused $c_{1,0}$ for simplicity. We then integrate the charge density $\rho(z) = c_0 v(z)^2$ over $r_1$ and $r_2$ to obtain
\begin{equation} \label{eq:linear_profile_result}
\begin{aligned}
	n(r_3) & = \frac{2 \pi g}{5} \left[ \frac{2 c_0 \mu}{g} \left( 1 - \frac{r_3^2}{2 \mu} \right) \right]^\frac{5}{2} \\
	& \times \left[ 1 + \frac{45 c_{1,0}}{32 c_0 \mu^2} \left\{ \frac{5}{\left( 1 - \frac{r_3^2}{2 \mu} \right)^2} - \frac{1}{\left( 1 - \frac{r_3^2}{2 \mu} \right)^3} \right\} + \ldots \right].
\end{aligned}
\end{equation}
This expression is valid for $r_3 \in [0, R_{cl} \sqrt{1 - \delta_\alpha}]$ and can in principle improve the fitting to experimental data.

\subsection{Lessons from the complete large-charge dilaton EFT}

The rich dynamics introduced by the inhomogeneity of the trapped system forced us to be very cautious and pedantic in the construction of the effective action, culminating in Eq.~\eqref{eq:FullDilatonEFT}. The fundamental principles of \ac{eft} building had to be fully exploited and carefully discussed in order to achieve this. This construction turned out to be completely self-consistent when computing the large-charge conformal dimension $\Delta_Q$, where the newly introduced edge \ac{eft} operators served as counterterms for the edge divergences observed already in the previous section. This is a beautiful illustration of the interplay between a 'bulk' \ac{eft} and a complementary \ac{eft} supported on an extended object---here, the fluctuating edge of the cloud of particles. Many lessons learnt in this section can be applied in other contexts with similar dynamics, \emph{e.g.} the effective string theory approach to the confinement problem \cite{Hellerman:2016hnf}.

The use of the dilaton formalism for the purpose of \ac{eft} building with spontaneously broken $U(1)$ and conformal symmetries had two main advantages. First, it allowed for a rather clear and straightforward construction of dressing operators. Indeed, any microscopic model realizing this superfluid spontaneous symmetry breaking pattern has a massive mode gapped at the cutoff scale set by the chemical potential. This massive mode, upon integrating it out, effectively dresses every operator built out of low-energy degrees of freedom in the resulting theory in a way that makes every term marginal. This is, by design, precisely what our complete large-charge \ac{eft} captures, where the edge dressing rule invoked a slightly more subtle version of this argument. Moreover, the dilaton mode allowed us to explore the fate of the ground-state energy $\Delta_Q$ near criticality by softly breaking conformal symmetry explicitly. While there was no phenomenological consequence of this in $d=2$, the situation in $d=3$ was different: the structural dependence of $\Delta_Q$ changed and acquired an extra $\sqrt{Q} \log Q$ term that can in principle be detected when fitting against experimental data.

However, it is also probably clear by now that a dilaton-based construction of the effective theory introduces a lot of redundancy. We carefully discussed how to deal with this aspect, although the formalism calls for a description in terms of the low-energy degrees of freedom only. Additionally, the $\delta$-layer regularization, while very intuitive and physical, can easily be replaced by a more familiar regularization, namely, dimensional regularization. Equipped with a thorough understanding of the rich dynamics taking place in these systems, we move on to describe how to efficiently reproduce these results in the simplest possible setup: the Goldstone \ac{eft} with dimensional regularization.
\pagebreak
\section{Systematic analysis II: large-charge Goldstone EFT}
\label{sec:NR_EFT_systematicsII}
\subsection{Heuristics}

It is enlightening to summarize the results obtained in the previous two sections by invoking only heuristic arguments based on the existence of a macroscopic limit and \ac{eft} principles, so let us do so in a way that gives a clear physical picture.

The theories of interest are invariant under the nonrelativistic incarnation of the conformal group---the Schrödinger group---which always includes a $U(1)$ central extension that corresponds to particle-number conservation. 
The nonrelativistic state-operator correspondence states that the (positively charged) operator spectrum is mapped bijectively to the energy spectrum of the system subjected to an external harmonic potential $A_0(\vect{r}) = \frac{m \omega^2}{2 \hbar} \vect{r}^2$~\cite{Werner_2006,Nishida:2007pj,Goldberger:2014hca}, with trapping frequency $\omega$. For the purpose of the argument, it is convenient to set $\hbar = m = 1$ but keep track of $\omega$, so we have
\begin{equation}
  E_0 = \omega \cdot \Delta_Q \ ,
\end{equation}
where $\Delta_Q$ is the lowest operator dimension of charge $Q$, and $E_0$ is the energy of the charged ground state, for which the charge $Q$---the number of particles in the trap---is sourced by a chemical potential $\mu$. Borrowing intuition from quantum mechanical systems in confining potentials and turning points, it is clear that the particle density decreases as we recede from the center of the trap. The system possesses a characteristic energy $\mu$ to overcome the energy barrier set by $A_0(\vect{r})$, so the resulting droplet of trapped particles occupies a finite region of space whose radius is given by $R_{cl}=\frac{\sqrt{2\mu}}{\omega}$. However, the particle density at the center of the cloud, or far enough from its edge, cannot depend on $\omega$ as the system locally looks like the untrapped one. This is dictated by the existence of macroscopic limit, and the particle density is therefore set by dimensional analysis to scale with $\mu$ like $\rho \sim \mu^\frac{d}{2}$, so $Q\sim\left(\frac{\mu}{\omega}\right)^d$. The volume of the cloud, $\Omega_d R_{cl}^d$, is then seen to be proportional to the square-root of the charge: $\text{Vol} \sim \omega^{-\frac{d}{2}} \sqrt{Q}$.

Now, consider the charge and energy densities 
in the bulk, \emph{i.e.} sufficiently far from the edge:
\begin{equation}
\begin{aligned}
	\text{charge density} & \sim \frac{Q}{\mathrm{Vol}} \sim \omega^{d/2} \sqrt{Q}\ , \\
	\text{energy density} & \sim \frac{E_0}{\mathrm{Vol}} \sim {\omega^{d/2+1}} \Delta_Q/{\sqrt{Q}}.
\end{aligned}
\end{equation}
The existence of macroscopic limit implies that the limit  $Q \to \infty$, $\omega\to 0$ can be taken while keeping the charge and energy densities fixed. In turn, this corresponds to the following leading-order scaling in $Q\gg1$:
\begin{equation} \label{eq:DeltaQ_LOsimple}
    \Delta_Q \sim Q^{(d+1)/d} + \text{(subleading)}.
\end{equation}

Moreover, the bulk \ac{eft} is characterized by the \ac{ir} length scale $R_{cl}$ and the \ac{uv} length scale $\mu^{-\frac{1}{2}} \sim \rho^{-\frac{1}{d}}$. The derivative expansion is thus controlled by
\begin{equation} \label{eq:BulkRatio}
    \frac{\rho^{-\frac{1}{d}}}{R_{cl}} \sim Q^{-1/d},
\end{equation}
so large separation of scales requires $Q\gg1$. In fact, in a parity-invariant theory, corrections to Eq.~(\ref{eq:DeltaQ_LOsimple}) are controlled by the square of this ratio, so the structure of $\Delta_Q$ established up to this point seems to follow precisely the relativistic one.

However, an inhomogeneous system might generically have regions where separation of scales is no longer satisfied: a familiar example is a superfluid developing vortices. The superfluid density at the vortex core drops and a naive superfluid \ac{eft} description would break down. This mandates the characterization of the vortex by a lower-dimensional \ac{eft} with a new \ac{uv} cutoff scale associated with the typical size of the vortex core. In our case, the situation is similar: the ratio Eq.~\eqref{eq:BulkRatio} is no longer satisfactory in some specific region of the cloud, namely, close to the edge. In fact, in terms of $z = 1 - \frac{r^2}{R_{cl}^2}$, the largest fluctuations that the \ac{eft} can appropriately describe \emph{locally} is not quite $R_{cl}$, but $R_{cl} \cdot z$, otherwise these fluctuations would leak out of the cloud. Similarly, the local \ac{uv} scale is given by $\rho(z)^{-\frac{1}{d}} \sim (\mu \cdot z)^{-\frac{1}{2}}$, so the bulk \ac{eft} breaks down in the region where $R_{cl} \cdot z \sim \rho(z)^{-\frac{1}{d}}$, that is, when
\begin{equation}
    z \sim \delta := \left(\frac{\omega}{\mu}\right)^\frac{2}{3} \sim Q^{-\frac{2}{3d}}.
\end{equation}
To accommodate for this breakdown, one needs to account for the edge dynamics by building an additional \ac{eft} living on the corresponding codimension-one fluctuating manifold, which is essentially a sphere of radius $R_{cl}$. From the point of view of this lower-dimensional theory, the typical size---hence the \ac{ir} length scale---is again $R_{cl}$, and the \ac{uv} one is the previous local \ac{uv} scale frozen at $z=\delta$, so the derivative expansion is expected to be controlled by
\begin{equation}
    \frac{\rho(\delta)^{-\frac{1}{d}}}{R_{cl}} \sim \delta,
\end{equation}
which is indeed what was computed in Section~\ref{sec:NR_EFT_systematicsI}. The edge \ac{eft} operators must be marginal operators built out of bulk primaries projected on the 'Dirichlet' region $\rho(z) = 0$, so their expectation values in the charged ground state cannot depend on $z$ except for a Dirac-delta factor $\delta(z)$. The leading one is built out of merely $\rho$ and its derivatives, and it is rather straightforward to see that it can only be
\begin{equation}
    \delta\left(\rho^\frac{2}{d}\right) \left[ \rho^{\frac{4}{d}-2} (\partial_i \rho)^2 \right]^\frac{d+4}{3},
\end{equation}
thereby contributing the leading edge contribution to $\Delta_Q$: a term proportional to $Q^\frac{2d-1}{3d}$. An entirely new structure therefore arises from the boundary dynamics of the system, as investigated in \cite{Hellerman:2020eff,Pellizzani:2021hzx} and reviewed in the previous section. The renormalization of edge divergences also hints at the possible presence of logarithms in the large-charge expansion of $\Delta_Q$. Finally, while we merely considered the leading-order relation between $\mu$ and $Q$, a careful treatment reveals that, upon replacing $\mu$ by $Q$ in the expansion, yet another series of terms emerges, starting at $Q^\frac{d-5}{3d}$, although its coefficients are not independent Wilsonian coefficients.

These series of heuristic arguments therefore allows to gain an intuitive understanding of the following structure, first revealed in~\cite{Pellizzani:2021hzx} by the author and explained in the previous section:
\begin{equation}
\begin{aligned}
	\Delta_Q & = Q^\frac{d+1}{d} \left[ a_1 + \frac{a_2}{Q^\frac{2}{d}} + \frac{a_3}{Q^\frac{4}{d}} + \dots \right] \\
	& + Q^\frac{2d-1}{3d} \left[ b_1 + \frac{b_2}{Q^\frac{2}{3d}} + \frac{b_3}{Q^\frac{4}{3d}} + \dots \right] \\
    & + Q^\frac{d-5}{3d} \left[ d_1 + \frac{d_2}{Q^\frac{2}{3d}} + \frac{d_3}{Q^\frac{4}{3d}} + \dots \right] \\
	& + \text{\(\log(Q)\) enhancements.}
\end{aligned}
\end{equation}
In this section, we discuss this in detail and recover these results from the \ac{nlsm}---or 'large-charge Goldstone \ac{eft}'---perspective using \ac{dimreg} and we clarify when the $\log(Q)$ terms occur.

Note that keeping track of $\omega$ helped us sketch the previous heuristic arguments, but we now switch it off again and work with $\hbar=m=\omega=1$. Our first task is to reconstruct the full large-charge EFT solely in terms of the Goldstone mode $\chi$.

\subsection{The complete large-charge Goldstone EFT}

To make what follows self-contained, we sketch again the preliminary steps needed to establish the leading-order results, and then quickly turn to the discussion of subleading operators in both the bulk and the edge \acp{eft}. Without much surprise, the leading-order term in the Lagrangian is of the form~\cite{Son:2005rv,schroedinger,Kravec:2018qnu}:
\begin{equation}%
\label{eq:LO-action}
	\mathcal{L}_{LO} = c_0 U^{\frac{d}{2}+1},
\end{equation}
where we recall that
\begin{equation}
	U = \dot \chi - \frac{1}{2} r^2 - \frac{1}{2} (\del_i \chi)^2\ .
\end{equation}
Indeed, $U$ has dimension 2, so $U^{\frac{d}{2}+1}$ is marginal. The harmonic trap $A_0(r) = \frac{1}{2} r^2$ (with $r = \abs{\vect{r}}$) may be regarded as being coupled to $\dot\chi$, as imposed by general coordinate invariance~\cite{Son:2005rv}. In the superfluid ground state, the Goldstone mode $\chi$ acquires the expectation value $\ev{\chi} = \mu \cdot t$. For comparison, the equivalent Lagrangian in the relativistic case would read $\mathcal{L} = c_0 (\del \chi)^{d+1}$~\cite{Hellerman:2015nra,Monin:2016jmo}.

In the charged ground state, $U$ thus has the following simple expectation value:
\begin{equation}
	\ev{U} = \mu \cdot z,
\end{equation}
where the dimensionless coordinate $z = 1 - \frac{r^2}{R_{cl}^2}$ has already been introduced earlier, and $R_{cl} = \sqrt{2 \mu}$ is the radius of the cloud. The ground-state charge density is
\begin{equation}
	\ev{ \rho }  \stackrel{LO}{=} \ev{ \pdv{\mathcal{L}_{LO}}{\dot \chi} } = \ev{\pdv{\mathcal{L}_{LO}}{ U}} \sim  \ev{U^{d/2}} \sim (\mu \cdot z)^{d/2},
\end{equation}
which is seen to decrease away from the origin ($z=1$), and vanishes when $z=0$, \emph{i.e.,} $r = R_{cl}$. The total charge $Q$ is obtained by integrating the charge density over the volume of the cloud, thereby relating it to the chemical potential:
\begin{equation} \label{eq:mu_LO}
	\mu \stackrel{LO}{=}  \frac{1}{\sqrt{2 \pi}} \left[ \frac{\Gamma(d+1)}{\Gamma\left( \frac{d}{2} + 2 \right) c_0} \right]^{1/d} Q^{1/d} =: \xi Q^{1/d}.
\end{equation}
Similarly, the (leading-order) ground-state energy---and, therefore, the dimension of the lowest operator of charge $Q$---can be obtained upon integrating the energy density over the volume of the cloud, which may be expressed in terms of $Q$ as
\begin{equation} 
\label{eq:DeltaQ_LO}
	\Delta_Q \stackrel{LO}{=} \frac{d}{d+1} \xi Q^{(d+1)/d}\ .
\end{equation}

We also summarize the discussion of the previous section regarding the general construction of dressing operators for systems that realize spontaneous breaking of the $U(1)$ and conformal symmetries, \emph{cf.} also \cite{Hellerman:2020eff} for the original discussion. First, we know that, besides $U$ and its derivatives, the only operator allowed by general coordinate invariance with a non-vanishing expectation value in the superfluid ground state is
\begin{equation}
	Z = \nabla^2 A_0 - \frac{1}{d} (\nabla^2 \chi)^2\ ,
\end{equation}
with $\ev{ Z } = d$. Therefore, in the bulk, all nontrivial operators are composite operators made out of integer powers of $U$, $(\del_i U)^2$ and $Z$, which are then dressed to marginality with appropriate dressing rules, established as follows. In a given region of the system, let $\{\mathcal{O}_i\}_i$ denote the set of all scalar non-composite primaries (of dimension $\Delta_i$) that have nonzero finite expectation value in the charged ground state---so $U$, $(\del_i U)^2$ and $Z$ in the bulk, and $(\del_i U)^2$ and $Z$ at the edge---and let
\begin{equation}
    \left[\prod_i \mathcal{O}_i^{p_i} \right]^\frac{2}{\sum_i p_i \Delta_i}
\end{equation}
be a generic composite operator whose dimension we chose to normalize to 2. The correct dressing operator in this given region of the system is the one with the highest power-law dependence on $\mu$. It is simple enough to find that the dressing operators in the bulk and at the edge are  given by
\begin{equation}
\begin{aligned}
    \mathcal{D}_{bulk} & := U, \\
    \mathcal{D}_{edge} & := \left[(\partial_i U)^2\right]^\frac{1}{3}.
\end{aligned}
\end{equation}

As mentioned above, the primary operators $\mathcal{O}_\Delta$ that need to be dressed are scalars of the form $U^l (\partial_i U)^{2m} Z^n$ with $l,m,n\in\mathbb{N}$. When it comes to building bulk \ac{eft} operators out of them, there is no need to include powers of $U$ as the dressing rule takes care of them, while for edge \ac{eft} operators, the same applies to $(\partial_i U)^2$ and, moreover, powers of $U$ cannot appear as their expectation values depend on $z$. The edge dressing rule is a dressed projection of $\mathcal{O}_\Delta$ onto the edge achieved by including an operator-valued Dirac-delta functional $\delta(U)$---which turns into $\frac{1}{\mu} \delta(z)$ in the charged ground state. The exhaustive list of allowed operators is therefore given by
\begin{equation} \label{eq:EFTopList}
\begin{aligned}
	\Op{m}{n} & := (\del_i U)^{2m} Z^n U^{\frac{d}{2}+1-(3m+2n)} \\
    & = \left(\frac{(\partial_i U)^2}{U^3}\right)^m \left(\frac{Z}{U^2}\right)^n U^{\frac{d}{2}+1}, \\
    \Z{p} & := Z^p \cdot \delta(U) \cdot \left[ (\partial_i U)^2 \right]^\frac{d+4(1-p)}{6}.
\end{aligned}
\end{equation}
The complete large-charge \ac{eft} thus reads
\begin{equation}
    \mathcal{L}(\chi) =
    U^{\frac{d}{2}+1} \sum_{m,n} c_{m,n} \, \left(\frac{(\partial_i U)^2}{U^3}\right)^m \left(\frac{Z}{U^2}\right)^n +
    \delta(U) \sum_p \kappa_p \, Z^p \left[ (\partial_i U)^2 \right]^\frac{d+4(1-p)}{6},
\end{equation}
where the $c_{m,n}$'s and the $\kappa_p$'s are real dimensionless Wilsonian coefficients, and $c_{0,0}$ denotes $c_0$ of the leading-order action described above.

\subsection{Large-charge conformal dimension}

We now perform a quick analysis of the leading contributions of the operators listed in Eq.~\eqref{eq:EFTopList} to the charge
\begin{equation}
    Q = \int_\text{cloud} \dd{\vect{r}} \ev{\pdv{\mathcal{L}}{U}}
\end{equation}
and the ground-state energy,
\begin{equation}
    \Delta_Q = \int_\text{cloud} \dd{\vect{r}} \ev{\pdv{\mathcal{L}}{U} \dot\chi - \mathcal{L}}.
\end{equation}
Recall that, at leading order in $\mu$ in the charged ground state, we have $\ev{U} = \mu \cdot z$, $\ev{(\partial_i U)^2} = 2\mu(1-z)$ and $\ev{Z} = d$. Moreover, the measure of integration over the cloud is given in terms of $z$ as $\frac{(2\pi\mu)^\frac{d}{2}}{\GammaF{\frac{d}{2}}} \int_0^1 \dd{z} (1-z)^{\frac{d}{2}-1}$. Every resulting integral is of the form of a Beta function,
\begin{equation}
    B(x, y) = \int_0^1 \dd{z} z^{x-1} (1-z)^{y-1} = \frac{\Gamma(x) \Gamma(y)}{\Gamma(x+y)},
    \qquad \text{Re}(x), \text{Re}(y) > 0.
\end{equation}
We shall invoke analytic continuation of the latter to perform the integrals explicitly.

We use the same superscript as the corresponding operator to denote its contribution. In the case of bulk operators, $\mathcal{O}^{(m,n)}_{bulk}$, we obtain
\begin{equation} \label{eq:QMN}
\begin{aligned}
	Q^{(m,n)} & := c_{m,n} \int_{\textnormal{cloud}} \dd{\vect{r}} \, \left\langle \frac{\partial \Op{m}{n}}{\partial U}\right\rangle \\
    & = c_{m,n} \frac{2^m d^n (2 \pi)^\frac{d}{2}}{\Gamma\left(\frac{d}{2}\right)} \mu^{d-2(m+n)} \cdot \frac{d + 2 - (6 m + 4 n)}{2} \int_0^1 \dd{z} \, (1 - z)^{\frac{d}{2}-1+m} z^{\frac{d}{2}-(3m+2n)} \\
    & = c_{m,n} \frac{2^m d^n (2 \pi)^\frac{d}{2}}{\Gamma\left(\frac{d}{2}\right)} \frac{\Gamma\left( \frac{d}{2} + m \right) \Gamma\left( \frac{d}{2} + 2 - (3 m + 2 n) \right)}{\Gamma(d + 1 - 2 (m + n))} \mu^{d-2(m+n)},
\end{aligned}
\end{equation}
and
\begin{equation} \label{eq:DeltaMN}
\begin{aligned}
	\Delta^{(m,n)}  & = c_{m,n} \int_{\textnormal{cloud}} \dd{\vect{r}} \, \left\langle \frac{\partial \Op{m}{n}}{\partial U} \dot\chi - \Op{m}{n} \right\rangle \\
	& = c_{m,n} \frac{2^m d^n (2 \pi)^\frac{d}{2}}{\Gamma\left(\frac{d}{2}\right)} \mu^{d+1-2(m+n)} \cdot \int_0^1 \dd{z} \, (1 - z)^{\frac{d}{2}-1+m} z^{\frac{d}{2}-(3m+2n)} \left[ \frac{d + 2 - (6 m + 4 n)}{2} - z \right] \\
	& = c_{m,n} \frac{2^m d^n (2\pi)^\frac{d}{2}}{\Gamma\left(\frac{d}{2}\right)} \frac{\Gamma\left( \frac{d}{2} + m \right) \Gamma\left( \frac{d}{2} + 2 - (3 m + 2 n) \right) (d - 2 (m + n))}{\Gamma(d+2-2(m+n))} \cdot \mu^{d+1-2(m+n)}.
\end{aligned}
\end{equation}
Note that the $(0, 0)$-contribution gives back Eqs.~\eqref{eq:mu_LO} and \eqref{eq:DeltaQ_LO}. Similarly, the contributions of edge operators are as follows:
\begin{equation} \label{eq:QP}
\begin{aligned}
	Q^{(p)} & := \kappa_p \int_{\textnormal{cloud}} \dd{\vect{r}}  \left\langle \frac{\partial \Z{p}}{\partial U} \right\rangle \\
	& = \kappa_p \frac{2^\frac{d+4(1-p)}{6} d^p (2 \pi)^\frac{d}{2}}{\Gamma\left(\frac{d}{2}\right)} \mu^\frac{2 d - 4 - 2 p}{3}  \int_0^1 \dd{z} \, \delta'(z) (1 - z)^\frac{2d-1-2p}{3} \\
	& = \kappa_p \frac{2^\frac{d+4(1-p)}{6} d^p (2 \pi)^\frac{d}{2}}{\Gamma\left(\frac{d}{2}\right)} \frac{2 d - 1 - 2 p}{3} \cdot \mu^\frac{2 d - 4 - 2 p}{3},
\end{aligned}
\end{equation}
and
\begin{equation} \label{eq:DeltaP}
\begin{aligned}
	\Delta^{(p)} & := \kappa_p \int_{\textnormal{cloud}} \dd{\vect{r}} \, \left\langle \frac{\partial \Z{p}}{\partial U} \dot\chi - \Z{p} \right\rangle \\
	& = \kappa_p \frac{2^\frac{d+4(1-p)}{6} d^p (2 \pi)^\frac{d}{2}}{\Gamma\left(\frac{d}{2}\right)} \mu^\frac{2d-1-2p}{3} \cdot \int_0^1 \dd{z} \, (1 - z)^\frac{2d-1-2p}{3} \left[ \delta'(z) - \delta(z) \right] \\
	& = \kappa_p \frac{2^\frac{d+4(1-p)}{6} d^p (2 \pi)^\frac{d}{2}}{\Gamma\left(\frac{d}{2}\right)} \frac{2 d - 4 - 2 p}{3} \cdot \mu^\frac{2d-1-2p}{3},
\end{aligned}
\end{equation}
where a prime refers to a derivative with respect to $z$, and $\delta'(z)$ is defined via integration by parts.

Based on the above results, it will be convenient to identify the leading \(\mu\)-scaling of the \emph{integrated} operators in the conformal dimension with the following notation:
\begin{equation} \label{eq:MuScalings}
\begin{aligned}
    \mudim{\Op{m}{n}} & := d+1-2(m+n), \\
    \mudim{\Z{p}} & := \frac{2(d -p)-1}{3}.
\end{aligned}
\end{equation}

The goal is to analyze in detail the physical content hidden in Eqs~\eqref{eq:QMN} and \eqref{eq:DeltaMN} using \ac{dimreg}, whereby these expressions are analytically continued in $d$, with the dimension $d$ shifted $2\varepsilon\ll1$ away from its physical value. This regularization scheme is dimensionless in that it does not introduce any additional scale, hence every power-law divergence found in $\delta$-layer regularization will be absent here, and the corresponding edge \ac{eft} operators used to renormalize them will simply give finite contributions without undergoing renormalization. In fact, \ac{dimreg} is a neat way to isolate and analyse \emph{logarithmic} divergences, which necessarily coincide with the ones observed in $\delta$-layer regularization. Here, they stem from simple poles of gamma functions in the numerator of Eqs~\eqref{eq:QMN} and \eqref{eq:DeltaMN}, and the renormalization procedure is thus rather straightforward to implement. After reviewing the structure of the large-charge expansion, we shall discuss this procedure in detail.

\subsection{Structure of the large-charge expansion}

So far, Eqs~\eqref{eq:QMN} and \eqref{eq:QP} yield the following structure (we omit coefficients for simplicity):
  \begin{equation}
    Q(\mu) = \mu^d \times
    \begin{bmatrix}
		1 & + & \mu^{-2} & + & \mu^{-4} & + & (\log) \\
		~ & + & \mu^{-\frac{d+4}{3}} & + & \mu^{-\frac{d+6}{3}} & + & \ldots
    \end{bmatrix}.
  \end{equation}
  In the first line, we indicated that contributions to $Q(\mu)$ might not just be power-laws in $\mu$, but might also contain logarithms, as alluded to before. This only happens in even $d$. Inverting this expression to find $\mu$ as an expansion in $Q$ gives rise to the following multi-layer expansion:
  \begin{equation}
    \mu(Q) = Q^\frac{1}{d} \times
    \begin{bmatrix}
		1 & + & Q^{-\frac{2}{d}} & + & Q^{-\frac{4}{d}} & + & (\log) \\
		~ & + & Q^{-\frac{d+4}{3d}} & + & Q^{-\frac{d+6}{3d}} & + & (\log) \\
		~ & + & Q^{-\frac{2d+8}{3d}} & + & Q^{-\frac{2d+10}{3d}} & + & \ldots
    \end{bmatrix}.
  \end{equation}
  When $d$ is odd, the third line really generates terms that are not already present in the first two, and its presence is due to the mixing of the second line in $Q(\mu)$ with itself upon inverting the expression. Similarly, Eqs.~\eqref{eq:DeltaMN} and \eqref{eq:DeltaP} imply, schematically, that
  \begin{equation}
    \begin{aligned}
      \Delta_{Q(\mu)} & = \mu^{d+1} \times
             \begin{bmatrix}
               1 & + & \mu^{-2} & + & \mu^{-4} & + & (\log) \\
               ~ & + & \mu^{-\frac{d+4}{3}} & + & \mu^{-\frac{d+6}{3}} & + & \ldots
             \end{bmatrix} \\
           & + \mu^0 \hspace{3.5mm} \times \left[ 1 + \mu^{-\frac{2}{3}} + \mu^{-\frac{4}{3}} + \text{(log)} \right].
    \end{aligned}
  \end{equation}
  As we shall see later, logarithms in the first line only appear in even $d$ and for terms with total $\mu$-scaling $> 0$. In the second line, we also anticipated the contributions of quantum corrections, the leading of which will be analysed in the next section and scales like $\mu^0$ or $\log \mu$. However, upon inserting the solution for $\mu(Q)$, we might generate terms with (multiple) logarithms even with total $Q$-scaling $\leq 0$. The final result has the form 
  \begin{equation} \label{eq:DeltaQ_Structure}
    \begin{aligned}
      \Delta_Q & = Q^\frac{d+1}{d} \times
             \begin{bmatrix}
               1 & + & Q^{-\frac{2}{d}} & + & Q^{-\frac{4}{d}} & + & (\log) \\
               ~ & + & Q^{-\frac{d+4}{3d}} & + & Q^{-\frac{d+6}{3d}} & + & (\log) \\
               ~ & + & Q^{-\frac{2d+8}{3d}} & + & Q^{-\frac{2d+10}{3d}} & + & \ldots
             \end{bmatrix} \\
           & + Q^0 \hspace{4mm} \times \left[ 1 + Q^{-\frac{2}{3d}} + Q^{-\frac{4}{3d}} + \text{(log)} \right].
    \end{aligned}
  \end{equation}
  Of course, the scaling of terms in these asymptotic series sometimes coincide, and not all terms are associated with a new Wilsonian coefficients.

\subsubsection{Example: $d = 2$}
  For the sake of clarity, let us see what happens in $d = 2$ up to $\mathcal{O}(Q^0)$. We consider
  \begin{equation}
    \mathcal{L} = \Op{0}{0} + c_{1,0} \Op{1}{0} + c_{0,1} \Op{0}{1} + \kappa_0 \Z{0} + \kappa_1 \Z{1},
  \end{equation}
  where we normalize $c_0$ away for convenience. As can be seen from Eq.~\eqref{eq:QMN}, the operator $\Op{1}{0}$ gives a singular contribution to the total charge, so we use \ac{dimreg} in $d = 2 + 2 \varepsilon$ to first find
  \begin{equation}
    Q(\mu) = 2 \pi \mu^2 \left[ 1 + \frac{4 c_{1,0} \log\mu + 2 \kappa_0^{ren.}}{\mu^2} + \frac{2^\frac{4}{3} \kappa_1}{3 \mu^\frac{8}{3}} + \mathcal{O}\left( \mu^{-\frac{10}{3}} \right) \right].
  \end{equation}
  The result has been written in terms of the renormalized edge coefficient $\kappa_0^{ren.}$ introduced via
  \begin{equation}
    \kappa_0 = \kappa_0^{ren.} - \left( \frac{1}{\varepsilon} + 1 + \gamma + \log 2\pi \right) c_{0,1} - c_{0,1},
  \end{equation}
  where, on top of the pole, we also absorbed some finite pieces for later convenience. Correspondingly, we get
  \begin{equation}
    Q(\mu) = 2 \pi \mu^2 \left[ 1 + \frac{4 c_{1,0} \log\mu + 2 \kappa_0^{ren.}}{\mu^2} + \frac{2^\frac{4}{3} \kappa_1}{3 \mu^\frac{8}{3}} + \mathcal{O}\left( \mu^{-\frac{10}{3}} \right) \right]
  \end{equation}
  and therefore,
  \begin{equation}
    \mu(Q) = \left( \frac{Q}{2 \pi} \right)^\frac{1}{2} \left[ 1 - 2\pi \frac{c_{1,0} \log\frac{Q}{2\pi} + \kappa_0^{ren.}}{Q} - \frac{(32 \pi^4)^\frac{1}{3} \kappa_1}{3 Q^\frac{4}{3}} + \mathcal{O}\left( Q^{-\frac{5}{3}} \right) \right].
  \end{equation}
  It is also rather straightforward to see that
  \begin{equation}
    \Delta_{Q(\mu)} = \frac{4 \pi}{3} \mu^3 \left[ 1 + \frac{6 c_{1,0}}{\mu^2} - \frac{2^\frac{4}{3} \kappa_1}{\mu^\frac{8}{3}} + \mathcal{O}\left( \mu^{-3} \right) \right].
  \end{equation}
  Finally, upon inserting the solution for $\mu(Q)$, we find
  \begin{equation}
    \begin{aligned}
        \Delta_Q & = \frac{2 Q^\frac{3}{2}}{3 \sqrt{2 \pi}} \left[ 1 - \frac{6 \pi ( c_{1,0} \log Q + \tilde\kappa_0^{ren.})}{Q} - \frac{3 (32 \pi^4)^\frac{1}{3} \kappa_1}{Q^\frac{4}{3}} + \mathcal{O}\left( Q^{-\frac{3}{2}} \right) \right] \\
        & = \frac{2 Q^\frac{3}{2}}{3 \sqrt{2 \pi}} - \sqrt{8\pi} Q^\frac{1}{2} (c_{1,0} \log Q + \tilde\kappa_0^{ren.}) - (2^{13} \pi^5)^\frac{1}{6} \kappa_1 Q^\frac{1}{6} + \mathcal{O}\left( Q^0 \right),
      \end{aligned}
    \end{equation}
    where $\tilde\kappa_0^{ren.}$ is the same as $\kappa_0^{ren.}$ up to finite terms. The presence of the $\sqrt{Q} \log Q$ term was first observed in~\cite{Kravec:2018qnu} for \(d=2\). The full renormalization procedure was subsequently presented in~\cite{Hellerman:2020eff}.
  
A couple of comments are in order. In Eqs~\eqref{eq:QMN} and \eqref{eq:DeltaMN}, the simple poles come from $\GammaF{\frac{d}{2}+2-(3m+2n)}$ in the numerator but, additionally, there is a factor of $(d-2(m+n))$ multiplying this gamma function in $\Delta^{(m,n)}$. So in the presence of the operator $\Op{1}{0}$ in $d=2$, a pole---which is turned into $\log\mu$ after renormalization---appears in $Q(\mu)$, but not in $\Delta_{Q(\mu)}$. One may therefore wonder whether $\Delta_Q$ will have logarithmic enhancements in the large-charge expansion. The answer is positive, as one has to first \emph{express $\mu$ as a function of Q}, thereby transferring all logarithmic enhancement to the final expression. The upshot is that we do not need to worry about the factor of $(d - 2 (m + n))$, and we can concentrate on the poles of the gamma functions in Eq.~\eqref{eq:DeltaMN}. Another observation that can be made from the above computation is that, again due to the fact that we need to invert the expression $Q(\mu)$, the conformal dimension $\Delta_Q$ would eventually include a term of the form
  \begin{equation}
    Q^\frac{3}{2} \left( \frac{c_{1,0} \log Q}{Q} \right)^2 = \frac{c_{1,0}^2 \log^2 Q}{Q^\frac{1}{2}},
  \end{equation}
had we pushed the expansion further. This is the first term in the series starting with $Q^\frac{d+1}{d} Q^{-\frac{2d+8}{3d}} = Q^\frac{d-5}{3d}$ in Eq.~\eqref{eq:DeltaQ_Structure}, and we see that in $d=2$, it gets a double logarithmic enhancement.

\subsection{Logarithmic terms from dimensional regularization}
\label{sec:logarithmic-terms}

In Eq.~\eqref{eq:DeltaMN}, we evaluated the leading-order classical (\emph{i.e.} tree-level) contribution of an insertion of the operator \(\Op{m}{n}\) to the large-charge conformal dimension to be
\begin{equation} \label{eq:DeltaMN_bis}
    \Delta^{(m,n)} = C_{m,n}^d \frac{\Gamma\left( \frac{d}{2} + 2 - (3 m + 2 n) \right)}{\Gamma\left( d + 2 - 2 (m + n) \right)} \cdot \mu^{d+1-2(m+n)},
\end{equation}
where $C_{m,n}^d$ is a $\mu$-independent constant that incorporates the Wilsonian coefficient and numerical factors. 

The gamma function has no zeros, but has single poles when the argument is a non-positive integer:
\begin{equation}
   \Gamma(-k + \varepsilon) = \frac{(-1)^k}{k!} \pqty{\frac{1}{\varepsilon} + H_k - \gamma} + \order{\varepsilon},
   \qquad k = 0,1, \ldots,
\end{equation}
where \(H_k = \sum_{l=1}^k \tfrac{1}{l}\) is the \(k\)-th harmonic number. We emphasize again that, compared to the $\delta$-layer regularization discussed in the previous section, the analytic continuation in terms of gamma functions captures the same physical content while dramatically simplifying the identification of logarithmic terms, leaving us with the simpler task of identifying the location of the poles and demonstrating that there always exists an appropriate counterterm for each of them.

\subsubsection{Location of the poles}

Let us enumerate which values of \(m\) and \(n\) give rise to poles, and at what order in \(\mu\) they enter the large charge conformal dimension. From the expression of \(\Delta^{(m,n)}\) in Eq.~\eqref{eq:DeltaMN_bis}, we see that there is a pole every time the argument of the gamma function in the numerator is a non-positive integer, and the argument of the gamma function in the denominator is a positive integer (\emph{i.e.,} every time the numerator exhibits a pole that is not canceled by a pole in the denominator):
\begin{equation}
  \label{eq:bulk-divergence-conditions}
  \begin{cases}
    \frac{d}{2} + 2 - (3 m + 2 n) = 0, -1, -2, \dots \\
     d + 2 - 2 (m + n) = 1, 2, 3, \dots
  \end{cases}
\end{equation}

\begin{figure} [!h]
  \centering
  \includegraphics{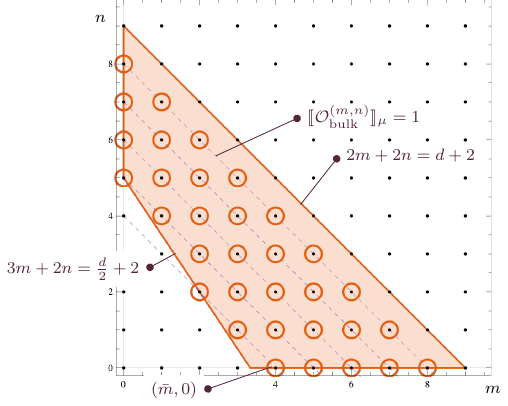}
  \caption{Divergent bulk operators \(\Op{m}{n}\). The operators corresponding to lattice points between the lines \( 2 m + 2 n = d + 2 \) and \(3m + 2n =\frac{d}{2} + 2\) have a simple pole and need to be regularized (shaded region).
    All the operators on a line \(m + n  = \text{const.}\) have the same \(Q\)-scaling (dashed lines).
  The operator \(\Op{\bar m}{0}\) has the highest \(Q\)-scaling.}
  \label{fig:Lines-and-dots}
\end{figure}

\begin{table}
  \centering
  \begin{tabular}{p{1em}lcc}
    \toprule
    \(d\) & \((m,n)\) & \(\mudim{\Op{m}{n}}\)  \\
    \midrule
    \(2\) & \((1,0)\) & \(1\) & \ \raisebox{-.6\height}{\includegraphics[width=3em]{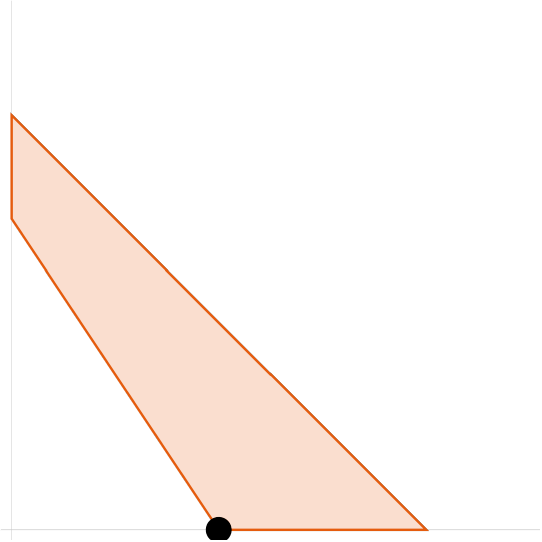} } \\[1em]
    \(4\) & \((2,0)\), \((1,1)\), \((0,2)\) & \(1\) & \raisebox{-.6\height}{\includegraphics[width=3em]{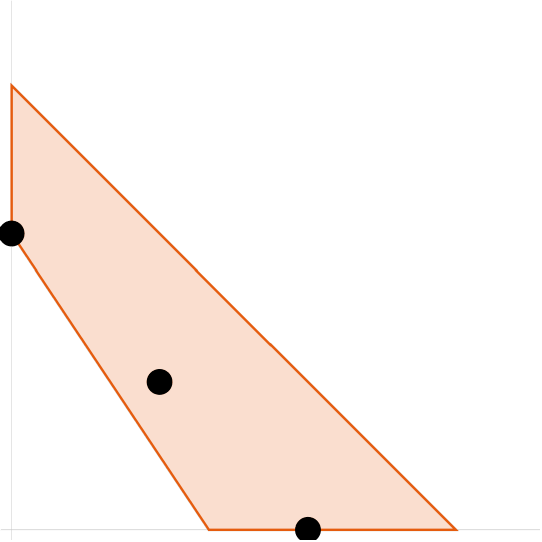}} \\[1em]
    \(6\) & \((2,0)\), \((1,1)\) & \(3\) & \multirow{2}{*}{\includegraphics[width=3em]{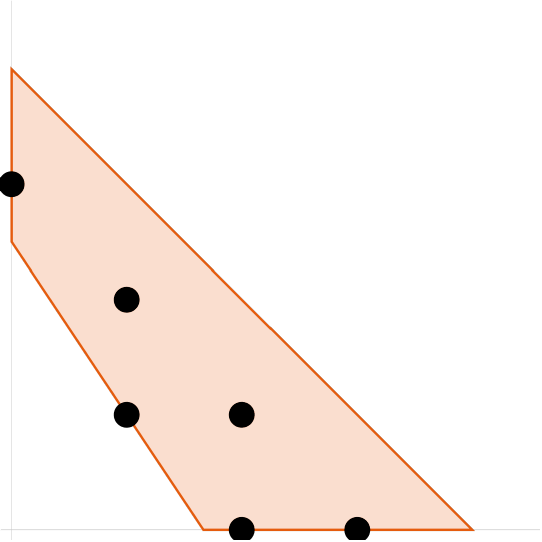}} \\ 
          & \((0,3)\), \((1,2)\), \((2,1)\), \((3,0)\) & \(1\) \\[1em]
    \(8\) & \((2,0)\) & \(5\) & \multirow{3}{*}{\includegraphics[width=3em]{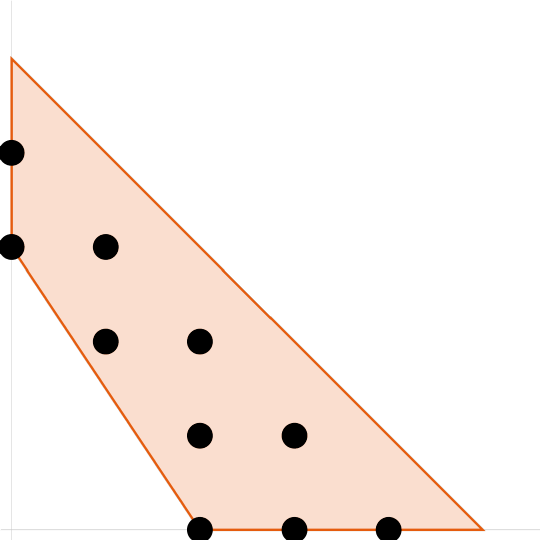}} \\
          & \((0,3)\), \((1,2)\), \((2,1)\), \((3,0)\) & \(3\) \\
          & \((0,4)\), \((1,3)\), \((2,2)\), \((3,1)\), \((4,0)\) & \(1\)\\
    \bottomrule
  \end{tabular}
  \caption{Divergent bulk operators  in different dimensions.}
  \label{tab:divergent-bulk-operators}
\end{table}

The solutions to these equations for the first few values of \(d\) are given in Table~\ref{tab:divergent-bulk-operators} (see also Figure~\ref{fig:Lines-and-dots}).
The first condition can only be satisfied in even $d$.\footnote{It is not clear to us if there is a simple or intuitive interpretation of this general rule.} The second condition can be rewritten using the \(\mu\)-scaling of the operator \(\Op{m}{n}\) as \(\mudim{\Op{m}{n}} + 1 = 1, 2, 3 \dots\)
However, since \(d \) must be even, the $\mu$-scaling must be odd and positive:
\begin{equation} \label{eq:evenD_logScaling}
  \mudim{\Op{m}{n}} = d + 1 - 2 (m + n ) = 1, 3, 5, \dots
\end{equation}
We have thus uncovered the following general property of the large-charge expansion of $\Delta_Q$: \emph{terms may acquire logarithmic enhancement from tree-level insertions only when $d$ is even and when their $Q$-scaling is positive}. The largest possible such scaling corresponds to the operator \(\Op{\bar m}{0}\), where \((\bar m,0)\) is the lattice point closest to the line \(\frac{d}{2} + 2 - 3 m - 2n= 0\) (see Figure~\ref{fig:Lines-and-dots}):
\begin{equation}
  \bar m = \left\lceil \frac{d}{6} + \frac{2}{3} \right\rceil\ ,
\end{equation}
so that the $\mu$-scaling of any other divergent bulk operator must satisfy
\begin{equation}
  1 \leq \mudim{\Op{m}{n}} \leq \mudim{\Op{\bar m}{0}} = d + 1 - 2  \left\lceil \frac{d}{6} + \frac{2}{3}   \right\rceil\ .
\end{equation}

\subsubsection{Dimensional regularization}

Dimensional regularization is then easily implemented: we simply replace $d$ by $d+2\varepsilon$ in Eq.~\eqref{eq:DeltaMN_bis}. When the conditions given in Eq.~\eqref{eq:bulk-divergence-conditions} are satisfied (in which $d$ remains an integer of course), we have
\begin{equation}
\begin{aligned}
    \Delta^{(m,n)}
    & = C_{m,n}^{d+2\varepsilon} \frac{\Gamma(-k+\epsilon)}{\Gamma(d+2-2(m+n)+\varepsilon)} \cdot \mu^{d+1-2(m+n)+2\varepsilon} \\
    & = \frac{(-1)^k C_{m,n}^d}{k! \Gamma(d+2-2(m+n))} \left( \frac{1}{\varepsilon} + 2 \log\mu + \text{(finite)} \right) \cdot \mu^{d+1-2(m+n)}
\end{aligned}
\end{equation}
where
\begin{equation} \label{eq:K}
    k := -\left[\frac{d}{2} + 2 - \left(3m+2n\right) \right] \in \mathbb{N} \qquad \text{under conditions \eqref{eq:bulk-divergence-conditions}}.
\end{equation}
As per our previous discussion in $d=2$, the coefficient $C_{m,n}^d$ may have been equal to zero from the beginning. But regardless of this, a $\log Q$ enhancement occurs whenever the conditions \eqref{eq:bulk-divergence-conditions} are satisfied, stemming from the relationship between $\mu$ and $Q$, meaning that the large-charge conformal dimension contains a term of the form
\begin{equation}
    Q^\frac{d+1-2(m+n)}{d} \log Q
\end{equation}
(as well as a term $Q^\frac{d+1-2(m+n)}{d}$ without enhancement). Moreover, looking at Figure~\ref{fig:Lines-and-dots}, it should be clear that there may be several divergent operators with fixed value of $m+n$ (the dashed lines) giving rise to the above contribution. We shall denote the corresponding cumulative singular contribution to $\Delta_Q$ as
\begin{equation} \label{eq:DeltaSing}
    \Delta^{(m,n)}_{sing.} := \frac{b}{\varepsilon} \cdot Q^\frac{d+1-2(m+n)}{d},
\end{equation}
where $b$ is the resulting coefficient. Removing such a divergence is our final task.

\subsubsection{Counterterms and renormalization}

Consider a bulk \ac{eft} operator $\Op{m}{n}$ whose contribution to the large-charge conformal dimension is singular and gives rise to Eq.~\eqref{eq:DeltaSing}. For any such operator, we want to show that there exists an edge \ac{eft} operator with the same $\mu$-scaling, that is, there exists $p\in\mathbb{N}$ such that
\begin{equation}
\label{eq:edge-regulation}
    \mudim{\Z{p}} = \mudim{\Op{m}{n}} \ .
\end{equation}
But using Eq.~\eqref{eq:MuScalings}, if $d$, $m$ and $n$ satisfy the conditions of Eq.~\eqref{eq:bulk-divergence-conditions}, the above equation becomes
\begin{equation}
\label{prelimConditionForBulkAndEdgeMapping}
  p 
  = n - \left[\frac{d}{2} + 2 - \left(3m+2n\right) \right] = n + k \in \mathbb{N},
\end{equation}
where $k$ was introduced in Eq.~\eqref{eq:K} above. We have therefore showed that (possibly cumulative) contributions from singular bulk \ac{eft} operators are always cured by the presence of a corresponding edge \ac{eft} operator. The coefficient of the latter gets renormalized through the relation
\begin{equation}
    \kappa_0 = \kappa_0^{ren.} - \frac{\text{cst}}{\varepsilon} + \text{(finite)},
\end{equation}
where the constant is fixed in a way that cancels the singular term, Eq.~\eqref{eq:DeltaSing}, as was done in the $d=2$ example before.

\subsection{Lessons from the large-charge Goldstone EFT}

This concludes the renormalization of singularities, leading to non-trivial tree-level logarithmic enhancements that have no counterpart in relativistic \acp{cft} at large charge.
Similar terms, however, do appear when one considers parity-violating CFT \cite{Cuomo:2021qws} or spinning operators \cite{Cuomo:2017vzg,Cuomo:2019ejv}.

It is striking how simple the analysis turned out to be using the large-charge Goldstone \ac{eft} and \ac{dimreg}. Of course, this is the culmination of the results discussed in previous sections, from which a lot of intuitive and quantitative understanding was gained. In particular, we proposed in the introduction a compact, heuristic argument that captures the main features of the trapped critical system. Moreover, we leveraged the power of \ac{dimreg} to isolate the tree-level logarithmic enhancements.

The agreement between the predictions of Sections~\ref{sec:NR_EFT_systematicsI} and \ref{sec:NR_EFT_systematicsII} is a very non-trivial cross-check of the various formalisms and schemes used in this chapter. The next natural step is therefore to use the large-chare Goldstone \ac{eft} with dimensional regularization to compute the large-charge expansion of further observables, akin to what has been initiated in \cite{Kravec:2018qnu,Kravec:2019djc} for higher-point functions and phases featuring vortices due to the presence of a large amount of angular momentum.

The results and methods that we found are a thorough and, in some sense, exhaustive extension of the seminal work \cite{Son:2005rv}.
\pagebreak
\section{Quantum corrections}%
\label{sec:NR_EFT_quantum}

In Sections~\ref{sec:NR_EFT_systematicsI} and \ref{sec:NR_EFT_systematicsII}, we thoroughly discussed the structure of the large-charge expansion of $\Delta_Q$ as given by \emph{tree-level} contributions, and we alluded to the quantum corrections at the beginning of Section~\ref{sec:NR_EFT_systematicsII}. It is however time to be more explicit about the latter.

The leading quantum contribution to the conformal dimension $\Delta_Q$ enters at order $Q^0$. This comes from the Casimir energy of the fluctuations over the ground state, whose spectrum is given by~\cite{Kravec:2018qnu}
\begin{align}
  E^d_{n,l} &= \sqrt{\frac{4 n}{d} (n + l + d - 1) + l}\ ,  & n, l &= 0, 1, \dots ,
\end{align}
with multiplicities $M^{d-1}_l$ enumerated by the modes on the \(S^{d-1}\) sphere:
\begin{equation}
  M^{d-1}_l = \pqty{ 2l + d -2} \frac{\Gamma(l+d-2)}{\Gamma(l+1) \Gamma(d-1)}\ .
\end{equation}
The Casimir energy is then given by\footnote{The prime indicates that the zero mode $n = l = 0$ is excluded.}
\begin{equation}
  \Ecasimir{d} = \frac{1}{2} \sideset{}{'}\sum_{n,l=0}^{\infty} M^{d-1}_l E^d_{n,l}\ .
\end{equation}
This expression is clearly divergent and needs to be regulated. This can typically be addressed in dimensional regularization using $\zeta$-function machinery. We are therefore facing a very similar, albeit physically different, problem as before. Indeed, there must exist a prescription to get rid of such a divergence in a way that leaves our physical observable $\Delta_Q$ scheme-independent. Just as before, since \ac{dimreg} is a dimensionless scheme, the only possible output is either a finite $Q^0$ contribution, or a singular one in the form of a simple pole in $\varepsilon$. The possibility of having $Q^0$ \emph{tree-level} contributions then has radically different consequences in these two scenarios. In the former case, tree-level and quantum contributions would mix, so quantum corrections (at least the leading one) would not have a distinct signature in the large-charge expansion, while in the latter case, the presence of tree-level $Q^0$ contributions is \emph{required} and serves as a counterterm to renormalize the divergence, akin to edge \ac{eft} operators serving as counterterms for edge singularities. The output is thus expected to be a $Q^0 \log Q$ contribution, which may possibly mix with $Q^0 \log Q$ tree-level contributions, if any.

In fact, as can be seen from our previous analysis of the structure of $\Delta_Q$, Eq.~\eqref{eq:DeltaQ_Structure}, tree-level contributions \emph{cannot contribute at $\order{Q^0}$ in even dimensions}. So the Casimir energy in even dimensions had better not be singular, otherwise this would signal a dramatic inconsistency. We shall however show that, in $d=2$, the situation is safe: the Casimir energy is simply given by a finite number. Since this is the only $Q^0$ contribution, it is a model-independent prediction valid across the whole superfluid large-charge universality class. In odd dimensions, there is always at least one operator contributing to $\order{Q^0}$, and we find that, at least in $d=3$, the Casimir energy is singular. The outcome in this case is indeed a $Q^0 \log Q$ quantum contribution and, since there is no tree-level logarithmic enhancement in odd dimensions, this term is again a universal prediction for the superfluid large-charge universality class.

Note that the Casimir energy contains higher-order corrections in \(1/Q\), of course, which generically will have to be regulated and can give logarithms multiplied by negative powers of $Q$ in the conformal dimension. In any case, they will have to be regulated by bulk or edge operators that are not divergent (only bulk operators with positive \(\mu\)-scaling can be divergent), so we do not face situations in which a divergence in the Casimir energy is to be regulated by a bulk operator that must, itself, be regulated by an edge term. This excludes the presence of double logarithms coming from one-loop effects in the expansion.

\subsection{The universal \texorpdfstring{$Q^0$}{} quantum contribution in \texorpdfstring{$d=2$}{}}

In $d=2$ the Casimir energy may be written as $E^{(d=2)}_{\text{Cas}} = E_2(-1/2)$, where
\begin{equation}
  E_2(s) = \frac{1}{2} \sum_{\substack{n=0,l = -\infty \\ (n,l) \neq (0,0)}}^\infty  \pqty{ 2 \pqty{n + \frac{1}{2} } \pqty{n + \abs{l} + \frac{1}{2} } - \frac{1}{2} }^{-s}.
\end{equation}
We are going to massage this expression until we recover multivariate zeta functions,
\begin{equation}
  \zeta(s_1, s_2) := \sum_{n_1 > n_2 \ge 1 } \frac{1}{n_1^{s_1} n_2^{s_2}},
\end{equation}
that can be analytically continued to a meromorphic function on \(\setC^2\)~\cite{Goncharov:2001iea}.

\begin{figure}
  \centering
  \tikzset{mystyle/.style={shape=circle,fill=black,scale=0.3}}
\tikzset{withtext/.style={fill=white}}

  \begin{tikzpicture}[scale=.5]
    \foreach \x in {0,...,8}
    \foreach \y in {1,...,4}
    {
      \node[mystyle] at (\x,\y){};
    }
    \foreach \x in {0,...,8}
    \foreach \y in {-4,...,-1}
    {
      \node[mystyle] at (\x,\y){};
    }
    \foreach \x in {1,...,8}
    {
      \node[mystyle] at (\x,0){};
    }
    \draw[-latex] (-.5,0) -- (8.5,0);
    \node at (9,0) {\(n\)};
    \draw[-latex] (0,-4.5) -- (0,4.5);
    \node at  (0,5) {\(l\)};
    \draw[smooth,red,thick] (.75,4.5) -- (.75,.75) -- (8.5, .75);
    \node[red] at (-.75,4.5) {\(I\)};
    \draw[smooth,red,thick] (.25,4.5) -- (.25,.75) -- (-.25, .75) -- (-.25,4.5);
    \node[red] at (8.5,4.5) {\(II\)};
    \draw[smooth,red,thick] (8.5,.25) -- (.75,.25) -- (.75, -.25) -- (8.5,-.25);
    \node[red] at (8.75,-.75) {\(III\)};
  \end{tikzpicture}
  \caption{Decomposition of the summing region. The summand is symmetric under the exchange \(l \to - l\).}
  \label{fig:summing-regions}
\end{figure}
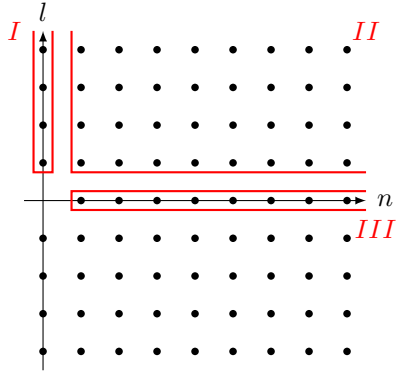

First we use the binomial expansion
\begin{equation}
  E_2(s) = \frac{1}{2^{s+1}} \sum_{n,l} \sum_{k=0}^{\infty} \binom{-s}{k} \pqty{-\frac{1}{4} }^k \pqty{n + \frac{1}{2} }^{-s-k} \pqty{n + l + \frac{1}{2} }^{-s -k} = \sum_{n,l,k} e_{n,l,k}(s)
\end{equation}
and separate the summing region into three parts (see figure~\ref{fig:summing-regions}):
\begin{align}
  I &= \set{(n,l) | n = 0, l \in \setN^*} \\
  II &= \set{ (n,l) | n \in \setN^* , l \in \setN^*} \\
  III &= \set{ (n,l) | n \in \setN^*, l = 0}.
\end{align}
The sum becomes
\begin{equation}
  E_2(s) = 2 E_2^I (s) + 2 E_2^{II} (s) + E_2^{III}(s),
\end{equation}
where
\begin{align}
  E_2^I(s) &= \sum_{l=1}^\infty \sum_{k = 0}^\infty e_{n,l,k}(s) = \frac{1}{2^{s+1}} \sum_{k=0}^\infty \binom{-s}{k}(-1)^k 2^{s - k} \zeta_H(s +k| \tfrac{3}{2}), \\
  E_2^{III}(s) &= \sum_{n=1}^\infty \sum_{k = 0}^\infty e_{n,l,k}(s) = \frac{1}{2^{s+1}} \sum_{k=0}^\infty \binom{-s}{k}\pqty{- \frac{1}{4} }^k  \zeta_H(2s +2k| \tfrac{3}{2}).
\end{align}
We made use of the Hurwitz zeta function:
\begin{equation}
  \zeta_H(s| x) = \sum_{n = 0}^\infty \frac{1}{(s + x)^s} = \sum_{k=0}^\infty \binom{-s}{k} \pqty{x - 1}^{k} \zeta(s+k) .
\end{equation}
In the region \(II = \set{n > 1, l >1}\) we can write the sum in terms of the multivariate Hurwitz zeta function~\cite{murty2006multiple}
\begin{equation}
  \zeta_H(s_1, s_2 | x_1, x_2) = \sum_{n_1 > n_2 \ge 1 } \frac{1}{(n_1 + x_1)^{s_1} (n_2 + x_2)^{s_2}} 
\end{equation}
as
\begin{equation}
  E_2^{II}(s) = \frac{1}{2^{s+1}} \sum_{k=0}^\infty \binom{-s}{k} \pqty{-\frac{1}{4} }^k \zeta_H(s+k, s+k| \tfrac{1}{2},\tfrac{1}{2})
\end{equation}
or then apply the binomial expansion twice and write
\begin{equation}
  E_2^{II}(s) = \frac{1}{2^{s+1}} \sum_{k, j_1, j_2 = 0}^\infty \binom{-s}{k} \pqty{-\frac{1}{4} }^k \binom{-s-k}{j_1} \binom{-s-k}{j_2} \pqty{\frac{1}{2}}^{j_1 + j_2} \zeta(s +k +j_2, s+k +j_1) .
\end{equation}
The factor in front of the zeta function in the sum is symmetric under the exchange \(j_1 \leftrightarrow j_2\), so we can use the reflection identity~\cite{Euler:1776}
\begin{equation}
  \zeta(s_1, s_2) + \zeta(s_2, s_1) = \zeta(s_1) \zeta(s_2) - \zeta(s_1 + s_2)
\end{equation}
and rewrite \(E_2^{II}(s)\) as
\begin{multline}
  E_2^{II}(s) =  \frac{1}{2^{s+2}} \sum_{k = 0}^\infty \binom{-s}{k}  \pqty{-\frac{1}{4} }^k \Bigg[ \zeta_H(s+k|\tfrac{3}{2})^2 \\ - \sum_{j_1,j_2 = 0 }^\infty \binom{-s-k}{j_1} \binom{-s-k}{j_2} \frac{1}{2^{j_1 + j_2}} \zeta(2s +2 k +j_1 + j_2) \Bigg] . 
\end{multline}
The sums in \(E_2^{II}(s)\) and \(E_2^{III}(s)\) have poles in \(s = -1/2\), respectively with residue \(\sqrt{2}/16\) and \(-\sqrt{2}/8\), because they both include \(\zeta(1)\).
The fact that the two poles cancel in \(E_2(s)\) is a nice confirmation of our chosen regularization.
The final series converges numerically very rapidly, and we find
\begin{equation}
  E^{(d=2)}_{\text{Cas}} = -0.294159\dots,
\end{equation}
which is the promised universal $Q^0$ contribution in $d=2$.


\subsection{The universal \texorpdfstring{$Q^0 \log Q$}{}  quantum contribution in \texorpdfstring{$d=3$}{}}

Let us first set the stage in general odd dimension. As argued at the beginning of this section, the Casimir energy in $d+2\varepsilon$ may be expected to develop a simple pole in $\varepsilon$ when $d$ is odd. Let us therefore suppose that
\begin{equation}
  \Ecasimir{d+2\varepsilon} = \frac{E_{-1}}{\varepsilon} + E_0 + \order{\varepsilon}.
\end{equation}
Then, we can use bulk operators \(\Op{m}{n}\) satisfying \(m+n = (d+1)/2\) as counterterms in \ac{dimreg}. We can write their cumulative leading-order contribution to $\Delta_Q$ as\footnote{For convenience, we abuse slightly the notation of Section~\ref{sec:NR_EFT_systematicsII} and reuse $C_{m,n}^d$ to denote here the cumulative contribution.}
\begin{equation}
\begin{aligned}
  \Delta^{(m,n)} = C_{m,n}^{d} + \order{\varepsilon}.
\end{aligned}
\end{equation}
The individual Wilsonian coefficients of these operators can be renormalized arbitrarily as
\begin{equation}
    c_{m,n} = c_{m,n}^{\text{ren}} + \frac{c_{m,n}^{(-1)}}{\varepsilon},
\end{equation}
where $c_{m,n}^{\text{ren}}$ and $c_{m,n}^{(-1)}$ are regular, provided the following overall condition is satisfied:
\begin{equation}
    C_{m,n}^{d+2\varepsilon} = \eval{C_{m,n}^{d+2\varepsilon}}_{c_{m,n}\to c_{m,n}^{\text{ren}}} - \frac{E_{-1}}{\varepsilon}\ .
\end{equation}
Reminiscent of the tree-level edge divergences, the result is that the divergence in the Casimir energy is traded for a logarithm:
\begin{equation}
\begin{aligned}
    \lim_{\varepsilon \to 0}(\Ecasimir{d+2\varepsilon} + \Delta^{(m,n)}) &=  \text{const.} - 2 E_{-1} \log(\mu) \\
    &= \text{const.} - \frac{2}{d} E_{-1} \log(Q)\ . 
    \label{eq:general-one-loop-log}
\end{aligned}
\end{equation}
Since there are no tree-level logarithmic enhancements in odd dimensions, the term proportional to $E_{-1}$ above is universal, similar to what was discussed in the relativistic context~\cite{Cuomo:2020rgt}.

Let us focus on $d=3$ from now on, and compute $E_{-1}$ explicitly. The Casimir energy is
\begin{equation}
    \Ecasimir{3+2\varepsilon} = \frac{1}{2} \sideset{}{'}\sum_{n,l=0}^{\infty} M^{2+2\varepsilon}_l E^{3+2\varepsilon}_{n,l}\ .
\end{equation}
In the limit \(\varepsilon \to 0\), the multiplicity is
\begin{equation}
  M_l^{2 + 2\varepsilon} = \pqty{2l + 1} + 2\left[ \pqty{2l +1} H_l - 2l \right] \varepsilon + \order{\varepsilon^2}\ ,
\end{equation}
where \(H_l\) is again the \(l\)-th harmonic number.
We are interested in the behavior around the pole, which is completely determined by the large-\(l\) behavior of the sum.
We can thus further expand
\begin{equation}
\begin{aligned}
    M_l^{2 + 2\varepsilon}
    & = \pqty{2 l + 1} + 2\pqty{ \pqty{2l+1} \pqty{\log(l) + \gamma - 1} + \order{\frac{1}{l}} } \varepsilon \\
    & = \pqty{2 l + 1 + 4 \varepsilon} \pqty{l^{2\varepsilon} + \pqty{\gamma -1} 2\varepsilon} + \order{\frac{\varepsilon}{l} }.
\end{aligned}
\end{equation}
Up to terms that vanish in the \(\varepsilon \to 0\) limit, the Casimir energy is given by
\begin{equation}
    \Ecasimir{3+ 2\varepsilon} = \frac{1}{2} \sideset{}{'}\sum_{n,l=0}^{\infty} \pqty{2l + 1} l^{2\varepsilon} \pqty{ \pqty{n + \frac{3}{4} } \pqty{l + n +\frac{5}{4} } - \frac{15}{16} }^{1/2} + \text{regular for \(\varepsilon \to 0\)~.}
\end{equation}
Applying the binomial expansion four times in a row, we can rewrite the sum in terms of the Mordell--Tornheim zeta function $\zeta_{MT}(s_1, s_2, s_3)$, defined as~\cite{mordell1958evaluation,tornheim1950harmonic,matsumoto2003mordell}
  \begin{equation}
    \zeta_{MT}(s_1, s_2, s_3) \equiv \sum_{n,l=1}^\infty l^{-s_1} n^{-s_2} (n + l)^{-s_3}~.
  \end{equation}
  We obtain:
  \begin{multline}
    \Ecasimir{3+ 2\varepsilon} = \left( \frac{3}{4} \right)^{-\sfrac{1}{2}} \sum_{k,k_1,k_2,k_3=0}^\infty \binom{\sfrac{1}{2}}{k} \binom{1}{k_1} \binom{\sfrac{1}{2}-k}{k_2} \binom{\sfrac{1}{2}-k}{k_3} 
	 \left( -\frac{15}{16} \right)^k \left( -\frac{1}{2} \right)^{k_1} \left( -\frac{1}{4} \right)^{k_2} \left( -\frac{3}{4} \right)^{k_3} \\
    \times \zeta_{MT}(k_1 -2 \varepsilon - 1 ,  k + k_2 - \sfrac{1}{2} , k + k_3 - \sfrac{1}{2}) + \text{regular.}
  \end{multline}
  The analytic structure of this special function is known~\cite{matsumoto2002analytic,tsumura2005mordell,matsumoto_value-relations_2008}, and from the expansion
\begin{equation}
\begin{aligned}
    \zeta_{MT}(s_1, s_2, s_3)
    & = \frac{\Gamma(s_2 + s_3 -1) \Gamma(1 - s_2)}{\Gamma(s_3)} \zeta(s_1 + s_2 + s_3 -1) \\
    & + \sum_{k=0}^{M-1} \binom{-s_3}{k} \zeta(s_1 + s_3 + k) \zeta(s_2 - k) + \text{regular,}
\end{aligned}
\end{equation}

we see that there are poles with unit residue if\footnote{Note that when $s_2$ is a positive (non-zero) integer, the expression admits two singular pieces that cancel each other.}
\begin{equation}
  \begin{cases}
    s_1 + s_3 = 1, 0, -1, -2, \dots \\
    s_2 + s_3 =  1, 0, -1, -2, \dots \\
    s_1 + s_2 + s_3 = 2~.
  \end{cases}
\end{equation}
In our case, only the third channel in this set can produce singularities, which arise for
\begin{equation}
  k_1 + k_2 + k_3 + 2 k -2 \varepsilon = 4~.  
\end{equation}
Summing all the contributions, we find the final result
\begin{equation}
  \Ecasimir{3 +2 \varepsilon} = -\frac{1}{ 2\sqrt{3} \varepsilon} + \text{regular.}
\end{equation}
Using the general formula in Eq.~\eqref{eq:general-one-loop-log}, we thus find that the conformal dimensions have a universal \(Q^0 \log(Q)\) term given by
\begin{equation}
\label{Q0LogQin3d}
  \eval{\Delta_Q}_{Q^0}  = \frac{1}{3\sqrt{3}} \log(Q) + \text{const.}
\end{equation}
It would be interesting to extend these kind of computations to higher dimensions or subleading order in the quantum corrections. This is left for future work.
\pagebreak
\section*{Closing word}
\addcontentsline{toc}{section}{Closing word}

Let us contemplate what we have achieved so far. In Section~\ref{sec:NR_EFT_simple}, we discussed how Coleman's dilaton-dressing prescription could help us understand the essential features of the nonrelativistic large-charge expansion by cooking up a simple class of large-charge dilaton \acp{eft}. It became immediately apparent that the nonrelativistic case offers a much richer analytical structure than its relativistic counterpart. This emerges from the inhomogeneity of the system, as mandated by the presence of an external potential. The latter is itself introduced to implement the state-operator correspondence and, in turn, offers a direct contact point with experimental data. The correct implementation of a complete large-charge \ac{eft} is therefore relevant both from the 'high-energy perspective' of collecting \ac{cft} data---falling broadly under the name \emph{conformal bootstrap}---and the 'low-energy perspective' of studying emergent collective behaviours of condensed matter systems such as the trapped unitary Fermi gas.

Even though the analysis of Section~\ref{sec:NR_EFT_simple} was somewhat rough, it clearly indicated the power of the large-charge \ac{eft} and its ability to capture features beyond any known results so far. In section~\ref{sec:NR_EFT_systematicsI}, we therefore undertook a thorough and exhaustive construction of the latter, still using Coleman's dilaton-dressing prescription and implementing the so-called $\delta$-layer regularization. We established the precise interplay between the bulk \ac{eft} and the complementary edge \ac{eft}. The latter is mandated by the breakdown of the former close to the edge of the cloud of particles, and a clear physical renormalization procedure was established. For a particle physicist, this kind of effective theory construction might seem exotic, but spontaneous symmetry breaking of spacetime symmetries due to a finite density (that is, condensed matter) state is rather common. These techniques are now even being applied in the context of confinement to effective string theories. Many lessons learnt here can therefore be lifted to various other situations. Importantly, we also implemented a small breaking of conformal symmetry via a small dilaton mass, which was the original motivation for the use of the dilaton formalism. The phenomenologically observable consequence of this is the presence of an otherwise absent $\sqrt{Q} \log Q$ term in $d=3$.\footnote{In this spirit, near-unitarity of the Fermi gas was for example discussed in~\cite{Escobedo:2009bh}.} As the precision of experiments performed on the unitary Fermi gas improve, it might eventually be possible to fit our proposed profile for the ground-state energy---including several subleading corrections---against these data, thereby extracting universal parameters beyond the leading one, that is, the famous Bertsch parameter.

In Section~\ref{sec:NR_EFT_systematicsII}, we first took some time to carefully analyse the intricate 'multi-series' expansion of $\Delta_Q$, as the bulk and the edge \acp{eft} contribute different kinds of expansions. We then went on to build the complete large-charge \ac{eft}, to be regarded as the extension of the Son \& Wingate \ac{eft} \cite{Son:2005rv}. Based on this, we elucidated the conditions for the appearance of logarithmic enhancements in the large-charge expansion of the large-charge conformal dimension $\Delta_Q$ stemming from tree-level insertions. Dimensional regularization provided a conformally invariant renormalization scheme that dramatically simplified this analysis, as compared to the preceding section: power-law divergences in $\delta$-layer regularization are simply absent after analytic continuation in $d$, and \ac{dimreg} therefore only captures the remaining simple poles that eventually turn into $\log Q$ enhancements. In odd dimension, we founds that such enhancements, caused by tree-level insertions, are absent. In even spatial dimension $d$, however, we found that such contributions can appear and are always multiplied by positive powers of $Q$. The corresponding simple poles are appropriately renormalized away thanks to the presence of edge \ac{eft} operators whose tree-level contributions have the same $Q$-scaling. We have provided a table of the particular $Q$-scalings where such logarithmic enhancements can occur in various even spatial dimensions (Table~\ref{tab:divergent-bulk-operators}); this pattern of classical \(\log(Q)\) enhancements is illustrated in Figure~\ref{fig:Lines-and-dots}.

Finally, in Section~\ref{sec:NR_EFT_quantum}, we have also analyzed the appearance of $Q^0$ or $Q^0 \log(Q)$ terms coming from the one-loop Casimir energy contribution to the ground-state energy. Using dimensional regularization to make the analysis tractable, we have found that the situation is in some sense the opposite as the one of tree-level insertions:
\begin{itemize} [itemsep=5pt,partopsep=0pt,topsep=0pt,parsep=0pt]
    \item In even spatial dimension, tree-level insertions generically give rise to logarithmic enhancements, but never contribute to order $Q^0$, whereas the leading quantum correction does not enjoy such an enhancement and therefore simply scales like $Q^0$. The latter does not depend on any Wilsonian coefficient: it appears as a number and is therefore a universal prediction valid across the whole superfluid large-charge universality class.
    \item In odd spatial dimension, tree-level insertions never acquire logarithmic enhancements, and there is always at least one operator contributing at $\order{Q^0}$. The latter serves as a counterterm for the Casimir energy, which is always divergent, resulting in a universal $Q^0 \log Q$ contribution. This is therefore the \emph{unique} logarithmic enhancement in odd dimension.
\end{itemize}
The fate of these leading quantum corrections is therefore completely analogous to the large-charge expansion in $(d+1)$-dimensional relativistic \ac{cft}, which were fully appreciated in~\cite{Cuomo:2020rgt}. We performed the explicit computation of the one-loop Casimir energy in $d=2$ and $d=3$. With sufficient precision, experimental data could potentially provide real-life checks of these predictions, akin to Regge intercepts.

\medskip
Let us wrap up. Via the nonrelativistic state-operator map, charged operators correspond to finite density states in a harmonic trap. We investigated the class of \acp{nrcft} whose large-charge sector is effectively described by a superfluid state in the trap, which is particularly relevant for the case of the unitary Fermi gas and critical anyons. Specifically, we exploited this effective description to extend all known results about the expansion of the conformal dimension of the lightest charged operator, up to and including the leading quantum corrections entering at order $Q^0$, thereby uncovering a rich structure of logarithmic enhancements. This was based on the recent treatment of divergences at the edge of the physical cloud of trapped particles discussed in~\cite{Hellerman:2020eff}, which we reviewed and extended. We also accounted for a small dilaton mass deformation in order to explore the near-conformal regime, and illustrated the full procedure in $d=2$ and $d=3$ cases. Whenever we found it appropriate, we commented on the connections with the ultracold atom literature and computed some new corrections, as in Eq.~\eqref{eq:linear_profile_result} for the doubly integrated density. This fruitful direction of research remains to be explored systematically. See also \cite{Son:2005rv,Escobedo:2009bh}.

Let us finally mention the seminal works \cite{Son:2008ye,Balasubramanian:2008dm} paving the way toward a gravity/\ac{nrcft} correspondence (see also \cite{Bekaert_2012} for a different proposal). It would be fascinating to understand whether the large-charge sector of certain \acp{nrcft} can be described in a dual picture and how this would relate to a semiclassical description akin to the one used here. The results presented here draw upon the rich structure inherent to conformal field theories, while relying on the dramatic simplifications afforded by large-quantum-number perturbation theory. Of course, much of this technology traces its parentage to a number of endeavors emerging from string theory, see \emph{e.g.} \cite{Berenstein:2002jq, Alday:2007mf, Komargodski:2012ek, Fitzpatrick:2012yx, Giombi:2021zfb}. As already emphasized, if cold atom systems in the unitary regime can be studied with sufficient precision, they may provide a crucial terrestrial test-bed for this increasingly far-reaching arena of techniques in theoretical physics.
\chapter{Microscopic realizations} \label{chap:ExplicitReal}
\setcounter{section}{0}

As promised, we are now going to discuss microscopic models whose large-charge sectors fit into the superfluid large-charge \ac{eft} framework discussed in the previous chapter. In fact, we are going to focus on a specific large-$N$ model describing the unitary Fermi gas, based on previous studies~\cite{Nikolic:2007zz,Veillette:2007zz,Sachdev2011,Bekaert_2012}. The novelty lies in the fact that we consider the system in a trapping potential. In particular, we extract the value of the Wilsonian coefficients in the large-charge \ac{eft} at large-$N$ and discuss the fate of edge contributions in this particular microscopic description. This is presented in Section~\ref{sec:UFG_largeN}, which based on~\cite{Hellerman:2023myh} by the author and his collaborators.

In Section~\ref{sec:MoreMicro}, we very briefly mention further microscopic models whose large-charge sectors could be investigated using similar techniques. In particular, we include a short description of some properties of anyons relevant to the large charge program.
\pagebreak
\section{The unitary Fermi gas at large charge and large N}
\label{sec:UFG_largeN}

We shall study the unitary Fermi gas in an external trapping potential starting from a microscopic theory in the limit of large charge and large number of fermion flavors $N$. We shall present an algorithmic procedure for extracting the large-charge expansion of the conformal dimension $\Delta_Q$ of the lightest operator of charge $Q\gg1$. The approach we propose here does not rely on any uncontrolled assumption. It takes the form of a gradient expansion valid in the interior of the cloud of particles, that is, far enough from its edge where the particle density drops to zero. In the latter region, we shall present a separate discussion resulting in the first microscopic evaluation of the contribution coming from the edge dynamics. The results stemming from the microscopic theory are matched against the predictions of the superfluid \acs{eft} of the last chapter, including the form of the gap and the energy of the system in a harmonic trap. As usual, the latter maps, via the non-relativistic state-operator correspondence, to $\Delta_Q$. This allows us to extract the values of the first three Wilsonian coefficients at leading order in $N$.

\subsection{Introduction}%
\label{sec:Introduction}



We devoted part of the first chapter to the description of the connection between the unitary Fermi gas and the nonrelativistic large-charge program. Here, we want to verify this connection explicitly from a specific microscopic theory. In order to motivate the model we are going to use, let us repeat the essential features characterizing this system.

Most importantly, the unitary Fermi gas is an experimentally accessible example of a system with Schrödinger symmetry.
It can be realized in the laboratory~\cite{ketterle2008making} by preparing an atomic gas, typically made out of ${}^6$Li or ${}^{40}$K, in an optical trap whose shape is approximately quadratic. The gas is then cooled down to very low temperatures and superfluidity emerges. Recall that, crucially, the gas is dilute, meaning that the inter-atomic spacing is much larger than the inter-atomic interaction range, which implies that $s$-wave scattering largely dominates, and the details of the inter-atomic potential do not matter. From a theoretical perspective, this justifies modeling atom scattering by a two-body contact interaction, whose strength is the only relevant parameter in the system and is related to the $s$-wave scattering length $a$. Ultracold Fermi gases can therefore be described in terms of a local nonrelativistic quantum field theory. Additionally, the parameter $a$ can be experimentally tuned using an external magnetic field. Upon going from the repulsive ($a>0$) to the attractive ($a<0$) regimes---or vice versa---the $s$-wave scattering length hits the unitary point where it diverges and the system is desribed by a local \acs{nrcft}. From a theoretical perspective, turning on a chemical potential\footnote{One could also consider imbalanced gases, requiring distinct chemical potentials for both fermion species.} sources a charge $Q$, corresponding to the number of particles in the system.\footnote{Turning on a chemical potential should not be viewed as a deformation of the theory in an RG sense. The chemical potential does not run as it couples to the charge density in a way that prepares a finite density state, not a different theory in the vacuum.} As we shall see, the corresponding ground state indeed corresponds to a superfluid state. Experimentally, the charge $Q$ is always large---typically $10^5$ or more---such that qualitative features of the theoretical large-charge description could be compared to measurable observables, such as for instance the doubly-integrated axial density, determined in~\cite{PhysRevLett.92.120401}.

In this section, we shall be interested in comparing our results, stemming from a microscopic description controlled by an artificially large number $N$ of fermion flavors, to the structure imposed by the large-charge \acs{eft} and extracting the corresponding Wilsonian coefficients. We can take Eq.~\eqref{eq:eff-action-nlo} as reference point since we shall keep track of the dimensionful constants $\hbar$ and $m$, in which case the Lagrangian evaluated in the superfluid ground state reads
\begin{equation} \label{eq:EFT_Lagr_nnlo}
    \frac{\hbar^3}{m^{3/2}}\Lag_{\text{GS}} = c_0(\hbar\mu - V(\vect{r}))^{5/2} + c_1 \frac{\hbar^2}{m}\frac{(\nabla V(\vect{r}))^2}{\sqrt{\hbar\mu - V(\vect{r})}} - 9 c_2\frac{\hbar^2}{m} \Laplacian V(\vect{r}) \sqrt{\hbar\mu - V(\vect{r})} + \ldots
\end{equation}
In the absence of the trapping potential, the behavior of the unitary Fermi gas is captured by a single Wilsonian coefficient, namely, $c_0$. Recall that we related this coefficient to the Bertsch parameter $\xi$~\cite{Chang:2007zzd}---defined as the ratio of the energy density of the gas at unitarity to that of the free Fermi gas at the same density---as $c_0 = \frac{2^{5/2}}{15\pi^2\xi^{3/2}}$.

Of course, the physics of the trapped system is described in terms of the additional Wilsonian coefficients $c_1$, $c_2$, etc. whose contributions to the observables are suppressed by fractional power of the charge, as reviewed in the last chapter. For most purposes, we will not need to commit to a specific choice of trapping potential\footnote{With respect to the previous notation, $V(\vect{r}) = \hbar A_0(\vect{r})$.} $V(\vect{r})$, although the specific choice
\begin{equation}
	V(\vect{r}) = \frac{m \omega^2}{2} \abs{\vect{r}}^2
\end{equation}
makes contact, via the state-operator correspondence, with the large-charge conformal dimension $\Delta_Q$ we have investigated so far. In this case, accounting for the fact that we include $N$ fermion flavors, we expect the following structure:
\begin{equation}
	\frac{\Delta_Q}{N} = \frac{ E(Q)}{\omega N} = \frac{\sqrt{\xi}}{4} \left(\frac{3Q}{N}\right)^\frac{4}{3} - \sqrt{2}\pi^2 \xi \left(c_1 - \frac{9}{2}c_2\right) \left(\frac{3Q}{N}\right)^\frac{2}{3} + \order{\left(\frac{3Q}{N}\right)^\frac{5}{9}}.
\end{equation}

The computation of $\Delta_Q$ resembles in spirit that of the relativistic $O(N)$ model at large $N$ and large charge~\cite{Alvarez-Gaume:2019biu}\footnote{In~\cite{Giombi:2020enj}, the same results are obtained without invoking the state-operator correspondence, but by working in flat space directly. This requires to construct the large-charge operators explicitly in terms of microscopic degrees of freedom. This is significantly harder to achieve in the case of fermions, although it would anyway be interesting to investigate this possibility in the case of the unitary Fermi gas.}, which relies on the usual Stratonovich transform which makes the dependence on the original fermionic degrees of freedom Gaussian. The latter can thus be integrated out, resulting in a non-local effective action for the Stratonovich field $\sigma$. The inhomogeneity caused by the potential, however, adds a layer of technical complexity which requires more sophisticated techniques. As will soon become apparent, evaluating the functional determinant characterizing the Stratonovich effective action becomes a nontrivial problem in the presence of the potential. Even at leading order in $N$, this problem is tractable only when the charge is large, in which case a controlled gradient expansion can be implemented. This is a reliable prescription to extract the \emph{bulk} contributions as characterized in the last chapter. Edge contributions evade this description and require a separate discussion, which we include too.

\subsubsection{Summary of the results and comparison with the literature}

Throughout, we work exclusively at leading order in the large-$N$ expansion. This description is in some sense a field-theoretic version of mean-field theory which provides an explicit realization of the \acs{bcs} mechanism, whereby the collective field $\sigma$ acquires a non-trivial expectation value in the superfluid ground state:
\begin{equation}
    \sigma_0(\vect{r}) := \ev{\sigma(\tau,\vect{r})}
    \neq 0.
\end{equation}
This is associated with spontaneous breaking of the internal symmetries. In the homogeneous case, $\sigma$ would characterize the gap in the spectrum of fermionic excitations, so we shall often refer to it as the gap, even in the presence of an inhomogeneous potential. We find by explicit calculation that
\begin{equation}
  \sigma_0(\vect{r}) = y_0 \pqtybis{\mu \hbar - V(\vect{r})} + y_1 \frac{\hbar^2}{m} \frac{(\nabla V(\vect{r}))^2}{\pqtybis{\mu \hbar - V(\vect{r})}^2} + y_2 \frac{\hbar^2}{m} \frac{\Laplacian{V(\vect{r})}}{\mu \hbar - V(\vect{r})} + \dots,
\end{equation}
with $y_0 \approx 1.1622\dots$, $y_1\approx -0.00434691\dots$, and $y_2\approx -0.160794\dots$. As we shall see, this expansion is controlled by the large value of the charge. Moreover, when $V(\vect{r})$ is set to be the harmonic potential, we find
\begin{equation}%
  \frac{\Delta_Q}{N} =  0.8313 \left(\frac{Q}{N}\right)^{4/3} + 0.26315 \left(\frac{Q}{N}\right)^{2/3} + \order{\left(\frac{Q}{N}\right)^\frac{5}{9}},
\end{equation}
which yields the following values for the first three Wilsonian coefficients:
\begin{align}%
  c_0 &=  0.0841\dots, & c_1 &= 0.006577\dots, & c_2 & \approx 0.004872\dots \, .
\end{align}
This reproduces the mean-field value of the Bertsch parameter, namely, 
\begin{equation}
    \xi \approx 0.5906\dots \, .
\end{equation}
 
As mentioned, edge effects require some care even in this controlled microscopic setup. While the analytic computation of the leading edge correction remains out of reach, we are able to estimate its scaling and show that it indeed matches the $Q^\frac{5}{9}$ scaling expected from the discussion of the last chapter and~\cite{Son:2005rv}. This is a very non-trivial first-principle verification---actually, the first of its kind---of the fascinating edge dynamics.

The condensed matter literature provides a wealth of different approaches to the description of the unitary Fermi gas, and the \acs{bcs}--\acs{bec} crossover more generally. They often employ physically motivated approximation schemes which are however often not applicable systematically to subleading orders. The simpler problem of the Fermi gas without the trapping potential was addressed at large $N$ in~\cite{Nikolic:2007zz,Veillette:2007zz}. They however include subleading corrections in $N$ and describe  the finite-temperature case. Our leading-order result, \emph{i.e.} the value of the Bertsch parameter, agrees with theirs. This also coincides with the well-established mean-field result. A derivative expansion in the presence of the harmonic trap was discussed in the superfluid \acs{eft} and then used to extend mean-field results in~\cite{MANES20091136}. It would be interesting to compare more thoroughly their approach and ours, which are in qualitative agreement. A more direct comparison of our results can be made with the work of~\cite{Csordas:2010id} which uses similar techniques, although the role of the charge as the controlling parameter was perhaps not fully appreciated. We anyway find complete agreement \emph{e.g.} for the shape of the gap.
There are also numerical studies~\cite{Forbes:2012yp}, though these deal with the opposite regime of small charge, where the quantized nature of the problem is apparent, and it is not clear how to make a direct comparison between our respective results. See also~\cite{Chang_2007,Blume_2007}.


\subsubsection{Outline of the section}

The plan of this section is as follows. In Subsection~\ref{sec:Model}, we discuss the microscopic model and use standard large-$N$ techniques to derive the expression of the functional determinant characterizing the Stratonovich effective action. We then phrase the problem in terms of the corresponding heat kernel. In Subsection~\ref{sec:Bulk}, we set up a gradient expansion by introducing the so-called Wigner coordinates and the Moyal product, which allows us to compute the large-charge expansion of the heat kernel in the bulk up to \acs{nnlo}. In Subsection~\ref{sec:Edge}, we discuss the edge expansion. In Subsection~\ref{sec:Conclusions}, we give conclusions and an outlook.

\subsection{The model}%
\label{sec:Model}

Building upon~\cite{Nikolic:2007zz,Veillette:2007zz,Sachdev2011,Bekaert_2012}, we consider an interacting $(3+1)$-dimensional model of $N$ fermion flavors with two hyperfine ('spin') species $\psi_{\sigma a}$, where $a \in \{1, \ldots, N\}$ and $\sigma \in \{\uparrow, \downarrow\}$, with a unique chemical potential $\mu$, coupled to an external potential $V(\vect{r})$. We work in Euclidean time $\tau$ at zero temperature, \emph{i.e.}, on $\mathbb{R}_\tau \times \mathbb{R}^3$. The system is described by
\begin{equation}
  \label{eq:action}
   	S[\psi] = \int \dd\tau \dd{\vect{r}} \left[
    \bar \psi_{\sigma a} \left( \hbar \partial_\tau{} - \frac{\hbar^2}{2 m} \Laplacian{} - \hbar\mu(\vect{r}) \right) \psi_{\sigma a} + \frac{2 u_0}{N}
    \bar\psi_{\uparrow a} \bar\psi_{\downarrow a} \psi_{\downarrow b} \psi_{\uparrow b} \right],
\end{equation}
where $\bar{\psi}_{\sigma a},~\psi_{\sigma a}$ are Grassmann fields\footnote{$\bar{\psi}_{\sigma a}$ denotes the complex conjugate of $\psi_{\sigma a}$. At finite temperature $T=\beta^{-1}$, these fields would have antiperiodic boundary conditions along the thermal circle $S_{\hbar\beta}^1$, with Matsubara frequencies $\omega_n = (2 n + 1) \pi/(\beta \hbar)$ ($n \in \mathbb{Z}$)~\cite{kapusta_finite-temperature_1989}.} and 
\begin{equation}
	\mu(\vect{r}) := \mu - \frac{1}{\hbar} V(\vect{r}).
\end{equation}
Throughout, summation over repeating flavor and spin indices is assumed. We keep the potential generic for now.\footnote{Note that special forms of the potential could lead to accidental symmetries and identifications of operators in the effective action.} The bare coupling $u_0<0$ describes an attractive 4-fermion contact interaction which, after renormalization, is proportional to the s-wave scattering length $a$.  The latter diverges at unitarity and the system is described by a strongly interacting \acs{nrcft}. 

The action is manifestly invariant under a global $(U(1)\otimes SU(2))^N$ symmetry, where the $U(1)$ copies correspond to particle number conservation of every flavor. The full symmetry group is however larger. Indeed, while the kinetic term has an explicit $U(2N)$ symmetry, the interaction term written in the form
\begin{equation}\label{eq:CooperInteraction}
  \frac{u_{0}}{2N} \left( \Psi^{T} \Omega \Psi \right)^{\dagger }  \left( \Psi^{T} \Omega \Psi \right),
\end{equation}
where $\Psi =\left( \psi_{1\uparrow} ...\psi_{N\uparrow } ,\psi_{1\downarrow } ...\psi_{N\downarrow } \right)^{T}$ and \(\Omega = i \sigma_2 \otimes \Id_N\), has a $U(1) \otimes SP(2N, \mathbb{C})$ symmetry. Hence, the correct symmetry group is $U(1) \otimes Sp(2N)$, with the usual definition $Sp(2N) = Sp(2N, \mathbb{C}) \cap U(2N)$~\cite{Bekaert_2012}. The interaction term in the form \eqref{eq:CooperInteraction} above is a Cooper type of interaction, so \acs{ssb} of the global $U(1)$ symmetry is expected to take place at fixed charge. This is different from the Gross--Neveu model~\cite{Dondi:2022zna}, where the symmetry breaking is exponentially suppressed---and thus nonperturbative---at large $N$.

We now introduce the (Hubbard--)Stratonovich field $\sigma(\tau, \vect{r})$ in the Cooper channel\footnote{See~\cite{altland_simons_2010} for a discussion of the different channels.}, namely,
\begin{equation} \label{eq:StrataDef}
    \sigma(\tau, \vect{r}) = -\frac{2 u_0}{N} \sum_{a=1}^N \psi_{\downarrow a}(\tau, \vect{r}) \psi_{\uparrow a}(\tau, \vect{r}),
\end{equation}
resulting in
\begin{equation} \label{eq:Stratonovich-action}
\begin{aligned}
  S[\psi, \sigma]
  & = \int \dd\tau \dd{\vect{r}} \left[
  \bar \psi_{\sigma a} \left( \hbar \partial_\tau - \frac{\hbar^2}{2 m} \Laplacian{} - \hbar \mu(\vect{r}) \right) \psi_{\sigma a} - \frac{N}{2 u_0} \sigma^* \sigma -  
  \sigma \bar\psi_{\uparrow a} \bar \psi_{\downarrow a} - \sigma^* \psi_{\downarrow a} \psi_{\uparrow a} \right] \\
  & = \int \dd\tau \dd{\vect{r}} \, \left[ -\bar \Psi_a G^{-1}[\sigma] \Psi_a - \frac{N}{2 u_0} \sigma^* \sigma \right].
\end{aligned}
\end{equation}
In the second line, the Nambu representation was used, namely,
\begin{equation}
\Psi_a := \begin{pmatrix} \psi_{\uparrow a} \\ \bar\psi_{\downarrow a} \end{pmatrix},
\end{equation}
and $\bar{\Psi }_{a} := \Psi^{\dag}_{a} $, along with the inverse fermion propagator $G^{-1}[\sigma]$,
\begin{equation}
  \label{eq:inverse-propagator}
  G^{-1}[\sigma] := \begin{pmatrix}
                -\hbar \partial_\tau + \frac{\hbar^2}{2 m} \Laplacian{} + \hbar \mu(\tau, \vect{r}) & \sigma(\tau, \vect{r}) \\
                 \sigma(\tau, \vect{r})^* & -\hbar \partial_\tau - \frac{\hbar^2}{2 m} \Laplacian{} - \hbar \mu(\vect{r})
              \end{pmatrix}.
\end{equation}
The Stratonovich field $\sigma$ is bilinear in the fundamental fermionic degrees of freedom, and it is charged under the $U(1)$ global symmetry. Therefore, $\sigma$ is the order parameter. This is how the \acs{bcs} mechanism manifests itself in this microscopic description.

Since the fermions now appear only quadratically, they can be integrated out to obtain
\begin{equation}
  \label{eq:sigma-action}
  S[\sigma] = -\hbar N \Tr \log \left( \abs{G^{-1}[\sigma]} \right) - \frac{N}{2u_{0}} \int \dd\tau \dd{\vect{r}} \, \sigma(\tau, \vect{r})^* \sigma (\tau, \vect{r}) .
\end{equation}
The crucial observation is that $N$, when large, controls the saddle-point approximation of the corresponding path integral. The Stratonovich field can thus be decomposed into a saddle-point value \(\sigma_0\) plus quantum fluctuations suppressed by \(1/\sqrt{N}\):
\begin{equation}%
  \label{eq:SigmaSaddleAndFluct}
  \sigma(\tau, \vect{r}) = \sigma_{0}(\tau, \vect{r}) +\frac{1}{\sqrt{N}} \hat{\sigma}(\tau,\vect{r}),
\end{equation}
with
\begin{equation}
  \frac{\delta S [\sigma]}{\delta \sigma (\tau, \vect{r})}\Bigg|_{\sigma_0} = 0.
\end{equation}
This saddle equation is commonly referred to as the \emph{gap equation}. In practice, we shall solve it to find the $\sigma_0$-profile only at the very end, therefore treating $\sigma_0$ as a generic function.

Eq.~\eqref{eq:sigma-action} can be expanded in $\frac{1}{N}$, and we write
\begin{equation}
    S[\sigma] = N \cdot S_0 + \order{N^0},
\end{equation}
where $S_0 := \frac{1}{N} S[\sigma_0]$. Similarly, we shall use \(G_0 := G[\sigma_0]\). Since we shall always consider time-independent external potentials $V(\vect{r})$, we can write the inverse fermion propagator as the sum of a time-derivative and a position-dependent operator\footnote{It is sometimes called the \ac{bdg} operator.} $B(\vect{r})$:
\begin{align}\label{eq:BdG}
    G_0^{-1} = -\hbar \Id \cdot \del_\tau{} + B(\vect{r}),
   &&\text{where} && 
    B(\vect{r}) :=
    \begin{pmatrix}
        -h(\vect{r}) & \sigma_0(\vect{r}) \\
        \sigma_0(\vect{r})^* & h(\vect{r})
    \end{pmatrix},
\end{align}
and \(h(\vect{r})\) is the one-particle Hamiltonian 
\begin{equation}
	h(\vect{r}) = -\frac{\hbar^2}{2m} \Laplacian{} - \hbar \mu(\vect{r}).
\end{equation}
The eigenvalues of $B(\vect{r})$ are real and come in pairs since it is Hermitian. Moreover, the eigenvalues of the time-derivative piece in the inverse propagator $G_0^{-1}$ are given by $i \hbar \omega_n$, where $\omega_n$ ($n \in \mathbb{Z}$) are the Matsubara frequencies (see previous footnote). The sum over $n$ can be performed explicitly which, in the zero-temperature limit, yields
\begin{equation} \label{eq:ZeroT_TrLog}
    \lim_{\beta\to\infty} \Tr \log\left( \abs{G_0^{-1}} \right) = \frac{\beta}{2} \Tr(\abs{B}) .
\end{equation}
The factor of $\frac{1}{2}$ here has a direct physical interpretation, as it makes the distinction between the 'large' first-quantized Hilbert space, generated by all the modes and the 'physical' Hilbert space of positive-energy modes. This is characteristic of fermions.

One can obtain Eq.~\eqref{eq:ZeroT_TrLog} by employing spectral zeta functions as follows. Consider a differential operator $\mathcal{O}$ whose eigenvalues we label by a generic index $\alpha$ as $o_\alpha$. We then define
\begin{equation}
    \zeta_\mathcal{O}(s) := \Tr( \abs{\mathcal{O}}^{-s}) = \sum_{\alpha} \abs{o_\alpha}^{-s}.
\end{equation}
With this, it is straightforward to show that
\begin{equation}
    \lim_{\beta\to\infty} \zeta_{G_0^{-1}}(s)
    = \frac{\beta \Gamma( \frac{s-1}{2} )}{2 \sqrt{\pi} \Gamma( \frac{s}{2} )} \zeta_B(s-1)
    = \frac{\beta \Gamma( \frac{s-1}{2} )}{2 \sqrt{\pi} \Gamma( \frac{s}{2} )} \zeta_{B^2}\left( \frac{s-1}{2} \right),
\end{equation}
so that
\begin{equation}
    \lim_{\beta\to\infty} \Tr \log( \abs{G_0^{-1}} ) = -\lim_{\beta\to\infty} \eval*{ \frac{d}{d s} \zeta_{G_0^{-1}}(s) }_{0^+}  = \frac{\beta}{2} \zeta_B(-1) = \frac{\beta}{2} \zeta_{B^2}\left( -\frac{1}{2} \right).
\end{equation}
We give the result in terms of the positive-definite operator $B^2$ for computational convenience. In order to compute its spectral zeta function, we compute the heat kernel $K_{B^2}(\vect{r}_1, \vect{r}_2; \proptime)$ associated with $B^2$, which is a 2-by-2 matrix satisfying\footnote{It is common practice to use $\tau$ to denote the \emph{proper time} of the heat kernel problem. We adopt this notation too, hoping it will not create confusion with the Euclidean time.}
\begin{equation} \label{eq:HeatK_Problem}
\begin{dcases}
 \left( \del_{\proptime}{} + B^2(\vect{r}_1) \right) K_{B^2}(\vect{r}_1, \vect{r}_2; \proptime), = 0\\
     \lim_{\proptime\to0^+} K_{B^2}(\vect{r}_1, \vect{r}_2; \proptime) =\deltaop(\vect{r}_1 - \vect{r}_2) \cdot \Id_2.
\end{dcases}
\end{equation}
Formally, this means that $K_{B^2}(\vect{r}_1, \vect{r}_2; \proptime) = \expval{\vect{r}_1| e^{-B^2 \proptime} |\vect{r}_2}$. To avoid later confusion, we denote the Dirac $\delta$-function with a hat, and $\Id_2$ is the 2-by-2 identity matrix. In turn, the coincident-point limit of the heat kernel allows to compute $\zeta_{B^2}$ through its Mellin transform as
\begin{equation} \label{eq:ZetaMellin}
  \zeta_{B^2}(s)
  = \frac{1}{\Gamma(s)} \int_0^\infty \frac{\dd{\proptime}}{\proptime} \proptime^s \int \dd{\vect{r}} \Tr\left(K_{B^2}(\vect{r}, \vect{r}; \proptime)\right),
\end{equation}
where $\Tr$ now just denotes for the matrix trace.

This expression turns the computation of the one-loop determinant---that is, the $\Tr\log(\abs{G_0^{-1}})$ term---into the (Euclidean) quantum mechanical problem of a particle with Hamiltonian \(B^2\). In the following subsection, we shall demonstrate that this is a natural starting point for setting up the large-charge expansion of the functional determinant when the system is inhomogeneous due to the presence of an external potential.

Before proceeding, it is convenient to write Eqs.~\eqref{eq:HeatK_Problem} in terms of dimensionless quantities, which we shall generically denote with a bar. The goal at this point is to express the problem in an exact form suggestive of a perturbative resolution, so no approximation is being made yet. Our description is valid for the class of potentials $V(\vect{r})$ that vanishes at the origin and that (classically) confines the particles to a finite region of space. Correspondingly, there is a characteristic length scale $R_{cl}$ associated with the smallest distance from the origin such that $\mu(\vect{r}) = 0$, \emph{i.e.}, where the particle density classically vanishes. Since we eventually are interested in computing conformal dimensions, one can consider the harmonic potential
$V(\vect{r}) = \frac{m \omega^2}{2} \vect{r}^2$ for concreteness. In this specific case, $R_{cl}$ is the radius of the spherical cloud of particles:
\begin{equation} \label{eq:R_cl}
    R_{cl} := \sqrt{\frac{2 \hbar \mu}{m \omega^2}}.
\end{equation}
The generic length scale $R_{cl}$ allows us to define dimensionless coordinates as
\begin{equation}
    \bar{\vect{r}} := \frac{\vect{r}}{R_{cl}}.
\end{equation}
Moreover, upon defining $\bar B(\bar{\vect{r}}) := \frac{1}{\hbar \mu} B(\vect{r})$ and $K_{\bar B^2}(\bar{\vect{r}}_1, \bar{\vect{r}}_2; \bar\proptime) := R_{cl}^3 K_{B^2}(\vect{r}_1, \vect{r}_2; \proptime)$ with $\bar\proptime := (\hbar \mu)^2 \proptime$, we obtain
\begin{equation} \label{eq:HeatK_Problem_DimLess}
\begin{dcases}
     \left( \del_{\bar\proptime}{} + \bar B^2(\bar{\vect{r}}_1) \right) K_{\bar B^2}(\bar{\vect{r}}_1, \bar{\vect{r}}_2; \bar\proptime) = 0, \\
    \lim_{\bar\proptime\to0^+} K_{\bar B^2}(\bar{\vect{r}}_1, \bar{\vect{r}}_2; \bar\proptime) = \deltaop(\bar{\vect{r}}_1 - \bar{\vect{r}}_2) \cdot \Id_2 ,
\end{dcases}
\end{equation}
where every quantity is now dimensionless. Explicitly, they are given by
\begin{equation}
    \bar B(\bar{\vect{r}}) =
    \begin{pmatrix}
        -\bar h(\bar{\vect{r}}) & \bar\sigma(\bar{\vect{r}}) \\
        \bar\sigma(\bar{\vect{r}})^* & \bar h(\bar{\vect{r}})
    \end{pmatrix},
\end{equation}
with $\bar\sigma(\bar{\vect{r}}) := \frac{\sigma_0(\vect{r})}{\hbar \mu}$, $\bar h(\bar{\vect{r}}) := \frac{h(\vect{r})}{\hbar \mu} = -\frac{\hbar}{2 m \mu R_{cl}^2} \Laplacian_{\bar{\vect{r}}} + \bar V(\bar{\vect{r}}) - 1$ and $\bar V(\bar{\vect{r}}) := \frac{1}{\hbar \mu} V(\vect{r})$.\footnote{If we consider a harmonic potential, we simply have $\bar V(\bar{\vect{r}}) = \bar{\vect{r}}^2$.}

\subsection{Setting up the bulk expansion}%
\label{sec:Bulk}

We have seen that the computation of the effective action at leading order in $N$ reduces to evaluating the trace of the absolute value (or the square) of the \acs{bdg} operator (cf.~Eq.~\eqref{eq:ZeroT_TrLog}).
In the absence of a confining potential, this is a standard calculation~\cite{Nikolic:2007zz,Veillette:2007zz}. 
In the presence of a confining potential, the problem is no longer translationally invariant and the particles are confined to a spherical region at the edge of which the particle density drops to zero.

In this subsection, we show how to perturbatively compute the free energy in the presence of a potential when the particle number is large, thereby providing the first explicit verification of the predictions of nonrelativistic large-charge \acs{eft}.
The underlying controlling parameter of this perturbative computation is the particle density, which naturally breaks down close to the edge of the particle cloud.
However, \acs{eft} considerations not only tell us that this had to be anticipated, but also that 'exotic' contributions enter the large-charge expansion of the free energy. While the setup in this subsection is not suited to capture these contributions, we discuss in subsection~\ref{sec:Edge} how to perform the matching between the expansion in the bulk of the cloud with the solution close to the edge, and find further agreement with the \acs{eft} predictions.

\subsubsection{Wigner coordinates and Moyal product}
\label{sec:wigner-moyal}

The lack of translational invariance can be addressed through the introduction of so-called mixed, or Wigner, coordinates~\cite{Wigner:1932eb,rammer_2007}.
The idea is to first write \(\bar B(\bar{\vect{r}})\) as a bilocal operator via
\begin{equation}
    \bar B(\bar{\vect{r}}_1, \bar{\vect{r}}_2) := \deltaop(\bar{\vect{r}}_1 - \bar{\vect{r}}_2) \bar B(\bar{\vect{r}}_1) = \begin{pmatrix}
        -\bar h(\bar{\vect{r}}_1, \bar{\vect{r}}_2) & \bar\sigma(\bar{\vect{r}}_1) \deltaop(\bar{\vect{r}}_1 - \bar{\vect{r}}_2) \\
        \bar\sigma(\bar{\vect{r}}_1)^* \deltaop(\bar{\vect{r}}_1 - \bar{\vect{r}}_2) & \bar h(\bar{\vect{r}}_1, \bar{\vect{r}}_2)
    \end{pmatrix},
\end{equation}
where
\begin{equation}
    \bar h(\bar{\vect{r}}_1, \bar{\vect{r}}_2) := -\frac{\hbar}{2 m \mu R_{cl}^2} \left( \Laplacian_{\bar{\vect{r}}_1}  \deltaop(\bar{\vect{r}}_1 - \bar{\vect{r}}_2) \right) + \left( \bar V(\bar{\vect{r}}_1) - 1 \right) \deltaop(\bar{\vect{r}}_1 - \bar{\vect{r}}_2), 
\end{equation}
and then perform a Fourier transform.  Introducing relative and center-of-mass positions
\begin{equation}
\begin{cases}
    \bar{\vect{r}}_{ij} := \bar{\vect{r}}_i - \bar{\vect{r}}_j, \\
    \bar{\vect{R}}_{ij} := \frac{\bar{\vect{r}}_i + \bar{\vect{r}}_j}{2},
\end{cases}
\end{equation}
a bilocal Fourier transform of a function $A(\bar{\vect{r}}_1, \bar{\vect{r}}_2)$ can be defined as
\begin{equation}
    a(\bar{\vect{R}}, \bar{\vect{p}}) := \int \dd{\bar{\vect{r}}} e^{-\frac{i\bar{\vect{p}}\cdot\bar{\vect{r}}}{\varepsilon}} A\left(\bar{\vect{R}} + \frac{\bar{\vect{r}}}{2}, \bar{\vect{R}} - \frac{\bar{\vect{r}}}{2}\right),
\end{equation}
and the inverse transformation is
\begin{equation}
    A(\bar{\vect{r}}_i, \bar{\vect{r}}_j) = \int \frac{\dd{\bar{\vect{p}}}}{(2 \pi \varepsilon)^3} e^{\frac{i\bar{\vect{p}}\cdot\bar{\vect{r}}_{ij}}{\varepsilon}} a(\bar{\vect{R}}_{ij}, \bar{\vect{p}}).
\end{equation}
We choose to put a bar on the momentum as well to indicate that it is dimensionless. In some sense, this transform allows one to disentangle the 'microscopic' dynamics of the system (associated with $\bar{\vect{r}}_{ij}$) from the 'macroscopic' properties (associated with $\bar{\vect{R}}_{ij}$) resulting from the external potential~\cite{rammer_2007}. 
For the moment, $\varepsilon$ is an arbitrary real parameter which will be assigned physical meaning soon. 

From this definition of the Fourier transform, it is easy to show the following general statement~\cite{Groenewold:1946kp,Moyal:1949sk}. If $C(\bar{\vect{r}}_1, \bar{\vect{r}}_2)$ is related to two other bilocal functions $A$ and $B$ by
\begin{equation}
	C(\bar{\vect{r}}_1, \bar{\vect{r}}_3) = \int \dd{\bar{\vect{r}}_2} A(\bar{\vect{r}}_1, \bar{\vect{r}}_2) B(\bar{\vect{r}}_2, \bar{\vect{r}}_3),
\end{equation}
then
\begin{equation}
	c(\bar{\vect{R}}, \bar{\vect{p}}) = a(\bar{\vect{R}}, \bar{\vect{p}}) \star b(\bar{\vect{R}}, \bar{\vect{p}}),
\end{equation}
where $\star$ is the Moyal product, defined by
\begin{equation}
\begin{aligned}
    a(\bar{\vect{R}}, \bar{\vect{p}}) \star b(\bar{\vect{R}}, \bar{\vect{p}})
    & := a(\bar{\vect{R}}, \bar{\vect{p}}) \exp[\frac{i \varepsilon}{2} \pqtybis{\cev{\del}_{\bar{\vect{R}}} \vec{\del}_{\bar{\vect{p}}} - \cev{\del}_{\bar{\vect{p}}} \vec{\del}_{\bar{\vect{R}}}}] b(\bar{\vect{R}}, \bar{\vect{p}}) \\
    & = \sum_{k=0}^\infty \left( \frac{i \varepsilon}{2} \right)^k \frac{1}{k!} \poisson{a(\bar{\vect{R}}, \bar{\vect{p}}), b(\bar{\vect{R}}, \bar{\vect{p}})}_k,
\end{aligned}
\end{equation}
with what we shall call the \emph{$k$-Poisson bracket}:
\begin{equation} \label{eq:KPoisson_def}
    \poisson{a(\bar{\vect{R}}, \bar{\vect{p}}), b(\bar{\vect{R}}, \bar{\vect{p}})}_k := a(\bar{\vect{R}}, \bar{\vect{p}}) \pqtybis{\cev{\del}_{\bar{\vect{R}}} \vec{\del}_{\bar{\vect{p}}} - \cev{\del}_{\bar{\vect{p}}} \vec{\del}_{\bar{\vect{R}}}}^k b(\bar{\vect{R}}, \bar{\vect{p}}).
\end{equation}
Of course, the $k=1$ case is just the normal Poisson bracket.\footnote{
Explicit computations involved in this section contain at most the $k=2$ case, which can also be put in the following form. Given the Poisson bivector $\Pi = i\sigma_2 \otimes \Id_d$ ($d = 3$ in our case) and the derivative $\del_I$ ($I = 1, \ldots, 2d$) acting on phase-space coordinates $(\bar{\vect{R}}, \bar{\vect{p}})$, and $\del_{IJ} := \del_I \del_J$, we have
\begin{align}
  \poisson{a, b}_1 &=  \Pi^{IJ} \del_I a \, \del_J b , \\
  \poisson{a, b}_2 &=  \Pi^{IJ}\Pi^{LM}\del_{IL} a \, \del_{JM} b.
\end{align}
}

This formalism was originally introduced to describe quantum mechanics in phase space, and is adapted to solve our heat kernel problem, which takes the form
\begin{equation}
\begin{dcases}
	 \del_{\bar\proptime} k(\bar{\vect{R}}, \bar{\vect{p}}; \bar\proptime) + b(\bar{\vect{R}}, \bar{\vect{p}}; \proptime) \star b(\bar{\vect{R}}, \bar{\vect{p}}; \bar\proptime) \star k(\bar{\vect{R}}, \bar{\vect{p}}; \proptime) = 0, \\
    \lim_{\bar\proptime\to0^+} k(\bar{\vect{R}}, \bar{\vect{p}}; \bar\proptime) = \Id_2  ,
\end{dcases}
\end{equation}
where $k(\bar{\vect{R}}, \bar{\vect{p}}; \bar\proptime)$ denotes the Fourier transform of $K_{\bar B^2}(\bar{\vect{r}}_1, \bar{\vect{r}}_2; \bar\proptime)$, as defined above, and 
\begin{equation}
    b(\bar{\vect{R}}, \bar{\vect{p}}) =
    \begin{pmatrix}
        -\bar h(\bar{\vect{R}}, \bar{\vect{p}}) & \bar\sigma(\bar{\vect{R}}) \\
        \bar\sigma(\bar{\vect{R}})^* & \bar h(\bar{\vect{R}}, \bar{\vect{p}})
    \end{pmatrix},
\end{equation}
with the dimensionless phase-space Hamiltonian given by
\begin{equation} \label{eq:Hamiltonian_PhaseSpace}
    \bar h(\bar{\vect{R}}, \bar{\vect{p}}) = \frac{\hbar}{2 m \mu R_{cl}^2 \varepsilon^2} \bar{\vect{p}}^2 + \bar V(\bar{\vect{R}}) - 1.
\end{equation}
Note that the gap \(\bar\sigma\) is only a function of \(\bar{\vect{R}}\) because it encodes pointwise interactions.

Since, at leading order in $N$, we can identify $\frac{1}{\hbar\beta} S_0$ with the grand-canonical potential $\Omega$, its zero-temperature limit can be expressed in terms of the solution $k(\bar{\vect{R}}, \bar{\vect{p}}; \bar\proptime)$ to the above heat kernel problem as 
\begin{equation} \label{eq:S0_DimLessStuff}
\begin{aligned}
    \Omega(\mu)
    = -\frac{\hbar \mu}{2} \left[ \int_0^\infty \frac{\dd{\bar\proptime}}{\Gamma(s) \bar\proptime} \bar\proptime^s \int \frac{\dd{\bar{\vect{R}}} \dd{\bar{\vect{p}}}}{(2 \pi \varepsilon)^3} \Tr\left(k(\bar{\vect{R}}, \bar{\vect{p}}; \bar\proptime)\right) \right]_{s=-\frac{1}{2}} - \frac{(\hbar \mu)^2 R_{cl}^3}{2 u_0} \int \dd{\bar{\vect{R}}} \abs{\bar\sigma(\bar{\vect{R}})}^2,
\end{aligned}
\end{equation}
on top of which one needs to impose the gap equation. The charge is then given by
\begin{equation}
    Q = -\frac{1}{\hbar} \frac{\del\Omega(\mu)}{\del\mu},
\end{equation}
which can be inverted to find $\mu$ as a function of $Q$, and the (zero-temperature) free energy is
\begin{equation}
    F(Q) = \Omega(\mu(Q)) + \hbar \mu(Q) \cdot Q.
\end{equation}
In the specific case of a critical system in a harmonic trap $V(\vect{r}) = \frac{m \omega^2}{2} \vect{r}^2$ (\emph{i.e.}, $\bar V(\bar{\vect{R}}) = \bar{\vect{R}}^2$ and $\varepsilon = {\omega}/({2 \mu})$, cf.~Eq.~\eqref{eq:R_cl}), the free energy is related to the conformal dimension of the lightest operator of charge $Q$ as~\cite{Nishida:2007pj,Goldberger:2014hca}
\begin{equation} \label{eq:DeltaQ_vs_FreeEnergy}
    \Delta(Q) = \frac{1}{\hbar\omega} F^\text{crit.}(Q).
\end{equation}

\subsubsection{Large-charge expansion of the heat kernel}%
\label{sec:large-charge-BdG}

Within the large-$N$ expansion, this general construction is thus far exact, although the fact that the Moyal product can be expanded as an asymptotic series in $\varepsilon$ suggests a perturbative calculation of the heat kernel problem.
It is convenient to canonically normalize the momentum in the phase-space Hamiltonian in Eq.~(\ref{eq:Hamiltonian_PhaseSpace}) by fixing
\begin{equation} \label{eq:EpsilonFixed}
    \varepsilon = \sqrt{\frac{\hbar}{2 m \mu}} \frac{1}{R_{cl}},
\end{equation}
which is small if the (dominating) scale associated with the potential is smaller than the scale set by the charge density. This is the large-charge limit because, as we will show, $Q \sim \varepsilon^{-3}$. Our goal is to compute $\Omega(\mu)$ as an expansion in $\varepsilon$,
\begin{equation} \label{eq:S0_EpsilonExpand}
    \Omega(\mu) := \Omega_{LO}(\mu) + \varepsilon \cdot \Omega_{NLO}(\mu) + \varepsilon^2 \cdot \Omega_{NNLO}(\mu) + \cdots,
\end{equation}
up to quadratic order, and thus to compute the corresponding free energy to this order.\footnote{As a reminder, the expansion to this order covers the dynamics of the effective action~\eqref{eq:eff-action-nlo} to 'NLO', in the language of~\cite{Son:2005rv}. On general grounds we expect the linear piece in \(\varepsilon\) to vanish because \(\Omega(\mu)\) is the the saddle value of the function \(\Omega(\sigma)\), and \(\Omega_{NLO}(\mu) \) is the value of its first variation.}

This amounts to a semiclassical analysis of the Hamiltonian, Eq.~\eqref{eq:Hamiltonian_PhaseSpace}, which is consistent as long as the potential term is bigger than the kinetic term.
In the bulk region, the gap profile \(\sigma(\vect{r})\) does not vary rapidly on the scale defined by its own \acs{vev}:
\begin{equation}
  \label{eq:bulk-condition}
  \sigma(\vect{r}) \gg \frac{\hbar^2}{2 m}\frac{\pqtybis{\del_{\vect{r}} \sigma(\vect{r})}^2}{\sigma(\vect{r})^2}.
\end{equation}
In terms of dimensionless variables, this turns into
\begin{equation}
  \bar \sigma(\bar{\vect{R}}) \gg \epsilon^2 \frac{\pqtybis{\del_{\bar{\vect{R}}} \bar \sigma(\bar{\vect{R}})}^2}{\bar \sigma(\bar{\vect{R}})^2} ,
\end{equation}
which is satisfied as long as \(\bar \sigma(\bar{\vect{R}})\) and its derivative with respect to \(\bar{\vect{R}}\) are both of order one.
More precisely, this means that $1 - \bar V (\bar{\vect{R}})$ must be bigger than some new parameter $\delta$, which we will evaluate below.
The construction is analogous to the standard \acs{wkb} approximation, in which the expansion in $\hbar$ is valid in the bulk away from the turning points.
For the moment, we concentrate on the bulk region, where $1 - \bar V (\bar{\vect{R}}) > \delta$ (in boundary layer theory this is called the 'outer region' \cite{bender1999advanced}).
We will turn to the edge of the cloud (the 'inner region'), which necessitates a separate treatment, in the next subsection.

Having clarified the interval of validity of our approximation, we can use standard phase-space quantum mechanics. With our choice of $\varepsilon$~\eqref{eq:EpsilonFixed}, the one-particle Hamiltonian in phase space takes the form 
\begin{equation} \label{eq:Hamiltonian_PhaseSpaceBis}
    \bar h(\bar{\vect{R}}, \bar{\vect{p}}) = \bar{\vect{p}}^2 + \bar V(\bar{\vect{R}}) - 1.
\end{equation}
The dependence on $\varepsilon$ has been reabsorbed by the Fourier transform in the same manner that $\hbar$ does not appear in the phase-space Hamiltonian in quantum mechanics.
This expression shows explicitly how in Wigner coordinates the contribution of the external potential, which depends only on the center-of-mass coordinate \(\bar{\vect{R}}\), is separated from the kinetic part depending only on the (dual) relative coordinate \(\bar{\vect{p}}\).

In the large-charge regime $\varepsilon \ll 1$, it is then natural to make the following Ansätze for the gap and the heat kernel:
\begin{align} \label{eq:gap-expansion}
	\bar \sigma(\bar{\vect{R}}) & = \sum_{k=0}^\infty \varepsilon^k \Sigma_k(\bar{\vect{R}}), \\
	  k(\bar{\vect{R}}, \bar{\vect{p}}; \bar\proptime) & 
    = \sum_{j=0}^\infty \varepsilon^j
    \begin{pmatrix}
        \Kdiag_j(\bar{\vect{R}}, \bar{\vect{p}}; \bar\proptime) & \Koff_j(\bar{\vect{R}}, \bar{\vect{p}}; \bar\proptime) \\
        \Koff_j(\bar{\vect{R}}, \bar{\vect{p}}; \bar\proptime)^* & \Kdiag_j(\bar{\vect{R}}, \bar{\vect{p}}; \bar\proptime)
    \end{pmatrix}.
\end{align}
Note that the matrix trace $\Tr \left( k(\bar{\vect{R}}, \bar{\vect{p}}; \bar\proptime) \right) = 2 \cdot \sum_{j=0}^\infty \varepsilon^j \Kdiag_j(\bar{\vect{R}}, \bar{\vect{p}}; \bar\proptime)$ will bring a factor of 2 into Eq.~\eqref{eq:S0_DimLessStuff}.

In the following, we will need to solve the saddle-point condition \(\delta \Omega / \delta \sigma = 0\).
It is convenient to reformulate it in terms of \(\Sigma_0\) by the chain rule
\begin{equation}
  \frac{\delta}{\delta \Sigma_0} = \frac{\delta \sigma}{\delta \Sigma_0} \frac{\delta}{\delta \sigma} = \frac{\delta \bar \sigma}{\delta \Sigma_0} \frac{\delta \sigma}{\delta \bar \sigma} \frac{\delta}{\delta \sigma} = \hbar \mu \frac{\delta}{\delta \sigma} ,
\end{equation}
so that the saddle-point condition reads \(\delta \Omega / \delta \Sigma_0 = 0\).
All the information about the saddle (\emph{i.e.}, the values of the functions \(\Sigma_j\)) is contained in the variations with respect to \(\Sigma_0\).
The same information is also contained in the variations with respect to any of the \(\Sigma_j\), but this is suppressed in the \(\varepsilon\) expansion because
\begin{equation}
  \frac{\delta}{\delta \Sigma_j} = \frac{\delta \sigma}{\delta \Sigma_j} \frac{\delta}{\delta \sigma} = \frac{\delta \bar \sigma}{\delta \Sigma_j} \frac{\delta \sigma}{\delta \bar \sigma} \frac{\delta}{\delta \sigma} = \hbar \mu \varepsilon^j \frac{\delta}{\delta \sigma} .
\end{equation}
In the expansion of \(\Omega\), the function  \(\Sigma_j\) appears linearly at order \(\varepsilon^j\):
\begin{equation}
\begin{aligned}
  \Omega(\bar \sigma) & = \Omega_{LO}(\Sigma_0) + \pqtybis{ \Omega_{NLO}(\Sigma_0) +  \Sigma_1 \hat \Omega_{NLO}(\Sigma_0) } \varepsilon \\
  & +  \pqtybis{ \Omega_{NNLO}(\Sigma_0, \Sigma_1) +  \Sigma_2 \hat \Omega_{NNLO}(\Sigma_0)} \varepsilon^2 + \dots,
\end{aligned}
\end{equation}
and, by the equation above, \( \hat \Omega_{NLO}(\Sigma_0) = \Omega_{LO}'(\Sigma_0) \), which vanishes at the saddle.
It follows that, apart from the leading order, the saddle-point value  \(\expval{\Omega}\) at order \(j\) does not depend on the saddle value \(\expval{\Sigma_j}\).

The phase-space Hamiltonian is quadratic in the variables $\bar{\vect{R}}$ and $\bar{\vect{p}}$, and it is easy to express the product \(b(\bar{\vect{R}}, \bar{\vect{p}}) \star b(\bar{\vect{R}}, \bar{\vect{p}})\) in closed form,
\begin{equation}
\begin{aligned}
    b(\bar{\vect{R}}, \bar{\vect{p}}) \star b(\bar{\vect{R}}, \bar{\vect{p}})
    & =
    \begin{pmatrix}
        h(\bar{\vect{R}}, \bar{\vect{p}})^2 + \abs{\bar\sigma(\bar{\vect{R}})}^2 - \frac{\varepsilon^2}{2} \Laplacian_{\bar{\vect{R}}}(\bar V(\bar{\vect{R}}))
        & \kern-10pt 2 i \varepsilon \bar{\vect{p}} \cdot \nabla_{\bar{\vect{R}}} \bar \sigma(\bar{\vect{R}})
        \\
        -2i \varepsilon \bar{\vect{p}} \cdot \nabla_{\bar{\vect{R}}} \bar \sigma(\bar{\vect{R}})
        & \kern-10pt h(\bar{\vect{R}}, \bar{\vect{p}})^2 + \abs{\bar\sigma(\bar{\vect{R}})}^2 - \frac{\varepsilon^2}{2} \Laplacian_{\bar{\vect{R}}}(\bar V(\bar{\vect{R}}))
  \end{pmatrix}  \\
  & =
  \sum_{i=0}^\infty \varepsilon^i
  \begin{pmatrix} 
        \bbdiag_i(\bar{\vect{R}}, \bar{\vect{p}}) & \bboff_i(\bar{\vect{R}}, \bar{\vect{p}}) \\
        \bboff_i^*(\bar{\vect{R}}, \bar{\vect{p}}) & \bbdiag_i(\bar{\vect{R}}, \bar{\vect{p}})
  \end{pmatrix},
\end{aligned} 
\end{equation}
where we have identified, order by order,
\begin{equation}
\begin{dcases}
    \bbdiag_0(\bar{\vect{R}}, \bar{\vect{p}})  := h(\bar{\vect{R}}, \bar{\vect{p}})^2 + \abs{\Sigma_0(\bar{\vect{R}})}^2, \\
    \bbdiag_{i}(\bar{\vect{R}}, \bar{\vect{p}})  := \sum_{k=0}^i \Sigma_k^*(\bar{\vect{R}}) \Sigma_{i-k}(\bar{\vect{R}}) - \frac{\delta_{i2}}{2} \Laplacian_{\bar{\vect{R}}}(\bar V(\bar{\vect{R}})) & \text{for $i = 1,2,\dots $,}
\end{dcases}
\end{equation}
and
\begin{equation}
\begin{cases}
    \bboff_0(\bar{\vect{R}}, \bar{\vect{p}}) := 0, \\
    \bboff_i(\bar{\vect{R}}, \bar{\vect{p}})  := 2i \bar{\vect{p}} \cdot \nabla_{\bar{\vect{R}}} \Sigma_{i-1}(\bar{\vect{R}}) & \text{for $i=1,2,\dots$.}
\end{cases}
\end{equation}
With this, the heat kernel equation can be understood hierarchically. Expanding the Moyal product in powers of $\varepsilon$ with the $k$-Poisson bracket notation introduced in Eq.~\eqref{eq:KPoisson_def}, and dropping the arguments to avoid cluttering the notation, the order-$\varepsilon^n$ heat kernel problem becomes
\begin{equation} \label{eq:HeatK_Problem_OrderN}
\begin{dcases}
 \del_{\bar\proptime} \Kdiag_n + \sum_{j=0}^n \sum_{k=0}^{n-j} \frac{i^k}{2^k k!} \left[ \poisson{\bbdiag_{n-j-k}, \Kdiag_j}_k + \poisson{\bboff_{n-j-k}, \Koff_j^*}_k \right] = 0, \\
\del_{\bar\proptime} \Koff_n + \sum_{j=0}^n \sum_{k=0}^{n-j} \frac{i^k}{2^k k!} \left[ \poisson{\bbdiag_{n-j-k}, \Koff_j}_k + \poisson{\bboff_{n-j-k}, \Kdiag_j}_k \right] = 0,
\end{dcases}
\end{equation}
with \acs{ic} $\lim\limits_{\bar\proptime\to 0} \Kdiag_n = \delta_{0n}$ and $\lim\limits_{\bar\proptime\to 0} \Koff_n = 0$. Note that each line contains a finite number of contributions and, for a given $n$, the two equations are decoupled, since $\bboff_0 = 0$. We investigate the order $\varepsilon^0$ contributions in the next subsection, and we will elaborate further on this system of equations when we reach subleading contributions.

\subsection{Explicit computations in the bulk}
\subsubsection{Leading order}%
\label{sec:leading-order-bulk}

The Moyal product at leading order is just a pointwise product and $\bar\sigma$ therefore commutes with the momentum at this order. It can therefore be treated as effectively constant. Said differently, at large charge, the position-dependent $\mu(\vect{r})$ can be regarded as a slowly varying function, and the computation is formally the same as the one without potential. It follows that, by dimensional analysis, the leading-order gap \(\Sigma_0(\bar{\vect{R}})\) must be proportional to the effective chemical potential $\mu(\vect{r})$, \emph{i.e.}
\begin{equation} \label{eq:Sigma0_LO}
   \Sigma_0(\bar{\vect{R}}) = y_0 \pqtybis{ 1 - \bar V(\bar{\vect{R}})},
   \quad \text{with} \quad y_0 \in \mathbb{R}.
\end{equation}
Additionally, this also implies that $y_0$ should reproduce the standard mean-field result for the Bertsch parameter in the absence of an external potential~\cite{Eagles:1969zz} (see also, \emph{e.g.},~\cite{Veillette:2007zz}).

The contributions at leading order stem from the $n=0$ (and therefore $j = k = 0$) terms in Eq.~\eqref{eq:HeatK_Problem_OrderN}, for which the heat kernel equations read
\begin{equation}
  \begin{cases}
    \del_{\bar\proptime} \Kdiag_0 + \bbdiag_0 \Kdiag_0 = 0, \\
    \del_{\bar\proptime} \Koff_0 + \bbdiag_0 \Koff_0 = 0 ,
  \end{cases}
\end{equation}
with initial condition $\Kdiag_0(\bar{\vect{R}},\bar{\vect{p}}; 0) = 1$ and $\Koff_0(\bar{\vect{R}},\bar{\vect{p}}; 0) = 0$. We thus find that 
$\Kdiag_0 = e^{-\bbdiag_0 \bar\proptime}$ and $\Koff_0 = 0$,
and Eq.~\eqref{eq:S0_DimLessStuff} at leading order becomes
\begin{equation} \label{eq:Omega_LO}
  \begin{aligned}
    \Omega_{LO}(\mu)
    ={}& -\hbar \mu \left[ \int_0^\infty \frac{\dd{\bar\proptime}}{\Gamma(s) \bar\proptime} \bar\proptime^s \int \frac{\dd{\bar{\vect{R}}} \dd{\bar{\vect{p}}}}{(2\pi \varepsilon)^3} e^{ -\pqtybis{\pqtybis{ 1 - \bar V(\bar{\vect{R}}) -  \bar{\vect{p}}^2 }^2 + \abs{\Sigma_0(\bar{\vect{R}})}^2 } \bar\proptime } \right]_{s=-1/2} \\ 
    &  - \frac{(\hbar \mu)^2 R_{cl}^3}{2 u_0} \int \dd{\bar{\vect{R}}} \abs{\Sigma_0(\bar{\vect{R}})}^2.
  \end{aligned}
\end{equation}
The $s$-wave scattering length is related to the renormalized coupling $u$ as $a = \frac{m u}{2 \pi \hbar^2}$~\cite{Strinati:2018wdg}. As mentioned, it diverges at the Feshbach resonance, indicating that the system has reached criticality. In the scheme we employ, $u=u_0$ and the unitary regime is thus reached as $u_0\to\infty$.
\footnote{
The regularization typically used in the \acs{bcs} literature goes as follows. Integrating Eq.~\eqref{eq:Omega_LO} over $\dd{\bar\tau}$ yields
\begin{equation} \label{eq:OmegaLO_a_la_BCS}
    \Omega_{LO}(\mu) = - \hbar \mu R_{cl}^3 \int \dd{\bar{\vect{R}}} \left[ \frac{\hbar \mu}{2 u_0} \abs{\Sigma_0(\bar{\vect{R}})}^2 
     + \left( \frac{2 m \mu}{\hbar} \right)^\frac{3}{2} \int \frac{\dd{\bar{\vect{p}}}}{(2 \pi)^3} \sqrt{ \pqtybis{1 - \bar V(\bar{\vect{R}}) - \bar{\vect{p}}^2}^2 + \abs{\Sigma_0(\bar{\vect{R}})}^2 } \right],
\end{equation}
where $\varepsilon$ was replaced by its explicit expression given in Eq.~\eqref{eq:EpsilonFixed}. The integrand diverges for large values of $\bar{\vect{p}}$ as
\begin{equation}
    \bar{\vect{p}}^2 \sqrt{\pqtybis{1 - \bar V(\bar{\vect{R}}) - \bar{\vect{p}}^2}^2 + \abs{\Sigma_0(\bar{\vect{R}})}^2 } = -\bar{\vect{p}}^2 \pqtybis{ 1 - \bar V(\bar{\vect{R}}) - \bar{\vect{p}}^2} + \frac{1}{2} \abs{\Sigma_0(\bar{\vect{R}})}^2 + \order*{1/\bar{\vect{p}}^2}.
\end{equation}
The first term is the result of the free theory, in which $\Sigma_0 = 0$, which can be absorbed in the normalization of the partition function. The second term, on the other hand, combines with the bare coupling $u_0$ to define the renormalized coupling as
\begin{equation}
    \frac{1}{u} := \frac{1}{u_0} + \left( \frac{2 m \mu}{\hbar} \right)^{\frac{3}{2}} \frac{1}{\hbar \mu} \int \frac{\dd{\bar{\vect{p}}}}{(2 \pi)^3} \frac{1}{\bar{\vect{p}}^2} = \frac{1}{u_0} + \int \frac{\dd{\vect{p}}}{(2 \pi \hbar)^3} \frac{2 m}{\vect{p}^2}.
\end{equation}
We restored the dimensionful momentum $\vect{p} := (2 m \hbar \mu)^{\frac{1}{2}} \bar{\vect{p}}$ in the second expression for a more direct comparison with the standard result (see \emph{e.g.}~\cite{Veillette:2007zz,Strinati:2018wdg}).
}

\newcommand*{\hypF}{\prescript{}{2}{F}_{1}}
Our goal is to express the result in the form of an integral over space. By construction, it is a priori only well-defined for \( 1- \bar V(\bar{\vect{R}}) < \delta\), but by consistency, the leading-order contribution cannot have edge divergences, as we shall see. Performing the integral over \(\bar\proptime\) first, we get integrals of the form
\begin{equation}\label{eq:polesThatNeedReg}
I_{m,n}(y) = - \eval*{ \frac{\Gamma(n-1/2)}{4 \sqrt{2}\pi^{7/2}} \int_0^{\infty} \dd{q} q^{2+m} \pqtybis{\pqtybis{q^2 - 1}^2 + y^2}^{1/2-n}}_{\text{reg}},
\end{equation}
which are well-defined for \(n > 1/2 \) and \(-3 < m < 4n -5\). They can however be analytically continued in terms of hypergeometric functions \(\hypF\) as
\begin{multline}
  I_{m,n}(y) = - \frac{y^{5/2 + m/2 - 2 n}}{2 (2 \pi)^{7/2}} \Bigg[ \Gamma(-\tfrac{3+m}{4}) \Gamma(n - \tfrac{5+m}{4})\hypF(- \tfrac{m+1}{4}, n - \tfrac{m+5}{4}, \tfrac{1}{2},-\tfrac{1}{y^2 })  \\
  + \frac{2}{y}  \Gamma(-\tfrac{5+m}{4}) \Gamma(n - \tfrac{3+m}{4})\hypF(\tfrac{m-1}{4}, n - \tfrac{m+3}{4}, \tfrac{3}{2},-\tfrac{1}{y^2 }) \Bigg] .
\end{multline}
They satisfy the recursion relation
\begin{equation}
  \frac{d}{dy}I_{m,n}(y) = -2 y I_{m,n+1}(y).
\end{equation}

\begin{figure}
  \centering
  \includegraphics[width=.5\textwidth]{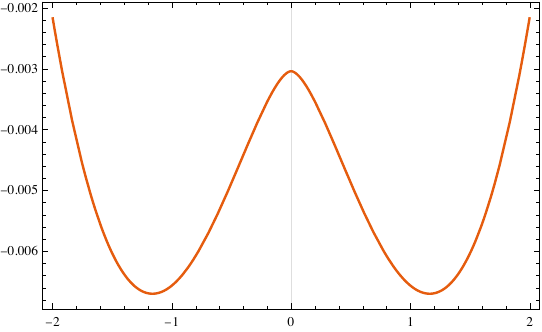}
  \caption{The function $-I_{0,0}(y)$ appearing in the effective action at leading order. It has a minimum for a non-zero value of the parameter \(y\) that controls the expectation value of the gap, which yields spontaneous symmetry breaking.}
  \label{fig:SSB}
\end{figure}

The leading-order contribution thus becomes
\begin{equation} \label{eq:OmegaLO}
\begin{aligned}
    \Omega_{LO}(\mu)
    & = -4 \pi \hbar \mu R_{cl}^3 \left( \frac{m \mu}{\hbar} \right)^{\frac{3}{2}}  \int \dd{\bar{\vect{R}}} I_{0,0}\pqtybis{\tfrac{\Sigma_{0}(\bar{\vect{R}})}{1- V(\bar{\vect{R}})} } \left( 1 - \bar V(\bar{\vect{R}}) \right)^{\frac{5}{2}} - \frac{(\hbar \mu)^2 R_{cl}^3}{2 u_0} \int \dd{\bar{\vect{R}}} \abs{\Sigma_0(\bar{\vect{R}})}^2\\
    & = -4 \pi \frac{m^{\frac{3}{2}}}{\hbar^3}  \int \dd{\vect{R}} I_{0,0}\pqtybis{\tfrac{\Sigma_{0}(\bar{\vect{R}})}{1- V(\bar{\vect{R}})} } \left( \hbar \mu - V(\vect{R}) \right)^{\frac{5}{2}} - \frac{(\hbar \mu)^2 R_{cl}^3}{2 u_0} \int \dd{\bar{\vect{R}}} \abs{\Sigma_0(\bar{\vect{R}})}^2,
\end{aligned}
\end{equation}
where
\begin{equation}
  I_{0,0}(y) = - \frac{y^{5/2}}{2 (2 \pi)^{7/2}} \left[ \Gamma(- \tfrac{5}{4}) \Gamma(\tfrac{3}{4}) \hypF(- \tfrac{5}{4}, -\tfrac{1}{4}, \tfrac{1}{2}; - \tfrac{1}{y^2}) + \frac{2}{y} \Gamma(- \tfrac{3}{4}) \Gamma(\tfrac{5}{4}) \hypF(- \tfrac{3}{4}, \tfrac{1}{4}, \tfrac{3}{2}; - \tfrac{1}{y^2}) \right] . 
\end{equation}
We then need to further impose the minimization condition set by the saddle equation which, at this order, reads
\begin{equation}
  \label{eq:leading-saddle}
    \Sigma_{0}(\bar{\vect{R}}) I_{0,1}\pqtybis{\tfrac{\Sigma_{0}(\bar{\vect{R}})}{1- V(\bar{\vect{R}})}} = 0.
\end{equation}
It admits the solution anticipated in Eq.~\eqref{eq:Sigma0_LO} with (see Figure~\ref{fig:SSB})
\begin{equation}
    y_0 \approx 1.1622\dots
\end{equation}

At large \(N\) and for a large chemical potential $\mu$, there is therefore an explicit realization of the Cooper mechanism: the Stratonovich field \(\sigma(\tau, \vect{r})\) acquires a non-trivial \acs{vev} which spontaneously breaks the $U(1)$ symmetry. The resulting state is a superfluid, allowing for an explicit computation of the Wilsonian coefficients characterizing the first few terms in the large-charge \ac{eft}. Here, we can extract
\begin{equation}
  c_0 = 4\pi\, I_{0,0}(y_0)  \approx 0.0841\dots,
\end{equation}
which, using Eq.~\eqref{eq:C0_VS_Bertsch}, allows us to extract the value of the Bertsch parameter:
\begin{equation}\label{eq:number-Bertsch}
    \xi \approx 0.5906\dots .
\end{equation}
In the absence of an external potential, this value was obtained via mean-field theory in the celebrated work~\cite{Eagles:1969zz}, and agrees with the large-$N$ computations in~\cite{Veillette:2007zz}, in which the next-to-leading order correction in $\frac{1}{N}$ was also computed.

\pagebreak
Next, recall that in Eq.~\eqref{eq:OmegaLO}, $R_{cl}^3 (2 m \mu / h)^\frac{3}{2} = \varepsilon^{-3} \gg 1$ was identified with the large-charge regime. This is justified a posteriori by the fact that the leading dependence of the charge on the chemical potential at criticality is given by
\begin{equation}
    Q_{LO} = -\frac{1}{\hbar} \frac{\del \Omega_{LO}^\text{crit.}(\mu)}{\del \mu}
    = \varepsilon^{-3} \cdot \frac{5 c_0}{2} \int \dd{\bar{\vect{R}}} \left( 1 - \bar V(\bar{\vect{R}}) \right)^\frac{3}{2} \gg 1.
\end{equation}
Specializing to the case of the external harmonic potential $V(\vect{r}) = \frac{m \omega^2}{2} \vect{r}^2$---\emph{i.e.}, $\bar V(\bar{\vect{R}}) = \bar{\vect{R}}^2$)---with $\omega \ll \mu$, we find
\begin{equation}
    Q_{LO} = \left( \frac{\mu}{\omega} \right)^3 \frac{5 \pi^2 c_0}{2^\frac{5}{2}} = \left( \frac{\mu}{\omega} \right)^3 \frac{1}{3 \xi^\frac{3}{2}} = \left( \frac{\mu}{\omega} \right)^3 \cdot 0.734\ldots
\end{equation}
Using Eq.~\eqref{eq:DeltaQ_vs_FreeEnergy}, the leading dependence of the conformal dimension of the large-charge operator on the charge $Q$ is therefore given by
\begin{equation}
    \Delta_Q \stackrel{LO}{=} \frac{3}{4} \left[ \frac{5 \pi^2 c_0}{2^\frac{5}{2}} \right]^{-3} Q^\frac{4}{3} = \frac{3^\frac{4}{3}}{4} \sqrt{\xi} Q^\frac{4}{3} = Q^\frac{4}{3} \cdot 1.893\ldots
\end{equation}

\subsubsection{Subleading corrections}%

By inspection of Eq.~\eqref{eq:HeatK_Problem_OrderN}, the components of the heat kernel are seen to be of the form
\begin{equation}
\begin{dcases}
    \Kdiag_n(\bar{\vect{R}}, \bar{\vect{p}}; \bar\proptime) & := e^{-\bbdiag_0(\bar{\vect{R}}, \bar{\vect{p}}) \bar\proptime} \sum_{l=0}^{n+1} \bar\proptime^l \Kdiag_{n,l}(\bar{\vect{R}}, \bar{\vect{p}}), \\
    \Koff_n(\bar{\vect{R}}, \bar{\vect{p}}; \bar\proptime) & := e^{-\bbdiag_0(\bar{\vect{R}}, \bar{\vect{p}}) \bar\proptime} \sum_{l=0}^{n+1} \bar\proptime^l \Koff_{n,l}(\bar{\vect{R}}, \bar{\vect{p}}),
\end{dcases}
\end{equation}
where $\Kdiag_{n,0}(\bar{\vect{R}}, \bar{\vect{p}}) = \delta_{0n}$ and $\Koff_{n,0}(\bar{\vect{R}}, \bar{\vect{p}}) = 0$. That is, these components are polynomials in $\bar\proptime$ of degree (at most) $n+1$, multiplied by an overall factor of $e^{-\bbdiag_0 \bar\proptime}$.


With this, the trace in Eq.~\eqref{eq:S0_DimLessStuff} can be decomposed explicitly order by order in \(\varepsilon\), and the integrals over momentum and proper time can be performed and expressed in terms of the functions \(I_{m,n}\).
At criticality,
\begin{equation}
\begin{aligned}
    \frac{\Omega_{N^nLO}^\text{crit.}(\mu)}{\hbar \mu R_{cl}^3 (2 m \mu / \hbar)^\frac{3}{2}}
     = -\sum_{l=0}^{n+1} \int \dd{\bar{\vect{R}}} \left[ \int \frac{\dd{\bar\proptime}}{\Gamma(s) \bar\proptime} \bar\proptime^{s+l} \int \frac{ \dd{\bar{\vect{p}}}}{(2 \pi)^3} \Kdiag_{n,l}(\bar{\vect{R}}, \bar{\vect{p}}) \cdot e^{-\bbdiag_0(\bar{\vect{R}}, \bar{\vect{p}}) \bar\proptime} \right]_{s=-\frac{1}{2}} ,
\end{aligned}
\end{equation}
which is subject to the gap equation,
\begin{equation}
  \frac{\delta \Omega_{N^nLO}^\text{crit.}}{\delta \Sigma_0} = 0.
\end{equation}
Once the above saddle equation is accounted for, this recipe gives the order-$\varepsilon^n$ \emph{bulk} contribution to the zero-temperature limit of the grand-canonical potential at criticality (and one still needs to account for the edge contributions).
\subsubsection{Next-to-leading order}%
\label{sec:next-to-leading-order-bulk}

At the next order in \(\varepsilon\), \emph{i.e.}, $n = 1$ in Eq.~\eqref{eq:HeatK_Problem_OrderN}, we use the fact that $\poisson{\bbdiag_0, \Kdiag_0} = 0$, so that the heat kernel equations at this order can be put into the form
\begin{equation}
    \begin{cases}
        \del_{\bar\proptime} \hat \Kdiag_1 + \bbdiag_1 = 0, \\
        \del_{\bar\proptime} \hat \Koff_1 + \bboff_1 = 0,
    \end{cases}
\end{equation}
with
\begin{equation}
	\begin{cases}
		\hat \Kdiag_1(\bar{\vect{R}},\bar{\vect{p}}; 0) = 0, \\
		\hat \Koff_1(\bar{\vect{R}},\bar{\vect{p}}; 0) = 0,
	\end{cases}
\end{equation}
where the hat notation means
\begin{align}
    \Kdiag_i & := \Kdiag_0 \cdot \hat \Kdiag_i, &  \Koff_i & := \Kdiag_0 \cdot \hat \Koff_i.
\end{align}
Thus, the first correction to the leading-order heat kernel is
\begin{equation}
    K_1(\bar{\vect{R}}, \bar{\vect{p}}; \bar\proptime) = -\bar\proptime
    e^{-\bbdiag_0(\bar{\vect{R}}, \bar{\vect{p}}) \bar\proptime}
    \begin{pmatrix}
    \Sigma_{0}(\bar{\vect{R}})^* \Sigma_{1}(\bar{\vect{R}}) + \Sigma_{1}(\bar{\vect{R}})^* \Sigma_{0}(\bar{\vect{R}})
    & 2 i \bar{\vect{p}} \nabla_{\bar{\vect{R}}} \Sigma_{0}(\bar{\vect{R}})
    \\
    -2 i \bar{\vect{p}} \nabla_{\bar{\vect{R}}} \Sigma_{0}(\bar{\vect{R}})^*
    & \Sigma_{0}(\bar{\vect{R}})^* \Sigma_{1}(\bar{\vect{R}}) + \Sigma_{1}(\bar{\vect{R}})^* \Sigma_{0}(\bar{\vect{R}})
    \end{pmatrix},
\end{equation}
and we have
 $\bbdiag_1 = \Sigma_0^* \Sigma_1 + \Sigma_1^* \Sigma_0$, so we obtain
\begin{equation}
  \frac{\Omega_{NLO}^\text{crit.}(\mu)}{\hbar \mu R_{cl}^3 \left( 2 m \mu / \hbar \right)^\frac{3}{2}} = - 4 \pi \int \dd{\bar{\vect{R}}} \left( \Sigma_0(\bar{\vect{R}})^* \Sigma_1(\bar{\vect{R}}) + \Sigma_1(\bar{\vect{R}})^* \Sigma_0(\bar{\vect{R}}) \right) I_{0,1}\pqtybis{\tfrac{\abs{\Sigma_{0}(\bar{\vect{R}})}}{1- V(\bar{\vect{R}})}}.
\end{equation}
The gap equation is
\begin{equation}
  \Sigma_{1}(\bar{\vect{R}}) \pqtybis{I_{0,1}\pqtybis{\tfrac{\abs{\Sigma_{0}(\bar{\vect{R}})}}{1- V(\bar{\vect{R}})}} - 2 \frac{ \Sigma_{0}(\bar{\vect{R}})^2 }{1- V(\bar{\vect{R}})}  I_{0,2}\pqtybis{\tfrac{\abs{\Sigma_{0}(\bar{\vect{R}})}}{1- V(\bar{\vect{R}})} } } = -2 \Sigma_{1}(\bar{\vect{R}}) \pqtybis{1- V(\bar{\vect{R}})}  y_0^2 I_{0,2}(y_0) = 0,
\end{equation}
which admits the single solution $\Sigma_1 = 0$.
We find that the first corrections in \(\varepsilon\) to the gap and to the grand potential both vanish, in agreement with the \acs{eft} prediction:
\begin{equation}
  \Omega_{NLO}(\mu) = 0 .
\end{equation}

\subsubsection{Next-to-next-to-leading order}%
\label{sec:next-to-next-to-leading-order-bulk}

At second order in \(\varepsilon\), or \acs{nnlo}, the heat kernel conditions become
\begin{equation}
  \begin{cases}
    \del_{\bar\proptime} \hat \Kdiag_2 + \bbdiag_2 - \pqtybis{ \bbdiag_1^2 + \bboff_1^2 + \ppoisson{\bbdiag_0, \bbdiag_0} } \bar\proptime + \frac{1}{8} \Pi_{IJ} \Pi_{LM} \del_{IL} \bbdiag_0 \del_J \bbdiag_0 \del_M \bbdiag_0 \bar\proptime^2 = 0, \\
    \del_{\bar\proptime} \hat \Koff_2 + \bboff_2 \bar\proptime + \bbdiag_1 \bboff_1 \bar\proptime^2 = 0,
  \end{cases}
\end{equation}
and  the \acs{nnlo} action takes the form
\begin{equation}
\begin{aligned}
 \frac{\Omega_{NNLO}^\text{crit.}(\mu)}{\hbar \mu R_{cl}^3 \left( 2 m \mu / \hbar \right)^\frac{3}{2}} = {}& \int \dd{\bar{\vect{R}}} \frac{4 \pi}{9 \pqtybis{1-\bar V}^{5/2}} \biggl( 4 (5 I_{2,3}-3 I_{0,3}) (\Sigma_0)^2 (\nabla \Sigma_0)^2 \\
  &\kern-50pt +\pqtybis{1-\bar V}^3 ((9 I_{0,1}-18 I_{0,2}+48 I_{2,2}-8 I_{2,3}-30 I_{4,2}+8 (3 I_{4,3}-3 I_{6,3}+I_{8,3})) \Laplacian \bar V\\
  &\kern-50pt -18 I_{0,1} ((\Sigma_1)^2+2 \Sigma_0 \Sigma_2))+2 \pqtybis{1-\bar V}^2 ((9 I_{0,2}-9 I_{2,2}+4 (I_{2,3}-2 I_{4,3}+I_{6,3})) (\nabla \Sigma_0)^2 \\
  &\kern-50pt +(9 I_{0,2}-15 I_{2,2}+4 (I_{2,3}-2 I_{4,3}+I_{6,3})) \Sigma_0 \Laplacian \Sigma_0\\
  &\kern-50pt +3 (3 I_{0,2}-2 I_{0,3}-3 I_{2,2}+6 I_{2,3}-6 I_{4,3}+2 I_{6,3}) (\nabla \bar V)^2) \\
  &\kern-50pt +12 \pqtybis{1-\bar V} \Sigma_0 (3 I_{0,2} \Sigma_0 (\Sigma_1)^2+2 (I_{0,3}-2 I_{2,3}+I_{4,3}) \nabla \bar V \nabla  \Sigma_0)\biggr) .
  \end{aligned}
\end{equation}
As expected, the coefficient of \(\Sigma_2\) is proportional to  \(\Sigma_0 I_{0,1}\), the leading-order variation in Eq.~\eqref{eq:leading-saddle}, and vanishes on shell.

The gap equation is obtained by varying with respect to \(\Sigma_0\):
\begin{equation}
  \frac{\delta}{\delta \Sigma_0} \Omega = \left[ \frac{\del}{\del \Sigma_0} - \nabla \frac{\del}{\del \nabla \Sigma_0} + \Laplacian \frac{\del}{\del \Laplacian  \Sigma_0} \right] \Omega = 0,
\end{equation}
and, using the relations between the \(I_{m,n}\) ,
\begin{equation}
	I_{m,n}(y) = (n-1/2) \pqtybis{I_{m+4,n+1} -2 I_{m+2, n+1} + (1+y^2) I_{m,n+1}},
\end{equation}
we find the profile
\begin{equation} \label{eq:StrataNNLO}
  \sigma_0(\tau, \vect{r}) = y_0 \pqtybis{\hbar\mu - V(\vect{r})} + y_1 \frac{\hbar^2}{m} \frac{(\nabla V(\vect{r}))^2}{\pqtybis{\hbar\mu - V(\vect{r})}^2} + y_2 \frac{\hbar^2}{m} \frac{\Laplacian{V(\vect{r})}}{\hbar\mu - V(\vect{r})} + \dots,
\end{equation}
with 
\begin{equation}
    y_1\approx -0.004347\dots, \qquad y_2\approx -0.1608\dots.
\end{equation}
This provides the first non-trivial correction to the profile of the Stratonovich field.
We stress once more that this expression is only valid up to a distance \(\delta\) from the edge at \(\bar{\vect{R}} = 1\).
We will discuss in the next subsection what happens closer to the edge of the cloud, where we will need to merge our small-gradient expansion with the perturbation theory of the boundary dynamics, subject to appropriate matching conditions.

As for the effective action, we obtain\footnote{Notably, we can express the action in terms of rational
coefficients of $I_{0,0}(y_0)$.}
\begin{equation}
\begin{aligned}
  S =  \frac{4\pi}{\hbar^3}\beta m^{3/2} I_{0,0}(y_0) \int \dd{\vect{r}} \biggl( 
  \pqtybis{\hbar\mu - V(\vect{r})}^{5/2} + \frac{5}{64} \frac{\hbar^2}{m} \frac{(\nabla V)^2}{\sqrt{\hbar\mu - V(\vect{r})}} - \frac{25}{48} \frac{\hbar^2}{m}  \Laplacian V(\vect{r}) \sqrt{\hbar\mu - V(\vect{r})} 
  \biggr).
\end{aligned}
\end{equation}
Comparing with Eq.~\eqref{eq:EFT_Lagr_nnlo}, we find the following Wilsonian coefficients:
\begin{align}%
  \label{eq:result-c1c2}
  c_1 &= \frac{5 \pi}{16}I_{0,0}(y_0) \approx 0.006577\dots, & c_2 &= \frac{25 \pi}{108} I_{0,0}(y_0) \approx 0.004872\dots .
\end{align}
This gives the final result for the scaling dimension of the lowest operator of charge $Q$
\begin{equation}%
  \label{eq:resultDelta}
  \frac{\Delta_Q}{N} =  0.8313 \pqtybis{\frac{Q}{N} }^{4/3} + 0.2631 \pqtybis{\frac{Q}{N} }^{2/3} + \dots
\end{equation}
The value for $c_2$ satisfies the bound of~\cite{Son:2005rv} on the transverse response function:  The transverse response itself must be negative for a stable condensate, requiring that $c_2 > 0$.  The fact that $2c_1 + 3c_2 > 0$ also places us in the regime were, in the zero-temperature, low-momentum limit, in the absence of an external potential or boundary dynamics, $1 \to 2$ phonon splitting is forbidden.  

It is important to stress that the form that we find for the gap and the Lagrangian density is precisely the one expected based on locality and dimensional arguments: as long as we only look at the bulk, the problem has two scales, $\mu$ and $\nabla V$, so any physical quantity of dimension \(\Delta(G)\), in the limit of large $\mu$, must take the form
\begin{equation}
	G(\mu, \vect r) = \pqtybis{ \hbar \mu-V(\vect r)}^{\Delta(G)/2} \mathscr{F}\pqtybis{\tfrac{\hbar}{m^{1/2}}\tfrac{ \nabla V(\vect r)}{(\hbar \mu-V(\vect r))^{3/2}}}. 
\end{equation}

\subsection{The edge expansion}%
\label{sec:Edge}

In the previous subsection, an asymptotic expansion of the bulk contributions to the free energy in terms of the small parameter $\varepsilon \sim \omega/\mu$ was obtained. Upon matching these contributions against the bulk large-charge \acs{eft}, we were able to extract the corresponding Wilsonian coefficients at leading order in $\frac{1}{N}$. This expansion is however only valid in the bulk, up to some distance $\delta$ away from the point where the particle density vanishes at leading order in $\varepsilon$. This is therefore a boundary-layer problem, see for instance~\cite{bender1999advanced}.

The procedure to tackle this kind of problem is the following. The domain of a differential equation is split into two (or more) regions. In each region, an asymptotic expansion of the solution is found and then matched over an intermediate region where both approximations are valid. When exact solutions are known across multiple regions, one must \emph{patch} these solutions smoothly at transition points.  Crucially, this is to be distinguished from cases where only asymptotic expansions are known in different regions, wherein one must instead perform a \emph{matching}. Said differently, given two such solutions supported in different regions, one needs to find an overlapping region (the \emph{intermediate limit}) in which the two asymptotic expansions have the same functional form. The possibility to find such a matching imposes a constraint on the thickness $\delta$ of the boundary layer. Roughly speaking, the boundary layer has to be sufficiently thick so that the corresponding asymptotic solution (controlled by $\delta$) is rich enough to permit a matching with the functional form coming from the outer (bulk) solution.

In practice, $\delta$ is fixed by a dominant-balance argument in which two or more of the terms of the one-particle Hamiltonian must be of the same order, allowing us to retain enough information. This characterizes the so-called \emph{distinguished limit}. For simplicity, we shall consider spherically-symmetric potentials only. Correspondingly, the boundary layer is a spherical shell around $\bar r := \abs{\bar{\vect{r}}} = 1$, so we introduce the \emph{inner} variable $\bar u$ as
\begin{equation}
  \bar{r} = 1 - \bar u \delta.
\end{equation}
In the limit of \(\delta \to 0\), the metric is approximately flat in terms of a radial and two orthogonal directions.
To see that, start from \(\setR^3\) in polar coordinates,
\begin{equation}
  \dd{s}^2 = \dd{\bar{r}}^2 + \bar{r}^2 \pqtybis{ \frac{\dd{z}^2}{1- z^2} + \pqtybis{1 - z^2} \dd{\phi}^2} ,
\end{equation}
with \(z \in (-1,1)\) and \(\phi \in (0, 2 \pi)\). Rescaling \(z = \bar x \delta\) and \(\phi = \bar y \delta\), so that \(\bar x \in(- 1/\delta, 1/\delta)\) and \(\bar y \in (0, 2 \pi/\delta)\), the metric becomes
\begin{equation}
  \dd{s}^2 = \delta^2 \pqtybis{ \dd{\bar u} + \dd{\bar x}^2 + \dd{\bar y}^2} = \delta^2 \pqtybis{\dd{\bar u} + \dd{\bar x_{\perp}}^2}  + \order{\delta}
\end{equation}
and the Laplacian with respect to the dimensionless coordinate $\bar{r}$ is approximately given by
\begin{equation}
  \Laplacian_{\bar r}{} = \frac{1}{\delta^2} \pqtybis{\Laplacian_{\perp}{} + \frac{\dd^{2}{}}{\dd{u^{2}}}} + \order{1/\delta}.
\end{equation}

Since $1 - \bar V(1) = 0$ by definition of $R_{cl}$, the dimensionless Hamiltonian $\bar h(\bar r)$ becomes
\begin{equation}
	\eval*{\bar h(\bar r)}_{\bar r =1 - \bar u \delta} = -\frac{\varepsilon^2}{\delta^2}\pqtybis{ \Laplacian_{\perp}{} + \frac{\dd^{2}{}}{\dd{u^{2}}}} - \bar u \delta \del_{\bar r} \bar V(1) + \order{(\bar u \delta)^2}.
\end{equation}
The only possible dominant balance is obtained when the first two terms are of the same order:
\begin{equation}
	\frac{\varepsilon^2}{\delta^3} = \order{1}.
\end{equation}
The distinguished limit therefore corresponds to the double-scaling limit $\varepsilon \to 0$, $\delta \to 0$, such that $\varepsilon^2/\delta^3=1$, and we obtain
\begin{equation}
  \eval*{\bar h(\bar r)}_{\bar r =1 - \bar u \delta} = -\delta \cdot \pqtybis{ \Laplacian_{\perp}{} + \frac{\dd^{2}{}}{\dd{u^{2}}} + \alpha \bar u } + \order{(\bar u \delta)^2},
\end{equation}
with \(\alpha := \bar V'(1)\). In particular, for the harmonic oscillator \(\alpha = 2\).
This is the Airy operator, corresponding to the linearization of the confining potential close to the turning point.

What concerns the gap profile, in contrast to the bulk condition in Eq.~\eqref{eq:bulk-condition}, the edge condition is that \(\sigma\) varies on the same scale as the one fixed by its own \acs{vev}
\begin{align}
   \sigma(\vect{r}) \lessapprox \frac{\hbar^2}{2 m}\frac{\pqtybis{\del_{\vect{r}} \sigma(\vect{r})}^2}{\sigma(\vect{r})^2}.
\end{align}
In the edge region%
\footnote{In the usual boundary theory language, the edge is the \emph{inner region} and the bulk is the \emph{outer region}.}%
, the Stratonovich field is expected to be an expansion in \(\delta\),
\begin{equation}
	\eval*{ \bar\sigma(\bar u)}_{\text{edge}} = \bar\sigma_{\text{edge}}^{(1)}(\bar u) \delta + \bar\sigma_{\text{edge}}^{(2)}(\bar u) \delta^2  + \dots,
\end{equation}
and the above condition thus becomes
\begin{equation}
    \bar \sigma_{\text{edge}}^{{1}}(\bar u) \lessapprox \frac{\epsilon^2}{\delta^3} \frac{\pqtybis{\del_{\bar u}  \bar \sigma_{\text{edge}}^{{1}}(\bar u)}^2}{ \bar \sigma_{\text{edge}}^{{1}}(\bar u)^2},
\end{equation}
which is satisfied when \(\bar \sigma_{\text{edge}}^{{1}}(\bar u)\) and its \(\bar u\) derivatives are of order \(\order{1}\). The boundary condition for \(\bar \sigma_{\text{edge}}\) imposes that the known result from the bulk is matched in the region where both the bulk and edge approximations are consistent.

In the spherically-symmetric case, the bulk expansion of the Stratonovich field, Eq.~\eqref{eq:StrataNNLO}, becomes
\begin{equation}
  \eval*{\bar\sigma(\bar{r})}_{\text{bulk}} = y_0 \pqtybis{1 - \bar V(\bar{r})} + \frac{2}{\varepsilon^2} \left[ y_1 \frac{\bar V'(\bar{r})^2}{\pqtybis{1 - \bar V(\bar{r})}^2} + y_2 \frac{\bar V'(\bar{r})}{1 - \bar V(\bar{r})} \right] + \dots
\end{equation}
For the harmonic potential, this becomes, in the intermediate limit $\delta \to 0$, $\bar u \to \infty$ with $ \bar u \delta \to 0$,
\begin{equation}
    \eval*{ \bar \sigma(1 - \bar u \delta)}_{\text{bulk}} = \delta \cdot \pqtybis{2 y_0 \bar u +\frac{y_1}{\bar u^2} + \dots } + \order{\delta^2},
\end{equation}
which also scales like $\delta$ at leading order. To be able to perform the matching, the edge solution $\bar\sigma_{\text{edge}}$ must have the same functional form for $\delta \to 0$:
\begin{equation}
	\sigma_{\text{edge}}^{(1)}(\bar u) \underset{\bar u \to \infty}{\sim} 2 y_0 \bar u +\frac{y_1}{\bar u^2} + \dots .
\end{equation}
This allows us to write the expression of the \acs{bdg} operator close to the boundary:
\begin{equation}
  B = \hbar \mu \delta \bar B^{(1)}(\bar u) + \order{\mu \delta^2} = 
    \hbar \mu \delta \begin{pmatrix}
         \Laplacian_{\perp}{} + \frac{\dd^{2}{}}{\dd{u^{2}}} + 2 \bar u & \sigma_{\text{edge}}^{(1)}(\bar u) \\
      \sigma_{\text{edge}}^{(1)}(\bar u)^* &  - \Laplacian_{\perp}{} - \frac{\dd^{2}{}}{\dd{u^{2}}} - 2 \bar u \end{pmatrix} + \dots .
\end{equation}
To compute the edge contribution to the free energy, we thus need to solve the gap equation in the following form:
\begin{equation}
  \label{eq:8}
  \begin{dcases}
    \frac{\delta \Tr{\abs{B}}}{\delta \sigma_{\text{edge}}^{(1)}(\bar u)} = 0 \\
    \sigma_{\text{edge}}^{(1)}(\bar u) \underset{\bar u \to \infty}{\sim} 2 y_0 \bar u +\frac{y_1}{\bar u^2} + \dots \\
    \sigma_{\text{edge}}^{(1)}(\bar u) \xrightarrow[\bar u \to -\infty]{}0 ,
  \end{dcases}
\end{equation}
to obtain the saddle profile \(\sigma_{\text{edge}}^{(1)}(\bar u)\), and then evaluate the corresponding value of the trace of the operator \(\abs{\bar B^{(1)}}\), 
\begin{equation}
  F_{\text{edge}} = \frac{\hbar \mu \delta }{2} \Tr{\abs{\bar B^{(1)}(\bar u)}}_{\expval{\sigma_{\text{edge}}^{(1)}(u)}} + \dots ,
\end{equation}
where the trace is understood over the edge region. This is a more challenging problem than the bulk calculation discussed previously, since there is no longer a small parameter controlling a perturbative expansion. Hence, a quantitative analysis is beyond the scope of the present work.

Rather, we shall \emph{estimate} the order in \(\mu\) and \(\omega\) at which this contribution enters the expression of the free energy. The first observation is that the trace is divergent and requires regulation. In fact, at first sight, the Airy linear potential might seem to imply a continuous spectrum. However, we are ultimately computing the trace of the absolute value \(\abs{\bar B^{(1)}} = \pqtybis{(B^{(1)})^2}^{1/2}\), and \((B^{(1)})^2\) is bounded from below and increases without bound in the \(u\) direction, so that its spectrum is manifestly discrete. A standard zeta-function regulator can therefore be employed, and the edge contribution to the free energy can be written as the Mellin transform of the heat kernel at coincident points
\begin{equation}
  \Tr{\abs{\bar B^{(1)}(\bar u)}} = \eval*{\frac{1}{ \Gamma(s)} \int \frac{\dd{\bar\proptime}}{\bar\proptime} \bar\proptime^s \Tr[e^{- \abs{\bar B^{(1)}} \bar\proptime}] }_{s=-1} ,
\end{equation}
which is well-defined. Inverting the order of integration, we can write the free energy as the integral of a local density,
\begin{equation}
  \bar b^{(1)}(\bar x, \bar y, \bar u) = \eval*{\frac{1}{\Gamma(s)}\int \frac{\dd{\bar\proptime}}{\bar\proptime} \bar\proptime^s \expval{\bar x, \bar y, \bar u|e^{- \abs{\bar B^{(1)} } \bar\proptime}|\bar x, \bar y, \bar u} }_{s=-1} ,
\end{equation}
which is parametrically of order \(\order{1}\). We are therefore in the position to extract the parametric dependence of \(F\):
\begin{equation}
  \begin{aligned}
    F_{\text{edge}} &= \frac{\hbar \mu \delta }{2} \Tr{\abs{\bar B^{(1)}}} = \frac{\hbar \mu \delta }{2} \int_{-1/\delta}^{1/\delta} \dd{\bar x} \int_0^{2\pi/\delta} \dd{\bar y} \int_0^1 \dd{\bar u} \bar b^{(1)}(\bar x, \bar y, \bar u) 
    \\
    &= \hbar \order*{\frac{\mu}{\delta}} = \hbar \order*{\mu^{4/3} R_{\text{cl}}^{2/3}}.
  \end{aligned}
\end{equation}
Note that we used the fact that the integral over \(\bar b^{(1)}\) is of order \(1/\delta^2\). In the case of the harmonic trap, we find
\begin{equation}
  \eval*{ F_{\text{edge}}}_{\text{ht}} = \frac{\hbar^{4/3}}{m^{1/3}} \order*{ \frac{\mu^{5/3}}{\omega^{2/3}} } + \dots ,
\end{equation}
which precisely matches the leading-order edge contribution in the large-charge \ac{eft}.

\subsection{Conclusion}%
\label{sec:Conclusions}

In this section, we studied a large-$N$ description of the unitary Fermi gas in a trapping potential at zero temperature, where $N$ denotes the number of Fermion flavors. Computations were analytically tractable in the large charge regime. This constitutes a first-principle approximation at the microscopic level, providing a foundation for the Thomas--Fermi approximation at leading order. We presented a clean algorithmic procedure for computing the Wilsonian coefficients of the bulk large-charge \ac{eft}, requiring no other inputs beyond dimensional and symmetry arguments and basic techniques of quantum mechanics.

Previous attempts involved a variety of approximation techniques, which we put here on a solid theoretical basis, whereby the large-$N$ limit turns the one-loop calculation into a justified approximation, and the large-charge regime allows to implement a gradient expansion of the functional determinant resulting from integrating out the original microscopic \acs{dof}.
In this way, we found a consistent microscopic derivation that confirms the \acs{eft} prediction for the form of the bulk operators, their contribution to the energy, and the form of the gap equation. Though the bulk expansion breaks down near the droplet edge, we were able to estimate the parametric dependence of the leading-order edge contributions to the energy starting from first principles, confirming the large-charge \acs{eft} prediction. We reserve the precise determination of the numerical coefficient for future work.

The natural next step would be to carefully connect the qualitative features of the large-charge expansion with experimental observations. As a first step, one might consider the doubly-integrated axial density~\cite{PhysRevLett.92.120401}. Conversely, it would be valuable to obtain experimental data (of the density profile, perhaps) with sufficient resolution to probe the concrete predictions set forth here, particularly in the regime beyond the reach of the Thomas--Fermi approximation.

\pagebreak
\section{Further microscopic models} \label{sec:MoreMicro}

An important task in the nonrelativistic large-charge program is to collect further evidence that a large set of theories indeed belong to the superfluid universality class, and to investigate the possibility of realizing other phases of matter. In the former case, despite our large-$N$ analysis of the last subsection, a \emph{quantitative} evaluation of edge effects is still lacking and should be a priority for future research projects based on microscopic realizations. We discuss the most natural starting point below. What concerns other phases of matter, we also provide a short, though surprising discussion of the relevance of anyons in this context.

\subsection{More realizations of the unitary Fermi gas}

We set $\hbar=m=1$ for simplicity, and discuss weakly-coupled models of the unitary Fermi gas close to 2 and 4 dimensions presented in~\cite{Nishida:2007pj} (see also \cite{Nikolic:2007zz}). Similar to the large-$N$ model, we first consider the so-called 'two-channel model' of Feshbach resonance:\footnote{One could introduce separate chemical potentials and distinct masses for the two fermion species.}
\begin{equation}
    S[\psi, \Phi] = \int\dd{\tau}\dd{\vect{r}} \left[
  \bar\psi_{\sigma} \left( \partial_\tau - \frac{\nabla^2}{2 m} - \mu(\vect{r}) \right) \psi_{\sigma} + \Phi^* \left( \partial_\tau - \frac{\nabla^2}{4m} - 2\mu(\vect{r}) \right) \Phi - g \Phi \bar\psi_{\uparrow} \bar \psi_{\downarrow} - g \Phi^* \psi_{\downarrow} \psi_{\uparrow} \right],
\end{equation}
with $d=4-\epsilon$ ($\epsilon>0$). The composite bosonic field $\Phi$, with mass $2m$, is dynamical (as opposed to the Stratonovich field) and represents rigid pairs of fermions. This is the critical action obtained in \acs{dimreg}, where the coupling is tuned to its critical value,
\begin{equation}
    g^2 = 8 \pi^2 \epsilon.
\end{equation}
This is critical value is computed when $\mu(\vect{r})$ is absent and is the exact result to all orders in $\epsilon$. This model therefore provides a simple weakly-coupled microscopic description of fermions at unitarity close to four dimensions. We expect that a gradient expansion---controlled by the charge $Q\gg1$---may be formulated around the homogeneous solution, as was done in the previous large-$N$ analysis. Generally speaking, the $\epsilon$-expansion is under slightly better analytic control than the large-$N$ expansion due to its proximity to the free theory. It is therefore reasonable to hope that the analysis of edge effects could be improved with respect to our previous results. 

A weakly-coupled approach to the unitary Fermi gas comes with a bonus: there is, in fact, another description as we approach two dimensions from above. Concretely, in $d=2+\tilde\epsilon$ ($\tilde\epsilon>0$), the critical model is simply given by~\cite{Nishida:2007pj,Nikolic:2007zz}
\begin{equation}
    \tilde{S}[\psi] = \int\dd{\tau} \dd{\vect{r}} \left[ \bar\psi_{\sigma} \left( \partial_\tau - \frac{1}{2m} \nabla^2 - \mu(\vect{r}) \right) \psi_{\sigma} - \tilde{g}^2 \bar\psi_{\uparrow} \bar\psi_{\downarrow} \psi_{\downarrow} \psi_{\uparrow} \right],
\end{equation}
in terms of the fixed-point coupling,
\begin{equation}
    \tilde{g}^2 = 2 \pi \tilde{\epsilon}.
\end{equation}
Both of these weakly-coupled descriptions have been investigated in the absence of the harmonic trap, and it is desirable to include it. The analysis should be phrased in terms of a double-scaling limit $\epsilon\to0$, $Q\to\infty$, keeping $\epsilon Q$ fixed (and similarly for $\tilde\epsilon$), as guided by the equivalent result in the relativistic case~\cite{Badel:2019oxl,Badel:2019khk}. A difficulty to overcome is the explicit construction of the large-charge operators in terms of the underlying fermionic degrees of freedom in these theories.

\subsection{Anyons} \label{sec:anyons}

Certain $(2+1)$-dimensional systems host quasiparticles with fractional statistics interpolating between those of fermions and those of bosons in terms of the (rational) phase $\theta \in [-\pi,\pi]$ acquired upon exchanging two identical particles: these are so-called anyons~\cite{wilczek1982magnetic,wilczek1982remarks,wu1984multiparticle,Chen:1989xs,chou1991multianyon,li1992thomas,basu1992class,sporre1992four,sen1992anyons}. The topic was initiated about 40 years ago by Wilczek~\cite{wilczek1982magnetic,wilczek1982remarks}, and it was quickly realized that they were central to the description of the fractional quantum Hall effect~\cite{laughlin1983anomalous,halperin1984statistics,arovas1984fractional}. Abelian anyons have been experimentally detected for the first time very recently~\cite{Bartolomei_2020,nakamura2020direct}. Their applications now range from non-invertible symmetries\footnote{\emph{E.g.} in the context of Adler-Bell-Jackiw (\textsc{ABJ}) anomalies~\cite{Cordova:2022ieu}, which are present in the Standard model~\cite{Choi:2022jqy}.} to quantum computing~\cite{Kitaev:1997wr,Aghaee_2023}.\footnote{Non-Abelian anyons are indeed ideal candidates for the preparation of qubits as their topological properties ('braiding') make them, in principle, resistant to quantum noise.}

Of central importance to us is the fact that they have a hidden $SL(2,\mathbb{R})$ symmetry which makes an \acs{nrcft} description appropriate. Conveniently, they are in fact often studied in a harmonic potential. Originally, the study of many-anyon states was phrased in terms of the quantum many-body wavefunction, see~\cite{date2003classicalquantummechanicsanyons} for a review. A more recent field-theoretic discussion appeared in~\cite{Nishida:2007pj,Doroud:2015fsz,Doroud:2016mfv}, see also~\cite{Fradkin_2013}. Before presenting a microscopic description of anyons, we discuss some striking properties of the $Q$-anyon ground-state conformal dimension, which we denote by $\Delta_{Q,\theta}$.


\begin{figure} [h]
    \centering
    \includegraphics[width=0.6\textwidth]{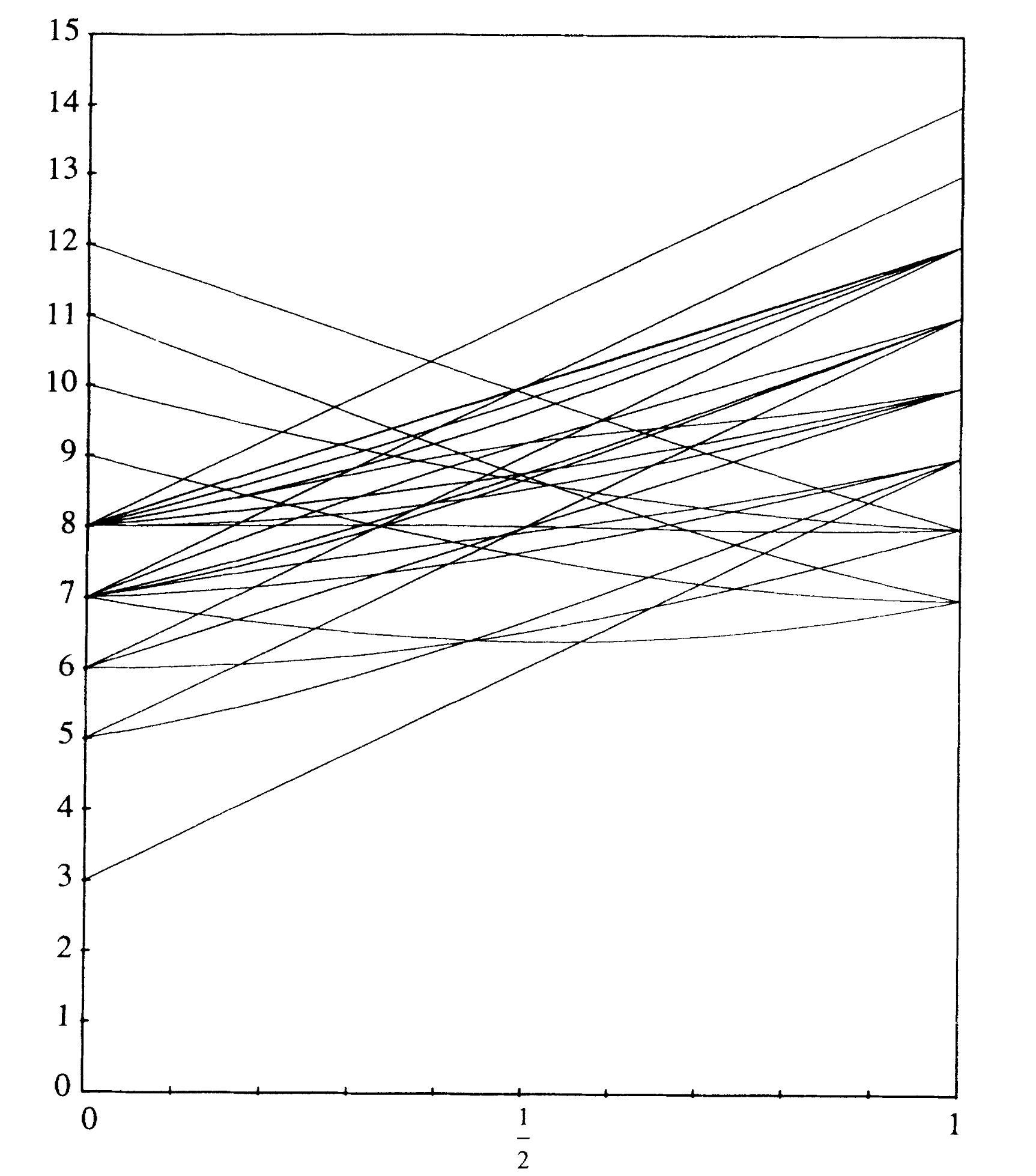}
    \caption{Picture taken from~\cite{sporre1992four}. Energy spectrum (vertical axis, in units of $\omega$) of a four-anyon system ($Q=4$) as a function of $\frac{\theta}{\pi}$ (horizontal axis), the parameter that interpolates between the bosonic behaviour ($\theta\approx0$) and the fermionic one ($\theta\approx\pi$). The strict limits $\theta=0,\pi$ correspond to free bosons and fermions, respectively. For $Q=4$, level crossing occurs around $\frac{\theta}{\pi} = 0.56$. This spectrum was obtained numerically.}
    \label{fig:FourAnyons}
\end{figure}

Starting from the bosonic end---that is, $\theta\approx0$, see Figure~\ref{fig:FourAnyons}---the ground state energy of a system of $Q$ anyons in a harmonic potential grows linearly with $\theta$ (the so-called linear solution):
\begin{equation} \label{eq:Nanyon_bos}
    \Delta_{Q,bos} = Q + \frac{Q(Q-1)}{2\pi} \theta.
\end{equation}
This conformal dimension is in fact \emph{exact}, as it is secretly protected by a hidden supersymmetry~\cite{Doroud:2015fsz}! In this supersymmetric description, the $Q$-anyon ground state is a nonrelativistic chiral primary operator: it is a \acs{bps} operator in disguise. Its conformal dimension is given exactly by $\Delta_{N,bos} = \frac{3}{2} R - J$ in terms of the $R$-charge $R=\frac{2(\theta-1)}{3\theta}Q$ and the angular momentum $J=-Q^2 \theta$. At large charge $Q\gg1$, this is---to the best of our knowledge---the only nonrelativistic known example of a large-charge state that evades the usual scaling behaviour. This provides a prototypical example for the moduli universality class of \acp{nrcft}, which moreover bears great phenomenological and theoretical importance.

Perhaps even more interesting, however, is the fact that there is level crossing: there exists a critical value $\theta_c$ above which the above state is in fact no longer the ground state, but an excited one. Remarkably, the large-$Q$ ground state slightly above $\theta_c$ is believed to have a conformal dimension given by~\cite{chitra1992ground}
\begin{equation}
    \Delta_{Q,ferm} = \sqrt{\frac{\theta}{\pi}} Q^\frac{3}{2} + \ldots
\end{equation}
at leading order, thereby matching the 'natural' scaling $Q^\frac{3}{2}$.\footnote{Note that, as $\theta$ approaches $\pi$, the system is described by a free Fermi gas, and the fact that $\Delta_{Q,\pi}$ scales like $Q^\frac{3}{2}$ can be seen from a simple counting argument (see~\cite{Komargodski:2021zzy} for the relativistic case).} This tells us that $\theta_c = \order{Q^{-\frac{1}{2}}}$ at large charge, making level crossing occur close to the bosonic end.\footnote{Note that a similar situation appeared in the (relativistic) supersymmetric model studied in~\cite{Hellerman:2015nra,Sharon:2020mjs}.} This said, at large but finite charge, this provides a rather unique crossover between two large-charge universality classes, parametrized by $\theta$. Of course, this large-charge scaling alone does not allow to conclude that the system is described by a conformal superfluid state at large charge. By now, however, there is compelling evidence that this is indeed the case~\cite{Chen:1989xs,sen1992anyons}.

In field-theoretic terms, anyons are described in terms of a Chern-Simons gauge field $a=(a_0,\vect{a})$ at level $k=\frac{\pi}{\theta}$ coupled to a nonrelativistic scalar field $\phi$ in $(2+1)$ dimensions~\cite{Nishida:2007pj}:\footnote{The model is expressed here in real time.}
\begin{equation}
\begin{aligned}
    S[a,\phi] & = \int\dd{t} \dd{\vect{r}} \left[ \frac{1}{4\theta} \partial_t \vect{a} \times \vect{a} - \frac{1}{2\theta} a_0 \nabla \times \vect{a} - \frac{1}{2\xi} (\nabla \cdot \vect{a})^2 \right. \\
    & \left. \hspace{16mm} + i\phi^*(\partial_t + i a_0) \phi - \frac{1}{2} \abs{(\nabla - i \vect{a}) \phi}^2 - \frac{v}{4} (\phi^* \phi)^2 \right].
\end{aligned}
\end{equation}
The Coulomb gauge corresponds to $\xi = 0$, and $v$ is fixed at one of its fixed-point values (repulsive and attractive, respectively):
\begin{equation}
    v = \pm \frac{2}{\abs{k}}.
\end{equation}
It would be very interesting to perform a microscopic computation of the large-charge conformal dimension starting from this model. We expect such a computation to be tractable at least close to the fermionic end~\cite{Chen:1989xs}, where a double-scaling limit $(\pi-\theta)\to0$, $Q\to\infty$, $(\pi-\theta) Q$ fixed should be appropriate. Computing $\Delta_{Q,\theta}$ for arbitrary $\theta$ is perhaps too formidable a task, although one may hope to make some progress in a large-$N$ generalization of the above model.

In the supersymmetric theory, it would also be interesting to consider other pieces of \acs{cft} data, akin to those associated with extremal correlators in the relativistic case. The simplicity of the latter however crucially relied on the \acs{ope} of chiral primaries being non-singular, which might be too much to ask for in this case. Finally, it would be interesting to discuss what the large-charge superfluid \acs{eft} describing the many-anyon ground state above $\theta_c$ really is. We expect it to be a nonrelativistic version of the parity-violating \acsp{cft} at large charge in $(2+1)$-dimensions studied in~\cite{Cuomo:2021qws}. Here, the relevant \acs{eft} would presumably be based on~\cite{Hoyos:2013eha,Moroz:2015cft,Du:2020gqf}, describing chiral superfluids.\footnote{In~\cite{Kravec:2018qnu}, a parity-violating term was identified in $(2+1)$ dimensions, but not included in the large-charge \ac{eft}. Whether this is enough to capture the many-anyon ground-state dynamics needs to be clarified.}

\pagebreak
\section*{Closing word}
\addcontentsline{toc}{section}{Closing word}

In this chapter, we performed an explicit computation of the first few terms in the large-charge expansion of $\Delta_Q$, including the leading edge contribution. This result was successfully matched against the structure corresponding to the large-charge \ac{eft} discussed in the previous chapter, which allowed us to extract the first three Wilsonian coefficients. This stems from a microscopic description of the unitariy Fermi gas in $(3+1)$ dimensions, in which the number of fermion flavors was taken to be artificially large. To the best of our knowledge, this is the first first-principle evaluation of the contribution coming from the edge dynamics. A quantitative computation of the corresponding Wilsonian coefficients remains however challenging and requires further investigation.

In the last section, we also proposed alternative avenues to address similar issues, starting from different microscopic models. In particular, we mentioned the weakly-coupled descriptions of the unitary Fermi gas around 2 and 4 dimensions. We then gave a very rough description of the fascinating dynamics of anyons in a trap. While this is arguably a mature topic, we believe that its relevance to the nonrelativistic large-charge program deserves some attention. We mentioned several possible entry points in this rich topic from a large-charge perspective.
\chapter*{Outlook}
\addcontentsline{toc}{chapter}{Outlook}

In this thesis, we reviewed the effort of the author and his collaborators to significantly improve the state of the art of the nonrelativistic large-charge program. For the benefit of the reader, we provided in the first chapter a comprehensive overview of the large-charge landscape in both relativistic and nonrelativistic \acp{cft}. We believe that this helps understand both the scope of the achievements established in~\cite{Orlando:2020idm,Pellizzani:2021hzx,Hellerman:2021qzz,Hellerman:2023myh} and the remaining challenges. We now discuss the most important open questions.~

First, despite the fact that this thesis focuses on nonrelativistic \acp{cft}, we mentioned in Section~\ref{sec:Intro_LargeQ_rel} a couple of challenges faced in the relativistic large-charge program.
\begin{itemize} [itemsep=5pt,partopsep=0pt,topsep=0pt,parsep=0pt]
    \item A better understanding of the large-charge versus large-spin phase diagram would be desirable, regarding in particular the nature of the phase transitions between them, as well as possible intermediate (excited) phases, accounting \emph{e.g.} for the Tkachenko modes.\footnote{\emph{Cf.}~\cite{Cuomo:2021qws} where these modes are mentioned in the large-charge context.}
    \item For instance, performing an explicit large-$N$ computation in one of the vortex phases by identifying the corresponding large-charge operator and solving the related saddle equation would already be a very non-trivial result.
    \item Since the large-spin behaviour is universal, it would be interesting to identify the characteristic feature in the \ac{cft} data that makes the theory enter the phase diagram of a specific universality class as the spin decreases.
    \item A systematic discussion of the Fermi sphere state at large-charge from an \ac{eft} perspective is lacking. The natural starting point for this is the \ac{eft} built in~\cite{Delacretaz:2022ocm}.
\end{itemize}

Regarding the core of the topic, that is, nonrelativistic conformal field theories at large charge, there is a number of open questions and challenges that one can reasonably hope to solve in the light of our results. Let us also reiterate that, regardless of the large-charge program, many formal ('bootstrap-like') developments in \acp{nrcft} are within reach, and it has become pressing to undertake such an investigation (see \emph{e.g.}~\cite{Raviv-Moshe:2024yzt}).

This said, let us discuss large-charge oriented challenges.
\begin{itemize} [itemsep=5pt,partopsep=0pt,topsep=0pt,parsep=0pt]
    \item A nonrelativistic version of the general macroscopic limit and its consequences for higher-point functions should be investigated systematically and compared with existing results~\cite{Kravec:2018qnu,Beane:2024kld}. In particular, the fate of theories with a moduli space, and supersymmetric \acp{nrcft} in general, should be given more attention. Perhaps this could also guide further investigations in the context of nonrelativistic holography.
    \item We mentioned some nonrelativistic Monte-Carlo simulations of $\Delta_Q$~\cite{Chang_2007,Blume_2007,Endres:2012cw,Yin_2015}. Making contact with these simulations and trying to improve or generalize them is important, given the rapid progress that has been made in the relativistic case\cite{Banerjee:2017fcx,Banerjee:2019jpw,Banerjee:2021bbw,Cuomo:2023mxg,Rong:2023owx}.
    \item The study of vortex phases and of the whole large-charge versus large-spin nonrelativistic phase diagram should be revisited in the light of the edge \ac{eft} and the recent work~\cite{Cuomo:2022kio}. Finding a bootstrap argument for a Regge-like regime when the spin is the dominating quantum number would be a fantastic achievement in itself.
    \item It would be interesting to generalize the computation of the Casimir energy performed in Section~\ref{sec:NR_EFT_quantum} to higher-dimensional cases, and to compute subleading quantum corrections.
    \item It is important to extend the set of microscopic realizations of the nonrelativistic superfluid universality class. We mentioned in Section~\ref{sec:MoreMicro} some simple models where this could reasonably be achieved using double-scaling limits. In particular, obtaining quantitative predictions for the edge contributions, at least the leading one, would be a major achievement. In the case of anyons, a large-$N$ model would perhaps allow to keep the phase factor $\theta$ arbitrary, thereby probing level crossing and the transition between two large-charge universality classes.
    \item Relatedly, figuring out the correct large-charge \ac{eft} describing the many-anyon ground state above $\theta_c$ is an important task, which would rely on the construction of chiral two-dimensional nonrelativistic superfluid \acp{eft} presented in~\cite{Hoyos:2013eha,Moroz:2015cft,Du:2020gqf}. This discussion would closely resemble the relativistic one~\cite{Cuomo:2021qws}.
    \item The case of superconformal anyons also deserves more attention, as there might be a class of simple correlators, akin to the extremal ones in the relativistic case, where large-charge techniques could prove very powerful.
    \item Finally, a more concrete comparison of the large-charge predictions with experimental data is important. In this regard, experiments involving fewer atoms are more sensitive to subleading corrections. It would be exciting to check whether the leading edge contribution---or even the universal leading quantum correction---could be extracted from the numerical fit obtained from (future) high-precision experiments.
    \item A related recent topic, going under the name of \emph{unnuclear physics}~\cite{Hammer:2021zxb,Chowdhury:2023ahp}, is another suggestion of applied \ac{nrcft}. This scenario is typically approximately realized in certain nuclear reactions emitting neutrons. In this context, it would be interesting to relate large-charge higher-point functions to phenomenological observations, see~\cite{Beane:2025tum}.
\end{itemize}

With this, we hope that we were able to convey the idea that the nonrelativistic large-charge program, being still in its infancy, offers a wide range of fascinating directions for future research, some of which have recently been unlocked thanks to the progress reviewed in this thesis.


\backmatter
\cleardoublepage
\phantomsection

\printbibliography

@article{Hagen:1972pd,
    author = "Hagen, C.R.",
    title = "{Scale and conformal transformations in galilean-covariant field theory}",
    doi = "10.1103/PhysRevD.5.377",
    journal = "Phys. Rev. D",
    volume = "5",
    pages = "377--388",
    year = "1972"
}

@article{Goldberger:2014hca,
    author = "Goldberger, Walter D. and Khandker, Zuhair U. and Prabhu, Siddharth",
    title = "{OPE convergence in non-relativistic conformal field theories}",
    eprint = "1412.8507",
    archivePrefix = "arXiv",
    primaryClass = "hep-th",
    doi = "10.1007/JHEP12(2015)048",
    journal = "JHEP",
    volume = "12",
    pages = "048",
    year = "2015"
}

@article{Golkar:2014mwa,
    author = "Golkar, Siavash and Son, Dam T.",
    title = "{Operator Product Expansion and Conservation Laws in Non-Relativistic Conformal Field Theories}",
    eprint = "1408.3629",
    archivePrefix = "arXiv",
    primaryClass = "hep-th",
    reportNumber = "EFI-14-27",
    doi = "10.1007/JHEP12(2014)063",
    journal = "JHEP",
    volume = "12",
    pages = "063",
    year = "2014"
}

@article{Hartnoll:2008vx,
    author = "Hartnoll, Sean A. and Herzog, Christopher P. and Horowitz, Gary T.",
    title = "{Building a Holographic Superconductor}",
    eprint = "0803.3295",
    archivePrefix = "arXiv",
    primaryClass = "hep-th",
    reportNumber = "NSF-KITP-08-38, PUPT-2261",
    doi = "10.1103/PhysRevLett.101.031601",
    journal = "Phys. Rev. Lett.",
    volume = "101",
    pages = "031601",
    year = "2008"
}

@article{Hartnoll:2008kx,
    author = "Hartnoll, Sean A. and Herzog, Christopher P. and Horowitz, Gary T.",
    title = "{Holographic Superconductors}",
    eprint = "0810.1563",
    archivePrefix = "arXiv",
    primaryClass = "hep-th",
    doi = "10.1088/1126-6708/2008/12/015",
    journal = "JHEP",
    volume = "12",
    pages = "015",
    year = "2008"
}

@article{Hartnoll:2009sz,
    author = "Hartnoll, Sean A.",
    editor = "Uranga, A. M.",
    title = "{Lectures on holographic methods for condensed matter physics}",
    eprint = "0903.3246",
    archivePrefix = "arXiv",
    primaryClass = "hep-th",
    doi = "10.1088/0264-9381/26/22/224002",
    journal = "Class. Quant. Grav.",
    volume = "26",
    pages = "224002",
    year = "2009"
}

@article{Esposito:2016ria,
    author = "Esposito, Angelo and Garcia-Saenz, Sebastian and Penco, Riccardo",
    title = "{First sound in holographic superfluids at zero temperature}",
    eprint = "1606.03104",
    archivePrefix = "arXiv",
    primaryClass = "hep-th",
    doi = "10.1007/JHEP12(2016)136",
    journal = "JHEP",
    volume = "12",
    pages = "136",
    year = "2016"
}

@article{Arav:2017plg,
    author = "Arav, Igal and Hason, Itamar and Oz, Yaron",
    title = "{Spontaneous Breaking of Non-Relativistic Scale Symmetry}",
    eprint = "1702.00690",
    archivePrefix = "arXiv",
    primaryClass = "hep-th",
    doi = "10.1007/JHEP10(2017)063",
    journal = "JHEP",
    volume = "10",
    pages = "063",
    year = "2017"
}

@article{Polchinski:1992ed,
	Archiveprefix = {arXiv},
	Author = {Polchinski, Joseph},
	Eprint = {hep-th/9210046},
	Primaryclass = {hep-th},
	Reportnumber = {NSF-ITP-92-132, UTTG-20-92, C92-06-03.1},
	Slaccitation = {%%CITATION = HEP-TH/9210046;%%},
	Title = {{Effective field theory and the Fermi surface}},
	Year = {1992}}

@article{Berenstein:2002jq,
	Archiveprefix = {arXiv},
	Author = {Berenstein, David Eliecer and Maldacena, Juan Martin and Nastase, Horatiu Stefan},
	Doi = {10.1088/1126-6708/2002/04/013},
	Eprint = {hep-th/0202021},
	Journal = {JHEP},
	Pages = {013},
	Primaryclass = {hep-th},
	Slaccitation = {%%CITATION = HEP-TH/0202021;%%},
	Title = {{Strings in flat space and pp waves from N=4 superYang-Mills}},
	Volume = {0204},
	Year = {2002}}

@article{Gross:2002su,
    author = "Gross, David J. and Mikhailov, Andrei and Roiban, Radu",
    title = "{Operators with large R charge in N=4 Yang-Mills theory}",
    eprint = "hep-th/0205066",
    archivePrefix = "arXiv",
    reportNumber = "NSF-ITP-02-36, ITEP-TH-17-02",
    doi = "10.1006/aphy.2002.6293",
    journal = "Annals Phys.",
    volume = "301",
    pages = "31--52",
    year = "2002"
}

@article{Srednicki:1994mfb,
    author = "Srednicki, Mark",
    title = "{Chaos and Quantum Thermalization}",
    eprint = "cond-mat/9403051",
    archivePrefix = "arXiv",
    doi = "10.1103/PhysRevE.50.888",
    journal = "Phys. Rev. E",
    volume = "50",
    month = "3",
    year = "1994"
}

@article{Dyer:2015zha,
	Archiveprefix = {arXiv},
	Author = {Dyer, Ethan and Mezei, M{\'a}rk and Pufu, Silviu S. and Sachdev, Subir},
	Eprint = {1504.00368},
	Primaryclass = {hep-th},
	Reportnumber = {PUPT-2479},
	Slaccitation = {%%CITATION = ARXIV:1504.00368;%%},
	Title = {{Scaling dimensions of monopole operators in the $\mathbb{CP}^{N_b - 1}$ theory in $2+1$ dimensions}},
	Year = {2015}}

@article{Komargodski:2011vj,
	Archiveprefix = {arXiv},
	Author = {Komargodski, Zohar and Schwimmer, Adam},
	Doi = {10.1007/JHEP12(2011)099},
	Eprint = {1107.3987},
	Journal = {JHEP},
	Pages = {099},
	Primaryclass = {hep-th},
	Slaccitation = {%%CITATION = ARXIV:1107.3987;%%},
	Title = {{On Renormalization Group Flows in Four Dimensions}},
	Volume = {1112},
	Year = {2011}}

@article{Fitzpatrick:2012yx,
	Archiveprefix = {arXiv},
	Author = {Fitzpatrick, A. Liam and Kaplan, Jared and Poland, David and Simmons-Duffin, David},
	Doi = {10.1007/JHEP12(2013)004},
	Eprint = {1212.3616},
	Journal = {JHEP},
	Pages = {004},
	Primaryclass = {hep-th},
	Slaccitation = {%%CITATION = ARXIV:1212.3616;%%},
	Title = {{The Analytic Bootstrap and AdS Superhorizon Locality}},
	Volume = {1312},
	Year = {2013}}

@article{Fardelli:2024heb,
    author = "Fardelli, Giulia and Fitzpatrick, A. Liam and Li, Wei",
    title = "{Holography and Regge Phases with $U(1)$ Charge}",
    eprint = "2403.07079",
    archivePrefix = "arXiv",
    primaryClass = "hep-th",
    month = "3",
    year = "2024"
}

@article{Dyer:2013fja,
	Archiveprefix = {arXiv},
	Author = {Dyer, Ethan and Mezei, M{\'a}rk and Pufu, Silviu S.},
	Eprint = {1309.1160},
	Primaryclass = {hep-th},
	Reportnumber = {MIT-CTP-4495},
	Slaccitation = {%%CITATION = ARXIV:1309.1160;%%},
	Title = {{Monopole Taxonomy in Three-Dimensional Conformal Field Theories}},
	Year = {2013}}

@article{Karananas:2021bqw,
    author = "Karananas, Georgios K. and Monin, Alexander",
    title = "{More on the operator-state map in nonrelativistic CFTs}",
    eprint = "2109.03836",
    archivePrefix = "arXiv",
    primaryClass = "hep-th",
    reportNumber = "LMU--ASC 39/21",
    doi = "10.1103/PhysRevD.105.065008",
    journal = "Phys. Rev. D",
    volume = "105",
    number = "6",
    pages = "065008",
    year = "2022"
}

@article{Gerchkovitz:2016gxx,
    author = "Gerchkovitz, Efrat and Gomis, Jaume and Ishtiaque, Nafiz and Karasik, Avner and Komargodski, Zohar and Pufu, Silviu S.",
    title = "{Correlation Functions of Coulomb Branch Operators}",
    eprint = "1602.05971",
    archivePrefix = "arXiv",
    primaryClass = "hep-th",
    reportNumber = "PUPT-2500",
    doi = "10.1007/JHEP01(2017)103",
    journal = "JHEP",
    volume = "01",
    pages = "103",
    year = "2017"
}

@article{Gimenez-Grau:2020jrx,
    author = "Gimenez-Grau, Aleix and Liendo, Pedro",
    title = "{Bootstrapping Coulomb and Higgs branch operators}",
    eprint = "2006.01847",
    archivePrefix = "arXiv",
    primaryClass = "hep-th",
    reportNumber = "DESY-20-099, DESY 20-099",
    doi = "10.1007/JHEP01(2021)175",
    journal = "JHEP",
    volume = "01",
    pages = "175",
    year = "2021"
}

@article{Ferrenberg_2018,
   title={Pushing the limits of Monte Carlo simulations for the three-dimensional Ising model},
   volume={97},
   ISSN={2470-0053},
   url={http://dx.doi.org/10.1103/PhysRevE.97.043301},
   DOI={10.1103/physreve.97.043301},
   number={4},
   journal={Physical Review E},
   publisher={American Physical Society (APS)},
   author={Ferrenberg, Alan M. and Xu, Jiahao and Landau, David P.},
   year={2018},
   month=apr }

@article{Zhu:2022gjc,
    author = "Zhu, Wei and Han, Chao and Huffman, Emilie and Hofmann, Johannes S. and He, Yin-Chen",
    title = "{Uncovering Conformal Symmetry in the 3D Ising Transition: State-Operator Correspondence from a Quantum Fuzzy Sphere Regularization}",
    eprint = "2210.13482",
    archivePrefix = "arXiv",
    primaryClass = "cond-mat.stat-mech",
    doi = "10.1103/PhysRevX.13.021009",
    journal = "Phys. Rev. X",
    volume = "13",
    number = "2",
    pages = "021009",
    year = "2023"
}

@article{Hu:2023xak,
    author = "Hu, Liangdong and He, Yin-Chen and Zhu, W.",
    title = "{Operator Product Expansion Coefficients of the 3D Ising Criticality via Quantum Fuzzy Spheres}",
    eprint = "2303.08844",
    archivePrefix = "arXiv",
    primaryClass = "cond-mat.stat-mech",
    doi = "10.1103/PhysRevLett.131.031601",
    journal = "Phys. Rev. Lett.",
    volume = "131",
    number = "3",
    pages = "031601",
    year = "2023"
}

@article{Han:2023yyb,
    author = "Han, Chao and Hu, Liangdong and Zhu, W. and He, Yin-Chen",
    title = "{Conformal four-point correlators of the three-dimensional Ising transition via the quantum fuzzy sphere}",
    eprint = "2306.04681",
    archivePrefix = "arXiv",
    primaryClass = "cond-mat.stat-mech",
    doi = "10.1103/PhysRevB.108.235123",
    journal = "Phys. Rev. B",
    volume = "108",
    number = "23",
    pages = "235123",
    year = "2023"
}

@article{Zhou:2023qfi,
    author = "Zhou, Zheng and Hu, Liangdong and Zhu, W. and He, Yin-Chen",
    title = "{SO(5) Deconfined Phase Transition under the Fuzzy-Sphere Microscope: Approximate Conformal Symmetry, Pseudo-Criticality, and Operator Spectrum}",
    eprint = "2306.16435",
    archivePrefix = "arXiv",
    primaryClass = "cond-mat.str-el",
    doi = "10.1103/PhysRevX.14.021044",
    journal = "Phys. Rev. X",
    volume = "14",
    number = "2",
    pages = "021044",
    year = "2024"
}

@article{Hu:2023ghk,
    author = "Hu, Liangdong and He, Yin-Chen and Zhu, W.",
    title = "{Solving conformal defects in 3D conformal field theory using fuzzy sphere regularization}",
    eprint = "2308.01903",
    archivePrefix = "arXiv",
    primaryClass = "cond-mat.stat-mech",
    doi = "10.1038/s41467-024-47978-y",
    journal = "Nature Commun.",
    volume = "15",
    number = "1",
    pages = "3659",
    year = "2024"
}

@article{Hofmann:2023llr,
    author = "Hofmann, Johannes Stephan and Goth, Florian and Zhu, Wei and He, Yin-Chen and Huffman, Emilie",
    title = "{Quantum Monte Carlo simulation of the 3D Ising transition on the fuzzy sphere}",
    eprint = "2310.19880",
    archivePrefix = "arXiv",
    primaryClass = "cond-mat.str-el",
    doi = "10.21468/SciPostPhysCore.7.2.028",
    journal = "SciPost Phys. Core",
    volume = "7",
    pages = "028",
    year = "2024"
}

@article{Han:2023lky,
    author = "Han, Chao and Hu, Liangdong and Zhu, W.",
    title = "{Conformal Operator Content of the Wilson-Fisher Transition on Fuzzy Sphere Bilayers}",
    eprint = "2312.04047",
    archivePrefix = "arXiv",
    primaryClass = "cond-mat.str-el",
    month = "12",
    year = "2023"
}

@article{Zhou:2023fqu,
    author = "Zhou, Zheng and Gaiotto, Davide and He, Yin-Chen and Zou, Yijian",
    title = "{The $g$-function and Defect Changing Operators from Wavefunction Overlap on a Fuzzy Sphere}",
    eprint = "2401.00039",
    archivePrefix = "arXiv",
    primaryClass = "hep-th",
    doi = "10.21468/SciPostPhys.17.1.021",
    journal = "SciPost Phys.",
    volume = "17",
    pages = "021",
    year = "2024"
}

@article{Hu:2024pen,
    author = "Hu, Liangdong and Zhu, W. and He, Yin-Chen",
    title = "{Entropic $F$-function of 3D Ising conformal field theory via the fuzzy sphere regularization}",
    eprint = "2401.17362",
    archivePrefix = "arXiv",
    primaryClass = "hep-th",
    month = "1",
    year = "2024"
}

@article{Zhou:2024dbt,
    author = "Zhou, Zheng and Zou, Yijian",
    title = "{Studying the 3d Ising surface CFTs on the fuzzy sphere}",
    eprint = "2407.15914",
    archivePrefix = "arXiv",
    primaryClass = "hep-th",
    month = "7",
    year = "2024"
}

@article{Dedushenko:2024nwi,
    author = "Dedushenko, Mykola",
    title = "{Ising BCFT from Fuzzy Hemisphere}",
    eprint = "2407.15948",
    archivePrefix = "arXiv",
    primaryClass = "hep-th",
    month = "7",
    year = "2024"
}

@article{gupta1996critical,
  title={Critical exponents of the 3-D Ising model},
  author={Gupta, Rajan and Tamayo, Pablo},
  journal={International Journal of Modern Physics C},
  volume={7},
  number={03},
  pages={305--319},
  year={1996},
  publisher={World Scientific}
}

@article{Rattazzi:2008pe,
	Archiveprefix = {arXiv},
	Author = {Rattazzi, Riccardo and Rychkov, Vyacheslav S. and Tonni, Erik and Vichi, Alessandro},
	Doi = {10.1088/1126-6708/2008/12/031},
	Eprint = {0807.0004},
	Journal = {JHEP},
	Pages = {031},
	Primaryclass = {hep-th},
	Slaccitation = {%%CITATION = ARXIV:0807.0004;%%},
	Title = {{Bounding scalar operator dimensions in 4D CFT}},
	Volume = {0812},
	Year = {2008}}

@article{Hellerman:2015nra,
	Archiveprefix = {arXiv},
	Author = {Hellerman, Simeon and Orlando, Domenico and Reffert, Susanne and Watanabe, Masataka},
	Doi = {10.1007/JHEP12(2015)071},
	Eprint = {1505.01537},
	Journal = {JHEP},
	Pages = {071},
	Primaryclass = {hep-th},
	Slaccitation = {%%CITATION = ARXIV:1505.01537;%%},
	Title = {{On the CFT Operator Spectrum at Large Global Charge}},
	Volume = {12},
	Year = {2015}}

@article{Alvarez-Gaume:2016vff,
	Archiveprefix = {arXiv},
	Author = {Alvarez-Gaume, Luis and Loukas, Orestis and Orlando, Domenico and Reffert, Susanne},
	Doi = {10.1007/JHEP04(2017)059},
	Eprint = {1610.04495},
	Journal = {JHEP},
	Pages = {059},
	Primaryclass = {hep-th},
	Reportnumber = {CERN-TH-2016-221},
	Slaccitation = {%%CITATION = ARXIV:1610.04495;%%},
	Title = {{Compensating strong coupling with large charge}},
	Volume = {04},
	Year = {2017}}

@article{Hellerman:2017efx,
      author         = "Hellerman, Simeon and Kobayashi, Nozomu and Maeda,
                        Shunsuke and Watanabe, Masataka",
      title          = "{A Note on Inhomogeneous Ground States at Large Global
                        Charge}",
      journal        = "JHEP",
      volume         = "10",
      year           = "2019",
      pages          = "038",
      doi            = "10.1007/JHEP10(2019)038",
      eprint         = "1705.05825",
      archivePrefix  = "arXiv",
      primaryClass   = "hep-th",
      reportNumber   = "IPMU17-0081, CALT-TH-2017-24",
      SLACcitation   = "%%CITATION = ARXIV:1705.05825;%%"
}

@article{Monin:2016jmo,
	Archiveprefix = {arXiv},
	Author = {Monin, Alexander and Pirtskhalava, David and Rattazzi, Riccardo and Seibold, Fiona K.},
	Doi = {10.1007/JHEP06(2017)011},
	Eprint = {1611.02912},
	Journal = {JHEP},
	Pages = {011},
	Primaryclass = {hep-th},
	Slaccitation = {%%CITATION = ARXIV:1611.02912;%%},
	Title = {{Semiclassics, Goldstone Bosons and CFT data}},
	Volume = {06},
	Year = {2017}}

@unpublished{sridipDiablerets,
    author = {Pal, Sridip},
    title = {Demystifying the State-Operator correspondence in NRCFT},
    note = {"Large Charge aux Diablerets" workshop, July 2022}
}

@article{Kolekar:2018sba,
    author = "Kolekar, Kedar S. and Narayan, K.",
    title = "{AdS$_2$ dilaton gravity from reductions of some nonrelativistic theories}",
    eprint = "1803.06827",
    archivePrefix = "arXiv",
    primaryClass = "hep-th",
    doi = "10.1103/PhysRevD.98.046012",
    journal = "Phys. Rev. D",
    volume = "98",
    number = "4",
    pages = "046012",
    year = "2018"
}

@article{Meyer:2017zfg,
    author = "Meyer, Adiel and Oz, Yaron and Raviv-Moshe, Avia",
    title = "{On Non-Relativistic Supersymmetry and its Spontaneous Breaking}",
    eprint = "1703.04740",
    archivePrefix = "arXiv",
    primaryClass = "hep-th",
    doi = "10.1007/JHEP06(2017)128",
    journal = "JHEP",
    volume = "06",
    pages = "128",
    year = "2017"
}

@article{Jafferis:2017zna,
	Archiveprefix = {arXiv},
	Author = {Jafferis, Daniel and Mukhametzhanov, Baur and Zhiboedov, Alexander},
	Eprint = {1710.11161},
	Primaryclass = {hep-th},
	Slaccitation = {%%CITATION = ARXIV:1710.11161;%%},
	Title = {{Conformal Bootstrap At Large Charge}},
	Year = {2017}}

@article{Hellerman:2017sur,
	Archiveprefix = {arXiv},
	Author = {Hellerman, Simeon and Maeda, Shunsuke},
	Doi = {10.1007/JHEP12(2017)135},
	Eprint = {1710.07336},
	Journal = {JHEP},
	Pages = {135},
	Primaryclass = {hep-th},
	Reportnumber = {IPMU17-0143, CALT-TH-2017-059},
	Slaccitation = {%%CITATION = ARXIV:1710.07336;%%},
	Title = {{On the Large $R$-charge Expansion in ${\mathcal N} = 2$ Superconformal Field Theories}},
	Volume = {12},
	Year = {2017}}

@article{Hellerman:2017veg,
	Archiveprefix = {arXiv},
	Author = {Hellerman, Simeon and Maeda, Shunsuke and Watanabe, Masataka},
	Doi = {10.1007/JHEP10(2017)089},
	Eprint = {1706.05743},
	Journal = {JHEP},
	Pages = {089},
	Primaryclass = {hep-th},
	Reportnumber = {IPMU-17-0015, CALT-TH-2017-032},
	Slaccitation = {%%CITATION = ARXIV:1706.05743;%%},
	Title = {{Operator Dimensions from Moduli}},
	Volume = {10},
	Year = {2017}}

@article{Banerjee:2017fcx,
	Archiveprefix = {arXiv},
	Author = {Banerjee, Debasish and Chandrasekharan, Shailesh and Orlando, Domenico},
	Doi = {10.1103/PhysRevLett.120.061603},
	Eprint = {1707.00711},
	Journal = {Phys. Rev. Lett.},
	Pages = {061603},
	Primaryclass = {hep-lat},
	Slaccitation = {%%CITATION = ARXIV:1707.00711;%%},
	Title = {{Conformal dimensions via large charge expansion}},
	Volume = {120},
	Year = {2018}}

@article{Cuomo:2023mxg,
    author = "Cuomo, Gabriel and Lopes, J. M. Viana Parente and Matos, Jos\'e and Oliveira, J\'ulio and Penedones, Joao",
    title = "{Numerical tests of the large charge expansion}",
    eprint = "2305.00499",
    archivePrefix = "arXiv",
    primaryClass = "hep-lat",
    doi = "10.1007/JHEP05(2024)161",
    journal = "JHEP",
    volume = "05",
    pages = "161",
    year = "2024"
}

@article{Cuomo:2017vzg,
	Archiveprefix = {arXiv},
	Author = {Cuomo, Gabriel and de la Fuente, Anton and Monin, Alexander and Pirtskhalava, David and Rattazzi, Riccardo},
	Eprint = {1711.02108},
	Primaryclass = {hep-th},
	Slaccitation = {%%CITATION = ARXIV:1711.02108;%%},
	Title = {{Rotating superfluids and spinning charged operators in conformal field theory}},
	Year = {2017}}

@article{Komargodski:2012ek,
	Archiveprefix = {arXiv},
	Author = {Komargodski, Zohar and Zhiboedov, Alexander},
	Doi = {10.1007/JHEP11(2013)140},
	Eprint = {1212.4103},
	Journal = {JHEP},
	Pages = {140},
	Primaryclass = {hep-th},
	Slaccitation = {%%CITATION = ARXIV:1212.4103;%%},
	Title = {{Convexity and Liberation at Large Spin}},
	Volume = {11},
	Year = {2013}}

@article{Watanabe:2013uya,
	Archiveprefix = {arXiv},
	Author = {Watanabe, Haruki and Brauner, Tom{\'a}{\v s} and Murayama, Hitoshi},
	Doi = {10.1103/PhysRevLett.111.021601},
	Eprint = {1303.1527},
	Journal = {Phys. Rev. Lett.},
	Number = {2},
	Pages = {021601},
	Primaryclass = {hep-th},
	Reportnumber = {BI-TP-2013-02, IPMU13-0056},
	Slaccitation = {%%CITATION = ARXIV:1303.1527;%%},
	Title = {{Massive Nambu-Goldstone Bosons}},
	Volume = {111},
	Year = {2013}}

@article{Callan:1969sn,
	Author = {Callan, Jr., Curtis G. and Coleman, Sidney R. and Wess, J. and Zumino, Bruno},
	Doi = {10.1103/PhysRev.177.2247},
	Journal = {Phys. Rev.},
	Pages = {2247-2250},
	Slaccitation = {%%CITATION = PHRVA,177,2247;%%},
	Title = {{Structure of phenomenological Lagrangians. 2.}},
	Volume = {177},
	Year = {1969}}

@article{Coleman:1969sm,
	Author = {Coleman, Sidney R. and Wess, J. and Zumino, Bruno},
	Doi = {10.1103/PhysRev.177.2239},
	Journal = {Phys. Rev.},
	Pages = {2239-2247},
	Slaccitation = {%%CITATION = PHRVA,177,2239;%%},
	Title = {{Structure of phenomenological Lagrangians. 1.}},
	Volume = {177},
	Year = {1969}}

@book{Rychkov:2016iqz,
	Archiveprefix = {arXiv},
	Author = {Rychkov, Slava},
	Doi = {10.1007/978-3-319-43626-5},
	Eprint = {1601.05000},
	Isbn = {9783319436258, 9783319436265},
	Primaryclass = {hep-th},
	Reportnumber = {CERN-TH-2016-012},
	Series = {SpringerBriefs in Physics},
	Slaccitation = {%%CITATION = ARXIV:1601.05000;%%},
	Title = {{EPFL Lectures on Conformal Field Theory in D $>=$ 3 Dimensions}},
	Url = {https://inspirehep.net/record/1415968/files/arXiv:1601.05000.pdf},
	Year = {2016}}

@article{Tan_2008a,
   title={Energetics of a strongly correlated Fermi gas},
   volume={323},
   ISSN={0003-4916},
   url={http://dx.doi.org/10.1016/j.aop.2008.03.004},
   DOI={10.1016/j.aop.2008.03.004},
   number={12},
   journal={Annals of Physics},
   publisher={Elsevier BV},
   author={Tan, Shina},
   year={2008},
   month=dec, pages={2952–2970} }

@article{Tan_2008b,
   title={Large momentum part of a strongly correlated Fermi gas},
   volume={323},
   ISSN={0003-4916},
   url={http://dx.doi.org/10.1016/j.aop.2008.03.005},
   DOI={10.1016/j.aop.2008.03.005},
   number={12},
   journal={Annals of Physics},
   publisher={Elsevier BV},
   author={Tan, Shina},
   year={2008},
   month=dec, pages={2971–2986} }

@article{Tan_2008c,
   title={Generalized virial theorem and pressure relation for a strongly correlated Fermi gas},
   volume={323},
   ISSN={0003-4916},
   url={http://dx.doi.org/10.1016/j.aop.2008.03.003},
   DOI={10.1016/j.aop.2008.03.003},
   number={12},
   journal={Annals of Physics},
   publisher={Elsevier BV},
   author={Tan, Shina},
   year={2008},
   month=dec, pages={2987–2990} }

@article{Braaten:2008uh,
    author = "Braaten, Eric and Platter, Lucas",
    title = "{Exact Relations for a Strongly Interacting Fermi Gas from the Operator Product Expansion}",
    eprint = "0803.1125",
    archivePrefix = "arXiv",
    primaryClass = "cond-mat.other",
    doi = "10.1103/PhysRevLett.100.205301",
    journal = "Phys. Rev. Lett.",
    volume = "100",
    pages = "205301",
    year = "2008"
}

@article{Braaten:2008bi,
    author = "Braaten, Eric and Kang, Daekyoung and Platter, Lucas",
    title = "{Exact Relations for a Strongly-interacting Fermi Gas near a Feshbach Resonance}",
    eprint = "0806.2277",
    archivePrefix = "arXiv",
    primaryClass = "cond-mat.other",
    doi = "10.1103/PhysRevA.78.053606",
    journal = "Phys. Rev. A",
    volume = "78",
    pages = "053606",
    year = "2008"
}

@article{Alday:2007mf,
	Archiveprefix = {arXiv},
	Author = {Alday, Luis F. and Maldacena, Juan Martin},
	Doi = {10.1088/1126-6708/2007/11/019},
	Eprint = {0708.0672},
	Journal = {JHEP},
	Pages = {019},
	Primaryclass = {hep-th},
	Slaccitation = {%%CITATION = ARXIV:0708.0672;%%},
	Title = {{Comments on operators with large spin}},
	Volume = {11},
	Year = {2007}}

@article{Hellerman:2018xpi,
	Archiveprefix = {arXiv},
	Author = {Hellerman, Simeon and Maeda, Shunsuke and Orlando, Domenico and Reffert, Susanne and Watanabe, Masataka},
	Doi = {10.1007/JHEP12(2019)047},
	Eprint = {1804.01535},
	Journal = {JHEP},
	Pages = {047},
	Primaryclass = {hep-th},
	Reportnumber = {CALT-TH-2018-014, IPMU18-0059},
	Slaccitation = {%%CITATION = ARXIV:1804.01535;%%},
	Title = {{Universal correlation functions in rank 1 SCFTs}},
	Volume = {12},
	Year = {2019}}

@article{schroedinger,
	Archiveprefix = {arXiv},
	Author = {Favrod, Samuel and Orlando, Domenico and Reffert, Susanne},
	Doi = {10.1007/JHEP12(2018)052},
	Eprint = {1809.06371},
	Journal = {JHEP},
	Pages = {052},
	Primaryclass = {hep-th},
	Slaccitation = {%%CITATION = ARXIV:1809.06371;%%},
	Title = {{The large-charge expansion for Schr{\"o}dinger systems}},
	Volume = {12},
	Year = {2018}}

@article{Hellerman:2018sjf,
	Archiveprefix = {arXiv},
	Author = {Hellerman, Simeon and Kobayashi, Nozomu and Maeda, Shunsuke and Watanabe, Masataka},
	Eprint = {1804.06495},
	Primaryclass = {hep-th},
	Reportnumber = {IPMU18-0069},
	Slaccitation = {%%CITATION = ARXIV:1804.06495;%%},
	Title = {{Observables in Inhomogeneous Ground States at Large Global Charge}},
	Year = {2018}}

@article{Bourget:2018obm,
	Archiveprefix = {arXiv},
	Author = {Bourget, Antoine and Rodriguez-Gomez, Diego and Russo, Jorge G.},
	Doi = {10.1007/JHEP05(2018)074},
	Eprint = {1803.00580},
	Journal = {JHEP},
	Pages = {074},
	Primaryclass = {hep-th},
	Slaccitation = {%%CITATION = ARXIV:1803.00580;%%},
	Title = {{A limit for large $R$-charge correlators in $\mathcal{N}=2$ theories}},
	Volume = {05},
	Year = {2018}}

@article{delaFuente:2018qwv,
	Archiveprefix = {arXiv},
	Author = {De La Fuente, Anton},
	Doi = {10.1007/JHEP08(2018)041},
	Eprint = {1805.00501},
	Journal = {JHEP},
	Pages = {041},
	Primaryclass = {hep-th},
	Slaccitation = {%%CITATION = ARXIV:1805.00501;%%},
	Title = {{The large charge expansion at large $N$}},
	Volume = {08},
	Year = {2018}}

@article{delaFuente:2020yua,
    author = "de la Fuente, Anton and Zosso, Jann",
    title = "{The large charge expansion and AdS/CFT}",
    eprint = "2005.06169",
    archivePrefix = "arXiv",
    primaryClass = "hep-th",
    doi = "10.1007/JHEP06(2020)178",
    journal = "JHEP",
    volume = "06",
    pages = "178",
    year = "2020"
}

@article{Nakayama:2015hga,
    author = "Nakayama, Yu and Nomura, Yasunori",
    title = "{Weak gravity conjecture in the AdS/CFT correspondence}",
    eprint = "1509.01647",
    archivePrefix = "arXiv",
    primaryClass = "hep-th",
    reportNumber = "CALT-TH-2015-045, IPMU15-0145, UCB-PTH-15-07",
    doi = "10.1103/PhysRevD.92.126006",
    journal = "Phys. Rev. D",
    volume = "92",
    number = "12",
    pages = "126006",
    year = "2015"
}

@article{Ishii:2024pnh,
    author = "Ishii, Takaaki and Nakayama, Yu",
    title = "{Convexity restoration from hairy black hole in Einstein-Maxwell-charged scalar system in AdS}",
    eprint = "2402.04552",
    archivePrefix = "arXiv",
    primaryClass = "hep-th",
    reportNumber = "RUP-24-2, YITP-24-11",
    doi = "10.1007/JHEP05(2024)197",
    journal = "JHEP",
    volume = "05",
    pages = "197",
    year = "2024"
}

@article{Bergman:1993kq,
	Archiveprefix = {arXiv},
	Author = {Bergman, O. and Lozano, G.},
	Doi = {10.1006/aphy.1994.1013},
	Eprint = {hep-th/9302116},
	Journal = {Annals Phys.},
	Pages = {416-427},
	Primaryclass = {hep-th},
	Reportnumber = {MIT-CTP-2182, IC-93-184},
	Slaccitation = {%%CITATION = HEP-TH/9302116;%%},
	Title = {{Aharonov-Bohm scattering, contact interactions and scale invariance}},
	Volume = {229},
	Year = {1994}}

@article{Nishida:2007pj,
	Archiveprefix = {arXiv},
	Author = {Nishida, Yusuke and Son, Dam T.},
	Doi = {10.1103/PhysRevD.76.086004},
	Eprint = {0706.3746},
	Journal = {Phys. Rev.},
	Pages = {086004},
	Primaryclass = {hep-th},
	Reportnumber = {INT-PUB-07-16},
	Slaccitation = {%%CITATION = ARXIV:0706.3746;%%},
	Title = {{Nonrelativistic conformal field theories}},
	Volume = {D76},
	Year = {2007}}

@article{Beccaria:2018xxl,
	Archiveprefix = {arXiv},
	Author = {Beccaria, Matteo},
	Eprint = {1809.06280},
	Primaryclass = {hep-th},
	Slaccitation = {%%CITATION = ARXIV:1809.06280;%%},
	Title = {{On the large R-charge $\mathcal N=2$ chiral correlators and the Toda equation}},
	Year = {2018}}

@article{Kravec:2018qnu,
	Archiveprefix = {arXiv},
	Author = {Kravec, S. M. and Pal, Sridip},
	Doi = {10.1007/JHEP02(2019)008},
	Eprint = {1809.08188},
	Journal = {JHEP},
	Pages = {008},
	Primaryclass = {hep-th},
	Slaccitation = {%%CITATION = ARXIV:1809.08188;%%},
	Title = {{Nonrelativistic Conformal Field Theories in the Large Charge Sector}},
	Volume = {02},
	Year = {2019}}

@article{Banerjee:2019jpw,
	Archiveprefix = {arXiv},
	Author = {Banerjee, Debasish and Chandrasekharan, Shailesh and Orlando, Domenico and Reffert, Susanne},
	Doi = {10.1103/PhysRevLett.123.051603},
	Eprint = {1902.09542},
	Journal = {Phys. Rev. Lett.},
	Number = {5},
	Pages = {051603},
	Primaryclass = {hep-lat},
	Slaccitation = {%%CITATION = ARXIV:1902.09542;%%},
	Title = {{Conformal dimensions in the large charge sectors at the O(4) Wilson-Fisher fixed point}},
	Volume = {123},
	Year = {2019}}

@article{Cuomo:2024psk,
    author = "Cuomo, Gabriel and He, Yin-Chen and Komargodski, Zohar",
    title = "{Impurities with a cusp: general theory and 3d Ising}",
    eprint = "2406.10186",
    archivePrefix = "arXiv",
    primaryClass = "hep-th",
    month = "6",
    year = "2024"
}

@article{Cuomo:2019ejv,
	Archiveprefix = {arXiv},
	Author = {Cuomo, Gabriel},
	Eprint = {1906.07283},
	Primaryclass = {hep-th},
	Slaccitation = {%%CITATION = ARXIV:1906.07283;%%},
	Title = {{Superfluids, vortices and spinning charged operators in 4d CFT}},
	Year = {2019}}

@article{Kravec:2019djc,
	Archiveprefix = {arXiv},
	Author = {Kravec, S. M. and Pal, Sridip},
	Doi = {10.1007/JHEP05(2019)194},
	Eprint = {1904.05462},
	Journal = {JHEP},
	Pages = {194},
	Primaryclass = {hep-th},
	Slaccitation = {%%CITATION = ARXIV:1904.05462;%%},
	Title = {{The Spinful Large Charge Sector of Non-Relativistic CFTs: From Phonons to Vortex Crystals}},
	Volume = {05},
	Year = {2019}}

@article{Alvarez-Gaume:2019biu,
      author         = "Alvarez-Gaume, Luis and Orlando, Domenico and Reffert,
                        Susanne",
      title          = "{Large charge at large N}",
      journal        = "JHEP",
      volume         = "12",
      year           = "2019",
      pages          = "142",
      doi            = "10.1007/JHEP12(2019)142",
      eprint         = "1909.02571",
      archivePrefix  = "arXiv",
      primaryClass   = "hep-th",
      SLACcitation   = "%%CITATION = ARXIV:1909.02571;%%"
}

@article{Badel:2019oxl,
	Archiveprefix = {arXiv},
	Author = {Badel, Gil and Cuomo, Gabriel and Monin, Alexander and Rattazzi, Riccardo},
	Eprint = {1909.01269},
	Primaryclass = {hep-th},
	Slaccitation = {%%CITATION = ARXIV:1909.01269;%%},
	Title = {{The Epsilon Expansion Meets Semiclassics}},
	Year = {2019}}

@article{Arias-Tamargo:2019xld,
	Archiveprefix = {arXiv},
	Author = {Arias-Tamargo, G. and Rodriguez-Gomez, D. and Russo, J. G.},
	Eprint = {1908.11347},
	Primaryclass = {hep-th},
	Slaccitation = {%%CITATION = ARXIV:1908.11347;%%},
	Title = {{The large charge limit of scalar field theories and the Wilson-Fisher fixed point at $\epsilon=0$}},
	Year = {2019}}

@article{Watanabe:2019pdh,
	Archiveprefix = {arXiv},
	Author = {Watanabe, Masataka},
	Eprint = {1909.01337},
	Primaryclass = {hep-th},
	Slaccitation = {%%CITATION = ARXIV:1909.01337;%%},
	Title = {{Accessing Large Global Charge via the $\epsilon$-Expansion}},
	Year = {2019}}

@article{Grassi:2019txd,
	Archiveprefix = {arXiv},
	Author = {Grassi, Alba and Komargodski, Zohar and Tizzano, Luigi},
	Eprint = {1908.10306},
	Primaryclass = {hep-th},
	Slaccitation = {%%CITATION = ARXIV:1908.10306;%%},
	Title = {{Extremal Correlators and Random Matrix Theory}},
	Year = {2019}}

@article{Grassi:2024bwl,
    author = "Grassi, Alba and Iossa, Cristoforo",
    title = "{Matrix models for extremal and integrated correlators of higher rank}",
    eprint = "2408.07391",
    archivePrefix = "arXiv",
    primaryClass = "hep-th",
    month = "8",
    year = "2024"
}

@article{Rothstein:2017niq,
    author = "Rothstein, Ira Z. and Shrivastava, Prashant",
    title = "{Symmetry Obstruction to Fermi Liquid Behavior in the Unitary Limit}",
    eprint = "1712.07797",
    archivePrefix = "arXiv",
    primaryClass = "cond-mat.str-el",
    doi = "10.1103/PhysRevB.99.035101",
    journal = "Phys. Rev. B",
    volume = "99",
    number = "3",
    pages = "035101",
    year = "2019"
}

@article{Orlando:2019skh,
    author = "Orlando, Domenico and Reffert, Susanne and Sannino, Francesco",
    archivePrefix = "arXiv",
    doi = "10.1103/PhysRevD.101.065018",
    eprint = "1909.08642",
    journal = "Phys.\ Rev.\ D",
    number = "6",
    pages = "065018",
    primaryClass = "hep-th",
    reportNumber = "CP3-Origins-2019-36 DNRF90",
    title = "{Near-Conformal Dynamics at Large Charge}",
    volume = "101",
    year = "2020"
}

@article{Antipin:2022dsm,
    author = "Antipin, Oleg and Bersini, Jahmall and Sannino, Francesco and Torres, Mat\'\i{}as",
    title = "{The analytic structure of the fixed charge expansion}",
    eprint = "2202.13165",
    archivePrefix = "arXiv",
    primaryClass = "hep-th",
    reportNumber = "RBI-ThPhys-2022-6, CERN-TH-2022-049",
    doi = "10.1007/JHEP06(2022)041",
    journal = "JHEP",
    volume = "06",
    pages = "041",
    year = "2022"
}

@article{Antipin:2021rsh,
    author = "Antipin, Oleg and Bersini, Jahmall and Sannino, Francesco and Wang, Zhi-Wei and Zhang, Chen",
    title = "{More on the weak gravity conjecture via convexity of charged operators}",
    eprint = "2109.04946",
    archivePrefix = "arXiv",
    primaryClass = "hep-th",
    reportNumber = "RBI-ThPhys-2021-35",
    doi = "10.1007/JHEP12(2021)204",
    journal = "JHEP",
    volume = "12",
    pages = "204",
    year = "2021"
}

@article{Brown:2023why,
    author = "Brown, Augustus and Wen, Congkao and Xie, Haitian",
    title = "{Generating functions and large-charge expansion of integrated correlators in \ensuremath{\mathscr{N}} = 4 supersymmetric Yang-Mills theory}",
    eprint = "2303.17570",
    archivePrefix = "arXiv",
    primaryClass = "hep-th",
    reportNumber = "QMUL-PH-23-03",
    doi = "10.1007/JHEP07(2023)129",
    journal = "JHEP",
    volume = "07",
    pages = "129",
    year = "2023"
}

@article{Brown:2024yvt,
    author = "Brown, Augustus and Galvagno, Francesco and Wen, Congkao",
    title = "{All-loop Heavy-Heavy-Light-Light correlators in $ \mathcal{N} $ = 4 super Yang-Mills theory}",
    eprint = "2407.02250",
    archivePrefix = "arXiv",
    primaryClass = "hep-th",
    reportNumber = "QMUL-PH-24-12",
    doi = "10.1007/JHEP10(2024)171",
    journal = "JHEP",
    volume = "10",
    pages = "171",
    year = "2024"
}

@article{Badel:2019khk,
	Archiveprefix = {arXiv},
	Author = {Badel, Gil and Cuomo, Gabriel and Monin, Alexander and Rattazzi, Riccardo},
	Doi = {10.1016/j.physletb.2020.135202},
	Eprint = {1911.08505},
	Journal = {Phys. Lett.},
	Pages = {135202},
	Primaryclass = {hep-th},
	Slaccitation = {%%CITATION = ARXIV:1911.08505;%%},
	Title = {{Feynman diagrams and the large charge expansion in $3-\varepsilon$ dimensions}},
	Volume = {B802},
	Year = {2020}}

@book{Coleman:1988aos,
	Author = {Sidney Coleman},
	Publisher = {Cambridge University Press},
	Title = {Aspects Of Symmetry},
	Year = {1988}}

@article{Moshe:2003xn,
	Archiveprefix = {arXiv},
	Author = {Moshe, Moshe and Zinn-Justin, Jean},
	Doi = {10.1016/S0370-1573(03)00263-1},
	Eprint = {hep-th/0306133},
	Journal = {Phys. Rept.},
	Pages = {69-228},
	Primaryclass = {hep-th},
	Slaccitation = {%%CITATION = HEP-TH/0306133;%%},
	Title = {{Quantum field theory in the large N limit: A Review}},
	Volume = {385},
	Year = {2003}}

@article{tHooft:1973alw,
	Author = {'t Hooft, Gerard},
	Doi = {10.1016/0550-3213(74)90154-0},
	Journal = {Nucl. Phys.},
	Note = {[,337(1973)]},
	Pages = {461},
	Reportnumber = {CERN-TH-1786},
	Slaccitation = {%%CITATION = NUPHA,B72,461;%%},
	Title = {{A Planar Diagram Theory for Strong Interactions}},
	Volume = {B72},
	Year = {1974}}

@article{Hellerman:2020eff,
    author = "Hellerman, Simeon and Swanson, Ian",
    title = "{Droplet-Edge Operators in Nonrelativistic Conformal Field Theories}",
    eprint = "2010.07967",
    archivePrefix = "arXiv",
    primaryClass = "hep-th",
    reportNumber = "IPMU20-0107",
    month = "10",
    year = "2020"
}

@article{Argurio:2020jcq,
    author = "Argurio, Riccardo and Hoyos, Carlos and Musso, Daniele and Naegels, Daniel",
    title = "{Gapped dilatons in scale invariant superfluids}",
    eprint = "2006.11047",
    archivePrefix = "arXiv",
    primaryClass = "hep-th",
    month = "6",
    year = "2020"
}

@article{Cuomo:2020gyl,
    author = "Cuomo, Gabriel and Esposito, Angelo and Gendy, Emanuele and Khmelnitsky, Andrei and Monin, Alexander and Rattazzi, Riccardo",
    title = "{Gapped Goldstones at the cut-off scale: a non-relativistic EFT}",
    eprint = "2005.12924",
    archivePrefix = "arXiv",
    primaryClass = "hep-th",
    reportNumber = "DESY-20-086",
    doi = "10.1007/JHEP02(2021)068",
    journal = "JHEP",
    volume = "02",
    pages = "068",
    year = "2021"
}

@article{Euler:1776,
	Author = {Leonhard Euler},
	Journal = {Novi Commentarii academiae scientiarum Petropolitanae},
	Pages = {140-186},
	Title = {Meditationes circa singulare serierum genus},
	Volume = {20},
	Year = {1776}}

@inproceedings{murty2006multiple,
  title={Multiple Hurwitz zeta functions},
  author={Murty, M Ram and Sinha, Kaneenika},
  booktitle={Proceedings of Symposia in Pure Mathematics},
  volume={75},
  pages={135},
  year={2006},
  organization={Providence, RI; American Mathematical Society; 1998}
}

@article{Goncharov:2001iea,
    author = "Goncharov, A.B.",
    title = "{Multiple polylogarithms and mixed Tate motives}",
    eprint = "math/0103059",
    archivePrefix = "arXiv",
    month = "3",
    year = "2001"
}

@article{Gaume:2020bmp,
    author = "Gaum\'e, Luis \`Alvarez and Orlando, Domenico and Reffert, Susanne",
    title = "{Selected Topics in the Large Quantum Number Expansion}",
    eprint = "2008.03308",
    archivePrefix = "arXiv",
    primaryClass = "hep-th",
    month = "8",
    year = "2020"
}

@article{Escobedo:2009bh,
    author = "Escobedo, Miguel Angel and Mannarelli, Massimo and Manuel, Cristina",
    title = "{Bulk viscosities for cold Fermi superfluids close to the unitary limit}",
    eprint = "0904.3023",
    archivePrefix = "arXiv",
    primaryClass = "cond-mat.quant-gas",
    reportNumber = "UB-ECM-PF-09-12",
    doi = "10.1103/PhysRevA.79.063623",
    journal = "Phys. Rev. A",
    volume = "79",
    pages = "063623",
    year = "2009"
}

@article{Orlando:2020yii,
    author = "Orlando, Domenico and Reffert, Susanne and Sannino, Francesco",
    title = "{Charging the Conformal Window}",
    eprint = "2003.08396",
    archivePrefix = "arXiv",
    primaryClass = "hep-th",
    month = "3",
    year = "2020"
}

@article{Antipin:2020abu,
    author = "Antipin, Oleg and Bersini, Jahmall and Sannino, Francesco and Wang, Zhi-Wei and Zhang, Chen",
    title = "{Charging the $O(N)$ model}",
    eprint = "2003.13121",
    archivePrefix = "arXiv",
    primaryClass = "hep-th",
    reportNumber = "RBI-ThPhys-2020-09, CP3-Origins-2020-04 DNRF90, NCTS-TH/2005",
    doi = "10.1103/PhysRevD.102.045011",
    journal = "Phys. Rev. D",
    volume = "102",
    number = "4",
    pages = "045011",
    year = "2020"
}

@article{Antipin:2021akb,
    author = "Antipin, Oleg and Bersini, Jahmall and Sannino, Francesco and Wang, Zhi-Wei and Zhang, Chen",
    title = "{Untangling scaling dimensions of fixed charge operators in Higgs theories}",
    eprint = "2102.04390",
    archivePrefix = "arXiv",
    primaryClass = "hep-th",
    reportNumber = "RBI-ThPhys-2021-8, CP3-Origins-2021-01 DNRF90",
    doi = "10.1103/PhysRevD.103.125024",
    journal = "Phys. Rev. D",
    volume = "103",
    number = "12",
    pages = "125024",
    year = "2021"
}

@article{Antipin:2020rdw,
    author = "Antipin, Oleg and Bersini, Jahmall and Sannino, Francesco and Wang, Zhi-Wei and Zhang, Chen",
    title = "{Charging non-Abelian Higgs theories}",
    eprint = "2006.10078",
    archivePrefix = "arXiv",
    primaryClass = "hep-th",
    reportNumber = "RBI-ThPhys-2020-22, CP3-Origins-2020-10 DNRF90",
    doi = "10.1103/PhysRevD.102.125033",
    journal = "Phys. Rev. D",
    volume = "102",
    number = "12",
    pages = "125033",
    year = "2020"
}

@article{Banerjee:2021bbw,
    author = "Banerjee, Debasish and Chandrasekharan, Shailesh",
    title = "{Subleading conformal dimensions at the O(4) Wilson-Fisher fixed point}",
    eprint = "2111.01202",
    archivePrefix = "arXiv",
    primaryClass = "hep-lat",
    doi = "10.1103/PhysRevD.105.L031507",
    journal = "Phys. Rev. D",
    volume = "105",
    number = "3",
    pages = "L031507",
    year = "2022"
}

@article{Balasubramanian:2008dm,
    author = "Balasubramanian, Koushik and McGreevy, John",
    title = "{Gravity duals for non-relativistic CFTs}",
    eprint = "0804.4053",
    archivePrefix = "arXiv",
    primaryClass = "hep-th",
    doi = "10.1103/PhysRevLett.101.061601",
    journal = "Phys. Rev. Lett.",
    volume = "101",
    pages = "061601",
    year = "2008"
}

@article{Bartolomei_2020,
   title={Fractional statistics in anyon collisions},
   volume={368},
   ISSN={1095-9203},
   url={http://dx.doi.org/10.1126/science.aaz5601},
   DOI={10.1126/science.aaz5601},
   number={6487},
   journal={Science},
   publisher={American Association for the Advancement of Science (AAAS)},
   author={Bartolomei, H. and Kumar, M. and Bisognin, R. and Marguerite, A. and Berroir, J.-M. and Bocquillon, E. and Plaçais, B. and Cavanna, A. and Dong, Q. and Gennser, U. and et al.},
   year={2020},
   month={4},
   pages={173–177}
}

@article{Brauner:2014aha,
    author = "Brauner, Tom\'a\v{s} and Watanabe, Haruki",
    title = "{Spontaneous breaking of spacetime symmetries and the inverse Higgs effect}",
    eprint = "1401.5596",
    archivePrefix = "arXiv",
    primaryClass = "hep-ph",
    doi = "10.1103/PhysRevD.89.085004",
    journal = "Phys. Rev. D",
    volume = "89",
    number = "8",
    pages = "085004",
    year = "2014"
}

@article{Aharony:2023ike,
    author = "Aharony, Ofer and Breitstein, Yacov-Nir",
    title = "{Tests of the Charge Convexity Conjecture in Caswell-Banks-Zaks theory}",
    eprint = "2305.08947",
    archivePrefix = "arXiv",
    primaryClass = "hep-th",
    doi = "10.1007/JHEP08(2023)044",
    journal = "JHEP",
    volume = "08",
    pages = "044",
    year = "2023"
}

@article{Cuomo:2020rgt,
    author = "Cuomo, Gabriel",
    title = "{A note on the large charge expansion in 4d CFT}",
    eprint = "2010.00407",
    archivePrefix = "arXiv",
    primaryClass = "hep-th",
    doi = "10.1016/j.physletb.2020.136014",
    journal = "Phys. Lett. B",
    volume = "812",
    pages = "136014",
    year = "2021"
}

@article{Mehen:1999nd,
    author = "Mehen, Thomas and Stewart, Iain W. and Wise, Mark B.",
    title = "{Conformal invariance for nonrelativistic field theory}",
    eprint = "hep-th/9910025",
    archivePrefix = "arXiv",
    reportNumber = "CALT-68-2242, UCSD-PTH-99-14",
    doi = "10.1016/S0370-2693(00)00006-X",
    journal = "Phys. Lett. B",
    volume = "474",
    pages = "145--152",
    year = "2000"
}

@article{Mehen:2007dn,
    author = "Mehen, Thomas",
    title = "{On non-relativistic conformal field theory and trapped atoms: Virial theorems and the state-operator correspondence in three dimensions}",
    eprint = "0712.0867",
    archivePrefix = "arXiv",
    primaryClass = "cond-mat.other",
    doi = "10.1103/PhysRevA.78.013614",
    journal = "Phys. Rev. A",
    volume = "78",
    pages = "013614",
    year = "2008"
}

@article{Cuomo:2021qws,
    author = "Cuomo, Gabriel and Delacretaz, Luca V. and Mehta, Umang",
    title = "{Large Charge Sector of 3d Parity-Violating CFTs}",
    eprint = "2102.05046",
    archivePrefix = "arXiv",
    primaryClass = "hep-th",
    reportNumber = "EFI-21-1",
    doi = "10.1007/JHEP05(2021)115",
    journal = "JHEP",
    volume = "05",
    pages = "115",
    year = "2021"
}

@article{Cuomo:2021ygt,
    author = "Cuomo, Gabriel",
    title = "{The OPE meets semiclassics}",
    eprint = "2103.01331",
    archivePrefix = "arXiv",
    primaryClass = "hep-th",
    month = "3",
    year = "2021"
}

@article{Dondi:2021buw,
    author = "Dondi, Nicola and Kalogerakis, Ioannis and Orlando, Domenico and Reffert, Susanne",
    title = "{Resurgence of the large-charge expansion}",
    eprint = "2102.12488",
    archivePrefix = "arXiv",
    primaryClass = "hep-th",
    month = "2",
    year = "2021"
}

@article{Dondi:2022wli,
    author = "Dondi, Nicola and Kalogerakis, Ioannis and Moser, Rafael and Orlando, Domenico and Reffert, Susanne",
    title = "{Spinning correlators in large-charge CFTs}",
    eprint = "2203.12624",
    archivePrefix = "arXiv",
    primaryClass = "hep-th",
    doi = "10.1016/j.nuclphysb.2022.115928",
    journal = "Nucl. Phys. B",
    volume = "983",
    pages = "115928",
    year = "2022"
}

@article{Giombi:2020enj,
    author = "Giombi, Simone and Hyman, Jonah",
    title = "{On the Large Charge Sector in the Critical $O(N)$ Model at Large $N$}",
    eprint = "2011.11622",
    archivePrefix = "arXiv",
    primaryClass = "hep-th",
    month = "11",
    year = "2020"
}

@article{Hellerman:2016hnf,
    author = "Hellerman, Simeon and Swanson, Ian",
    title = "{Boundary Operators in Effective String Theory}",
    eprint = "1609.01736",
    archivePrefix = "arXiv",
    primaryClass = "hep-th",
    reportNumber = "IPMU16-0117, CALT-TH-2016-025",
    doi = "10.1007/JHEP04(2017)085",
    journal = "JHEP",
    volume = "04",
    pages = "085",
    year = "2017"
}

@article{Hellerman:2020sqj,
    author = "Hellerman, Simeon and Maeda, Shunsuke and Orlando, Domenico and Reffert, Susanne and Watanabe, Masataka",
    title = "{S-duality and correlation functions at large R-charge}",
    eprint = "2005.03021",
    archivePrefix = "arXiv",
    primaryClass = "hep-th",
    month = "5",
    year = "2020"
}

@article{Hellerman:2021qzz,
    author = "Hellerman, Simeon and Orlando, Domenico and Pellizzani, Vito and Reffert, Susanne and Swanson, Ian",
    title = "{Nonrelativistic CFTs at Large Charge: Casimir Energy and Logarithmic Enhancements}",
    eprint = "2111.12094",
    archivePrefix = "arXiv",
    primaryClass = "hep-th",
    month = "11",
    year = "2021"
}

@article{Hellerman:2023myh,
    author = "Hellerman, Simeon and Krichevskiy, Daniil and Orlando, Domenico and Pellizzani, Vito and Reffert, Susanne and Swanson, Ian",
    title = "{The unitary Fermi gas at large charge and large N}",
    eprint = "2311.14793",
    archivePrefix = "arXiv",
    primaryClass = "hep-th",
    doi = "10.1007/JHEP05(2024)323",
    journal = "JHEP",
    volume = "05",
    pages = "323",
    year = "2024"
}

@article{Orlando:2020idm,
    author = "Orlando, Domenico and Pellizzani, Vito and Reffert, Susanne",
    title = {{Near-Schr\"odinger dynamics at large charge}},
    eprint = "2010.07942",
    archivePrefix = "arXiv",
    primaryClass = "hep-th",
    month = "10",
    year = "2020"
}

@article{PhysRevD.103.105026,
  title = {Charging the conformal window},
  author = {Orlando, Domenico and Reffert, Susanne and Sannino, Francesco},
  journal = {Phys. Rev. D},
  volume = {103},
  issue = {10},
  pages = {105026},
  numpages = {11},
  year = {2021},
  month = {5},
  publisher = {American Physical Society},
  doi = {10.1103/PhysRevD.103.105026},
  url = {https://link.aps.org/doi/10.1103/PhysRevD.103.105026}
}

@article{PhysRevD.101.065018,
  title = {Near-conformal dynamics at large charge},
  author = {Orlando, Domenico and Reffert, Susanne and Sannino, Francesco},
  journal = {Phys. Rev. D},
  volume = {101},
  issue = {6},
  pages = {065018},
  numpages = {6},
  year = {2020},
  month = {3},
  publisher = {American Physical Society},
  doi = {10.1103/PhysRevD.101.065018},
  url = {https://link.aps.org/doi/10.1103/PhysRevD.101.065018}
}

@article{Son:2008ye,
    author = "Son, D. T.",
    title = "{Toward an AdS/cold atoms correspondence: A Geometric realization of the Schrodinger symmetry}",
    eprint = "0804.3972",
    archivePrefix = "arXiv",
    primaryClass = "hep-th",
    reportNumber = "INT-PUB-08-08",
    doi = "10.1103/PhysRevD.78.046003",
    journal = "Phys. Rev. D",
    volume = "78",
    pages = "046003",
    year = "2008"
}

@article{Wilczek:1982wy,
    author = "Wilczek, Frank",
    title = "{Quantum Mechanics of Fractional Spin Particles}",
    reportNumber = "NSF-ITP-82-56",
    doi = "10.1103/PhysRevLett.49.957",
    journal = "Phys. Rev. Lett.",
    volume = "49",
    pages = "957--959",
    year = "1982"
}

@article{niederer1972maximal,
  title={Maximal kinematical invariance group of the free Schrodinger equation},
  author={Niederer, U},
  journal={Helv. Phys. Acta},
  volume={45},
  pages={802--810},
  year={1972}
}

@article{Valenzuela:2009gu,
    author = "Valenzuela, Mauricio",
    title = {{Higher Spin Symmetries of the Free Schr\"odinger Equation}},
    eprint = "0912.0789",
    archivePrefix = "arXiv",
    primaryClass = "hep-th",
    doi = "10.1155/2016/5739410",
    journal = "Adv. Math. Phys.",
    volume = "2016",
    pages = "5739410",
    year = "2016"
}

@inproceedings{Giombi:2016ejx,
    author = "Giombi, Simone",
    title = "{Higher Spin \textemdash{} CFT Duality}",
    booktitle = "{Theoretical Advanced Study Institute in Elementary Particle Physics}: {New Frontiers in Fields and Strings}",
    eprint = "1607.02967",
    archivePrefix = "arXiv",
    primaryClass = "hep-th",
    doi = "10.1142/9789813149441_0003",
    pages = "137--214",
    year = "2017"
}

@article{PhysRevLett.92.120401,
  title = {Crossover from a Molecular Bose-Einstein Condensate to a Degenerate Fermi Gas},
  author = {Bartenstein, M. and Altmeyer, A. and Riedl, S. and Jochim, S. and Chin, C. and Denschlag, J. Hecker and Grimm, R.},
  journal = {Phys. Rev. Lett.},
  volume = {92},
  issue = {12},
  pages = {120401},
  numpages = {4},
  year = {2004},
  month = {3},
  publisher = {American Physical Society},
  doi = {10.1103/PhysRevLett.92.120401},
  url = {https://link.aps.org/doi/10.1103/PhysRevLett.92.120401}
}

@misc{nakamura2020direct,
      title={Direct observation of anyonic braiding statistics at the $\nu$=1/3 fractional quantum Hall state}, 
      author={James Nakamura and Shuang Liang and Geoffrey C. Gardner and Michael J. Manfra},
      year={2020},
      eprint={2006.14115},
      archivePrefix={arXiv},
      primaryClass={cond-mat.mes-hall}
}

@article{Aharony:2021mpc,
    author = "Aharony, Ofer and Palti, Eran",
    title = "{On Convexity of Charged Operators in CFTs and the Weak Gravity Conjecture}",
    eprint = "2108.04594",
    archivePrefix = "arXiv",
    primaryClass = "hep-th",
    month = "8",
    year = "2021"
}

@article{Antipin:2021jiw,
    author = "Antipin, Oleg and Bersini, Jahmall and Sannino, Francesco and Wang, Zhi-Wei and Zhang, Chen",
    title = "{More on the cubic versus quartic interaction equivalence in the $O(N)$ model}",
    eprint = "2107.02528",
    archivePrefix = "arXiv",
    primaryClass = "hep-th",
    reportNumber = "RBI-ThPhys-2021-27",
    doi = "10.1103/PhysRevD.104.085002",
    journal = "Phys. Rev. D",
    volume = "104",
    pages = "085002",
    year = "2021"
}

@article{Arias-Tamargo:2019kfr,
    author = "Arias-Tamargo, Guillermo and Rodriguez-Gomez, Diego and Russo, Jorge G.",
    title = "{Correlation functions in scalar field theory at large charge}",
    eprint = "1912.01623",
    archivePrefix = "arXiv",
    primaryClass = "hep-th",
    reportNumber = "ICCUB-19-018",
    doi = "10.1007/JHEP01(2020)171",
    journal = "JHEP",
    volume = "01",
    pages = "171",
    year = "2020"
}

@article{Arias-Tamargo:2020fow,
    author = "Arias-Tamargo, Guillermo and Rodriguez-Gomez, Diego and Russo, Jorge G.",
    title = "{On the UV completion of the $O(N)$ model in $6-\epsilon$ dimensions: a stable large-charge sector}",
    eprint = "2003.13772",
    archivePrefix = "arXiv",
    primaryClass = "hep-th",
    doi = "10.1007/JHEP09(2020)064",
    journal = "JHEP",
    volume = "09",
    pages = "064",
    year = "2020"
}

@article{Beane:2024kld,
    author = "Beane, Silas R. and Orlando, Domenico and Reffert, Susanne",
    title = "{Exact evaluation of large-charge correlation functions in nonrelativistic conformal field theory}",
    eprint = "2403.18898",
    archivePrefix = "arXiv",
    primaryClass = "hep-th",
    reportNumber = "YITP-24-48,NT@UW-24-05",
    doi = "10.1103/PhysRevD.110.025011",
    journal = "Phys. Rev. D",
    volume = "110",
    number = "2",
    pages = "025011",
    year = "2024"
}

@article{Chang_2007,
   title={Unitary Fermi gas in a harmonic trap},
   volume={76},
   ISSN={1094-1622},
   url={http://dx.doi.org/10.1103/PhysRevA.76.021603},
   DOI={10.1103/physreva.76.021603},
   number={2},
   journal={Physical Review A},
   publisher={American Physical Society (APS)},
   author={Chang, S. Y. and Bertsch, G. F.},
   year={2007},
   month=aug }

@article{Blume_2007,
   title={Universal Properties of a Trapped Two-Component Fermi Gas at Unitarity},
   volume={99},
   ISSN={1079-7114},
   url={http://dx.doi.org/10.1103/PhysRevLett.99.233201},
   DOI={10.1103/physrevlett.99.233201},
   number={23},
   journal={Physical Review Letters},
   publisher={American Physical Society (APS)},
   author={Blume, D. and von Stecher, J. and Greene, Chris H.},
   year={2007},
   month=dec }

@article{Yin_2015,
   title={Trapped unitary two-component Fermi gases with up to ten particles},
   volume={92},
   ISSN={1094-1622},
   url={http://dx.doi.org/10.1103/PhysRevA.92.013608},
   DOI={10.1103/physreva.92.013608},
   number={1},
   journal={Physical Review A},
   publisher={American Physical Society (APS)},
   author={Yin, X. Y. and Blume, D.},
   year={2015},
   month=jul }

@article{Cuomo:2021cnb,
    author = "Cuomo, Gabriel and Mezei, M\'ark and Raviv-Moshe, Avia",
    title = "{Boundary conformal field theory at large charge}",
    eprint = "2108.06579",
    archivePrefix = "arXiv",
    primaryClass = "hep-th",
    doi = "10.1007/JHEP10(2021)143",
    journal = "JHEP",
    volume = "10",
    pages = "143",
    year = "2021"
}

@article{Cuomo:2022xgw,
    author = "Cuomo, Gabriel and Komargodski, Zohar and Mezei, M\'ark and Raviv-Moshe, Avia",
    title = "{Spin impurities, Wilson lines and semiclassics}",
    eprint = "2202.00040",
    archivePrefix = "arXiv",
    primaryClass = "hep-th",
    doi = "10.1007/JHEP06(2022)112",
    journal = "JHEP",
    volume = "06",
    pages = "112",
    year = "2022"
}

@article{Krishnan:2024wrc,
    author = "Krishnan, Abijith and Metlitski, Max A.",
    title = "{The Kondo impurity in the large spin limit}",
    eprint = "2408.12650",
    archivePrefix = "arXiv",
    primaryClass = "cond-mat.str-el",
    month = "8",
    year = "2024"
}

@article{Dupuis:2021yej,
    author = "Dupuis, \'Eric and Witczak-Krempa, William",
    title = "{Monopole hierarchy in transitions out of a Dirac spin liquid}",
    eprint = "2102.04885",
    archivePrefix = "arXiv",
    primaryClass = "cond-mat.str-el",
    month = "2",
    year = "2021"
}

@article{Dupuis:2021flq,
    author = "Dupuis, \'Eric and Boyack, Rufus and Witczak-Krempa, William",
    title = "{Anomalous dimensions of monopole operators at the transitions between Dirac and topological spin liquids}",
    eprint = "2108.05922",
    archivePrefix = "arXiv",
    primaryClass = "cond-mat.str-el",
    month = "8",
    year = "2021"
}

@article{Dupuis:2019uhs,
    author = "Dupuis, \'Eric and Paranjape, M. B. and Witczak-Krempa, William",
    title = "{Transition from a Dirac spin liquid to an antiferromagnet: Monopoles in a QED3-Gross-Neveu theory}",
    eprint = "1905.02750",
    archivePrefix = "arXiv",
    primaryClass = "cond-mat.str-el",
    doi = "10.1103/PhysRevB.100.094443",
    journal = "Phys. Rev. B",
    volume = "100",
    number = "9",
    pages = "094443",
    year = "2019"
}

@inproceedings{Dupuis:2019xdo,
    author = "Dupuis, \'Eric and Paranjape, M. B. and Witczak-Krempa, William",
    title = "{Monopole Operators and Their Symmetries in QED3-Gross\textendash{}Neveu Models}",
    booktitle = "{11th International Symposium on Quantum Theory and Symmetries}",
    eprint = "1911.05802",
    archivePrefix = "arXiv",
    primaryClass = "cond-mat.str-el",
    doi = "10.1007/978-3-030-55777-5_31",
    month = "11",
    year = "2019"
}

@article{Giombi:2022gjj,
    author = "Giombi, Simone and Helfenberger, Elizabeth and Khanchandani, Himanshu",
    title = "{Long range, large charge, large N}",
    eprint = "2205.00500",
    archivePrefix = "arXiv",
    primaryClass = "hep-th",
    reportNumber = "PUPT-2633",
    doi = "10.1007/JHEP01(2023)166",
    journal = "JHEP",
    volume = "01",
    pages = "166",
    year = "2023"
}

@article{Giombi:2021zfb,
    author = "Giombi, Simone and Komatsu, Shota and Offertaler, Bendeguz",
    title = "{Large Charges on the Wilson Loop in $\mathcal{N}=4$ SYM: Matrix Model and Classical String}",
    eprint = "2110.13126",
    archivePrefix = "arXiv",
    primaryClass = "hep-th",
    reportNumber = "CERN-TH-2021-161",
    month = "10",
    year = "2021"
}

@article{Giombi:2022anm,
    author = "Giombi, Simone and Komatsu, Shota and Offertaler, Bendeguz",
    title = "{Large charges on the Wilson loop in $ \mathcal{N} $ = 4 SYM. Part II. Quantum fluctuations, OPE, and spectral curve}",
    eprint = "2202.07627",
    archivePrefix = "arXiv",
    primaryClass = "hep-th",
    reportNumber = "CERN-TH-2022-017",
    doi = "10.1007/JHEP08(2022)011",
    journal = "JHEP",
    volume = "08",
    pages = "011",
    year = "2022"
}

@article{Hellerman:2021duh,
    author = "Hellerman, Simeon",
    title = "{On the exponentially small corrections to ${\cal N} = 2$ superconformal correlators at large R-charge}",
    eprint = "2103.09312",
    archivePrefix = "arXiv",
    primaryClass = "hep-th",
    month = "3",
    year = "2021"
}

@article{Hellerman:2021yqz,
    author = "Hellerman, Simeon and Orlando, Domenico",
    title = "{Large R-charge EFT correlators in N=2 SQCD}",
    eprint = "2103.05642",
    archivePrefix = "arXiv",
    primaryClass = "hep-th",
    month = "3",
    year = "2021"
}

@article{Pal:2018idc,
    author = "Pal, Sridip",
    title = "{Unitarity and universality in nonrelativistic conformal field theory}",
    eprint = "1802.02262",
    archivePrefix = "arXiv",
    primaryClass = "hep-th",
    doi = "10.1103/PhysRevD.97.105031",
    journal = "Phys. Rev. D",
    volume = "97",
    number = "10",
    pages = "105031",
    year = "2018"
}

@article{Volovich:2009yh,
    author = "Volovich, Anastasia and Wen, Congkao",
    title = "{Correlation Functions in Non-Relativistic Holography}",
    eprint = "0903.2455",
    archivePrefix = "arXiv",
    primaryClass = "hep-th",
    reportNumber = "BROWN-HET-1579",
    doi = "10.1088/1126-6708/2009/05/087",
    journal = "JHEP",
    volume = "05",
    pages = "087",
    year = "2009"
}

@article{Komargodski:2021zzy,
    author = "Komargodski, Zohar and Mezei, M\'ark and Pal, Sridip and Raviv-Moshe, Avia",
    title = "{Spontaneously broken boosts in CFTs}",
    eprint = "2102.12583",
    archivePrefix = "arXiv",
    primaryClass = "hep-th",
    doi = "10.1007/JHEP09(2021)064",
    journal = "JHEP",
    volume = "09",
    pages = "064",
    year = "2021"
}

@article{Nishida:2010tm,
    author = "Nishida, Yusuke and Son, Dam Thanh",
    title = "{Unitary Fermi gas, epsilon expansion, and nonrelativistic conformal field theories}",
    eprint = "1004.3597",
    archivePrefix = "arXiv",
    primaryClass = "cond-mat.quant-gas",
    reportNumber = "MIT-CTP-4141, INT-PUB-10-017",
    doi = "10.1007/978-3-642-21978-8_7",
    journal = "Lect. Notes Phys.",
    volume = "836",
    pages = "233--275",
    year = "2012"
}

@article{Orlando:2021usz,
    author = "Orlando, Domenico and Reffert, Susanne and Schmidt, Tim",
    title = "{Following the flow for large N and large charge}",
    eprint = "2110.07616",
    archivePrefix = "arXiv",
    primaryClass = "hep-th",
    month = "10",
    year = "2021"
}

@article{Pellizzani:2021hzx,
    author = "Pellizzani, Vito",
    title = "{Operator spectrum of nonrelativistic CFTs at large charge}",
    eprint = "2107.12127",
    archivePrefix = "arXiv",
    primaryClass = "hep-th",
    month = "7",
    year = "2021"
}

@article{Sharon:2020mjs,
    author = "Sharon, Adar and Watanabe, Masataka",
    title = "{Transition of Large $R$-Charge Operators on a Conformal Manifold}",
    eprint = "2008.01106",
    archivePrefix = "arXiv",
    primaryClass = "hep-th",
    doi = "10.1007/JHEP01(2021)068",
    journal = "JHEP",
    volume = "01",
    pages = "068",
    year = "2021"
}

@article{Dondi:2022zna,
    author = "Dondi, Nicola and Hellerman, Simeon and Kalogerakis, Ioannis and Moser, Rafael and Orlando, Domenico and Reffert, Susanne",
    title = "{Fermionic CFTs at large charge and large N}",
    eprint = "2211.15318",
    archivePrefix = "arXiv",
    primaryClass = "hep-th",
    doi = "10.1007/JHEP08(2023)180",
    journal = "JHEP",
    volume = "08",
    pages = "180",
    year = "2023"
}

@article{Csordas:2010id,
    author = "Csordas, Andras and Almasy, Orsolya and Szepfalusy, Peter",
    title = "{Gradient corrections to the local density approximation for trapped superfluid Fermi gases}",
    eprint = "1009.4822",
    archivePrefix = "arXiv",
    primaryClass = "cond-mat.quant-gas",
    doi = "10.1103/PhysRevA.82.063609",
    journal = "Phys. Rev. A",
    volume = "82",
    pages = "063609",
    year = "2010"
}

@article{Hammer:2021zxb,
    author = "Hammer, Hans-Werner and Son, Dam Thanh",
    title = "{Unnuclear physics}",
    eprint = "2103.12610",
    archivePrefix = "arXiv",
    primaryClass = "nucl-th",
    doi = "10.1073/pnas.2108716118",
    journal = "Proc. Nat. Acad. Sci.",
    volume = "118",
    pages = "e2108716118",
    year = "2021"
}

@article{Chowdhury:2023ahp,
    author = "Chowdhury, Subham Dutta and Mishra, Ruchira and Son, Dam Thanh",
    title = "{Applied nonrelativistic conformal field theory: scattering-length and effective-range corrections to unnuclear physics}",
    eprint = "2309.15177",
    archivePrefix = "arXiv",
    primaryClass = "hep-th",
    month = "9",
    year = "2023"
}

@article{Chang:2007zzd,
    author = "Chang, S. Y. and Bertsch, G. F.",
    title = "{Unitary Fermi gas in a harmonic trap}",
    eprint = "physics/0703190",
    archivePrefix = "arXiv",
    doi = "10.1103/PhysRevA.76.021603",
    journal = "Phys. Rev. A",
    volume = "76",
    pages = "021603",
    year = "2007"
}

@article{Eagles:1969zz,
    author = "Eagles, D. M.",
    title = "{Possible Pairing without Superconductivity at Low Carrier Concentrations in Bulk and Thin-Film Superconducting Semiconductors}",
    doi = "10.1103/PhysRev.186.456",
    journal = "Phys. Rev.",
    volume = "186",
    pages = "456--463",
    year = "1969"
}

@article{Forbes:2012yp,
    author = "Forbes, Michael McNeil",
    title = "{The Unitary Fermi Gas in a Harmonic Trap and its Static Response}",
    eprint = "1211.3779",
    archivePrefix = "arXiv",
    primaryClass = "cond-mat.quant-gas",
    reportNumber = "INT-PUB-12-057",
    month = "11",
    year = "2012"
}

@article{Groenewold:1946kp,
    author = "Groenewold, H. J.",
    title = "{On the Principles of elementary quantum mechanics}",
    doi = "10.1016/S0031-8914(46)80059-4",
    journal = "Physica",
    volume = "12",
    pages = "405--460",
    year = "1946"
}

@article{Moser:2021bes,
    author = "Moser, Rafael and Orlando, Domenico and Reffert, Susanne",
    title = "{Convexity, large charge and the large-N phase diagram of the \ensuremath{\varphi}$^{4}$ theory}",
    eprint = "2110.07617",
    archivePrefix = "arXiv",
    primaryClass = "hep-th",
    doi = "10.1007/JHEP02(2022)152",
    journal = "JHEP",
    volume = "02",
    pages = "152",
    year = "2022"
}

@article{Moyal:1949sk,
    author = "Moyal, J. E.",
    title = "{Quantum mechanics as a statistical theory}",
    doi = "10.1017/S0305004100000487",
    journal = "Proc. Cambridge Phil. Soc.",
    volume = "45",
    pages = "99--124",
    year = "1949"
}

@article{Nikolic:2007zz,
    author = "Nikolic, Predrag and Sachdev, Subir",
    title = "{Renormalization-group fixed points, universal phase diagram, and 1/N expansion for quantum liquids with interactions near the unitarity limit}",
    eprint = "cond-mat/0609106",
    archivePrefix = "arXiv",
    doi = "10.1103/PhysRevA.75.033608",
    journal = "Phys. Rev. A",
    volume = "75",
    pages = "033608",
    year = "2007"
}

@article{Strinati:2018wdg,
    author = {Strinati, Giancarlo Calvanese and Pieri, Pierbiagio and R\"opke, Gerd and Schuck, Peter and Urban, Michael},
    title = "{The BCS\textendash{}BEC crossover: From ultra-cold Fermi gases to nuclear systems}",
    eprint = "1802.05997",
    archivePrefix = "arXiv",
    primaryClass = "cond-mat.quant-gas",
    doi = "10.1016/j.physrep.2018.02.004",
    journal = "Phys. Rept.",
    volume = "738",
    pages = "1--76",
    year = "2018"
}

@article{Veillette:2007zz,
    author = "Veillette, Martin Y. and Sheehy, Daniel E. and Radzihovsky, Leo",
    title = "{Large-N expansion for unitary superfluid Fermi gases}",
    eprint = "cond-mat/0610798",
    archivePrefix = "arXiv",
    doi = "10.1103/PhysRevA.75.043614",
    journal = "Phys. Rev. A",
    volume = "75",
    pages = "043614",
    year = "2007"
}

@article{Wigner:1932eb,
    author = "Wigner, Eugene P.",
    title = "{On the quantum correction for thermodynamic equilibrium}",
    doi = "10.1103/PhysRev.40.749",
    journal = "Phys. Rev.",
    volume = "40",
    pages = "749--760",
    year = "1932"
}

@book{bender1999advanced,
	author = {Bender, C.M. and Orszag, S.A.},
	doi = {10.1007/978-1-4757-3069-2},
	isbn = {9780387989310},
	lccn = {99044783},
	publisher = {Springer},
	title = {Advanced Mathematical Methods for Scientists and Engineers I: Asymptotic Methods and Perturbation Theory},
	year = {1999}}

@article{MANES20091136,
	author = {Juan L. Ma{\~n}es and Manuel A. Valle},
	doi = {https://doi.org/10.1016/j.aop.2009.01.003},
	issn = {0003-4916},
	journal = {Annals of Physics},
	number = {5},
	pages = {1136-1157},
	title = {Effective theory for the Goldstone field in the BCS--BEC crossover at T=0},
	url = {https://www.sciencedirect.com/science/article/pii/S0003491609000165},
	volume = {324},
	year = {2009}}

@book{rammer_2007,
	author = {Rammer, J{\o}rgen},
	doi = {10.1017/CBO9780511618956},
	place = {Cambridge},
	publisher = {Cambridge University Press},
	title = {Quantum Field Theory of Non-equilibrium States},
	year = {2007}}

@book{altland_simons_2010,
	author = {Altland, Alexander and Simons, Ben D.},
	doi = {10.1017/CBO9780511789984},
	edition = {2},
	place = {Cambridge},
	publisher = {Cambridge University Press},
	title = {Condensed Matter Field Theory},
	year = {2010}}

@article{Fuertes:2009ex,
    author = "Fuertes, Carlos A. and Moroz, Sergej",
    title = "{Correlation functions in the non-relativistic AdS/CFT correspondence}",
    eprint = "0903.1844",
    archivePrefix = "arXiv",
    primaryClass = "hep-th",
    reportNumber = "IFT-UAM-CSIC-09-13",
    doi = "10.1103/PhysRevD.79.106004",
    journal = "Phys. Rev. D",
    volume = "79",
    pages = "106004",
    year = "2009"
}

@article{Bekaert_2012,
	author = {Xavier Bekaert and Elisa Meunier and Sergej Moroz},
	doi = {10.1007/jhep02(2012)113},
	journal = {Journal of High Energy Physics},
	number = {2},
	publisher = {Springer Science and Business Media {LLC}},
	title = {Symmetries and currents of the ideal and unitary Fermi gases},
	url = {https://doi.org/10.1007%2Fjhep02%282012%29113},
	volume = {2012},
	year = 2012}

@Book{Sachdev2011,
  author     = {Sachdev, Subir},
  publisher  = {Cambridge University Press},
  title      = {Quantum Phase Transitions},
  year       = {2011},
  edition    = {2},
  doi        = {10.1017/CBO9780511973765},
  place      = {Cambridge},
}

@article{kapusta_finite-temperature_1989,
	title = {Finite-temperature field theory},
	volume = 15,
	url = {https://doi.org/10.1088/0954-3899/15/3/005},
	doi = {10.1088/0954-3899/15/3/005},
	number = 3,
	journal = {Journal of Physics G: Nuclear and Particle Physics},
	author = {Kapusta, J. I. and Landshoff, P. V.},
	month = {3},
	year = 1989,
	note = {Publisher: IOP Publishing},
	pages = {267--285},
}

@article{mordell1958evaluation,
  title={On the evaluation of some multiple series},
  author={Mordell, LJ},
  journal={Journal of the London Mathematical Society},
  volume={1},
  number={3},
  pages={368--371},
  year={1958},
  publisher={Wiley Online Library}
}

@article{tornheim1950harmonic,
  title={Harmonic double series},
  author={Tornheim, Leonard},
  journal={American Journal of Mathematics},
  volume={72},
  number={2},
  pages={303--314},
  year={1950},
  publisher={JSTOR}
}

@inproceedings{matsumoto2003mordell,
  title={On Mordell-Tornheim and other multiple zeta-functions},
  author={Matsumoto, Kohji},
  booktitle={Proceedings of the Session in Analytic Number Theory and Diophantine Equations, Bonner Math. Schriften},
  volume={360},
  pages={17},
  year={2003}
}

@article{tsumura2005mordell,
  title={On Mordell-Tornheim zeta values},
  author={Tsumura, Hirofumi},
  journal={Proceedings of the American Mathematical Society},
  volume={133},
  number={8},
  pages={2387--2393},
  year={2005}
}

@inproceedings{matsumoto2002analytic,
  title={On analytic continuation of various multiple zeta-functions},
  author={Matsumoto, Kohji},
  booktitle={Surveys in Number Theory: Papers from the Millennial Conference on Number Theory},
  pages={169},
  year={2002},
  organization={AK Peters/CRC Press}
}

@article{matsumoto_value-relations_2008,
	title = {On value-relations, functional relations and singularities of {Mordell}–{Tornheim} and related triple zeta-functions},
	volume = {132},
	issn = {0065-1036, 1730-6264},
	url = {http://journals.impan.pl/cgi-bin/doi?aa132-2-1},
	doi = {10.4064/aa132-2-1},
	language = {en},
	number = {2},
	urldate = {2021-09-22},
	journal = {Acta Arithmetica},
	author = {Matsumoto, Kohji and Nakamura, Takashi and Ochiai, Hiroyuki and Tsumura, Hirofumi},
	year = {2008},
	pages = {99--125},
}

@article{Werner_2006,
   title={Unitary gas in an isotropic harmonic trap: Symmetry properties and applications},
   volume={74},
   ISSN={1094-1622},
   url={http://dx.doi.org/10.1103/PhysRevA.74.053604},
   DOI={10.1103/physreva.74.053604},
   number={5},
   journal={Physical Review A},
   publisher={American Physical Society (APS)},
   author={Werner, Felix and Castin, Yvan},
   year={2006},
   month={11}
}

@article{Cuomo:2022kio,
    author = "Cuomo, Gabriel and Komargodski, Zohar",
    title = "{Giant Vortices and the Regge Limit}",
    eprint = "2210.15694",
    archivePrefix = "arXiv",
    primaryClass = "hep-th",
    doi = "10.1007/JHEP01(2023)006",
    journal = "JHEP",
    volume = "01",
    pages = "006",
    year = "2023"
}

@article{Lashkari:2016vgj,
    author = "Lashkari, Nima and Dymarsky, Anatoly and Liu, Hong",
    title = "{Eigenstate Thermalization Hypothesis in Conformal Field Theory}",
    eprint = "1610.00302",
    archivePrefix = "arXiv",
    primaryClass = "hep-th",
    doi = "10.1088/1742-5468/aab020",
    journal = "J. Stat. Mech.",
    volume = "1803",
    number = "3",
    pages = "033101",
    year = "2018"
}

@article{DAlessio:2015qtq,
    author = "D'Alessio, Luca and Kafri, Yariv and Polkovnikov, Anatoli and Rigol, Marcos",
    title = "{From quantum chaos and eigenstate thermalization to statistical mechanics and thermodynamics}",
    eprint = "1509.06411",
    archivePrefix = "arXiv",
    primaryClass = "cond-mat.stat-mech",
    doi = "10.1080/00018732.2016.1198134",
    journal = "Adv. Phys.",
    volume = "65",
    number = "3",
    pages = "239--362",
    year = "2016"
}

@article{Caron-Huot:2017vep,
    author = "Caron-Huot, Simon",
    title = "{Analyticity in Spin in Conformal Theories}",
    eprint = "1703.00278",
    archivePrefix = "arXiv",
    primaryClass = "hep-th",
    doi = "10.1007/JHEP09(2017)078",
    journal = "JHEP",
    volume = "09",
    pages = "078",
    year = "2017"
}

@article{Cuomo:2024fuy,
    author = "Cuomo, Gabriel and Rastelli, Leonardo and Sharon, Adar",
    title = "{Moduli Spaces in CFT: Large Charge Operators}",
    eprint = "2406.19441",
    archivePrefix = "arXiv",
    primaryClass = "hep-th",
    month = "6",
    year = "2024"
}

@article{Ivanovskiy:2024vel,
    author = "Ivanovskiy, Vyacheslav and Komatsu, Shota and Mishnyakov, Victor and Terziev, Nikolay and Zaigraev, Nikita and Zarembo, Konstantin",
    title = "{Vacuum Condensates on the Coulomb Branch}",
    eprint = "2405.19043",
    archivePrefix = "arXiv",
    primaryClass = "hep-th",
    month = "5",
    year = "2024"
}

@book{zwerger2011bcs,
  title={The BCS-BEC crossover and the unitary Fermi gas},
  author={Zwerger, Wilhelm},
  volume={836},
  year={2011},
  publisher={Springer Science \& Business Media}
}

@book{Schmitt:2014eka,
    author = "Schmitt, Andreas",
    title = "{Introduction to Superfluidity}: {Field-theoretical approach and applications}",
    eprint = "1404.1284",
    archivePrefix = "arXiv",
    primaryClass = "hep-ph",
    doi = "10.1007/978-3-319-07947-9",
    isbn = "978-3-319-07946-2, 978-3-319-07947-9",
    volume = "888",
    year = "2015"
}

@article{bardeen1957microscopic,
  title={Microscopic theory of superconductivity},
  author={Bardeen, John and Cooper, Leon N and Schrieffer, J Robert},
  journal={Physical Review},
  volume={106},
  number={1},
  pages={162},
  year={1957},
  publisher={APS}
}

@article{bardeen1957theory,
  title={Theory of superconductivity},
  author={Bardeen, John and Cooper, Leon N and Schrieffer, John Robert},
  journal={Physical review},
  volume={108},
  number={5},
  pages={1175},
  year={1957},
  publisher={APS}
}

@article{Hellerman:2013kba,
    author = "Hellerman, Simeon and Swanson, Ian",
    title = "{String Theory of the Regge Intercept}",
    eprint = "1312.0999",
    archivePrefix = "arXiv",
    primaryClass = "hep-th",
    reportNumber = "IPMU-13-0230",
    doi = "10.1103/PhysRevLett.114.111601",
    journal = "Phys. Rev. Lett.",
    volume = "114",
    number = "11",
    pages = "111601",
    year = "2015"
}

@article{Pestun:2016zxk,
    author = "Pestun, Vasily and others",
    title = "{Localization techniques in quantum field theories}",
    eprint = "1608.02952",
    archivePrefix = "arXiv",
    primaryClass = "hep-th",
    doi = "10.1088/1751-8121/aa63c1",
    journal = "J. Phys. A",
    volume = "50",
    number = "44",
    pages = "440301",
    year = "2017"
}

@article{Beisert:2010jr,
    author = "Beisert, Niklas and others",
    title = "{Review of AdS/CFT Integrability: An Overview}",
    eprint = "1012.3982",
    archivePrefix = "arXiv",
    primaryClass = "hep-th",
    reportNumber = "AEI-2010-175, CERN-PH-TH-2010-306, HU-EP-10-87, HU-MATH-2010-22, KCL-MTH-10-10, UMTG-270, UUITP-41-10",
    doi = "10.1007/s11005-011-0529-2",
    journal = "Lett. Math. Phys.",
    volume = "99",
    pages = "3--32",
    year = "2012"
}

@article{Aharony:1999ti,
    author = "Aharony, Ofer and Gubser, Steven S. and Maldacena, Juan Martin and Ooguri, Hirosi and Oz, Yaron",
    title = "{Large N field theories, string theory and gravity}",
    eprint = "hep-th/9905111",
    archivePrefix = "arXiv",
    reportNumber = "CERN-TH-99-122, HUTP-99-A027, LBNL-43113, RU-99-18, UCB-PTH-99-16, LBL-43113",
    doi = "10.1016/S0370-1573(99)00083-6",
    journal = "Phys. Rept.",
    volume = "323",
    pages = "183--386",
    year = "2000"
}

@article{Seiberg:2016gmd,
    author = "Seiberg, Nathan and Senthil, T. and Wang, Chong and Witten, Edward",
    title = "{A Duality Web in 2+1 Dimensions and Condensed Matter Physics}",
    eprint = "1606.01989",
    archivePrefix = "arXiv",
    primaryClass = "hep-th",
    doi = "10.1016/j.aop.2016.08.007",
    journal = "Annals Phys.",
    volume = "374",
    pages = "395--433",
    year = "2016"
}

@article{osheroff1972evidence,
  title={Evidence for a new phase of solid He 3},
  author={Osheroff, DD and Richardson, RC and Lee, DM},
  journal={Physical Review Letters},
  volume={28},
  number={14},
  pages={885},
  year={1972},
  publisher={APS}
}

@article{zwierlein2005vortices,
  title={Vortices and superfluidity in a strongly interacting Fermi gas},
  author={Zwierlein, Martin W and Abo-Shaeer, Jamil R and Schirotzek, Andre and Schunck, Christian H and Ketterle, Wolfgang},
  journal={Nature},
  volume={435},
  number={7045},
  pages={1047--1051},
  year={2005},
  publisher={Nature Publishing Group UK London}
}

@misc{date2003classicalquantummechanicsanyons,
      title={Classical and Quantum Mechanics of Anyons}, 
      author={G. Date and M. V. N. Murthy and Radhika Vathsan},
      year={2003},
      eprint={cond-mat/0302019},
      archivePrefix={arXiv},
      primaryClass={cond-mat},
      url={https://arxiv.org/abs/cond-mat/0302019}, 
}

@article{Doroud:2015fsz,
    author = "Doroud, Nima and Tong, David and Turner, Carl",
    title = "{On Superconformal Anyons}",
    eprint = "1511.01491",
    archivePrefix = "arXiv",
    primaryClass = "hep-th",
    doi = "10.1007/JHEP01(2016)138",
    journal = "JHEP",
    volume = "01",
    pages = "138",
    year = "2016"
}

@article{Doroud:2016mfv,
    author = "Doroud, Nima and Tong, David and Turner, Carl",
    title = "{The Conformal Spectrum of Non-Abelian Anyons}",
    eprint = "1611.05848",
    archivePrefix = "arXiv",
    primaryClass = "hep-th",
    doi = "10.21468/SciPostPhys.4.4.022",
    journal = "SciPost Phys.",
    volume = "4",
    number = "4",
    pages = "022",
    year = "2018"
}

@article{wu1984multiparticle,
  title={Multiparticle quantum mechanics obeying fractional statistics},
  author={Wu, Yong-Shi},
  journal={Physical Review Letters},
  volume={53},
  number={2},
  pages={111},
  year={1984},
  publisher={APS}
}

@article{chou1991multianyon,
  title={Multianyon spectra and wave functions},
  author={Chou, Chihong},
  journal={Physical Review D},
  volume={44},
  number={8},
  pages={2533},
  year={1991},
  publisher={APS}
}

@article{li1992thomas,
  title={Thomas-Fermi approximation for confined anyons},
  author={Li, Shuxi and Bhaduri, RK and Murthy, MVN},
  journal={Physical Review B},
  volume={46},
  number={2},
  pages={1228},
  year={1992},
  publisher={APS}
}

@article{basu1992class,
  title={Class of exact solutions for many-anyon quantum mechanics},
  author={Basu, Rahul and Date, G and Murthy, MVN},
  journal={Physical Review B},
  volume={46},
  number={5},
  pages={3139},
  year={1992},
  publisher={APS}
}

@article{sporre1992four,
  title={Four anyons in a harmonic well},
  author={Sporre, M and Verbaarschot, JJM and Zahed, I},
  journal={Physical Review B},
  volume={46},
  number={9},
  pages={5738},
  year={1992},
  publisher={APS}
}

@article{chitra1992ground,
  title={Ground state of many anyons in a harmonic potential},
  author={Chitra, R and Sen, Diptiman},
  journal={Physical Review B},
  volume={46},
  number={17},
  pages={10923},
  year={1992},
  publisher={APS}
}

@article{sen1992anyons,
  title={Anyons as perturbed bosons},
  author={Sen, Diptiman and Chitra, R},
  journal={Physical Review B},
  volume={45},
  number={2},
  pages={881},
  year={1992},
  publisher={APS}
}

@article{wilczek1982magnetic,
  title={Magnetic flux, angular momentum, and statistics},
  author={Wilczek, Frank},
  journal={Physical Review Letters},
  volume={48},
  number={17},
  pages={1144},
  year={1982},
  publisher={APS}
}

@article{wilczek1982remarks,
  title={Remarks on dyons},
  author={Wilczek, Frank},
  journal={Physical Review Letters},
  volume={48},
  number={17},
  pages={1146},
  year={1982},
  publisher={APS}
}

@article{laughlin1983anomalous,
  title={Anomalous quantum Hall effect: an incompressible quantum fluid with fractionally charged excitations},
  author={Laughlin, Robert B},
  journal={Physical Review Letters},
  volume={50},
  number={18},
  pages={1395},
  year={1983},
  publisher={APS}
}

@article{halperin1984statistics,
  title={Statistics of quasiparticles and the hierarchy of fractional quantized Hall states},
  author={Halperin, Bertrand I},
  journal={Physical Review Letters},
  volume={52},
  number={18},
  pages={1583},
  year={1984},
  publisher={APS}
}

@article{arovas1984fractional,
  title={Fractional statistics and the quantum Hall effect},
  author={Arovas, Daniel and Schrieffer, John R and Wilczek, Frank},
  journal={Physical review letters},
  volume={53},
  number={7},
  pages={722},
  year={1984},
  publisher={APS}
}

@inbook{Fradkin_2013, place={Cambridge}, title={Anyon superconductivity}, booktitle={Field Theories of Condensed Matter Physics}, publisher={Cambridge University Press}, author={Fradkin, Eduardo}, year={2013}, pages={414–431}}

@article{Cordova:2022ieu,
    author = "Cordova, Clay and Ohmori, Kantaro",
    title = "{Noninvertible Chiral Symmetry and Exponential Hierarchies}",
    eprint = "2205.06243",
    archivePrefix = "arXiv",
    primaryClass = "hep-th",
    doi = "10.1103/PhysRevX.13.011034",
    journal = "Phys. Rev. X",
    volume = "13",
    number = "1",
    pages = "011034",
    year = "2023"
}

@article{Kitaev:1997wr,
    author = "Kitaev, A. Yu.",
    title = "{Fault tolerant quantum computation by anyons}",
    eprint = "quant-ph/9707021",
    archivePrefix = "arXiv",
    doi = "10.1016/S0003-4916(02)00018-0",
    journal = "Annals Phys.",
    volume = "303",
    pages = "2--30",
    year = "2003"
}

@article{Aghaee_2023,
   title={InAs-Al hybrid devices passing the topological gap protocol},
   volume={107},
   ISSN={2469-9969},
   url={http://dx.doi.org/10.1103/PhysRevB.107.245423},
   DOI={10.1103/physrevb.107.245423},
   number={24},
   journal={Physical Review B},
   publisher={American Physical Society (APS)},
   author={Aghaee, Morteza and Akkala, Arun and Alam, Zulfi and Ali, Rizwan and Alcaraz Ramirez, Alejandro and Andrzejczuk, Mariusz and Antipov, Andrey E. and Aseev, Pavel and Astafev, Mikhail and Bauer, Bela and Becker, Jonathan and Boddapati, Srini and Boekhout, Frenk and Bommer, Jouri and Bosma, Tom and Bourdet, Leo and Boutin, Samuel and Caroff, Philippe and Casparis, Lucas and Cassidy, Maja and Chatoor, Sohail and Christensen, Anna Wulf and Clay, Noah and Cole, William S. and Corsetti, Fabiano and Cui, Ajuan and Dalampiras, Paschalis and Dokania, Anand and de Lange, Gijs and de Moor, Michiel and Estrada Saldaña, Juan Carlos and Fallahi, Saeed and Fathabad, Zahra Heidarnia and Gamble, John and Gardner, Geoff and Govender, Deshan and Griggio, Flavio and Grigoryan, Ruben and Gronin, Sergei and Gukelberger, Jan and Hansen, Esben Bork and Heedt, Sebastian and Herranz Zamorano, Jesús and Ho, Samantha and Holgaard, Ulrik Laurens and Ingerslev, Henrik and Johansson, Linda and Jones, Jeffrey and Kallaher, Ray and Karimi, Farhad and Karzig, Torsten and King, Evelyn and Kloster, Maren Elisabeth and Knapp, Christina and Kocon, Dariusz and Koski, Jonne and Kostamo, Pasi and Krogstrup, Peter and Kumar, Mahesh and Laeven, Tom and Larsen, Thorvald and Li, Kongyi and Lindemann, Tyler and Love, Julie and Lutchyn, Roman and Madsen, Morten Hannibal and Manfra, Michael and Markussen, Signe and Martinez, Esteban and McNeil, Robert and Memisevic, Elvedin and Morgan, Trevor and Mullally, Andrew and Nayak, Chetan and Nielsen, Jens and Nielsen, William Hvidtfelt Padkær and Nijholt, Bas and Nurmohamed, Anne and O’Farrell, Eoin and Otani, Keita and Pauka, Sebastian and Petersson, Karl and Petit, Luca and Pikulin, Dmitry I. and Preiss, Frank and Quintero-Perez, Marina and Rajpalke, Mohana and Rasmussen, Katrine and Razmadze, Davydas and Reentila, Outi and Reilly, David and Rouse, Richard and Sadovskyy, Ivan and Sainiemi, Lauri and Schreppler, Sydney and Sidorkin, Vadim and Singh, Amrita and Singh, Shilpi and Sinha, Sarat and Sohr, Patrick and Stankevič, Tomaš and Stek, Lieuwe and Suominen, Henri and Suter, Judith and Svidenko, Vicky and Teicher, Sam and Temuerhan, Mine and Thiyagarajah, Nivetha and Tholapi, Raj and Thomas, Mason and Toomey, Emily and Upadhyay, Shivendra and Urban, Ivan and Vaitiekėnas, Saulius and Van Hoogdalem, Kevin and Van Woerkom, David and Viazmitinov, Dmitrii V. and Vogel, Dominik and Waddy, Steven and Watson, John and Weston, Joseph and Winkler, Georg W. and Yang, Chung Kai and Yau, Sean and Yi, Daniel and Yucelen, Emrah and Webster, Alex and Zeisel, Roland and Zhao, Ruichen},
   year={2023},
   month=jun }

@article{Choi:2022jqy,
    author = "Choi, Yichul and Lam, Ho Tat and Shao, Shu-Heng",
    title = "{Noninvertible Global Symmetries in the Standard Model}",
    eprint = "2205.05086",
    archivePrefix = "arXiv",
    primaryClass = "hep-th",
    reportNumber = "YITP-SB-2022-21, MIT-CTP/5433",
    doi = "10.1103/PhysRevLett.129.161601",
    journal = "Phys. Rev. Lett.",
    volume = "129",
    number = "16",
    pages = "161601",
    year = "2022"
}

@article{ketterle2008making,
  title={Making, probing and understanding ultracold Fermi gases},
  author={Ketterle, Wolfgang and Zwierlein, Martin W},
  journal={La Rivista del Nuovo Cimento},
  volume={31},
  number={5},
  pages={247--422},
  year={2008},
  publisher={Springer}
}

@article{sa2008fermions,
  title={When fermions become bosons: Pairing in ultracold gases},
  author={S{\'a} de Melo, Carlos AR},
  journal={Physics Today},
  volume={61},
  number={10},
  pages={45--51},
  year={2008},
  publisher={AIP Publishing}
}

@article{Henkel:1993sg,
    author = "Henkel, Malte",
    title = "{Schrodinger invariance in strongly anisotropic critical systems}",
    eprint = "hep-th/9310081",
    archivePrefix = "arXiv",
    reportNumber = "OUTP-93-33-S, UGVA-DPT-1993-09-833",
    doi = "10.1007/BF02186756",
    journal = "J. Statist. Phys.",
    volume = "75",
    pages = "1023--1061",
    year = "1994"
}

@article{Coddington_2003,
   title={Observation of Tkachenko Oscillations in Rapidly Rotating Bose-Einstein Condensates},
   volume={91},
   ISSN={1079-7114},
   url={http://dx.doi.org/10.1103/PhysRevLett.91.100402},
   DOI={10.1103/physrevlett.91.100402},
   number={10},
   journal={Physical Review Letters},
   publisher={American Physical Society (APS)},
   author={Coddington, I. and Engels, P. and Schweikhard, V. and Cornell, E. A.},
   year={2003},
   month=sep }

@article{Moroz:2018noc,
    author = "Moroz, Sergej and Hoyos, Carlos and Benzoni, Claudio and Son, Dam Thanh",
    title = "{Effective field theory of a vortex lattice in a bosonic superfluid}",
    eprint = "1803.10934",
    archivePrefix = "arXiv",
    primaryClass = "cond-mat.quant-gas",
    doi = "10.21468/SciPostPhys.5.4.039",
    journal = "SciPost Phys.",
    volume = "5",
    number = "4",
    pages = "039",
    year = "2018"
}

@article{Moroz:2019qdw,
    author = "Moroz, Sergej and Son, Dam Thanh",
    title = "{Bosonic Superfluid on the Lowest Landau Level}",
    eprint = "1901.06088",
    archivePrefix = "arXiv",
    primaryClass = "cond-mat.quant-gas",
    doi = "10.1103/PhysRevLett.122.235301",
    journal = "Phys. Rev. Lett.",
    volume = "122",
    number = "23",
    pages = "235301",
    year = "2019"
}

@article{Nguyen:2020yve,
    author = "Nguyen, Dung Xuan and Gromov, Andrey and Moroz, Sergej",
    title = "{Fracton-elasticity duality of two-dimensional superfluid vortex crystals: defect interactions and quantum melting}",
    eprint = "2005.12317",
    archivePrefix = "arXiv",
    primaryClass = "cond-mat.quant-gas",
    doi = "10.21468/SciPostPhys.9.5.076",
    journal = "SciPost Phys.",
    volume = "9",
    pages = "076",
    year = "2020"
}

@article{Hoyos:2013eha,
    author = "Hoyos, Carlos and Moroz, Sergej and Son, Dam Thanh",
    title = "{Effective theory of chiral two-dimensional superfluids}",
    eprint = "1305.3925",
    archivePrefix = "arXiv",
    primaryClass = "cond-mat.quant-gas",
    reportNumber = "TAUP-2965-13, NT@UW-13-18, EFI-13-9",
    doi = "10.1103/PhysRevB.89.174507",
    journal = "Phys. Rev. B",
    volume = "89",
    number = "17",
    pages = "174507",
    year = "2014"
}

@article{Moroz:2015cft,
    author = "Moroz, Sergej and Hoyos, Carlos and Radzihovsky, Leo",
    title = "{Chiral p\ensuremath{\pm}ip superfluid on a sphere}",
    eprint = "1511.03502",
    archivePrefix = "arXiv",
    primaryClass = "cond-mat.quant-gas",
    reportNumber = "FPAU0-15-16",
    doi = "10.1103/PhysRevB.93.024521",
    journal = "Phys. Rev. B",
    volume = "93",
    number = "2",
    pages = "024521",
    year = "2016"
}

@article{Benjamin:2024kdg,
    author = "Benjamin, Nathan and Lee, Jaeha and Pal, Sridip and Simmons-Duffin, David and Xu, Yixin",
    title = "{Angular fractals in thermal QFT}",
    eprint = "2405.17562",
    archivePrefix = "arXiv",
    primaryClass = "hep-th",
    reportNumber = "CALT-TH 2024-021",
    month = "5",
    year = "2024"
}

@article{Chen:1989xs,
    author = "Chen, Yi-Hong and Wilczek, Frank and Witten, Edward and Halperin, Bertrand I.",
    editor = "Wilczek, Frank",
    title = "{On Anyon Superconductivity}",
    reportNumber = "IASSNS-HEP-89-27",
    doi = "10.1142/S0217979289000725",
    journal = "Int. J. Mod. Phys. B",
    volume = "3",
    pages = "1001",
    year = "1989"
}

@article{Du:2020gqf,
    author = "Du, Yi-Hsien and Mehta, Umang and Son, Thanh",
    title = "{Rotons in Anyon Superfluids}",
    eprint = "2012.07991",
    archivePrefix = "arXiv",
    primaryClass = "cond-mat.mes-hall",
    doi = "10.1007/JHEP03(2021)101",
    journal = "JHEP",
    volume = "21",
    pages = "101",
    year = "2020"
}

@article{Son:2005rv,
    author = "Son, D. T. and Wingate, M.",
    title = "{General coordinate invariance and conformal invariance in nonrelativistic physics: Unitary Fermi gas}",
    eprint = "cond-mat/0509786",
    archivePrefix = "arXiv",
    reportNumber = "INT-PUB-05-23",
    doi = "10.1016/j.aop.2005.11.001",
    journal = "Annals Phys.",
    volume = "321",
    pages = "197--224",
    year = "2006"
}

@article{Delacretaz:2022ocm,
    author = "Delacretaz, Luca V. and Du, Yi-Hsien and Mehta, Umang and Son, Dam Thanh",
    title = "{Nonlinear bosonization of Fermi surfaces: The method of coadjoint orbits}",
    eprint = "2203.05004",
    archivePrefix = "arXiv",
    primaryClass = "cond-mat.str-el",
    reportNumber = "EFI 22-3",
    doi = "10.1103/PhysRevResearch.4.033131",
    journal = "Phys. Rev. Res.",
    volume = "4",
    number = "3",
    pages = "033131",
    year = "2022"
}

@article{Abrikosov1956MagneticPO,
  title={Magnetic properties of superconductors of the second group},
  author={Alexei A. Abrikosov},
  journal={Journal of Experimental and Theoretical Physics},
  year={1956},
  url={https://api.semanticscholar.org/CorpusID:117984570}
}

@article{Watanabe_2008,
   title={Tkachenko modes in a superfluid Fermi gas at unitarity},
   volume={77},
   ISSN={1094-1622},
   url={http://dx.doi.org/10.1103/PhysRevA.77.021602},
   DOI={10.1103/physreva.77.021602},
   number={2},
   journal={Physical Review A},
   publisher={American Physical Society (APS)},
   author={Watanabe, Gentaro and Cozzini, Marco and Stringari, Sandro},
   year={2008},
   month=feb }

@article{Tkachenko1966OnVL,
  title={On Vortex Lattices},
  author={V. K. Tkachenko},
  journal={Journal of Experimental and Theoretical Physics},
  year={1966},
  url={https://api.semanticscholar.org/CorpusID:117777154}
}

@ARTICLE{1966JETP23.1049T,
       author = {{Tkachenko}, V.~K.},
        title = "{Stability of Vortex Lattices}",
      journal = {Soviet Journal of Experimental and Theoretical Physics},
         year = 1966,
        month = dec,
       volume = {23},
        pages = {1049},
       adsurl = {https://ui.adsabs.harvard.edu/abs/1966JETP...23.1049T}
}

@article{Tkachenko1969ElasticityOV,
  title={Elasticity of Vortex Lattices},
  author={V. K. Tkachenko},
  journal={Journal of Experimental and Theoretical Physics},
  year={1969},
  volume={29},
  pages={945},
  url={https://api.semanticscholar.org/CorpusID:230747266}
}

@article{Kasamatsu_2002,
   title={Giant hole and circular superflow in a fast rotating Bose-Einstein condensate},
   volume={66},
   ISSN={1094-1622},
   url={http://dx.doi.org/10.1103/PhysRevA.66.053606},
   DOI={10.1103/physreva.66.053606},
   number={5},
   journal={Physical Review A},
   publisher={American Physical Society (APS)},
   author={Kasamatsu, Kenichi and Tsubota, Makoto and Ueda, Masahito},
   year={2002},
   month=nov }

@inproceedings{Hartman:2022zik,
    author = "Hartman, Thomas and Mazac, Dalimil and Simmons-Duffin, David and Zhiboedov, Alexander",
    title = "{Snowmass White Paper: The Analytic Conformal Bootstrap}",
    booktitle = "{Snowmass 2021}",
    eprint = "2202.11012",
    archivePrefix = "arXiv",
    primaryClass = "hep-th",
    month = "2",
    year = "2022"
}

@inproceedings{Poland:2022qrs,
    author = "Poland, David and Simmons-Duffin, David",
    title = "{Snowmass White Paper: The Numerical Conformal Bootstrap}",
    booktitle = "{Snowmass 2021}",
    eprint = "2203.08117",
    archivePrefix = "arXiv",
    primaryClass = "hep-th",
    reportNumber = "CALT-TH 2022-013",
    month = "3",
    year = "2022"
}

@article{Fischer_2003,
   title={Vortex States of Rapidly Rotating Dilute Bose-Einstein Condensates},
   volume={90},
   ISSN={1079-7114},
   url={http://dx.doi.org/10.1103/PhysRevLett.90.140402},
   DOI={10.1103/physrevlett.90.140402},
   number={14},
   journal={Physical Review Letters},
   publisher={American Physical Society (APS)},
   author={Fischer, Uwe R. and Baym, Gordon},
   year={2003},
   month=apr }

@article{Guo_2020,
   title={Supersonic Rotation of a Superfluid: A Long-Lived Dynamical Ring},
   volume={124},
   ISSN={1079-7114},
   url={http://dx.doi.org/10.1103/PhysRevLett.124.025301},
   DOI={10.1103/physrevlett.124.025301},
   number={2},
   journal={Physical Review Letters},
   publisher={American Physical Society (APS)},
   author={Guo, Yanliang and Dubessy, Romain and de Herve, Mathieu de Goër and Kumar, Avinash and Badr, Thomas and Perrin, Aurélien and Longchambon, Laurent and Perrin, Hélène},
   year={2020},
   month=jan }

@article{Brauner:2020zov,
    author = "Brauner, Tomas",
    title = "{Remarks on relativistic scalar models with chemical potential}",
    eprint = "2009.08895",
    archivePrefix = "arXiv",
    primaryClass = "hep-th",
    month = "9",
    year = "2020"
}

@article{Ku_2012,
   title={Revealing the Superfluid Lambda Transition in the Universal Thermodynamics of a Unitary Fermi Gas},
   volume={335},
   ISSN={1095-9203},
   url={http://dx.doi.org/10.1126/science.1214987},
   DOI={10.1126/science.1214987},
   number={6068},
   journal={Science},
   publisher={American Association for the Advancement of Science (AAAS)},
   author={Ku, Mark J. H. and Sommer, Ariel T. and Cheuk, Lawrence W. and Zwierlein, Martin W.},
   year={2012},
   month=feb, pages={563–567} }

@article{Endres:2012cw,
    author = "Endres, Michael G. and Kaplan, David B. and Lee, Jong-Wan and Nicholson, Amy N.",
    title = "{Lattice Monte Carlo calculations for unitary fermions in a finite box}",
    eprint = "1203.3169",
    archivePrefix = "arXiv",
    primaryClass = "hep-lat",
    reportNumber = "INT-PUB-12-011, KEK-CP-266, RIKEN-QHP-16, UM-DOE-ER-40762-517",
    doi = "10.1103/PhysRevA.87.023615",
    journal = "Phys. Rev. A",
    volume = "87",
    number = "2",
    pages = "023615",
    year = "2013"
}

@article{Carlson:2011kv,
    author = "Carlson, J. and Gandolfi, Stefano and Schmidt, Kevin E. and Zhang, Shiwei",
    title = "{Auxiliary Field quantum Monte Carlo for Strongly Paired Fermions}",
    eprint = "1107.5848",
    archivePrefix = "arXiv",
    primaryClass = "cond-mat.quant-gas",
    doi = "10.1103/PhysRevA.84.061602",
    journal = "Phys. Rev. A",
    volume = "84",
    pages = "061602",
    year = "2011"
}

@article{zamolodchikov1987renormalization,
  title={Renormalization group and perturbation theory about fixed points in two-dimensional field theory},
  author={Zamolodchikov, AB},
  journal={Sov. J. Nucl. Phys.(Engl. Transl.);(United States)},
  volume={46},
  number={6},
  year={1987},
  publisher={LD Landau Institute of Theoretical Physics, Academy of Sciences of the USSR}
}

@article{Simmons-Duffin:2016wlq,
    author = "Simmons-Duffin, David",
    title = "{The Lightcone Bootstrap and the Spectrum of the 3d Ising CFT}",
    eprint = "1612.08471",
    archivePrefix = "arXiv",
    primaryClass = "hep-th",
    doi = "10.1007/JHEP03(2017)086",
    journal = "JHEP",
    volume = "03",
    pages = "086",
    year = "2017"
}

@article{Benjamin:2023qsc,
    author = "Benjamin, Nathan and Lee, Jaeha and Ooguri, Hirosi and Simmons-Duffin, David",
    title = "{Universal asymptotics for high energy CFT data}",
    eprint = "2306.08031",
    archivePrefix = "arXiv",
    primaryClass = "hep-th",
    reportNumber = "CALT-TH 2023-014, IPMU 23-0020",
    doi = "10.1007/JHEP03(2024)115",
    journal = "JHEP",
    volume = "03",
    pages = "115",
    year = "2024"
}

@article{Cremonesi:2022fvg,
    author = "Cremonesi, Stefano and Lanza, Stefano and Martucci, Luca",
    title = "{Semiclassics of three-dimensional SCFTs from holography}",
    eprint = "2202.06970",
    archivePrefix = "arXiv",
    primaryClass = "hep-th",
    doi = "10.1007/JHEP10(2022)111",
    journal = "JHEP",
    volume = "10",
    pages = "111",
    year = "2022"
}

@article{Chester:2017vdh,
    author = "Chester, Shai M. and Iliesiu, Luca V. and Mezei, Mark and Pufu, Silviu S.",
    title = "{Monopole Operators in $U(1)$ Chern-Simons-Matter Theories}",
    eprint = "1710.00654",
    archivePrefix = "arXiv",
    primaryClass = "hep-th",
    reportNumber = "PUPT-2537",
    doi = "10.1007/JHEP05(2018)157",
    journal = "JHEP",
    volume = "05",
    pages = "157",
    year = "2018"
}

@article{Creminelli:2022onn,
    author = "Creminelli, Paolo and Janssen, Oliver and Senatore, Leonardo",
    title = "{Positivity bounds on effective field theories with spontaneously broken Lorentz invariance}",
    eprint = "2207.14224",
    archivePrefix = "arXiv",
    primaryClass = "hep-th",
    doi = "10.1007/JHEP09(2022)201",
    journal = "JHEP",
    volume = "09",
    pages = "201",
    year = "2022"
}

@article{Metlitski:2008dw,
    author = "Metlitski, Max A. and Hermele, Michael and Senthil, T. and Fisher, Matthew P. A.",
    title = "{Monopoles in CP**(N-1) model via the state-operator correspondence}",
    eprint = "0809.2816",
    archivePrefix = "arXiv",
    primaryClass = "cond-mat.str-el",
    doi = "10.1103/PhysRevB.78.214418",
    journal = "Phys. Rev. B",
    volume = "78",
    pages = "214418",
    year = "2008"
}

@article{Moitra:2023yyc,
    author = "Moitra, Upamanyu",
    title = "{Newton vs Coulomb in AdS/CFT and the weak gravity conjecture}",
    eprint = "2305.08907",
    archivePrefix = "arXiv",
    primaryClass = "hep-th",
    doi = "10.1103/PhysRevD.109.L041903",
    journal = "Phys. Rev. D",
    volume = "109",
    number = "4",
    pages = "L041903",
    year = "2024"
}

@article{Vafa:2005ui,
    author = "Vafa, Cumrun",
    title = "{The String landscape and the swampland}",
    eprint = "hep-th/0509212",
    archivePrefix = "arXiv",
    reportNumber = "HUTP-05-A043",
    month = "9",
    year = "2005"
}

@article{Palti:2019pca,
    author = "Palti, Eran",
    title = "{The Swampland: Introduction and Review}",
    eprint = "1903.06239",
    archivePrefix = "arXiv",
    primaryClass = "hep-th",
    reportNumber = "MPP-2019-53",
    doi = "10.1002/prop.201900037",
    journal = "Fortsch. Phys.",
    volume = "67",
    number = "6",
    pages = "1900037",
    year = "2019"
}

@article{Arkani-Hamed:2006emk,
    author = "Arkani-Hamed, Nima and Motl, Lubos and Nicolis, Alberto and Vafa, Cumrun",
    title = "{The String landscape, black holes and gravity as the weakest force}",
    eprint = "hep-th/0601001",
    archivePrefix = "arXiv",
    reportNumber = "HUTP-05-A0057",
    doi = "10.1088/1126-6708/2007/06/060",
    journal = "JHEP",
    volume = "06",
    pages = "060",
    year = "2007"
}

@article{Harlow:2018tng,
    author = "Harlow, Daniel and Ooguri, Hirosi",
    title = "{Symmetries in quantum field theory and quantum gravity}",
    eprint = "1810.05338",
    archivePrefix = "arXiv",
    primaryClass = "hep-th",
    doi = "10.1007/s00220-021-04040-y",
    journal = "Commun. Math. Phys.",
    volume = "383",
    number = "3",
    pages = "1669--1804",
    year = "2021"
}

@article{Rong:2023owx,
    author = "Rong, Junchen and Su, Ning",
    title = "{From O(3) to Cubic CFT: Conformal Perturbation and the Large Charge Sector}",
    eprint = "2311.00933",
    archivePrefix = "arXiv",
    primaryClass = "hep-th",
    month = "11",
    year = "2023"
}

@article{Boyack:2023uml,
    author = "Boyack, Rufus and Delacr\'etaz, Luca and Dupuis, \'Eric and Witczak-Krempa, William",
    title = "{Conformal field theories in a magnetic field}",
    eprint = "2312.12546",
    archivePrefix = "arXiv",
    primaryClass = "hep-th",
    month = "12",
    year = "2023"
}

@article{Chester:2022wur,
    author = "Chester, Shai M. and Dupuis, \'Eric and Witczak-Krempa, William",
    title = "{Evidence for web of dualities from monopole operators}",
    eprint = "2210.12370",
    archivePrefix = "arXiv",
    primaryClass = "hep-th",
    doi = "10.1103/PhysRevD.108.L021701",
    journal = "Phys. Rev. D",
    volume = "108",
    number = "2",
    pages = "L021701",
    year = "2023"
}

@article{Borokhov:2002ib,
    author = "Borokhov, Vadim and Kapustin, Anton and Wu, Xin-kai",
    title = "{Topological disorder operators in three-dimensional conformal field theory}",
    eprint = "hep-th/0206054",
    archivePrefix = "arXiv",
    reportNumber = "CALT-68-2391",
    doi = "10.1088/1126-6708/2002/11/049",
    journal = "JHEP",
    volume = "11",
    pages = "049",
    year = "2002"
}

@article{Sharon:2023drx,
    author = "Sharon, Adar and Watanabe, Masataka",
    title = "{A counterexample to the CFT convexity conjecture}",
    eprint = "2301.08262",
    archivePrefix = "arXiv",
    primaryClass = "hep-th",
    doi = "10.1007/JHEP05(2023)202",
    journal = "JHEP",
    volume = "05",
    pages = "202",
    year = "2023"
}

@article{Caetano:2023zwe,
    author = "Caetano, Jo\~ao and Komatsu, Shota and Wang, Yifan",
    title = "{Large charge \textquoteright{}t Hooft limit of $ \mathcal{N} $ = 4 super-Yang-Mills}",
    eprint = "2306.00929",
    archivePrefix = "arXiv",
    primaryClass = "hep-th",
    reportNumber = "CERN-TH-2023-089",
    doi = "10.1007/JHEP02(2024)047",
    journal = "JHEP",
    volume = "02",
    pages = "047",
    year = "2024"
}

@article{Kravchuk:2018htv,
    author = "Kravchuk, Petr and Simmons-Duffin, David",
    title = "{Light-ray operators in conformal field theory}",
    eprint = "1805.00098",
    archivePrefix = "arXiv",
    primaryClass = "hep-th",
    reportNumber = "CALT-TH 2018-018",
    doi = "10.1007/JHEP11(2018)102",
    journal = "JHEP",
    volume = "11",
    pages = "102",
    year = "2018"
}

@article{luscher1975global,
  title={Global conformal invariance in quantum field theory},
  author={L{\"u}scher, M and Mack, G},
  journal={Communications in Mathematical Physics},
  volume={41},
  number={3},
  pages={203--234},
  year={1975}
}

@article{Paul:2023rka,
    author = "Paul, Hynek and Perlmutter, Eric and Raj, Himanshu",
    title = "{Exact large charge in $ \mathcal{N} $ = 4 SYM and semiclassical string theory}",
    eprint = "2303.13207",
    archivePrefix = "arXiv",
    primaryClass = "hep-th",
    doi = "10.1007/JHEP08(2023)078",
    journal = "JHEP",
    volume = "08",
    pages = "078",
    year = "2023"
}

@article{Choi:2025tql,
    author = "Choi, Jaehyeok and Lee, Eunwoo",
    title = "{Large charge operators at large spin from relativistically rotating vortices}",
    eprint = "2501.07198",
    archivePrefix = "arXiv",
    primaryClass = "hep-th",
    month = "1",
    year = "2025"
}

@article{Taylor:2008tg,
    author = "Taylor, Marika",
    title = "{Non-relativistic holography}",
    eprint = "0812.0530",
    archivePrefix = "arXiv",
    primaryClass = "hep-th",
    reportNumber = "ITFA-2008-48",
    month = "12",
    year = "2008"
}

@article{Taylor:2015glc,
    author = "Taylor, Marika",
    title = "{Lifshitz holography}",
    eprint = "1512.03554",
    archivePrefix = "arXiv",
    primaryClass = "hep-th",
    doi = "10.1088/0264-9381/33/3/033001",
    journal = "Class. Quant. Grav.",
    volume = "33",
    number = "3",
    pages = "033001",
    year = "2016"
}

@article{Delacretaz:2025ifh,
    author = "Delacr\'etaz, Luca V. and Chowdhury, Subham Dutta and Mehta, Umang",
    title = "{Symmetry and causality constraints on Fermi liquids}",
    eprint = "2501.02073",
    archivePrefix = "arXiv",
    primaryClass = "hep-th",
    month = "1",
    year = "2025"
}

@article{Baume:2020ure,
    author = "Baume, Florent and Heckman, Jonathan J. and Lawrie, Craig",
    title = "{6D SCFTs, 4D SCFTs, Conformal Matter, and Spin Chains}",
    eprint = "2007.07262",
    archivePrefix = "arXiv",
    primaryClass = "hep-th",
    doi = "10.1016/j.nuclphysb.2021.115401",
    journal = "Nucl. Phys. B",
    volume = "967",
    pages = "115401",
    year = "2021"
}

@article{Baume:2022cot,
    author = "Baume, Florent and Heckman, Jonathan J. and Lawrie, Craig",
    title = "{Super-spin chains for 6D SCFTs}",
    eprint = "2208.02272",
    archivePrefix = "arXiv",
    primaryClass = "hep-th",
    reportNumber = "DESY-22-132",
    doi = "10.1016/j.nuclphysb.2023.116250",
    journal = "Nucl. Phys. B",
    volume = "992",
    pages = "116250",
    year = "2023"
}

@article{Heckman:2024erd,
    author = "Heckman, Jonathan J. and Sharon, Adar and Watanabe, Masataka",
    title = "{6d Large Charge and 2d Virasoro Blocks}",
    eprint = "2409.05944",
    archivePrefix = "arXiv",
    primaryClass = "hep-th",
    month = "9",
    year = "2024"
}

@article{Karthik:2018rcg,
    author = "Karthik, Nikhil",
    title = "{Monopole scaling dimension using Monte-Carlo simulation}",
    eprint = "1808.08970",
    archivePrefix = "arXiv",
    primaryClass = "cond-mat.str-el",
    doi = "10.1103/PhysRevD.98.074513",
    journal = "Phys. Rev. D",
    volume = "98",
    number = "7",
    pages = "074513",
    year = "2018"
}

@article{Dondi:2024vua,
    author = "Dondi, Nicola Andrea and Sberveglieri, Giacomo",
    title = "{NLO in the large charge sector of the critical $O(N)$ model at large $N$}",
    eprint = "2409.06781",
    archivePrefix = "arXiv",
    primaryClass = "hep-th",
    month = "9",
    year = "2024"
}

@article{Pavaskar:2024pqf,
    author = "Pavaskar, Shashin and Rothstein, Ira Z.",
    title = "{First Principle Predictions for Cold Fermionic Gases Near Criticality via Critical Boson Dominance and Anomaly Matching}",
    eprint = "2409.18379",
    archivePrefix = "arXiv",
    primaryClass = "cond-mat.quant-gas",
    month = "9",
    year = "2024"
}

@article{Sen:2011cn,
    author = "Sen, Ashoke",
    title = "{State Operator Correspondence and Entanglement in $AdS_2/CFT_1$}",
    eprint = "1101.4254",
    archivePrefix = "arXiv",
    primaryClass = "hep-th",
    doi = "10.3390/e13071305",
    journal = "Entropy",
    volume = "13",
    pages = "1305--1323",
    year = "2011"
}

@article{Beane:2025tum,
    author = "Beane, Silas R. and Orlando, Domenico and Reffert, Susanne",
    title = "{Unnuclear matter at large-charge}",
    eprint = "2501.10505",
    archivePrefix = "arXiv",
    primaryClass = "nucl-th",
    reportNumber = "NT@UW-24-12,YITP-24-173",
    month = "1",
    year = "2025"
}

@article{Raviv-Moshe:2024yzt,
    author = "Raviv-Moshe, Avia and Zhong, Siwei",
    title = {{Impurities in Schr\"odinger field theories and s-wave resonance}},
    eprint = "2411.04040",
    archivePrefix = "arXiv",
    primaryClass = "cond-mat.quant-gas",
    month = "11",
    year = "2024"
}

@article{Brown:2024ajk,
    author = "Brown, Adam R. and Iliesiu, Luca V. and Penington, Geoff and Usatyuk, Mykhaylo",
    title = "{The evaporation of charged black holes}",
    eprint = "2411.03447",
    archivePrefix = "arXiv",
    primaryClass = "hep-th",
    month = "11",
    year = "2024"
}

@article{Baiguera:2024vlj,
    author = "Baiguera, Stefano and Harmark, Troels and Lei, Yang and Yan, Ziqi",
    title = "{Conformal Mapping of Non-Lorentzian Geometries in SU(1,2) Conformal Field Theory}",
    eprint = "2411.11951",
    archivePrefix = "arXiv",
    primaryClass = "hep-th",
    reportNumber = "NORDITA 2024-043",
    month = "11",
    year = "2024"
}

@article{Jensen:2014aia,
    author = "Jensen, Kristan",
    title = "{On the coupling of Galilean-invariant field theories to curved spacetime}",
    eprint = "1408.6855",
    archivePrefix = "arXiv",
    primaryClass = "hep-th",
    reportNumber = "YITP-SB-14-26",
    doi = "10.21468/SciPostPhys.5.1.011",
    journal = "SciPost Phys.",
    volume = "5",
    number = "1",
    pages = "011",
    year = "2018"
}

\addcontentsline{toc}{chapter}{Bibliography}

\end{document}